(69.0 k words inclusive; Figure 49; Table 16; Refs 638; Updated: 05/25/05)

# Size and confinement effect on nanostructures

Chang Q. Sun*

*School of Electrical and Electronic Engineering, Nanyang Technological University, Singapore 639798*

*Institute of Advanced Materials Physics and Faculty of Science, Tianjin University, 300072, P. R. China*

**Abstract**

This report presents a systematic understanding of the nature behind the unusual behavior of a nanosolid, and a surface as well, in mechanics, thermodynamics, acoustics, optoelectronics, magnetism, dielectrics, atomic diffusivity and chemical reactivity towards predictable design and controllable growth of nanostructured materials. A bond order-length-strength (BOLS) correlation mechanism has been developed, which has enabled the tunability of a variety of properties of a nanosolid in connection with surface to be consistently predicted and experimentally verified. The BOLS correlation indicates that the coordination number (CN) imperfection of an atom at site surrounding a defect or in the surface skin causes the remaining bonds of the lower-coordinated atom to contract spontaneously. The spontaneous bond contraction is associated with bond-strength gain or atomic potential well deppression, which localize electrons and enhance the density of charge, mass, and energy in the relaxed region. The enhancement of energy density in the relaxed region perturbs the Hamiltonian and the associated properties such as the band-gap width, core-level energy, Stokes shift (electron-phonon interaction), and dielectric susceptibility. On the other hand, bond order loss lowers generally the cohesive energy of the lower-coordinated atom from the value of a fully-coordinated atom, which dictates the thermodynamic process such as self-assembly growth, atomic vibration, thermal stability, and activation energies for atomic dislocation and diffusion. Consistency between predictions and observations evidences the enormous impact of atomic CN imperfection to the low-dimensional and disordered systems, including surface, amorphous and nanosolid states, and the validity and essentiality of the BOLS correlation.

* E-mail: ecqsun@ntu.edu.sg; URL: www.ntu.edu.sg/home/ecqsun/

Nomenclature

| | |
|---|---|
| $\omega$ | The angular frequency |
| $\mu$ | Atomic magnetic momentum |
| $\beta$ | Compressibility/extensibility |
| $\chi$ | Dielectric susceptibility |
| $\sigma$ | Surface stress/conductivity |
| $\Phi$ | Work function |
| $\varepsilon'_r$ | Imaginary part of dielectric constant |
| $\varepsilon_0$ | Dielectric permittivity of vacuum |
| $\eta_1$ | Specific heat per coordinate |
| $\eta_2$ | Thermal energy per coordinate for evaporating a molten atom |
| $\mu_B$ | Bohr magneton |
| $\gamma_{ij}$ | Atomic portion in the i th atomic shell over the entire solid of size $D_j$ |
| $\delta_K$ | Kobo gap |
| $\varepsilon_r$ | Real part of dielectric constant |
| $\mu_B$ | Bohr magneton |
| BBB | Bond-band-barrier correlation |
| BOLS | Bond-order-length-strength correlation |
| CN(z) | Coordination number |
| CNT | Carbon nanotube |
| $d_0$ | Atomic diameter or bond length |
| DFT | Density functional theory |
| $D_j$ | Diameter of jth spherical nanosolid |
| DOS | Density-of-state |
| $E_v$ | Core level energy |
| $E_v(1)$ | Single energy level of a statically isolated atom |
| $E_B$ | Atomic cohesive energy /vacancy formation energy |
| $E_b$ | Cohesive energy per bond |
| $E_G$ | Band gap |
| e-h | Electron-hole |
| e-p | Electron-phonon |
| GB | Grain boundary |
| HOPG | Highly oriented pyrolytic graphite |
| IHPR | Invere Hall-Petch relationship |
| $k_B$ | Boltzmann constant |
| $K_j$ | Dimensionless form of the radius of a sphere or the thickness of a plate |
| MC | Monte Carlo |
| MD | Molecular dynamics |
| $M_S$ | Saturation magnetization |
| P | Stress |
| p-Si | Porous silicon |
| $Q(K_j)$ | Measurable quantity of a nanosolid |
| QC | Quantum confinement |
| R | Radius/Resistance |
| RCL | Resistance–capacitance–inductance |
| RT | Room temperature |
| SIMS | Secondary ion mass spectroscopy |
| SMAT | Surface mechanical attrition treatment |
| SPB | Surface potential barrier |
| STE | Self-trapping exciton |
| STM/S | Scanning tunneling microscopy/spectroscopy |
| $T_C$ | Critical/Curie temperature |
| TEM | Transition electronic spectroscopy |
| $T_m$ | Melting point |
| W | Stokes shift |
| XAFS | X-ray absorption fine structure spectroscopy |
| XPS | X-ray photoelectron spectroscopy |
| XRD | X-ray diffraction |
| Y | Young's modulus |

# 1 Introduction

## 1.1 Scope

This report starts, in Section 1, with a brief overview on the progress in understanding the unusual behavior of a solid in the nanometer regime by highlighting some intriguing phenomena that challenge the currently reported efforts towards consistent insight into the mechanism behind, aiming at predictable design and controllable growth of nanostructures with desired functions. Section 2 will describe the original bond order-length-strength (*BOLS*) correlation mechanism for the consequences of the often-oversighted event of atomic coordination-number (*CN*) imperfection on the length and strength of the remaining bonds of the lower-coordinated atom at site surrounding a defect or in the surface skin of a couple of atomic layers thick. A brief summary is also given of the recent chemical-bond – valence-band – potential-barrier (*BBB*) correlation mechanism[1] for tetrahedron bond formation of O, N and C interacting with atoms in solid phase and its impact to surface chemistry and surface probing technologies, which provides some background knowledge of surface chemical passivation. A general scaling law expression is derived based on the shell structure for the shape-and-size dependence of a nanosolid. Sections 3 – 10 apply the BOLS correlation to the behavior of a nanosolid in various aspects. In each Section, efforts in experimental observations and theoretical modeling exercises will be summarized with appreciation and then are followed by the BOLS formulation on the particular property of concern. Correspondence between the shape-and-size and the elevated, or the suppressed, physical properties will be established and then compared with experimental observations. Agreement between predictions and experimental observations show consistently that atomic CN imperfection and its consequence on the binding energy density per unit volume in the relaxed surface skin or on the cohesive energy per lower-coordinated atom or their combination dictate the physical size effect. Section 11 summarizes the main contributions by showing the advantages and limitations of the BOLS premise. Recommendations are given on further extension of the BOLS correlation premise and the associated approaches towards new knowledge and innovative measures for practical applications.

## 1.2 Overview

A nanosolid, or so-called as nanoparticle, nanocluster, nanocrystallite, nanograin, etc., is defined as substances or devices that are in the shape of spherical dot, rod, thin plate, or any irregular shape smaller than $10^2$ nanometer across or substances consisting of such grains that are weakly linked. The nanosolid can also be foams or porous structures.[2] The substances may be composites, compounds, alloys, or elemental solids. The dot and rod may be solid or hollow inside. We prefer the term of nanosolid rather than adding a prefix of quantum as nanosolid may properly describe the state of solid and the scale of size in real space. The characteristics of a nanosolid is the high portion of surface atoms. For a spherical dot of one micrometer across the volume or number ratio of surface-to-volume is only 1% while for a dot of 10 nm size it is 25%. The surface-to-volume ratio reaches 100% when the solid is around one nm across or consisting of three atomic shells or less.

The discovery of nanosolid of various shapes, and its assemblies has been quite surprising and has thus generated enormously ever-increasing interest for scientific insights and technological thrusts. Properties of solids determined by their shapes and sizes are indeed fascinating and form the basis of the emerging field of nanoscience and nanotechnology that has been recognized as the key significance in science, technology, and economics in the 21st century. Nanoscale materials are offering a variety of novel features. New physical and chemical properties are expected to occur in such systems, arising from the large fraction of the low-coordinated atoms at the surface and the confinement of electrons to a rather small volume. From a more fundamental point of view, nanostructures bridge the gap between the behavior of an isolated atom and that of the corresponding bulk counterpart where interatomic interaction becomes dominance.

The study of nanocrystalline materials is an active area of research in physics, chemistry, and materials engineering and sciences.[3,4] The striking significance of miniaturizing a solid to nanometer scale is the tunability of the measurable quantities of the solid in all aspects and the quantized and resonant features in transport. In addition to the large surface-to-volume ratios, the surface and quantum effects take on a significance that is normally inconsequential for bulk materials. Varieties of physical properties such as mechanical strength, plasticity,[5] melting,[6,7,8] sintering and alloying ability,[9] diffusivity,[10] chemical reactivity,[11,12] as well as the mode of crystal growth (self-assembly) have been found dependent upon particle size. Property tunability also includes thermodynamics (critical temperatures for phase transition, liquidation, and evaporation), acoustics (lattice vibration), optics (photoemission and absorption), electronics (work function, energy level positions, electron-phonon coupling),[13,14] magnetism (magnetization tailoring or enhancement) and dielectric modulation. Surface passivated by electronegative additives such as C, N, and O, also affects the performance of the nanosolids.[15] A recent review[16] suggests that not only size-dependent phase transitions, but also chemical interaction between the core of the nanoparticle and its surfactant molecules are responsible for the observed x-ray absorption fine structure spectral (XAFS) changes, which can be explained when constructing detailed models of core-surfactant interaction. Besides, external stimuli such as thermally heating and mechanically stretching/pressing also cause abnormal responses from the nanosolid compared with the bulk counterpart.

Materials composed of nanosolids possess unusual features, leading to new phenomena that are indeed surprising.[17,18,19] For instances, the structural and electronic properties are modified near and at the surface, resulting in a breaking of lattice symmetry and broken bonds, giving rise to site-specific surface anisotropy, weakened exchange coupling, and surface spin disorder.[20] Moreover, spin-spin coupling at the interface and between the surface and the core magnetic structures can give rise to exchange anisotropy.[21] An individual defect-free silicon nanosphere with a diameter of 40 nm is measured roughly three-four times harder than bulk silicon (12 GPa) at the ambient temperature.[22] The hardness of nanocrystalline (nc)/amorphous (a) composites such as nc-TiN/a-$Si_3N_4$, nc-TiN/a-$Si_3N_4$/ and nc-$TiSi_2$, nc-$(Ti_{1-x}Al_x)N$/a-$Si_3N_4$, nc-TiN/$TiB_2$, nc-TiN/BN, approaches that of diamond.[23,24] Generally, mechanical strength of a particle increases inversely with the square root of its size and then becomes soft at sizes around 10 nm, which is termed as the inverse Hall-Petch relationship.[25] Ceramic blocks made of nanometric grains can be moulded into engine parts or other useful shapes without shattering during the process, as do ceramics made from larger particles. The sinterability of zeolite crystal increases at the ambient temperature as the solid size is reduced. On heating nanocrystallites of 40 to 80 nm at 80 ºC, solution-mediated transport results in additionally substantial crystal growth.[9] A carbon nanotube (CNT) is much stiffer than the bulk graphite[26,27] while a single-walled CNT melts at ~1600 K,[28] being 0.42 times the melting point ($T_m(\infty)$ = 3800 K) of graphite bulk. A CdS nanodot of ~2.5 nm across melts at 600 K[6] that is much lower than the bulk value of 1675K. The $T_m$ of other nanocrystals also changes with their sizes.[8,29]

Grains of semiconductors of a few nanometers across emit blue-shifted light than do slightly larger chunks of the same material.[30,31] The band gap ($E_G$) in CdSe can be tuned from deep red (1.7 eV) to green (2.4 eV) by simply reducing the solid diameter from 20 to 2 nm.[32] Both the $E_G$[33,34] and the core-level ($E_v$)[35] shift of nano-semiconductors increase whereas the dielectric susceptibility ($\chi$) decreases when the solid size is reduced. Without triggering the electron-phonon (e-p) interaction or electron-hole (e-h) production, scanning tunneling spectroscopy/microscopy (STS/M) measurement revealed that the $E_G$ expands from 1.1 to 3.5 eV when the Si nanorod diameter reduces from 7.0 to 1.3 nm.[36]

Magnetic nanocomposites also exhibit enhanced or tailored magnetic properties under various conditions.[37] At low temperatures, the saturation magnetization of a small solid is higher than that of the bulk but at the ambient temperatures, an opposite trend dominates. Nitriding happens at much lower temperature (300 °C for 9 hrs) to Fe surface covered with nanoparticles compared with nitriding of smooth Fe surface which occurs at 500 °C or higher for more than 48 hrs under atmospheric pressure of amonia.[38] The diffusivity of Ag to Au nanoparticle at the ambient temperature is much higher upon the particle size being reduced.[39] Decreasing the particle size of tin oxide particles in the range of 10-35 nm leads to an increase of the sensitivity to changing gas conditions.[40] The ductility of metallic nanowires such as Cu is ~ 50 times higher than that of the bulk counterparts.[5,41,42] Pure copper samples with a high density of nanoscale grains show a tensile strength about 10 times higher than that of conventional coarse-grained copper, while retaining an electrical conductivity comparable to that of pure copper.[43] If platinum is generated with a continuum network of nanometer-sized pores, they can generate reversible strain with amplitudes comparable to those of commercial materials through surface-charging effects under potentials of about 1 volt.[44] The conversion of an external electrical signal into a volume change, and hence mechanical force, known as actuation is of considerable importance in the development of small-scale devices.

As uncovered by Hu et al[45] nano-sized (27 nm) $SrTiO_3$ obtained by high-energy ball milling could lower substantially the sensoring operation temperature from 970 to 310 K, closing to the temperature of human body. The grain size increases whereas the sample resistivity decreases when the annealing temperature is increased.[46] Higher gas sensitivity of size-selected $SnO_2$ nanoparticles[47] and size-induced structural transformation and ionicity enhancement of $Cu_2O$ nanoparticles have been observed by Mehta and coworkers.[48] Introducing ferroelectric materials of different sizes into a photonic crystal could modulate its refractive index and hence the photonic gap, $E_G$, which is not only sensitive to the external stimuli such as temperature or electric field but also tunable by varying the particle sizes. Zhou et al[49,50] filled barium titanate ($BaTiO_3$) and lead lanthanum zirconate titanate into the silicon-dioxide colloid crystal matrix and found that near the ferroelectric phase-transition point of $BaTiO_3$ (100 to 150 °C), the photonic $E_G$ of the resulting assembly exhibits strong temperature dependence. At the Curie point ($T_C$), a 20-nm red shift of the $E_G$ has been detected. The photonic $E_G$ gradually shifts to longer wavelength with the increase of the applied electric field, suggesting that the refractive index increases with the applied voltage. The photonic $E_G$ tunability could be used not only for simple on-off switching, but also in devices requiring more localized control of light propagation. A comprehensive review given by Lu[51] showed that the thermal expansion coefficient, resistivity, and specific heat of metallic nanosolids/alloys increase with the inverse of solid size whereas the temperature coefficient of resistivity and the temperature of magnetic transition drop with solid size. Indeed, the increased surface-to-volume ratio has caused dramatic change of many physical properties, which are enlisted endless, as timely reviewed by many researchers.[52,53,54,55,56]

The size induced property change has inspired numbers outstanding theories in a certain aspect from various perspectives. For instance, the following models describe the size-induced blue shift in the photoluminescence (PL):

(i) *Quantum confinement* theory[57,58,59,60] suggests that the potential and kinetic energies of e-h pairs (or termed as exciton) are responsible for the intrinsic $E_G$ expansion, which dictates the PL blue shift of a semiconductor nanosolid.

(ii) *Free-exciton collision* model[61] proposes that during the PL measurement the excitation laser heats the free excitions that then collide with the boundaries of the nanometer-sized fragments. The PL blue shift originates from the activation of hot-phonon-assisted electronic transitions rather than from the effect of quantum confinement.

(iii) *Impurity luminescent center* model[62] assumes that different impurity centers in the solid takes responsibility for the PL blue shift. The density and types of impurity centers vary with particle size.

(iv) *Surface states and surface alloying* mechanism[63] considers that the extent of surface catalytic reaction and measurement temperature determine the PL blue shift and the passivation effect varies with the processing parameters and aging conditions.[64]

(v) *Inter-cluster interaction and oxidation*[65] argument also claims for the dominance of the PL blue shift.

The melting point of an isolated nanosolid, or a system with weakly linked nanoparticles, drops with solid size (called as supercooling), while the $T_m$ may rise (called as superheating) for an embedded nano-system due to the interfacial effect. Mechanisms for the $T_m$ elevation or suppression in the nanometer regime are highly disputed with either the Lindermann's criterion[66] of atomic vibration enhancement or Born's criterion[67] of shear modulus disappearance. Other models include (i) homogeneous melting and growth,[68,69] (ii) liquid shell nucleation,[70,71,72,73,74] (iii) liquid nucleation and growth,[70,75,76] (iv) lattice-vibrational instability,[29,77,78,79,80,81] (v) random fluctuation melting,[82] (vi) liquid-drop[83] and, (vii) the surface-phonon instability.[84,85,86]

### 1.3 Challenge

Overwhelming contribution has been made to the development of nanotechnology such as atomic imaging and manipulating, nanosolid synthesizing, functioning, and characterizing as well as structural patterning for device fabrication. However, consistent insight into the mechanism behind the nanosolid tunability remains yet infancy. For a single phenomenon, there are often numerous theories discussing from various perspectives. Reconciliation of all observations in a comprehensive yet straightforward way is a high challenge.

Predictable design and controllable growth of nanostructured materials or devices are foremost important to scientific and technological communities. One needs not only to understand the performance but also needs to know the origin, the trend, and the limitation of the changes and the interdependence of various properties in order to predict and control the process for fabricating materials and devices.

Furthermore, structural miniaturization provides us with an additional freedom that not only allows us to tune the properties of the solid by changing its shape and size, but also challenges us to gain quantitative information by making use of the new freedom, which is beyond traditional approaches.

### 1.4 Objective

In earlier 1990's, the practitioner found that it is essential for the surface bond to contract in decoding the very-low-energy electron diffraction (VLEED, in the energy range of 6.0 – 16.0 eV) data from O-Cu(001) surface reaction kinetics. This finding drove the practitioner to seek evidence from various sources and found that Goldschmidt[87] in 1927 and Pauling[88] in 1947 (see Appendix A) showed that the atomic CN-imperfection could cause the shrink of atomic size, or the contraction of the remaining bonds, of the lower-coordinated atoms. However, at that point of time, consequence of the spontaneous bond contraction on the bond energy was oversighted. The combination of VLEED practice and the premise of Goldschmidt and Pauling drove the practitioner to move into the field of low-dimensional system with a large portion of lower-coordinated atoms as characteristics.

Apparently, only atomic CN imperfection occurs at sites surrounding defects (point defects, voids, dislocations, etc) or at the flat surface skin of a bulk material or at the curved surface of a nanosolid, or in amorphous state with randomly distributed defects. Therefore, the effect of atomic CN imperfection and the increased portion of the lower-coordinated atoms of a nanosolid should dictate the property change. By consideration of the spontaneous effect on the system energy that determines the physical and chemical properties of a solid, the practitioner extends the Goldschmidt and Paulings' bond order-length premise to cover its consequences on the bond strength gain. Efforts have led to the currently reported BOLS correlation mechanism that has been intensively verified and widely applied in recent years of practice.

The objective of this report is to share with the community the consideration, the formulation, the verification and the application of the BOLS premise and to highlight the enormous impact of atomic CN imperfection in dictating the performance of a low-dimensional system and the disordered amorphous state. It is anticipated that the single and simple model could generalize as far as possible the shape and size effect on the imaginable and measurable quantities of a nanosolid. Understanding so far has formed encouraging impact to the physical behavior of a low-dimensional system such as interdependence of a surface and a nanosolid of various shapes. The encouraging progress made insofar includes:

(i) The unusual behavior of a surface and a nanosolid in mean lattice contraction,[89,90] mechanical strength,[91,92] phase transition,[93,94,95] thermal stability,[96] acoustic and optical phonons,[97,35,98] optoelectronics,[33,34,99,100] magnetism,[101] dielectrics,[102,103,104] and chemical reactivity[105] has been consistently predicted and experimentally verified with formulations depending on atomic CN imperfection and its consequences.

(ii) Most encouragingly, single energy levels of an *isolated* Si, Pd, Au, Ag and Cu atoms and their shift upon bulk and nanosolid formation have been quantified by matching predictions to the observed size-and-shape dependence of the XPS data. This attainment in turn enhances the capability of the XPS and provides an effective way of discriminating the contribution from intra-atomic trapping from the contribution of crystal binding to the specific electrons.[106,107] Attainment is beyond the scope of a combination of XPS and laser cooling that only measures the energy level separation of the slowly moving atoms/clusters in gaseous phase.[108,109]

(iii) Quantitative information about dimer vibration[97] and e-p interaction[110] has been elucidated by matching predictions to the measured shape and size dependence of Raman and photoemission/absorption spectra of Si and other III-V and II-VI compounds. The CN imperfection of different orders unifies the phase stability of ferromagnetic, ferroelectric and superconductive nanosolids.[111] In conjunction with the previous bond-band-barrier correlation mechanism,[1] the present approach allows us to distinguish the extent of oxidation[112] and contribution of surface passivation[113] to the dielectric susceptibility of porous silicon.

(iv) The bonding identities such as the length, strength, extensibility, and thermal and chemical stability,[114] in metallic monatomic chains (MCs)[115,116] and in the CNTs[117] have been determined. Understanding has been extended to the mechanical strength and ductility of metallic nanowires, and the inverse Hall-Petch relationship that shows the mechanical strength transition from hard to soft in the nanometer regime. Further investigation in this direction is still in progress.

For simplicity, the work will use the dimensionless form to express the relative change (%) of a detectable quantity and the dimensionless form of size $K_j$ (being the number of atoms lined along the radius of a sphere or cross the thin film) unless indicated otherwise throughout the course. The dimensionless approach also allows the generality of the formulations and minimizes the contribution from impurities and errors in measurement. Attempt is made to minimize and simplify numerical expressions and focus more on physical understanding.

# 2 Principles

## 2.1 BOLS correlation

### 2.1.1 Effects of lattice periodicity termination

- Barrier confinement - quantum uncertainty

The termination of the lattice periodicity in the surface normal direction has two effects. One is the creation of the surface potential barrier (SPB), work function, or contact potential, and the other is the reduction of the atomic CN. The CN of an atom in a highly curved surface is lower compared with the CN of an atom at a flat surface. For a negatively curved surface (such as the inner side of a pore or a bubble), the CN may be slightly higher. Therefore, from the atomic CN-imperfection point of view, there is no substantial difference in nature between a nanosolid, a nanopore, and a flat surface. This premise can be extended to the structural defects or defaults such as voids surrounding which atoms are suffer from CN imperfection.

The work function is expressed as: $\Phi = E_0 - E_F(\rho(E)^{2/3})$,[118] which is the energy separation between the vacuum level, $E_0$, and the Fermi energy, $E_F$. The $\Phi$ depends on the charge density ($\rho(E)$) in the surface region and energy dependent. The charge density varies with the valence states of the surface atoms. If dipoles form at the surface through reaction with electronegative elements such as nitrogen or oxygen, the $\Phi$ of a metal surface can be reduced (by ~1.2 eV).[119] However, if hydrogen-like bonds form at the surface, the $\Phi$ will restore to the original value or even higher because the metal dipoles donate the polarized electrons to the additional electronegative additives to form '+/dipole' at the surface.[1] The shape and the saturation degree of the SPB depend on the surface atomic-valence states[120] but the height of the SPB approaches to the muffin-tin inner-potential-constant of atoms inside the solid, $V_0$.[121] The real (elastic) and imaginary (inelastic) parts of the SPB take the following forms:[121,122]

$$
\begin{aligned}
\mathrm{Re}V(z) &= \begin{cases} \dfrac{-V_0}{1+A\exp[-B(z-z_0)]}, z \ge z_0 \text{ (a pseudo - Fermi - z function)} \\ \dfrac{1-\exp[\lambda(z-z_0)]}{4(z-z_0)}, \quad z < z_0 \text{ (the classical image potential)} \end{cases} \\
\mathrm{Im}V(z,E) &= \mathrm{Im}[V(z)\times V(E)] \\
&= \gamma \times \rho(z) \times \exp\left[\frac{E-\phi(E)}{\delta}\right] \\
&= \frac{\gamma \times \exp\left[\dfrac{E-\phi(\mathrm{E})}{\delta}\right]}{1+\exp\left[-\dfrac{z-z_1}{\alpha}\right]}
\end{aligned}
\tag{1}
$$

where A, B, γ, and δ are constants. α and λ describe the degree of saturation. $z_0$ is the origin of the image plane inside which electron occupies. ϕ(E), the energy-dependent local Φ(E) depends on the density of states ρ(E). The $\nabla^2[\mathrm{Re}V(z)] = -\rho(z) \propto \mathrm{Im}V(z)$ describes the spatial distribution of charges.

Figure 1 (link) One-dimensional SPB model showing that the real and imaginary parts are as functions of the distance z from the surface. The z - axis is directed into the crystal and $z_0$ is the origin for the image plane.[123]

The SPB is the intrinsic feature of a surface, which confines only electrons that are freely moving inside the solid. However, the SPB has nothing to do with the *strongly localized* electrons in deeper bands or with those form sharing electron pairs in a bond. According to the quantum uncertainty principle, reducing the dimension (D) of the space inside which energetic particles are moving increases the fluctuation, rather than the average value, of the momentum, p, or kinetic energy, $E_k$, of the moving particles:

$$
\begin{aligned}
&\Delta p \propto \hbar / D \\
&p = \bar{p} \pm \Delta p \\
&E_k = \bar{p}^2 / (2m)
\end{aligned}
\tag{2}
$$

where $\hbar$ being the Plank constant corresponds to the minimal quanta in energy and momentum spaces and m is the mass of the moving particle. Therefore, SPB confinement causes energy-rise of neither the freely moving carriers nor the localized ones. Therefore, the kinetic energies of carriers or e-h pairs do not change at all with solid dimension.

- Atomic CN imperfection

Figure 2 illustrates situations of atomic CN imperfection. The CN of an atom in the interior of a monatomic chain and an atom at the open end of a single-walled CNT is 2; while in the CNT wall, the CN is 3. For an atom in the fcc unit cell, the CN varies from site to site. The CN of an atom at the edge or corner differs from the CN of an atom in the plane or inside the unit cell. Atoms with deformed bond lengths or deviated angles in the CNT are the same as in amorphous states. The CN imperfection is referred to the standard value of 12 in the bulk irrespective of the bond nature or the crystal structure.[87,88] For example, the CN of an atom in diamond tetrahedron is the same as that in an fcc structure as a tetrahedron unit cell is an interlock of two fcc unit cells.

Figure 2 (Link) Atomic CN of (a) monatomic chain (z = 2); (b) single-walled CNT (z = 2, 3); and (c) an fcc unit cell (z varies from site to site).

### 2.1.2 BOLS formulation

- Bond order-length correlation

Goldschmidt[87] and Pauling[88] indicated that if the CN of an atom is reduced the ionic and the metallic radius of the atom would shrink spontaneously. Therefore, the CN-imperfection will shorten the remaining bonds of the lower-coordinated atom, which is independent of the nature of the specific chemical bond.[124] A 10% contraction of spacing between the first and second atomic surface layers has been detected in the liquid phase of Sn, Hg, Ga, and In.[125] As impurity has induced 8% bond contraction around the impurity (acceptor dopant As) at the Te sublattice in CdTe has also been observed using EXAFS (extended X-ray absorption fine structure) and XANES (X-ray absorption near edge spectroscopy).[126] The finding of impurity-induced bond contraction could provide important impact to an atomic scale understanding of the bond in a junction interface that has been puzzled for decades. Therefore, CN imperfection or impurity induced bond contraction is common.  It is reasonable to extend this initiative to an atom at a curved or a flat surface or at site surrounding a defect in amorphous or polycrystalline solid. Figure 3a shows the CN dependence of the bond contraction coefficient, $c_i(z_i)$. The solid curve formulates the Goldschmidt premise which states that an ionic radius contracts by 12%, 4%, and 3% if the CN of the atom reduces from 12 to 4, 6 and 8, respectively. Feibelman[127] has noted a 30% contraction of the dimer bond of Ti and Zr, and a 40% contraction of the dimer-bond of Vanadium, which is also in line with the formulation. The Goldschmidt-Feibelman contraction coefficient and the associated bond energy increase form the BOLS correlation mechanism that is formulated as:

$$\begin{cases} c_i(z_i) & = d_i/d_0 = 2/\{1+\exp[(12-z_i)/(8z_i)]\} & (BOLS-coefficient) \\ E_i & = c_i^{-m}E_b & (Single-bond-energy) \\ E_{B,i} & = z_iE_i & (atomic-cohesive-energy) \end{cases} \quad (3)$$

Subscript i denotes an atom in the i th atomic layer, which may be countered up to three from the outermost atomic layer to the center of the solid as no CN-reduction is expected for i > 3. The index m is a key parameter that represents the nature of the bond. For Au, Ag, Ni metals, m ≡ 1; for alloys and compounds m is around four; for C and Si, the m has been optimized to be 2.56[117] and 4.88,[35] respectively. The m value may vary if the bond nature evolves with atomic CN. If the surface bond expands in cases, we simply expand the $c_i$ from a value that is smaller than unity to greater, and the m from positive to negative to represent the spontaneous process of which system energy is minimized. The $c_i(z_i)$ should be anisotropic and depend on the effective CN rather than a certain order of CN. The $z_i$ also varies with the particle size due to the change of the surface curvature. Experience reveals that the $z_i$ takes the following values:[35]

$$z_1 = \begin{cases} 4(1-0.75/K_j) & curved-surface \\ 4 & flat-surface \end{cases} \quad (4)$$

Generally, $z_2 = 6$ and $z_3 = 8$ or 12 would be sufficient. The BOLS correlation illustrated in Figure 3 has nothing to do with the particular form of the pairing potential as the approach involves only atomic distance at equilibrium.

Figure 3 (link) illustration of the BOLS correlation. Solid curve in (a) is the contraction coefficient $c_i$ derived from the notations of Goldschmidt[87] (open circles) and Feibelman[127] (open square). As a spontaneous process of bond contraction, the bond energy at equilibrium atomic separation will rise in absolute energy, $E_i = c_i^{-m}E_b$. The *m* is a parameter that represents the nature of the bond. However, the atomic cohesive energy, $z_iE_i$, changes with both the m and $z_i$ values. (b) Atomic CN-imperfection modified pairing potential energy. CN imperfection causes the bond to contract from one unit (in $d_0$) to $c_i$ and the cohesive energy per coordinate increases from one unit to $c_i^{-m}$ unit. Separation between $E_i(T)$ and $E_i(0)$ is the thermal vibration energy. Separation between $E_i(T_{m,i})$ and $E_i(T)$ corresponds to melting energy per bond at T, which dominates the mechanical strength. $T_{m,i}$ is the melting point. The energy required to break the bond at T is $\eta_{1i}(T_{m,i} - T) + \eta_{2i}$.

From a study of interatomic distances of the C-C bonds in organic chemistry, Pauling derived the relation:[88]

$$r(1) - r(v/z_0) = 0.030\log(v/z_0) \quad \text{(nm)} \tag{5}$$

where r(1) is the radius of a single atom, or the length of a dimer bond. The r(v/z) is the radius of an s-fold bond and s = v/z, where v is the number of valency bonds and $z_0$ is the number of equivalent coordinate. As an illustration of the use of this relation, the radius of Ti has been computed as an hcp from the data of Ti as bcc. As a bcc, the radius of Ti is 0.1442 nm, and there are eight bonds of this length. The next closest bonds are six situated 0.1667 nm from any given Ti atom. These values are calculated from the known lattice parameter of 0.333 nm. The valence, v, of Ti is four. The problem is to determine what fraction of these bonds are associated with the eight near neighbours and with the six others removed from these. From eq (5),

$$r(1) - r(x/8) = 0.030\log(x/8) \tag{6}$$

and for the next nearest neighbours,

$$r(1) - r[(4-x)/6] = 0.030\log[(4-x)/6] \tag{7}$$

where x is the number of bonds associated with the eight near neighbours and (4-x) the number associated with the other six bonds. Subtracting (6) from (7) and using the value (0.1667 – 0.1442) obtained from the lattice parameter, one can find that x = 3.75 and the dimer bond length r(1) = 0.13435 nm which contracts by 0.00985 nm. For a CN = 12 in the hcp structure, the bond number v/z is 4/12, and the corresponding bond length is r(4/12) = 0.13435 - 0.03log(4/12) = 0.1486 nm. From Eq (5), one can also deduce the bond length of an atom with a reduced CN($z_i$):

$$r(v/z_i) = r(v/z_0) + 0.030\log(z_i/z_0) \tag{8}$$

Appendix A combines the Goldschmidt and Paulings' notations of electronegativity (η), metallic (ionic) valencies, and metallic (ionic) radii of the elements. Pauling's theory introduced here contains numerous assumptions and it is somewhat empirical in nature. Compared to the formulation (3), Pauling's notation is $d_0$ dependent and somewhat complicated, which gives:

$$c_i = 1 + 0.06\log(z_i/z_0)/d_0(v/z_0) \tag{9}$$

However, this notation does give some surprisingly good answers in certain cases, as commented by Sinnott.[124] Both eqs (3) and (9) should be valid but here we prefer relation (3), as it covers Feibelmen's notation and it is element ($d_0$) independent.

- Bond length-strength correlation

Figure 3b illustrates schematically the BOLS correlation using the pairing potential, u(r). When the CN of an atom is reduced, the equilibrium atomic distance will contract from one unit (in $d_0$) to $c_i$ and the cohesive energy of the shortened bond will increase in magnitude from one unit (in $E_b$) to $c_i^{-m}$. The solid and the broken u(r) curves correspond to the pairing potential of a dimer bond with and without CN imperfection. The bond length-strength correlation herein is consistent with the trend reported in Ref. [128] though the extents of bond contraction and energy enhancement therein vary from situation to situation. There are several characteristic energies in Figure 3b, which correspond to the facts:

(i) $T_{m,i}$ being the local melting point is proportional to the atomic cohesive energy, $z_iE_i(0)$,[96] per atom with $z_i$ coordinate.[129]

(ii) Separation between E = 0 and $E_i(T)$, or $\eta_{1i}(T_{m,i} - T) + \eta_{2i}$, corresponds to the cohesive energy per coordinate, $E_i$ at T, being required for the bond fracture under mechanical or thermal stimulus. $\eta_{1i}$ is the specific heat per coordinate.

(iii) The separation between E = 0 and $E_i(T_m)$, or $\eta_{2i}$, is $1/z_i$ fold energy that is required for atomization of an atom in molten state.

(iv) The spacing between $E_i(T)$ and $E_i(0)$ is the thermal vibration energy.

(v) The energy contributing to the mechanical strength is the separation between the $E_i(T_m)$ and the $E_i(T)$, as a molten phase is extremely soft and highly compressible.

Values of $\eta_{1i}$ and $\eta_{2i}$ can be determined with the known $c_i^{-m}$ and the bulk $\eta_{1b}$ and $\eta_{2b}$ values that have been determined for various crystal structures as given in Table 1.

Table 1 Relation between the bond energy and the $T_m$ of various structures. $E_b = \eta_{1b}T_m + \eta_{2b}$,[83] see Figure 4. $\eta_{2b} < 0$ for an fcc structure means that the energy required for breaking all the bonds of an atom in molten state is included in the term of $\eta_{1b}zT_m$ and therefore the $\eta_{2b}$ exaggerates the specific heat per CN.

| | fcc | bcc | Diamond structure |
|---|---|---|---|
| $\eta_{1b}$ ($10^{-4}$ eV/K) | 5.542 | 5.919 | 5.736 |
| $\eta_{2b}$ (eV) | -0.24 | 0.0364 | 1.29 |

Figure 4 (Link) Correlation between the cohesive energy per coordinate and the $T_m$ of different elements and crystal structures.[83] The data for cohesive energy per atom are taken from Refs. [130,131]

- Nanosolid potential well

As the relaxation (either contraction or expansion) is a spontaneous process, the binding energy of the relaxed bond will be lowered (rise in magnitude) to stabilize the system. The relaxed bond will be stronger. Bond expansion might happen but the system energy must be minimized, unless the relaxation is a process under external stimulus such as heating or stretching.

Figure 5 compares the quantum potential well of QC convention with that of BOLS correlation for a nanosolid. QC convention extends the monotrapping center potential of an isolated atom be rescaling the size. BOLS covers contribution from individual atoms with multi-trapping-center potential wells and the effect of atomic CN imperfection in the surface region. Atomic CN-imperfection induced bond contraction and the associated bond-strength gain deepens the potential well of trapping near the edge of the surface. Therefore, the density of charge, energy, and mass in the relaxed surface region are higher than other sites inside the solid. Consequently, surface stress that is in the dimension of energy density will increase in the relaxed region. Electrons in the relaxed region are more localized because of the deepened potential well of trapping, which lowers the work function and conductivity in the surface region, but enhances the angular momentum of the surface atoms.

Figure 5 (link) Schematic illustration of (a) conventional quantum well with a monotrapping center extended from that of a single atom, and (b) BOLS derived nanosolid potential with multi-trap centers and CN imperfection induced features. In the relaxed surface region, the density of charge, energy and mass will be higher than other sites due to atomic CN imperfection.[107]

## 2.2 Surface passivation

### 2.2.1 Bond-band-barrier correlation

Besides the effect of atomic CN imperfection, surface passivation by adsorbing electronegative elements also contributes to the behavior of a nanosolid. The catalytic-BBB (chemical-bond – valence-band – potential-barrier) correlation mechanism[1] indicates that it is necessary for an atom of oxygen, nitrogen, and carbon to hybridize its *sp* orbitals upon interacting with atoms in solid phase. Because of tetrahedron formation, non-bonding lone pairs, anti-bonding dipoles and hydrogen-like bonds are produced, which add corresponding features to the density-of-states (DOS) of the valence band of the host, as illustrated in Figure 6.[132] Bond forming also alters the sizes and valences of the involved atoms and causes a collective dislocation of these atoms. Alteration of atomic valences roughens the surface, giving rise to corrugations of surface morphology. Charge transportation not only alters the nature of the chemical bond but also produces holes below the $E_F$ and thus creates or enlarges the $E_G$.[133] In reality, the lone pair induced metal dipoles often direct into the open end of a surface due to the strong repulsive forces among the lone pairs and among the dipoles as well. This dipole orientation leads to the surface dipole layer with lowered $\Phi$. For a nitride tetrahedron, the single lone pair may direct into the bulk center, which produces ionic layer at the surface. The ionic surface network deepens the well depth, or increases the $\Phi$, as the host surface atoms donate their electrons to the electron acceptors. For carbide, no lone pair is produced but the weak antibonding feature exists due to the ion-induced polarization. However, hydrogen adsorption neither adds DOS features to the valence band nor expands the $E_G$ as hydrogen adsorption terminates the dangling bond at a surface, which minimizes the midgap impurity DOS of silicon, for instance.[134] Addition of light elements such as S and F is expected to produce dipoles but this anticipation is subject to confirmation.

Figure 6 (link) N and O induced *DOS* differences between a compound and the parent metal (upper) or the parent semiconductor (lower). The lone-pair polarized anti-bonding state lowers the $\Phi$ and the formation of bonding and anti-bonding generate holes close to $E_F$ of a metal or near the valence band edge of a semiconductor. For carbide, no lone pair features appear but the ion induced antibonding states will remain.

### 2.2.2 Significance

The validity of the BBB correlation premise and the associated approaches are testified by the following major progress. Interested readers may be referred to the recent reports.[1,135,136,137]

(i) The reaction kinetics of over 30 samples of O-derived phases on transition metals Cu, Co, Ag, and V, noble metals Rh, Ru, and Pd, and non-metallic diamond surfaces and of C/N on Ni(001) surface has been generalized using formulae of reactions with identification of individual atomic valences and bond forming kinetics at the surfaces.

(ii) The adsorbate-derived signatures of STM/S, LEED, XRD, UPS, XPS, thermal desorption spectroscopy, electron energy loss spectroscopy and Raman spectroscopy, have been unified in terms of atomic valence, bond geometry, valence DOS, bond strength and bonding kinetics. This achievement also enhances in turn the capacities of these probing techniques in revealing details of bond forming kinetics and its consequence on the behavior of the involved atoms and valence electrons.

(iii) Most encouragingly, a $Cu_3O_2$ bond geometry and its four-stage forming kinetics on the O-Cu(001) surface has been quantified by decoding LEED and STM data as: one bond forms first and then the other follows; the sp-orbital then hybridizes with creation of lone pairs that induce dipoles. The four-stage bonding kinetics holds generally true to other analyzed systems based on various observations.

(iv) It has been uncovered that formation of the basic tetrahedron, and consequently, the four-stage bond forming kinetics and the adsorbate-derived DOS features, are *intrinsically* common for all the analyzed systems though the patterns of observations may vary from situation to situation. What differs one surface phase from another in observations are: (a) the site selectivity of the adsorbate, (b) the order of the ionic bond formation and, (c) the orientation of the tetrahedron at the host surfaces. The valences of adsorbate, the scale and geometrical orientation of the host lattice and the electronegativity of the host elements determine these specific differences *extrinsically*.

(v) New knowledge derived from the BBB premise has enabled development of new measures or functional materials for blue-light emission,[138] photonic switch,[49,50] electron emission,[139,140] diamond-metal adhesion,[141] nitride self-lubrication,[142] magnetization modulation,[143] and other systems in applications.[144,145,146]

## 2.3 Shape-and-size dependency

### 2.3.1 Surface-to-volume ratio

It is easy to derive the volume or number ratio of a certain atomic layer, denoted i, to that of the entire solid as:

$$\gamma_{ij} = \frac{N_i}{N_j} = \frac{V_i}{V_j} = \frac{R_{i,out}^{\tau} - R_{i,in}^{\tau}}{R_{K_j,out}^{\tau} - R_{L,in}^{\tau}} \cong \begin{cases} \dfrac{\tau c_i}{K_j}, & K_j > 3 \\ 1, & K_j \le 3 \end{cases} \tag{10}$$

where $K_j = R_j/d_0$ is the dimensionless form of size, which is the number of atoms lined along the radius of a spherical dot, a rod, or cross the thickness of a thin plate. $\tau$ is the dimensionality of a thin plate ($\tau = 1$, and monatomic chain as well), a rod ($\tau = 2$) and a spherical dot ($\tau = 3$) of any size. $L$ is the number of atomic layers without being occupied by atoms in a hollow structure. For a solid system, $L = 0$; while for a hollow sphere or a hollow tube, $L < K_j$. For a hollow system, the $\gamma_{ij}$ should count both external and internal sides of the hollow system. $d_i = R_{i,out}-R_{i,in}$ is the thickness of the ith atomic shell. With reducing the particle size, performance of surface atoms will dominate because at the smallest size ($K_j \to 3$) $\gamma_1$ approaches unity. At $K_j = 1$, the solid will degenerate into an isolated atom. The definition of dimensionality herein differs from convention in transport considerations in which a nano-sphere is defined as zero-dimension (quantum dot), a rod as one dimension (quantum wire), and a plate two dimension (quantum well). Figure 7 illustrates the derivation of the surface-to-volume ratio. As the $K_j$ is an integer, property change will show quantized oscillation features at small particle sizes, which varies from structure to structure, as illustrated in Figure 7c.

Figure 7 (link) illustration of (a) the surface-to-volume ratio ($\gamma_{ij}$) of a nanosolid with involvement of CN-imperfection-induced bond contraction. (b) The $\gamma_{ij}$ drops from unity to infinitely small when the solid grows from atomic scale, $K_j = 1$, to infinitely large, $K_j = \infty$. At the lower end of the size limit, the solid degenerates into an isolated atom. (c) Number ratio shows oscillatory features for fcc and bcc small solids.

### 2.3.2 Scaling law

Generally, the mean relative change of a measurable quantity of a nanosolid containing $N_j$ atoms, with dimension $K_j$, can be expressed as $Q(K_j)$; and as $Q(\infty)$ for the same solid without CN-imperfection contribution. The $Q(K_j)$ relates to $Q(\infty) = Nq_0$ as follows:

$$Q(K_j) = (N_j - N_s)q_0 + N_s q_s = N_j q_0 + N_s(q_s - q_0) \tag{11}$$

The $q_0$ and $q_s$ correspond to the local density of $Q$ inside the bulk and in the surface region, respectively. $N_s = \sum N_i$ is the number of atoms in the surface atomic shells. Eq (11) leads to the immediate relation:

$$
\begin{aligned}
\frac{\Delta Q(K_j)}{Q(\infty)} &= \frac{Q(K_j)-Q(\infty)}{Q(\infty)} = \frac{N_s}{N_j}\left(\frac{q_s}{q_0}-1\right) \\
&= \sum_{i\le 3}\gamma_{ij}\left(q_i/q_0-1\right) \\
&= \sum_{i\le 3}\gamma_{ij}\left(\Delta q_i/q_0\right) = \Delta_{qj}
\end{aligned}
\tag{12}
$$

The weighting factor, $\gamma_{ij}$, represents the geometrical contributions from dimension ($K_j$, L) and dimensionality ($\tau$) of the solid, which determines the magnitude of change. The quantity $\Delta q_i/q_0$ is the origin of change. The $\sum_{i\le 3}\gamma_{ij}$ drops in a $K_j^{-1}$ fashion from unity to infinitely small when the solid dimension grows from atomic level to infinitely large. For a spherical dot at the lower end of the size limit, $K_j = 1.5$ ($K_j d_0 = 0.43$ nm for an Au spherical dot example), $\gamma_{1j} = 1$, $\gamma_{2j} = \gamma_{3j} = 0$, and $z_1 = 2$, which is identical in situation to an atom in a monatomic chain despite the geometrical orientation of the two interatomic bonds. Actually, the bond orientation is not involved in the modeling iteration. Therefore, the performance of an atom in the smallest nanosolid is a mimic of an atom in an MC of the same element without presence of external stimulus such as stretching or heating. At the lower end of the size limit, the property change of a nanosolid relates directly to the behavior of a single bond, being the case of a flat surface.

Generally, experimental observed size-and-shape dependence of a detectable quantity follows a scaling law. Equilibrating the scaling law with Eq (12), one has:

$$
Q(K_j)-Q(\infty) = \begin{cases} bK_j^{-1} & (measurement) \\ Q(\infty)\times\Delta_{qj} & (theory) \end{cases}
\tag{13}
$$

where the slope $b \equiv Q(\infty)\times\Delta_{qj}\times K_j \cong$ constant is the focus of various modeling pursues. The $\Delta_j \propto K_j^{-1}$ varies simply with the $\gamma_{ij}(\tau, K_j, c_i)$ if the functional dependence of $q(z_i, c_i, m)$ on the atomic CN, bond length, and bond energy is given.

Physical quantities of a solid can normally be categorized as follows:

(i) Quantities that are related directly to bond length, such as the mean lattice constant, atomic density, and binding energy.

(ii) Quantities that depend on the cohesive energy per discrete atom, $E_{B,i} = \sum_{zi} E_i = z_i E_i$, such as self-organization growth, thermal stability, Coulomb blockade, critical temperature for liquidation, evaporation and phase transition of a nanosolid and the activation energy for atomic dislocation, diffusion, and bond unfolding.[147]

(iii) Properties that vary with the binding energy density in the relaxed continuum region such as the Hamiltonian that determines the entire band structure and related properties such as band-gap, core-level energy, magnetization, phonon frequency. The binding energy density is proportional to the single bond energy $E_i$ because the number of bonds per circumferential area between neighboring atomic layers in the relaxed region does not change.

(iv) Properties that are contributed from the joint effect of the binding energy density and atomic cohesive energy such as mechanical strength, Young's modulus, surface energy, surface stress, extensibility and compressibility of a nanosolid, as well as the magnetic performance of a ferromagnetic nanosolid.

Therefore, if one knows the functional dependence of the $q$ on atomic separation or its derivatives, the size dependence of the quantity $Q$ is then definite. This approach means that one can design a nanomaterial with desired functions based on the prediction as such by simply tuning the shape and size of the solid.

2.4 Summary

We have addressed the event of atomic CN imperfection and its effect on the bond length and bond strength of the lower-coordinated atoms and the effect of chemical reaction with C, N, O, and H atoms. In using the BOLS correlation, one needs to consider the cohesive energy per bond and per discrete atom when we deal with thermally activated process such as phase transition and crystal growth. One also needs to consider the binding energy density in the continuum region when we deal with the Hamiltonian of the system that dictates the change of the entire band structure of a nanosolid. Some properties such as mechanical strength and magnetization, both atomic cohesion and energy density come into competition. As we know, the performance of a material is determined by the bond and band structures. Chemical reaction with electronegative additives also affects the bond length and the DOS in the valence band.[1] In the following sections, we will apply the BOLS premise to various topics towards consistent insight into the size and shape induced property change of nanosolids, aiming at predictable design and controllable growth of nanostructures with designed functions.

# 3 Surface and nanosolid densification

## 3.1 Nanosolid densification

### 3.1.1 Observations

For an isolated nanosolid or a complex consisting of highly dispersed nanosolids, the lattice constants are often measured to contract while for a nanosolid embedded in a matrix of different materials or passivated chemically, they may expand. For example, oxygen chemisorption could expand the first metallic interlayer by up to 10%-25% due to the penetration of the oxygen atoms into the interlayer spacing.[1] Lattice expansion induced $M_S$ suppression of Ni nanosolids in the diameter range of 6 – 27 nm have been observed due to surface oxidation that extends to 0.4 nm in depth.[148] Using extended XAFS, Montano et al [149] measured the nearest-neighbor distance for silver particles of 2.5 – 13 nm sizes isolated in solid argon and found a noticeable contraction of the nearest-neighbor atomic distance. Lamber et al [150] have measured the lattice-contraction of Pd particles of 1.4 – 5.0 nm sizes using LEED. Mi et al [151] resolved an offset of High-resolution TEM diffraction patterns of 14 nm FePt nanoparticles embedded in amorphous carbon matrix, which corresponds to 4% contraction of lattice constant. Yu et al [152] found using XRD that the mean lattice constants of Sn and Bi nano-particles contract with decreasing the particle size and the absolute amount of contraction of the c-axis lattice is more significant than that of the a-axis lattice. Using XRD, Reddy and Reddy[153] measured ~8 % contraction of the lattice parameter of a 12.5 nm sized ZnMnTe nanosolid. Extended XAFS investigation revealed that the Cu-Cu distance in copper nanosolids with 0.7 to 1.5 nm mean diameter contracts with size in a $D^{-1}$ way and the Cu-Cu dimer bond is reduced by 13% from 0.2555 to 0.221 nm of the 0.7 nm-sized particle.[154]

In comparison, an effective-medium theory approximation[155] suggested that the bond lengths of small (100–1000 atoms) Cu particles at various temperatures suffer only negligible changes. DFT calculations[156] suggest that the atomic distance of Ge and Si expands in the central sites while the bond length contracts in the surface edges, and therefore, the mean lattice constants of the whole Ge and Si nanosolids are smaller than the bulk values.

Mechanisms for the nanosolid densification are quite controversial. From mechanical point of view, the lattice contraction was ascribed as the hydrostatic pressure effect of the surface stress[157] and the intrinsic compressibility of the material,[149,158,159,160] while the central lattice expansion is expected to be consequences of surface oxidation. The mean lattice strain is also explained in terms of incorporation of impurities like hydrogen, carbon, and oxygen, or pseudomorphism in the case of crystalline supports.[150] Nanda et al [161] adopted a liquid-drop model to illustrate the lattice strain and indicated that the anisotropic lattice contraction in Bi and Sn arises from the anisotropy of both the compressibility and the thermal expansion coefficient of the corresponding bulk counterpart in the c and in the a axes. Jiang et al [162,163] developed a model for the size-induced lattice contraction based on the Laplace-Young equation and the size dependence of the solid-liquid interface energy. Yu et al [152] attributed such lattice variation to the super-saturation of the vacant lattice sites inside the particle. Reddy and Reddy[153] suggested that both the atomic density and the refractive index in the core region are higher than those near the edge of the surface. Both the atomic density and the refractive index slowly decrease as one moves away from the center to the surface of the solid. By examining distance between neighboring atoms of Ag, Cu, Ni, and Fe in different CNs, Kara and Rahman[164] found in DFT calculations that these elements follow a strong bond order-length correlation. Because of this correlation, the bond length between an atom and its neighbors would decrease with decreasing CN. Thus, the bond lengths of the dimers (2.53, 2.22, 2.15, and 2.02 Å, for Ag, Cu, Ni, and Fe, respectively[164]), are shorter than the nearest-neighboring atomic distances in their respective bulk by 12.5% for Ag, 13.2% for Cu, 13.6% for Ni, and 18.6% for Fe. Nevertheless, from numerical point of view, all the modeling arguments could fit the experimental data well despite different physical origins.

Actually, surface stress and surface energy result intrinsically from, rather than in, the bond contraction as no external pressure is applied to the surface during measurement. For instance, the compressibility, or the inverse of Young's modulus,

$$\beta = -\frac{1}{V}\left(\frac{\partial V}{\partial P}\right)\Bigg|_T = \left(-V\frac{\partial^2 u}{\partial V^2}\Bigg|_T\right)^{-1} = Y^{-1}$$

and the thermal expansion coefficient,

$$\alpha = \frac{1}{V}\left(\frac{\partial V}{\partial T}\right)\Bigg|_P, \tag{14}$$

are intrinsic properties of a solid and they depend functionally on the interatomic interaction and the atomic size. These measurable quantities describe the response of the lattice ($V \propto d^3$) to the external stimulus such as pressure, $\Delta P$, or temperature change, $\Delta T$.

$$\frac{\Delta V}{V} = 3\frac{\Delta d}{d} = \begin{cases}\beta \times \Delta P \\ \alpha \times \Delta T\end{cases}$$

The external stimulus simply provides a probe detecting the responses: compression or expansion. It may not be applicable to assume a constant compressibility or a constant thermal expansion coefficient in dealing with a nanometric solid. In fact, the surface stress and interfacial energy are derivatives of the binding energy that is enhanced at the surface by the spontaneous process. The idea of core atomic density increases with reducing particle size could hardly explain the lattice and property change of a nanosolid, as at the lower end of the size there no 'core' exists at all.

### 3.1.2 BOLS formulation

The contraction of the mean lattice constant of the entire solid originates from the CN-imperfection induced bond shortening in the surface region and the rise in the surface-to-volume ratio with decreasing solid size, which can be easily derived in a shell structure as follows:

$$\begin{aligned}\bar{d} &= \left[N_j d_0 + \sum_{i\le 3} N_i (c_i - 1) d_0\right]\Big/ N_j \\ &= d_0\left[1 + \sum_{i\le 3}\gamma_{ij}(c_i - 1)\right]\end{aligned} \tag{15}$$

Accordingly, the density of charge, bind energy, and mass in the relaxed region will increase. All the possible models for the mean lattice contraction of a nanosolid are summarized as follows:[161,162]

$$\frac{\Delta d(K_j)}{d_0(\infty)} = \begin{cases}\sum_{i\le 3}\gamma_{ij}(c_i - 1) = \Delta_d & (BOLS) \\ -(2\beta\sigma)/(3K_j) & (\text{liquid-drop}) \\ -(\beta\gamma_{s\text{-}1}d_0)^{1/2}/K_j & (\text{surface-stress})\end{cases} \tag{16}$$

where β is the compressibility and σ the surface stress of the corresponding bulk solid. $\gamma_{s-l} = (2d_0 S_{vib} H_m)/(3VR_j)$ is the solid-liquid interface energy, which is a function of the bulk melting enthalpy, $H_m$, and Molar volume, V, and the vibrational part of melting entropy, $S_{vib}$. All the models fit numerically well the measured data for solids larger than a critical size. The relative change in the mean lattice-constant of a particle in the present BOLS premise simply depends on the $\gamma_{ij}$ and the bond-contraction coefficient $c_i$ without needing other quantities that may vary with the solid size.

With a given shape and size and the known atomic diameters of the constituent atoms of a nanosolid, one can easily predict the lattice contraction of the nanosolid using Eq (16). In the particular ZnMnTe case[153] in which the nanosolid was assumed as a spherical dot with diameter D, the diameters of the constituent atoms are taken as 0.1306(Zn), 0.1379(Mn), and 0.1350(Te) nm, respectively, and the effective CN of the outermost three atomic layers are as 4, 6, and 8. Calculation results listed in Table 2 agree fairly well with the observation (~ 8% contraction for 12.5 nm sized ZnMnTe solid) and show that the lattice constant reaches its bulk value only when the solid dimension is sufficiently large. At the lower end of the size limit, the mean lattice contraction of the solid approaches to the extent of a dimer bond of the same atomic constituents. Furthermore, predictions based on the BOLS premise also agree with the observed trends of lattice contraction for ZnS:Mn films[165] and Sn and Bi nanoparticles.[152] Measurements[164] show that for the 2.0, 2.5, and 3.5 nm sized Ag crystals, the Ag-Ag atomic distance becomes shorter than the bulk one. For the 5.0 nm crystal, 60% of the atoms have the bulk value but 40% have a shorter atomic distance. Consistency between predictions and observations of lattice contraction for a number of metals has been achieved as shown in Figure 8. One can see that the average atomic distance for the three elements, Ag, Cu, and Ni, is shortened by as much as 1.6% – 2.0% for small nano-crystals and about 0.6% for relatively large ones, as compared to the bulk value. In the current approach using $K_j$ as the lateral axis, the observed anisotropy of Bi lattice contraction does not exist at all if one considers the relative change of c and a axis, Δa/a and Δc/c, instead of the absolute amount of variation.

Table 2 Prediction of the size-induced mean lattice contraction of the ZnMnTe nanosolid.

| D(nm) | 5 | 6 | 7 | 8 | 9 | 10 | 11 | 12 | 13 | 100 | 1000 |
|---|---|---|---|---|---|---|---|---|---|---|---|
| $\Delta d(K_j)/d_0(\infty)(\%)$ | -9.9 | -9.1 | -8.7 | -8.4 | -8.0 | -7.7 | -7.4 | -7.1 | -6.8 | -1.3 | -0.13 |

Figure 8 (link) Comparison of BOLS predations with measured size dependence of mean lattice contraction of (a) Al,[166] Ni,[164] Pd,[150] and Pt,[167] (b) Ag-01,[160] Ag-02,[164] Ag-03,[149] Au-01,[158] Au-02,[168] (c) Bi-a-01,[152] Bi-a-02,[169] Bi-c,[152] (d) Cu-01,[170] Cu-02,[171] Cu-03[164] nanoparticles. (e) Thickness dependence of lattice constant of $Pr_2O_3$ films on Si substrate.[172] No anisotropy presents when using Δa/a and Δc/c to calibrate the Bi lattice contraction.

### 3.1.3 Further confirmation

Structural deviations in nanoparticles relative to the bulk crystals are not well understood because they are hard to resolve experimentally, which is under theoretical debate. It is generally assumed in theoretical models of that they have bulk like interior structure.[173] Tight-binding calculations[174] suggested that surfaces of nanoparticles relax in a manner comparable to that of bulk surfaces. However, classical and quantum molecular dynamics simulations have suggested that disorder may pervade throughout nanoparticles.[175,176] Combining the pair distribution function (PDF) derived from wide-angle x-ray scattering and EXAFS analyses, Gilbert et al [177] investigated the intermediate-range order in 3.4 ± 0.3 -nm-diameter ZnS nanoparticles. They found the structural coherence loss over distances beyond 2 nm rather than at 3.4 nm and suggested the presence of structural disorder throughout the nanoparticles.

The PDF for real ZnS nanoparticles is distinct from that of ideal ZnS nanoparticles in the following respects, as shown in **Figure 9**:[177] (i) The first-shell PDF peak intensity is lower compared with that for the ideal case. (ii) PDF peak intensities at higher correlation distances diminish more rapidly than the ideal nanoparticle. (iii) PDF peak widths are broader in the real nanoparticle. (iv) PDF peak positions are shifted closer to the reference atom. The shift is more apparent at r = 1.0 nm and 1.4 nm (shortened by 0.008 and 0.02 nm, respectively), indicating a contraction of mean bond length of the nanoparticle. The Einstein vibration frequency of ZnS nanoparticles is estimated to increase from the bulk value of 7.12 ± 1.2 to 11.6 ± 0.4 THz, implying bond stiffening. It was suggested that structural disorder pervades throughout the nanosolid and that the observed stiffening is due to the structural disorder rather the measured 1% radial contraction of the solid.

Figure 9 (Link) Comparison of the pair distribution function of ZnS bulk solid, calculation for ideal nanosolid and the measurement for real nanosolid show the cohesive length loss of nanosolid. [177]

The findings of ZnS nanosolid straining and stiffening agree appreciably well with the BOLS prediction suggesting that the PDF intensity weakening or diminishing results from volume loss of high-order CN atoms. The PDF peak shifting and broadening arise from the broad range of bond contraction in the outermost two or three atomic layers and the non-uniformity of nanosolid sizes. As the XRD and the EXAFS collect statistic information from numerous nanosolids, one could hardly recognize the structure distortion arises either from the surface region or from the interior of the nanosolids. However, the diameter difference of (3.4 - 2.0) 1.4 nm corresponds to the scale of thickness of the outermost atomic capping and surface layers[178] (3 × 2× 0.255 nm) of which atoms are subject to CN imperfection. Compared with the PDF of an amorphous solid of which the structure coherence extends only to a couple of atomic distances,[134] the detected PDF coincides with the core size of the measured nanosolid. Therefore, the surface layers dominate the bond length distortion. As illustrated in next section, the strengthening of the shortened bonds is responsible for the stiffening of the entire nanosolid. The difference between a solid composed of nanoparticles and a solid in amorphous state is the distribution of atoms with CN imperfection. For the former, lower-coordinated atoms are located at the surface; for the latter, they distribute randomly inside the solid and the distribution is sensitive to processing conditions. In collecting statistic information, the low-CN atoms contribute identically irrespective of their locations. Therefore, BOLS correlation explains well the observations. Furthermore, it is anticipated that the PDF correlation length, or core size, increases with solid dimension and further verification is necessary.

## 3.2 Surface relaxation

### 3.2.1 Monolayer relaxation

There exists sufficient evidence for the bond contraction at surfaces (see samples in Table 3). For instance, about a 10% reduction of the first layer spacing ($d_{12}$) of the hcp($10\bar{1}0$) surface of Ru,[179] Co[180] and Re[181] has been detected using LEED and DFT approaches. The $d_{12}$ of the diamond (111) surface was reported to be ~30% smaller than the (111) spacing in the bulk with a substantial reduction of the surface energy.[182] It has been uncovered[183,135] with VLEED that the O-Cu bond contraction (from 4% to 12%) forms one of the four essential stages of the O-Cu(001) bond forming kinetics and, about 10% bond contraction for the O-Cu(110) surface is necessary.[184] TiCrN surface bond contraction (by 12% - 14%) was further confirmed by measuring the enhanced surface stress and Young's modulus of the TiCrN surfaces.[142] Theoretical calculations[185] confirmed that the interatomic distance drops significantly associated with cohesive energy per bond rise as the dimensionality decreases from three to one for Ag, Au and Cu nanosolids, as shown in Figure 10.

Figure 10 (Link) DFT calculation of the bond-length-strength relation for bulk fcc crystals and straight atomic chains.[128] The bond length contracts and the binding energy per bond in the chains are about 2–3 times that of the bulk fcc crystals, which complies with the current BOLS correlation albeit the absolute amounts of variation.

However, the $d_{12}$ of the Be(0001) and Mg(0001) surfaces and the dimer bonds of the II-b elements of Zn, Cd, and Hg have been reported to expand. With a reduction of Se grain size from 70 to 13 nm, the *a* lattice was found to expand by 0.3%, but *c* spacing decreases slightly, the unit-cell volume increases by about 0.7% at *D* =13 nm.[186] The reported expansion appears not in line with notations of Goldschmidt and Pauling who emphasized that the global bond contraction depends only on the reduction of the atomic CN and it is independent of the bond nature and the constituent elements (appendix A).

Table 3 Bond length relaxation for typical covalent, metallic and ionic solids and its effect on the physical properties of the corresponding solid or surface. Where $d_0$, and $d_l$ is the bond lengths for atoms inside the bulk and for atoms at the surface, respectively. The $c_l$ is the bond contraction coefficient that varies from source to source.

| Bond nature | Medium | $c_l = d_l/d_0$ | Effect |
|---|---|---|---|
| Covalent | Diamond {111}[182] | 0.7 | Surface energy decrease |
| Metallic | hcp $(10\bar{1}0)$ surface of Ru,[179] Co,[180] and Re[181]<br>Fe-W, Fe-Fe [187]<br>Fe(310)[188],Ni(210)[189]<br><br>Al(001)[190]<br>Ni, Cu, Ag, Au, Pt and Pd dimer[128]<br>Ti, Zr[127]<br>V [127] | 0.9<br>0.9<br>0.88<br>0.88<br><br>0.85-0.9<br><br>0.7<br>0.6 | Atomic magnetic momentum enhancement by (25~27)%.<br>Cohesive energy rises by 0.3 eV per bond.<br>Single bond energy increases by 2 ~ 3 times. |
| Ionic | O-Cu(001)[183,135]<br>O-Cu(110)[184]<br>N-Ti/Cr[142] | 0.88-0.96<br>0.9<br>0.86-0.88 | 100% rise in hardness |
| Extraordinary cases | (Be, Mg)(0001) surface Zn, Cd, and Hg dimer bond,[127] Nb[191] | > 1.0 | No report is available about its effects on physical properties. |

### 3.2.2 Multilayer relaxation

In some numerical calculations and diffractional data optimizations, bond contraction/expansion has also been extended to more than 10 atomic layers for a number of clean metals though the physical ground is ambiguous.[192] For instance, calculations suggest that[193] atomic relaxations on the stepped Ag(410) and Cu(320) surfaces extend several layers into the bulk with non-uniform character in damping magnitudes of interlayer relaxations. The calculated contractions (with respect to the bulk) of 11.6, 5.3, and 9.9 % for the top three interlayer separations of Ag(410) are followed by lattice expansion of 2.1% and 6.7% for the subsequent two interlayer spacings. Investigations[194] of the temperature dependence of the first three interlayer spacings of Ag(110) using LEED and DFT, and molecular dynamics calculations suggest that the $d_{12}$ contracts by 8% at 133 K and by 0.2% at 673 K associated with Debye temperature rise from 150 ± 65 K to 170 ± 100 K compared with the bulk value of 225 K. For a Cu(320) surface, 13.6% and 9.2% contraction of the first two interlayers are followed by an expansion of 2.9%, and then an 8.8% contraction, and finally a 10.7% expansion for the subsequent three. The $d_{12}$ of Au(110) surface is reduced by 13.8%, the $d_{23}$ is expanded by 6.9%, and finally the $d_{34}$ is reduced by 3.2%.[195] LEED analysis for Ag(410) suggests[196] initially that there was no measurable relaxation of the interlayer spacings.

However, later analysis of the same LEED Ag(410) data shows 36% contraction for $d_{23}$ and 18% expansion for $d_{34}$. On the other hand, LEED measurements of Cu(320) revealed a 24% contraction for $d_{12}$ and 16% contraction for $d_{23}$, followed by 10% expansion for $d_{34}$. Therefore, physical constraints may be necessary to specify a unique solution from the numerous values given by pure mathematics. According to the BOLS correlation, the surface bond contracts by 12% if we take the effective CN of four.

### 3.3 Impact of bond contraction

The bond contraction at a surface has indeed enormous effects on various physical properties of a nanosolid. Besides the magnetic enhancement,[187-189] the relaxation of Al, Ag, Cu, and Pd surfaces has been found to lead to a shift in the frequencies of the surface states and to a change in the number and localization of the states.[197] For Ag nanocrystals, densification stiffens the atomic force constants by up to 120% when compared to that for Ag bulk.[164] The vibrational free energy and the heat capacity of the step and the terrace atoms on the Cu(711) surface are sensitive to the local atomic environment, and vibrational contribution to the excess free energy of the step atoms near room temperature is a significant fraction of the kink formation energy.[198] The Al(001) surface relaxation expands the bandwidth for the relaxed monolayer by 1.5 eV compared with the value for the bulk-truncated monolayer with 0.3 eV enhancement of atomic cohesive energy.[190] The lattice constant of Au nanoparticles capped in n-dodecanethiol contracts by 1.4%, 1.1% and 0.7% with an association of 0.36, 0.21, and 0.13 eV 4f-core-level shift for 1.6, 2.4, and 4.0 nm sized Au particles.[199] Therefore, the impact of atomic CN imperfection and the associated bond-strength gain is indeed enormous to the physical properties of a system with large portion of lower-coordinated atoms.

## 4 Mechanical strength

### 4.1 Surfaces

#### 4.1.1 Outstanding models

- Surface stress and surface energy

Surface stress (P) being the same in dimension to hardness (H) reflects the internal energy response to volume strain at a given temperature. Hardness is the ability of one material to resist being scratched or dented by another. The former often applies to elastic regime while the latter to inelastic deformation. Surface stress is an important concept, which links the microscopic bonding configuration at an interfacial region with its macroscopic properties,[200,201] such as the threshold of cold cathode field emission in carbon.[202] The stress also plays a central role in the thermodynamics and acoustics of solid surfaces. During the past decades, increasing interest has been paid to processes that are strongly influenced by surface stress effects such as reconstruction, interfacial mixing, segregation, and self-organization at solid surfaces. However, surface stress and hardness are not so easily determined at atomic scale, and there is no theory that could tell us how to arrange atoms to make a hard structure. Therefore, detailed knowledge underlying the atomistic processes of surface stress is yet lacking.[200,201]

Comparatively, surface energy ($\gamma$) representing a fundamental material property is normally defined as half the energy needed to cut a given crystal into two half crystals. As such, the surface energy naturally depends on the strength of the bonding and on the orientation of the surface plane. Despite its importance, the value of surface energy is difficult to determine experimentally. Most of the experiments are performed at high temperatures where surface tension of liquid is measured, which are extrapolated to zero Kelvin. This kind of experiments contains uncertainties of unknown magnitude[203] and corresponds to only $\gamma$ value of an isotropic crystal.[204] Documented data determined by the contact angle of metal droplets or from peel tests conflict one another, which can be induced by the presence of impurities or by mechanical contributions, such as dislocation slip or the transfer of material across the grain boundary.[205]

Numerical attempts has been made to calculate the $\gamma$ values of metals using *ab initio* techniques,[204,206,207] tight-binding (TB) parameterizations,[208] and semi-empirical methods.[209] The $\gamma$ values, work functions, and surface relaxation for the whole series of bcc and fcc 4d transition metals have been studied using the full-potential (FP) linear muffin-tin orbital (LMTO) method in conjunction with the local-spin density approximation to the exchange-correlation potential.[206] In the same spirit, the γ values and the work functions of most elemental metals including the light actinides have been carried out using the Green′s function with LMTO method.[207,210] Documented database is accompanied with a mean deviation of 10% for the 4d transition metals from FP methods.[204] In conjunction with the pair-potential model,[211] the database has been further extended to estimating the formation energy of monatomic steps on low-index surfaces for an ensemble of the bcc and fcc metals.[204]

- Bond broken argument

Besides, the traditional broken-bond model is often used to estimate the γ values of the transition and the noble metals with different facets.[195] The simplest approach to estimate the γ values at $T$ = 0 K is to determine the broken bond number $z_{(h\,k\,l)} = z_b - z_S$ for creating a surface area by cutting a crystal along a certain crystallographic plane with a Miller index *(hkl)* where $z_S$ is the *CN* of a surface atom and $z_b$ the corresponding bulk one. Galanakis et al [195] investigated the correlation between the broken bond and the surface energy of noble metals using two different full-potential *ab initio* techniques. They introduced a simple rule based on the number of broken nearest-neighbor bonds to determine the surface energies of Cu, Ag, and Au metals. The physical argument for the bond-broken rule is derived from a tight binding approximation, which relates the surface energy to the atomic cohesion energies. In a nearest-neighbor TB model, the γ value for a transition metal surface is given by,[195]

$$\gamma \cong \frac{W_S - W_B}{20} n_d \left(n_d - 10\right)$$

where $n_d$ is the number of d-electrons. $W_S$ and $W_B$ are the bandwidths for the surface and bulk DOS, which are assumed rectangular forms. The γ value of the Au(110) surface was calculated to be reduced by 6.5% compared with energy in the bulk region.[195]

Using Kelvin equation,[212] the relation between the atomic cohesive energy, $E_B$, and the activation energy, $E_A(N)$, for removing one atom from a nanosolid is given as,[213]

$$\begin{cases} \dfrac{P_s(D)}{P_s(\infty)} & = \exp\left(\dfrac{4\gamma m_a}{\rho_P k_B T D}\right), \quad (Kelvin) \\ E_A(N) & = E_B - \dfrac{2\pi d^2 \gamma}{3N^{1/3}}, \quad (E_A \sim E_B) \end{cases} \tag{17}$$

where $P_s(D)$ and $P_s(\infty)$ are the vapor pressure of the nanosolid and the corresponding bulk surface; $m_a$ is the atomic weight, $\rho_P$ is the mass density of the particle. Nanda et al [214] developed a method based on this approach to determine the γ value of Ag and PbS nanosolids by measuring the onset temperature ($T_{onset}$) of particle evaporation:

$$\frac{T_{onset}(K_j)}{T_{onset}(\infty)} = 1 - \frac{8\pi\gamma d_0^2}{3E_B K_j} \tag{18}$$

Using $E_B$ = 2.95 eV, and $d_0$ = 0.158 nm, a γ value of 7.37 $Jm^{-2}$ has been derived by analyzing the measured size dependent $T_{onset}$ of Ag free standing nanosolids, which is 5-6 times higher than that of the bulk value and this value may be over estimated.[215]

Multiplying the broken bond number with the cohesion energy per bond $E_b = E_B/z_b$ in the bulk for the non-spin-polarized atom at 0 K, γ is suggested to follow,[201]

$$\gamma = \left(1 - z_s/z_b\right)E_B = E_b\left(z_b - z_s\right) \tag{19}$$

the γ corresponds actually to the energy loss due to CN reduction, which is always lower than the bulk value. On the other hand, the broken-bond rule seems to contradict to the basic knowledge about the electronic structure since $E_B$ in general does not scale linearly with $z_S$. Nevertheless, the above estimation provides the order of magnitude of γ and shows a possible relationship between γ and atomic binding strength. Despite the absence of verification from experiments, such a rule has been used to describe the γ value of Al reasonablly.[195]

In the second-moment TB approximation, the width of the local DOS on an atomic scale with $z_S$, leads to an energy gain that is proportional to $\sqrt{z_S}$ due to the lowering of the occupied states.[216] Neglecting the repulsive terms, the energy per nearest neighbour is then proportional to $\sqrt{z_S}$ . By assuming that the total crystalline energy is a sum of contributions from all bonds of an atom, the surface energy is suggested to follow the relation:[216]

$$\gamma \cong \left(1-\sqrt{z_s/z_b}\right)E_B = E_b\left(\sqrt{z_b}-\sqrt{z_s}\right). \tag{20}$$

According to this model, the rearrangement of the electronic charge does not practically change the nature of the remaining bonds when one bond is broken. Thus, the energy needed to break a bond is independent of the surface orientation, so that the γ value is proportional to the square root of the number of the nearest-neighbour broken bonds.

Direct utilization of either eq (19) or (20) to estimate the γ value has been widely accepted. However, eq (19) fails to consider the variation of bond strength with *CN* reduction, while eq (20) is less complete because only attractive forces are taken into account.[206] Namely, eq (19) neglects while eq (20) overestimates the effect of relaxation on γ when the γ is directly related to the number of broken bonds. Neither of them could alone give satisfied predictions for γ values in comparison with the experimental and theoretical results.[206] To obtain a more general expression, Jiang et al [217] assumed that an average of eqs (19) and (20) could make up the deficiency of each. Thus, the γ values may be determined by averaging them without elaborating estimation on the relaxation energy. They also considered the contribution from the high-order CN to the bond energy and suggested a form,

$$\gamma = \frac{\left[2-z_s/z_b-\left(z_s/z_b\right)^{1/2}\right]+\lambda\left[2-z'_s/z'_b-\left(z'_s/z'_b\right)^{1/2}\right]}{2+2\lambda}E_B \tag{21}$$

where the prime denotes the next-nearest *CN* of the surface atoms and λ is the total bond strength ratio between a next-nearest neighbor and a nearest neighbor.[218] The approach improves substantially the agreement between predictions and experimental observations of a number of elemental surfaces.[217]

4.1.2 BOLS formulation

- Surface energy

Apparently, approaches mentioned above give inconsistent trends of surface energy compared with the bulk binding energy because of the different definitions. Actually, Kelvin's equation describes energy density in the surface region while the bond broken argument describes energy loss upon surface formation. It is reasonable that Kelvin's equation suggests an elevation while the broken bond arguments suggest a suppression of the surface energy. The fact of surface bond contraction and its effect on the bond energy may provide us with the possible physical origin for surface energy and interfacial stress, despite the effect of chemical passivation. The BOLS correlation relates the surface stress and surface energy directly to the binding energy density in the relaxed surface region, which become stronger due to atomic CN reduction. As the number of bonds in the relaxed region does not change apparently, and therefore, the surface energy, or the binding energy density in the relaxed region, should increase by:

$$\Delta\gamma_i/\gamma \cong c_i^{-(m+3)}-1>0 \tag{22}$$

For a defect-free metallic surface (m = 1), the surface energy is expected to increase by $0.88^{-4}$-1 ≅ 67% in magnitude, if we define the surface energy as the energy density in the relaxed outermost surface region.

- Young's modulus and surface stress

Young's modulus (Y) is the stress of a material divided by its strain in the elastic deformation region, meaning that how much the material yields for each pound of load put on it. The Y and P at a surface can be expressed as functions of the binding energy, $E_b$, volume, v, and the atomic distance, d. They share the same dimension:

$$\begin{cases} P &= -\left.\dfrac{\partial u(r)}{\partial v}\right|_{r=d} \propto E_b / d^3 \\ Y &= v\dfrac{\partial P}{\partial v} = v\left.\dfrac{\partial^2 u(r)}{\partial v^2}\right|_{r=d} \propto E_b / d^3 \end{cases} \tag{23}$$

The relative change of the local $Y_i$, $P_i$ and $\gamma_i$ shares a commonly dimensionless form:

$$\frac{\Delta Y_i}{Y} = \frac{\Delta P_i}{P} = \frac{\Delta \gamma_i}{\gamma} = c_i^{-(m+3)} - 1 \tag{24}$$

This relation implies that both the Y and the P at a surface be higher than the bulk values because of the rise of cohesive energy per shortened bond at temperatures, ideally, far below the melting point. One may note that the Young's modulus describes the elasticity while its inverse, or the extensibility/compressibility, covers both the elastic and plastic deformation. However, any of the elastic or plastic processes is related to the process of bond distortion with energy consumption (integrating the stress with respect to strain). No matter how complicated the actual process of deformation (with linear or nonlinear response) or recovery (reversible or irreversible) is, a specific process consumes a fixed portion of bond energy, and the exact portion for a specific process comes not into play in the numerical expression for the relative change. Therefore, relations (23) and (24) are valid for any substances in any phases such as gaseous, liquid, or solid state and any scale as well, as they are quantities of intensity that is dimensionless. It might not be appropriate to think about the stress of a single atom but instead the stress at a specific atomic sire.

There is massive evidence for the surface-enhanced mechanical strength of a thin film. An examination of the hardness (also stress) and the Young's modulus using nanoindentation revealed that the surface of the TiCrN thin film (2 μm thick) is 100% higher than its bulk value.[142] The same trend holds for amorphous carbon[219] and AlGaN surfaces.[220] Ni films also show maximal hardness in ~ 5 nm depth of the surface, though the peak maxima vary with the shapes of nanoindentor tips.[221] Low-dielectric thin film shows 2-3 times higher value of hardness and Young's modulus near the surface at the ambient temperature. When the measuring temperature rises, the stress drops and transits from tensile to compressive.[222]

Solving eq (24) with the measured value of $\Delta P / P = (50-25)/25 = 1$ in Figure 11 gives rise to the $c_i$ values of 0.860 and 0.883 associated with m = 3, and 4, respectively. Lower values of the P peaks are also observed in measurement, which arise from grain boundaries or from defects. Figure 12 shows the indentation profiles of N doped tetrahedron carbon films.[223] The thickness dependence of both the Y and the P follow the similar trend and that is independent of the nitrogen content, evidencing the validity of relation (24), though the Young's modulus is defined for the elastic deformation and the hardness is for plastic deformation. Figure 13a shows the indentation depth dependence of the hardness of Ag, Ni, Cu, Al, $\alpha_2$-TiAl, and γ-TiAl films.[224] The measured strain-gradient plasticity was explained in terms of the material characteristic length that depends functionally on quantities of the Burger's vector, the shear modulus, and the material reference stress. Figure 13b compares predictions with theoretically calculated thickness dependence of Young's modulus for Ni-I and Cu-I,[225] Ni-II, Cu-II, and Ag[226] thin films by summing Eq (24) over the top three atomic layers. Agreement is reached with $z_1 = 4$, $z_2 = 6$, and m = 1 for the pure metals. At the thinnest limit (two atomic layers), the Young's modulus of Cu is 100% higher than the bulk value, which agrees with that detected from the TiCrN surface.[142]

Figure 11 (link) Nanoindentation hardness-depth profile for TiCrN[142] and GaAlN[220] thin films. The peak shift corresponds to the surface roughness (Ra = 10 nm as confirmed by AFM).

Figure 12 (link) Nitrogen partial pressure dependence of (a) the hardness and (b) Young's modulus of ta-C:N films versus penetration depth measured using a continuous stiffness method.[223]

Figure 13 (link) (a) Comparison of indentation size effect. The hardness $H$ is normalized by its value, $H_{1000}$, at a depth of 1000 nm. [224] (b) Agreement between predictions and measured size dependence of the Young's modulus of *Cu*-I and *Ni*-I, Ni-II and Cu-II and Ag thin films.[225,226]

### 4.2 Nanospheres

By squeezing Si nanospheres of different sizes between a diamond-tipped probe and the sapphire surface, Gerberich et al [22] measured that a defect-free silicon nanosphere with a diameter of 40 nm is ~ 3 times (50 GPa) harder than bulk silicon (12 GPa). The smaller the sphere, the harder it was: spheres with a diameter of 100 nm had a hardness of around 20 GPa. For comparison, sapphire has a hardness of about 40 GPa, and diamond 90 GPa. The silicon nanospheres are comparable in hardness with materials such as nitrides and carbides, which typically have hardness values in the range of 30-40 GPa. Figure 14(a-c) shows the loading curves and the derived hardness of different particle sizes. Liu et al [227] measured, as shown in Figure 14(d), that the Y value of nano-grained steel increases from 218 to 270 GPa associated with lattice contraction from 0.2872 to 0.2864 nm in the grain-size range of 700 nm and lower.

Figure 14 (link) (a) Load–displacement curves for a Si particle with an original height of 50.3 nm exhibiting a strain reversal of approximately 34%. (b) Load–displacement curves for a particle with an original height of 92.7 nm exhibiting a strain reversal of approximately 19%. (c) Hardening curves for silicon nanospheres of various dimensions. The right-hand column lists the particle radii.[22] (d) Correlation between the Young's modulus and lattice constant of various planes of nanograined steel.[227]

The afore-discussed findings coincide with a recent curry of activity used in the nanotechnology community which has focused on the indentation size effect[22,228] and the mechanical behavior of small volumes.[229,230] Small volumes include mechanically milled iron powders with hardness of 8.4 GPa,[231] and nanocrystalline composite films of $TiN/Si_3N_4$ reportedly having hardness in the range of 20-100 GPa.[232] The hardness of Ti, Zr, and Hf carbide films on silicon substrate increases from bulk value of 18 GPa to 45 GPa when the film thickness decreases from 9000 to 300 nm.[233] Molecular dynamics simulations[234] of 105–106 atoms of nanocrystalline aluminum under an applied stress of 2–3 GPa suggest that the stress is substantially higher than what their normal bulk counterparts could sustain, which are believed to have unique properties at least partially due to their small length scales. These seemingly widely disparate material systems have a common thread in that line defects or dislocations in these refined microstructures are generated at very high pressures. In the process of testing, dislocations that are generated and squeezed closely together result in extremely high internal stress, which is suggested to be responsible for resisting plastic deformation in these fine microstructures. A single edge dislocation has an elastic shear stress field distribution that could produce a stress of nearly 3 GPa at a distance of one nm from the line. The consequences of these very high internal stresses have potential for the design of superhard materials. Evaluation of such materials becomes in question, as substrate, contact area, or pressure effects represent confounding aspects in measuring hardness.

The size dependent hardness and the Young' modulus enhancement at temperature far below the melting point of the addressed samples are in good accordance with the BOLS prediction. From eqs (23) and (24), one may realize that the factors that enhance the Y and P simultaneously are the shortened bond length and the associated bond-strength gain. Atoms in the outermost atomic layer, or at an interface, should dominate the hardness due to the atomic CN induced bond contraction and bond strength gain.

Applying m = 4.88 for Si to eq (24) gives immediately $c_1 = (15 \pm 1)\%$ and the corresponding $z_1 \approx 3.60 \pm 0.25$. The $z_1$ is slightly lower than $z_1 = 4$ for a flat surface and the $c_1$ is slightly greater than that of the TiCrN flat surface (13 ± 1)%. The bond contraction (from 0.263 nm to 0.23 nm) and band gap expansion (from 1.1 to 3.5 eV) of Si nanorod with the rod diameter reduction[36] evidences directly the BOLS formulation of the Young's modulus and the hardness of Si nanosolids. For steel, Liu et al [227] proposed a functional dependence of Y value on the lattice constant, which agrees with the BOLS prediction of $Y(D)/Y(\infty) = c_i^{-(m+3)}$ with m = 1 for metals. The $c_i$ is estimated to be 94.8% according to the measured maximal Y value of Liu et al , as given in Figure 14d.

### 4.3 Compounds and alloys

Blending different types of atoms in a solid could enhance the hardness of the solid preferably in amorphous state, so called as high entropy materials.[235] Gao et al [236] proposed a formula based on the concept of ionicity to predict the hardness of several compounds. Ionicity ($f_i$) is a measure of the degree of charge sharing: covalent bonds have the lowest ionicity, and ionic bonds have the highest. The hardness, or the activation energy required for plastic gliding, was related to the band gap $E_G$ and the hardness of covalent solid is given as:

$$\begin{cases} H_v = AN_aE_G = 556N_a d_0^{-2.5}\exp(-1.191 f_i)(GPa) \\ f_i = 1 - E_h^2/E_G^2 \end{cases}$$

where A is a constant coefficient and $N_a$ is the number of covalent bond per volume. The $E_G$ for a binary polar covalent system can be separated into both covalent or homopolar gap $E_h$ (= $39.74d_0^{-2.5}$)[237] and ionic or heteropolar gap C. The $d_0$ is the covalent bond length in angstrom. Philips,[238] Liu and Cohen,[239] and Korsunskii et al ,[240] have proposed a similar relationship for the bulk modulus $B$ of a compound solid, which follows:

$$B = N_c/4 \times (19.71 - 2.20 f_i) d_0^{-3.5} (Mbar)$$

$N_c$ is the nearest atomic CN. The parameter $f_i$ accounts for the reduction in $B$ arising from increased charge transfer. The value of $f_i$ = 0, 1 and 2 for groups IV, III-V, and II-VI solids in the periodic table. For a tetrahedral system, $N_c = 4$, otherwise, the $N_c$ is an average of atomic CN. For diamond, $f_i = 0$, d = 1.54 Å, and hence $B$ = 4.35 Mbar, compared with an average experimental value of 4.43 Mbar. This relationship was applied to *BN* and β-$Si_3N_4$ with corresponding prediction of $B$ = 3.69 and 2.68 Mbar. Litovchenko[241] also derived that the $E_G \propto d^{-2}$ and then the elastic modulus follows the relation of $B \propto d^{-5}$. These predictions stimulated tremendous interest of experimental pursue for superhard carbon nitride phase worldwide,[242,243,244,245] as the diameter of an N atom is shorter (0.14 ~0.148 nm) compared with the C-C diamond bond length of 0.154 nm.

Albeit the difference in the power index, -2.5, -3.5, -5, and –(m+3) in the current approach, all the expressions indicate that shorter and denser chemical bonds as well as smaller ionicity should favor hardness. In order to obtain a compound with large bulk modulus, one must find such a covalent compound that has both shorter bond length and smaller ionicity, and high compactness in atomic arrangement inside. Thus, the atomic CN-imperfection induced bond contraction should contribute directly to the hardness at the surface or sites surrounding defects. Therefore, a nanometer sized diamond is expected to be 100% ($0.88^{-5.56} - 1$) harder than the bulk nature diamond.

4.4 Inverse Hall-Petch relation

The mechanically strengthening with grain refinement of crystals with mean grain-size of 100 nm or bigger has been traditionally rationalized with the so-called T-unapparent Hall-Petch relationship (HPR)[246] that can be simplified in a dimensionless form:

$$P(x_j, T)/P(0, T) = 1 + A' x_j$$

The $P(x_j, T)$ is the yield strength (or flow stress), and the slope A′ is an adjustable parameter for data fitting. The $x_j = K_j^{-1/2}$ and $K_j$ is the dimensionless form of size. P(0, T) is the bulk strength measured at the same T. The dimensionless form should be processing condition independent.

The pile-up of dislocations at grain boundaries is envisioned as a key mechanistic process underlying the enhanced resistance to plastic flow from grain refinement. As the crystal is refined from the micrometer regime into the nanometer regime, this process invariably breaks down and the yield strength versus grain size relationship departs markedly from that seen at larger grain sizes. With further grain refinement, the yield stress peaks at a mean grain size in the order range of 10 nm or so in many cases. Further decrease in grain size can cause softening of the solid, and then the HPR slope turns from positive to negative in the nanometer range, which is called as the inverse Hall-Petch relationship (IHPR).[247]

There is a concerted global effort underway using a combination of novel processing routes, experiments and large-scale computations to develop deeper insights into these phenomena. It has been suggested that the grain boundaries consisting of lower-coordinated atoms contribute to the grain-boundary strengthening.[248] The strength maximum at a grain size of 10 ~ 15 nm for Cu nanosolid is attributed to a switch in the microscopic deformation mechanism from dislocation-mediated plasticity in the coarse-grain interior to grain boundary sliding in the nanocrystalline regime.[249] A significant portion of atoms resides in the grain boundaries and the plastic flow of the grain-boundary region is responsible for the unique characteristics displayed by such materials.[250] In the HPR regime, crystallographic slips in the grain interiors govern the plastic behavior of the polycrystalline; while in the IHPR regime, the plastic behavior is dominated by the grain boundaries. During the transition, both grain interiors and grain boundaries contribute competitively. The slope in the HPR is suggested to be proportional to the work required to eject dislocations from grain boundaries.[251] Wang et al [252] proposed two mechanisms that influence the effective stiffness and other mechanical properties of nanomaterials. One is the softening effect due to the distorted atomic structures and the increased atomic spacings in the interface region, and the other is the baffling effect due to the existence of boundary layers between the interface and the crystalline interior. The mechanical performance of a nanocrystallite depends on the competition between these two origins. Molecular-dynamics simulations suggest that the IHPR arises from sliding-accommodated grain-boundary diffusion creep.[253] The critical size depends strongly on the stacking-fault energy and the magnitude of the applied stress.[254] Unfortunately, an analytical form for the IHPR was absent until recently when Zhao et al [25] firstly modified the HPR by introducing the activation energy that can be related directly to the melting point, $T_m$, to the slope A′.[129]

Although there is a growing body of experimental evidence pointing to such unusual deformation responses in the nanometer regime, the underlying mechanisms are not fully understood. As pointed out recently by Kumar et al ,[255] the physical origin of the IHPR has been a long-standing puzzle and the factors that dominate the critical size at which the HPR transits are yet poorly known.

It is possible to incorporate the BOLS correlation to the IHPR by considering the competition between the effect of bond-order loss on cohesive energy and bond-strength gain on the mechanical strength in the continuum region. According to the BOLS correlation, the surface is harder at low temperature but melts easier. Separation between the operation temperature and the $T_m(K_j)$ will drop with solid size. The closer the $T_m(K_j)$-T, the softer the nanosolid. Further investigation on the size and temperature dependence of mechanical strength and extensibility of nanosolid and the factors dominating the critical size for different materials are in progress.

## 5 Thermal stability

### 5.1 Cohesive energy

#### 5.1.1 Defination

The cohesive energy of a solid ($E_{coh}$) is an important physical quantity to account for the binding strength of the crystal, which equals to the energy dividing the crystal into individually isolated atoms by breaking all the bonds of the solid. The $E_{coh}$ is given as: $E_{coh}(N_j) = N_jE_B = N_jz_bE_b$, if no atomic CN imperfection is considered. The cohesive energy for a single atom, $E_B$, is the sum of the single bond energy $E_b$ over the atomic *CN*, $E_B = z_bE_b$, (*or* $E_{Bi} = z_iE_i$ for the ith specific atom). As the heat required for loosening an atom is proportional to the atomic $E_B$ that varies with not only the atomic CN but also the CN reduction induced bond strength, the difference of the mean $E_B$ in different systems is responsible for the fall (undercooking) or rise (overheating) of the $T_m$ of a surface and a nanosolid. The $E_B$ is also responsible for other thermally activated behaviors such as phase transition, catalytic reactivity, crystal structural stability, alloy formation (segregation and diffusion), and stability of electrically charged particles (Coulomb explosion), as well as crystal growth and atomic diffusion, atomic gliding displacement that determine the ductility of nanosolids.

#### 5.1.2 Outstanding models

- Surface-area difference

The lower-coordinated surface atoms will be less thermally stable than those inside the bulk though the strength gain happens to the remaining bonds of the lower-coordinated atoms. For large bulk materials, effects of surface CN-imperfection is ignorable but for small particles, surface effects become dominant because of the appreciably large portion of such lower-coordinated atoms at the surface. One approach to determine the $E_{coh}$ of a nanosolid is to consider the difference between the surface area of a whole particle and the overall surface area of all the constituent atoms in isolated state.[256] For a spherical dot with radius $R_j$ and $N_j$ atoms of diameter $d_0$, the $E_{coh}$ equals to the energy required to generate the area difference, $\Delta S$, between the isolated $N_j$ atoms and the nanodot without changing the volume:

$$\begin{cases} N_j 4\pi(d_0/2)^3/3 = 4\pi(R_j)^3/3 & (volume-conservation) \\ \Delta S = \pi\left[N_j d_0{}^2 - (2R_j)^2\right] & (Surface-area-difference) \end{cases}$$

Let the surface energy per unit area at 0 K is $\gamma_0$, and then the overall $E_{coh}(N_j)$ is,

$$\begin{cases} E_{coh}(K_j) & = \gamma_0 \Delta S = \pi N_j d_0^2 \gamma_0 \left(1 - N_j^{-1/3}\right) \\ & = E_{coh}(\infty)\left(1 - \alpha/K_j\right) \end{cases}$$

$E_{coh}(\infty)$ is the cohesive energy of the $N_j$ atoms without the effect of atomic CN imperfection. The factor $\alpha$ varies with the shape and dimensionality of the solid. For a cube, the factor is 9/4;[256] for a spherical dot, it is 1/2.

Recent development of the model by Qi and coworkers[257] covers the situations of both isolated and embedded nanosolids by considering the contribution from interface/surface atom:

$$E_{B,s} = \left[E_B + 3\beta\left(E_B/2 + kE_m/2\right)\right]/4$$

and the mean atomic cohesive energy becomes,

$$\begin{aligned} E_B(K_j) &= E_B + \gamma_{ij}\left(E_{B,S} - E_B\right) \\ &= E_B + 3\gamma_{ij}\left[k\beta E_m - (2-\beta)E_B\right]/8 \end{aligned}$$

where $\beta$ is the ratio of the interface area to the whole surface area, *k* denotes the degree of cohesion between the nanocrystal and the matrix with $E_m$ atomic cohesive energy. For a nanocrystal wholly embedded in the matrix, $\beta$ = 1 and *k* = 1; for an isolated crystal, $\beta$ = 0 and *k* = 0. This model improves agreement between modeling calculations and measurement than the model that assumes the surface atomic cohesive energy to be $E_B/4$.[258]

- Atomic CN-difference

By considering the effect of CN imperfection, Tomanek et al [147] derived the $E_B$ for an individual atom denoted i:

$$E_{B,i} = \left(z_i/z_b\right)^{1/2} E_B(\infty) + E_R$$

where $z_b$ is the effective CN of an atom in the bulk. $E_R$ is the repulsive interactions that can be replaced by a hard-core potential. The $E_R$ is neglected at equilibrium distances. The mean $E_B$ in a nanosolid with $N_j$ atoms is obtained by summing all bonds over the $N_j$ atoms:

$$\left\langle E_B\left(N_j\right)\right\rangle = \sum_{<i,N_j>} \left(z_i\right)^{1/2} E_i / N_j$$

Based on the model for the latent heat, the size dependent cohesive energy or boiling heat can be derived,[259] which agrees reasonably with the experimental results of W and Mo nanosolids.

### 5.1.3 BOLS formulation

The BOLS correlation considers contribution from atoms in the shells near the surface edge. Using the same spherical-dot containing $N_j$ atoms with $N_S$ (sums the $N_i$ over the outermost two or three atomic layers) atoms at the surface shells, the average $<E_{coh}(N_j)>$, or $<E_B(N_j)>$ is,

$$\begin{aligned}
\left\langle E_{coh}\left(N_j\right)\right\rangle &= N_j z_b E_b + \sum_{i\le 3} N_i\left(z_i E_i - z_b E_b\right) \\
&= N_j E_B(\infty) + \sum_{i\le 3} N_i z_b E_b \left(z_{ib} E_{ib} - 1\right) \\
&= E_{coh}(\infty)\left[1 + \sum_{i\le 3} \gamma_{ij}\left(z_{ib} c_i^{-m} - 1\right)\right] \qquad = E_{coh}(\infty)\left(1 + \Delta_B\right) \\
or, \quad \left\langle E_B\left(N_j\right)\right\rangle &= E_B(\infty)\left(1 + \Delta_B\right)
\end{aligned} \tag{25}$$

where $E_{coh}(\infty) = N_j z_b E_b$ represents the ideal situation without *CN*-imperfection. The $z_{ib} = z_i/z_b$ is the normalized *CN* and $E_{ib} = E_i/E_b \cong c_i^{-m}$ is the binding energy per coordinate of a surface atom normalized by the bulk value. For an isolated surface, $\Delta_B < 0$; for an intermixed interface, $\Delta_B$ may change depending on the interfacial interaction. Summarizing all the models, one may find that the size dependence of $E_B$ can be numerically estimated in any of the scaling relationships:

$$\frac{\Delta E_B(K_j)}{E_B(\infty)} = \begin{cases} \sum_{i\le 3} \gamma_{ij}\left(z_{ib} c_i^{-m} - 1\right) = \Delta_B & (BOLS) \\ \sum_{i\le 3} \gamma'_{ij}\left[\left(z_{ib}\right)^{1/2} - 1\right] = \Delta'_B & (Atomic-CN-loss) \\ -\alpha/K_j = \delta_B & (Surf-area-difference) \end{cases} \tag{26}$$

where $\gamma_{ij} \sim \tau c_i/K_j$ and $\gamma'_{ij} \sim \tau_i/K_j$ are the surface-to-volume ratio in the corresponding descriptions. Figure 15 compares the modelling predictions with the measured size dependent $<E_B(K_j)>$ of Mo and W nanosolids.[260] From the viewpoint of numerical wise, one could hardly tell which model is preferred to others though physical indications of the compared models are entirely different.

Figure 15 (Lnik) Comparison of the modeling predictions with experimental results on the size dependent $E_B$ of Mo and W nanosolids.[260] Numerical agreement is nearly identical for the compared models albeit the different physical origins.

### 5.1.4 Atomic vacancy formation

As an elemental of structural defects, atomic vacancies or point defects are very important in materials, which have remarkable effect on the physical properties of a material such as electrical resistance, heat capacity, and mechanical strength, etc. An atomic vacancy formation needs energy to break all the bonds of the specific atom to its surroundings, which is the same to the atomic $E_B$ in the current BOLS iteration though deformation is involved upon atom evaporation. Nevertheless, the deformation costs no additional external energy. The vacancy volume should be greater than the atomic size due the effect of atomic CN-imperfection on atoms surrounding the vacancy. The $E_B$ is detectable but the experimental values are subject to accuracy. For instance, the $E_B$ of a Mo atom was measured to vary in a large range from 2.24 to 3.3 eV.[261] Sophisticated theoretical efforts also predict the $E_B$ of metals and alloys but the theories are rather complicated and not general to most metals accurately.[262]

- Brook's convention

In the 1950s, Brooks[263] developed a semi-empirical method to calculate the $E_B$ of bulk materials. In this method, the crystal is assumed isotropic, and the formation of vacancy is considered as equivalent to creating new surface, equal to the area of one unit cell, being approximately the spherical surface of the atomic volume. Meanwhile, it was assumed that the surface tension of the hole would squeeze the hole to contract in size by distorting the rest of the crystal elastically. Then the $E_B$ equals the minimum sum of the increased surface energy and the distortion energy. The $E_B$ for atomic vacancy formation inside a bulk solid is thus given as:

$$E_B = \pi d_0^3 \gamma_0 G (\gamma_0 + G d_0)^{-1} \tag{27}$$

$G$ is the shear modulus and $\gamma_0$ the surface energy per unit area surrounding the vacancy. Introducing the size effect to the $d_0$, $G$, and $\gamma_0$, the relative change of the mean $E_B$ in a nanoparticle becomes,

$$\frac{E_p - E_B}{E_B} = \frac{d_p^3}{d_0^3}\left(\frac{G d_0 + \gamma_0}{G d_p + \gamma_0}\right) - 1 \tag{28}$$

where $E_p$ and $d_p$ are the corresponding vacancy formation energy and mean atomic diameter of the nanosolid. Qi and Wong[264] extended Brook's approach to nanostructures by assuming that the G and the $\gamma_0$ of a nanosolid remain the bulk values. The key factor influencing the $E_p$ of a spherical dot of diameter D is the size dependent atom size. Supposing that a small size shrink of $\varepsilon D$ ($\varepsilon \ll 1$) results from the atomic size contraction, the surface energy variation $\Delta\gamma$ and the strain dependent elastic energy $f$ of the particle become,

$$\begin{aligned} \Delta\gamma &= \pi D^2 \left[(1-\varepsilon)^2 - 1\right]\gamma_0 \\ f &= \pi G D^3 \varepsilon^2 \end{aligned} \tag{29}$$

At equilibrium state, the total energy $F$, or the sum of $\Delta\gamma$ and $f$, is minimal, that is, $dF/d\varepsilon = 0$, and then the strain of the particle is:

$$\varepsilon = \left[1 + (G/\gamma_0) D\right]^{-1} \tag{30}$$

The average size $d_p$ of an atom shrinks due to the presence of G and $\gamma_0$, $d_p = d_0(1-\varepsilon)$.

- BOLS analysis

In comparison with the BOLS formulation, the bond strain can be expressed as:

$$\frac{\Delta d_p}{d_0} = \begin{cases} -\left[1 + (G/\gamma_0) D\right]^{-1} \cong -K_{dc}/(K_j + K_{dc}) & (Brook) \\ \sum_{i\le 3} \gamma_i (c_i - 1) = \Delta_d & (BOLS) \end{cases} \tag{31}$$

where $K_{dc} = \gamma_0/(2 d_0 G)$ is the critical value and $K_j$ remains its usual meaning of dimensionless size. Further simplification of Eq (28) leads to the atomic vacancy-formation energy in a nanometric system as given in comparison with the BOLS derivative:

$$\frac{\Delta E_B(K_j)}{E_B(\infty)} = \begin{cases} -(1+K_j/\alpha)^{-1} & \cong -K_{Ec}/(K_{Ec}+K_j) \\ \sum_{i\le 3}\gamma_{ij}\left(z_{ib}c_i^{-m}-1\right) & = \Delta_B \end{cases} \qquad (32)$$

where $\alpha = (2gd_0+3)/\left[2d_0\left(g^2d_0+g\right)\right]$ (~$10^{-1}$ level) and $g = G/\gamma_0$~10 $nm^{-1}$. $K_{Ec} = \alpha/(2d_0)$ is the critical value of $K_j$. For Pd and Au nanosolids, the critical $K_{Ec}$ and $E_{dc}$ values are calculated based on the given G and $\gamma_0$ bulk values as listed in Table 4. Figure 16 compares the predictions of the two models. At the lower end of the size limit ($K_j$ = 1.5), the particle contracts by 40% associated with 12% reduction of the $E_B$ according to Brook's convention.[264] In comparison, the BOLS correlation predicts a 25% bond contraction and 70% lower of the $E_B$ for the smallest size. Approximation based on Brook's relation seems to over-estimate the bond contraction and under-estimate the $E_B$ suppression because of the assumption of size independent G and $\gamma_0$. Actually, the atomic vacant hole should expand instead as the remaining bonds of the surrounding atoms will contract. If T ≠ 0, one has to consider the temperature dependence of surface energy, which is beyond the scope of current discussion focusing on the effect of size. One may note that $E_B$ varies from site to site due the difference of atomic CN imperfection at various locations of the solid.

Figure 16 (Link) Comparison of the bond (particle size) contraction and atomic vacation-formation energy derived from the BOLS premise and from Brook's approach for Pd and Au nanosolids.

Table 4 Shear modulus, surface energy and the calculated α values for Pd and Au.

| | $G(10^{10}\ N/m^2)$[265] | $\gamma_0(J/m^2)$[266] | α/nm | $K_{Ec}/K_{dc}$ |
|---|---|---|---|---|
| Pd | 4.36 | 2.1 | 0.104 | 0.1894/0.8770 |
| Au | 2.6 | 1.55 | 0.119 | 0.2066/1.035 |

## 5.2 Liquid-solid transition

### 5.2.1 Outstanding models

The melting behavior of a surface and a nanosolid has attracted tremendous interest of research both theoretically and experimentally for decades.[6] In many physical systems, surface melting and evaporating often occur at temperatures lower than the corresponding bulk values.[267,268] For substrate-supported nanosolids with relatively free surfaces, the $T_m$ decreases with decreasing particle size (supercooling). In contrast, as per the existing experimental evidence for embedded nanosolids, the $T_m$ can be lower than the bulk $T_m$ for some matrices. However, the same nanosolids embedded in some other matrices may exhibit superheating to temperatures higher than the bulk $T_m$. The $T_m$ suppression for free surface is attributed to the reduced degree of confinement, and hence the increased entropy of the molecules at the surface compared with atoms in the bulk, whereas, the $T_m$ elevation or depression of the embedded nanosolids depends on the coherency between the nanosolids and the embedding matrix.[269,270]

There have been a huge volume of database for surface and nanosolid melting point suppression.[271,272,273,274,275] The general trend of melting point suppression applies to small particles disregarding the composition. For instances, a photoelectron emission study[276] confirmed that the lithium (110) surface melting occurs 50 K below the bulk melting point (453.69 K). A thermal and temperature resolved XRD analysis revealed that the melting point of nanometer sized drugs (polymer) also drops (by 33 and 30 K for 11 nm sized griseofulvin and 7.5 sized nifedipine, respectively) in a 1/R fashion.[277] STM measurements of a reversible, temperature-driven structural surface phase transition of Pb/Si(111) nanoislands indicate that the transition temperature decreases with inverse of island and domain size and the phase transition has little to do with the processes of cooling or healing.[278]

Numerous models for the size effect on the nanosolid melting in terms of classical thermodynamics and atomistic/molecular dynamics.[29,68,69,70,71,73,74,75,76,82,85,279,280,281,282] In general, the size dependent $T_m(K_j)$ follows the empirical scaling relationship:

$$\frac{\Delta T_m(K_j)}{T_m(\infty)} = -\frac{K_C}{K_j}, \tag{33}$$

where $K_C$ is the critical limit at which the nanosolid melts completely. The physical meaning of the $K_C$ is profound, which is the focus of modeling pursues.

- Classical thermodynamics

Classical thermodynamic theories based on the surface Laplace and the Gibbs-Duhem equations[281] have derived that the $K_C$ follows the following relations depending on mechanisms behind:[279,282]

$$K_C = \frac{-2}{H_m(\infty)} \times \begin{cases} \sigma_{sv} - \sigma_{lv}(\rho_s/\rho_l)^{2/3}, & (HGM) \\ \sigma_{sl}(1 - K_0/K_j)^{-1} + \sigma_{lv}(1 - \rho_s/\rho_l), & (LSM) \\ [\sigma_{sl}, 3(\sigma_{sv} - \sigma_{lv}\,\rho_s/\rho_l)/2] & (LNG) \end{cases}$$

Where $H_m$ is the latent heat of fusion. $\rho$ and $\sigma$ are the mass density and the interfacial energy. Subscripts *s*, *l*, and *v* represent the phases of solid, liquid and vapor, respectively. The critical $R_C$ (= $K_C d_0$) is normally several nanometers. Expressions of the $K_C$ correspond to three outstanding models in terms of classical thermodynamics:

(i) The homogeneous melting and growth (*HMG*) model[68,69] considers the equilibrium between the entire solid and the entire molten particle, which suggests that the melt proceeds throughout the solid simultaneously.

(ii) The liquid shell nucleation (*LSN*) model[70,73] assumes that a liquid layer of $K_0$ thick is in equilibrium at the surface, which indicates that the surface melts before the core of the solid.

(iii) The liquid nucleation and growth (*LNG*) model[75,76] suggests that melting starts by the nucleation of liquid layer at the surface and moves into the solid as a slow process with definite activation energy.

- Atomistic models

However, models based on atomistic/molecular dynamics suggest that the critical $R_C$ follows:

$$R_C = \begin{cases} 5230 v_0 \gamma, & (v_0 = 4\pi d_0^3/3) \quad (Liquid-drop) \\ \alpha_m d_0, & (\alpha_m - constant) \quad (Surf-phonon) \\ R_j \times [\exp(-(\beta-1)/(N_j/N_S - 1)) - 1] \approx R_0\left(\frac{1-\beta}{1-R_0/R_j}\right), & (Surf-RSMD) \end{cases} \tag{34}$$

The liquid-drop model[83] relates the $T_m$ to the $E_{coh}$ of the entire particle of $N_j$ atoms. With surface involvement, the $E_{coh}$ equals the difference between the volume cohesive energy ($N_jE_B$) and the surface energy ($4\pi d_0^2 N_j^{2/3}\gamma$). The mean cohesive energy per atom with volume $v_0$ in the solid is: $E_B(R_j) = E_B - E_{B,S}N_j^{-1/3}$, where $E_{B,S} = 4\pi d_0^2\gamma$ is the cohesive energy for an atom at the surface. The relation between the $E_B$ and the $E_{B,S}$ is given empirically as, $E_{B,S} = 0.82E_B$.[283] Based on the Lindemann's criterion of melting, an expression for the $T_m$ of the bulk material is derived as:[284]

$$T_m(\infty) = nE_B f_e^2/(3k_B Z) \propto E_B \tag{35}$$

where $n$ is the exponent of the repulsive part of the interaction potential between constituent atoms, $Z$ is the valence of the atom, which is different from the atomic CN. Coefficient $f_e$ is the thermal expansion magnitude of an atom at $T_m$. At the $T_m$, the $f_e$ is around 3%.[285] The fact that the bulk $T_m$ varies linearly with the $E_B$ and hence with the $E_{B,S}$ agrees with the data measured for metals.[286] Therefore, the $T_m$ of a solid can be simply related to the mean atomic $<E_B(K_j)>$ of the solid. Replacing the $E_B$ with $E_B(K_j)$, Nanda et al [83] derived the liquid-drop model for the size dependent $T_m(K_j)$ based on the relation between the bulk $T_m$ and the cohesive energy per coordinate:

$$E_b(\infty) = \eta_{1b}T_m(\infty) + \eta_{2b} \tag{36}$$

where the constant $\eta_{2b}$ represents 1/z fold of enthalpy of fusion and atomization being required for evaporating an atom in molten state. $\eta_{1b}$ is the specific heat per coordinate in the bulk. The $\eta_{1b}$ and $\eta_{2b}$ values, as tabulated in Table 1 (see Section 2), for various structures and elements have been obtained from fitting experimental data.[83]

According to the liquid-drop model, the critical radius at which $T_m(K_C)$ is zero Kelvin is in the range of 0.34 (for Mn) -1.68 nm (for Ga). The liquid drop model underestimates the $T_m$ of Sn, Bi, In, and Pd nanosolids by 3% - 12%, due to the inter-cluster interaction and particle-substrate interaction.[83]

Surface-phonon instability model[84,287] suggests that the $T_m(K_j)$ varies with $T_m(\infty)$ and with the energy of intrinsic defects formation at the surface and that within the thermo dynamical limit (particle radius larger than about two nm), the cumulated effect of size reduction and high electronic excitation combine.[288]

Lattice-vibration instability (*RMSD*) model[77,78,79,80,81,85] was advanced based on the Lindemann's vibrational-lattice instability criterion. [66] The melting behavior of a nanosolid is related to the ratio ($\beta$) of the root-mean-square-displacement (*RMSD*, $\delta^2$) of an atom at the surface to the *RMSD* of an atom inside a spherical dot. As an adjustable parameter, the $\beta$ is assumed size independent:

$$\beta = \delta_s^2(D)/\delta_b^2(D) = \delta_s^2(\infty)/\delta_b^2(\infty)$$

The $K_C$ in the *RMSD* model is determined by $K_0 = \tau$ at which all the constituent atoms have surface features. This model indicates that if $\beta > 1$, the surface melts below the bulk $T_m$, and vice versa. The nanosolid melts at 0 K at the lower end of the size limit, $K_C = \tau$ shells.

- Superheating

In the case of embedded nanosolids, the coefficient of surface energy will be replaced by the interfacial energy if the surfaces are completely saturated with atoms of the surrounding matrix. Nanda et al [83] describe the superheating by introducing the ratio as perturbation of surface energy between the matrix and the embedded specimen,

$$\frac{\Delta T_m(K_j)}{T_m(\infty)} = -\frac{K_C}{K_j}\left(1 - \frac{\gamma_{Mat}}{\gamma}\right)$$

If the surface energy of the matrix $\gamma_{Mat} > \gamma$, the nanosolid melts at temperature that is higher than the bulk $T_m$. This expression matches the experimental data of Pb particles embedded in Al matrix but overestimates the $T_m$ for Indium particles embedded in Al matrix by some 10-20 K using the known $\gamma$ and $\gamma_{Mat}$ values. Based on the size-dependent magnitudes of the atomic vibrations, Jiang *et al* [8,85,] extended the $T_m(K_j)$ model for the superheating, according to which the superheating is possible if the diameter of the constituent atoms of the matrix is smaller than the atomic diameter in the embedded nanosolid. Therefore, both the superheating and undercooling of a nanosolid can be modeled simply by adjusting the $\beta$ value in the *RMSD* model. Superheating happens when $\beta < 1$, which means that the matrix confines the vibration of the interfacial atoms.

However, molecular dynamics simulations[289] suggest that atoms in the bulk interior of an isolated nanosolid melt prior to the surface and the surface melting occurs at relatively higher temperatures.

This prediction seems to be excluded by existing evidence but it is possible if the bond nature changes upon CN reduction, as discussed shortly for the superheating of smallest $Ge^+$ and Sn clusters.

The models of LSN, HMG and LNG suit only for the cases of $T_m$ suppression ($\Delta T_m < 0$) while the liquid-drop and the RMSD models cover both the undercooling of an isolated solid and the superheating of an embedded system. For particles larger than several nanometres, all the models work sufficient well in simulating the size-dependent melting despite the disputable mechanisms. However, a large number of independent parameters such as the latent heat of fusion, mass density and interfacial energy of different phases are to be considered when evaluating the melting behaviour of a nanosolid. Furthermore, these quantities are also size dependent.

### 5.2.2 BOLS formulation

It is known that the total energy of a pair of atoms (Section 2) can be expressed in a Taylor's series, which can be decomposed as energies of binding at 0 K, $E_b(r)$, and the thermal vibration energy, $E_V(T)$:

$$\begin{aligned} E_{total}(r,T) &= \sum_n \left( \frac{d^n u(r)}{n!dr^n} \right)_{r=d_0} (r-d_0)^n \\ &= u(d_0) + 0 + \left.\frac{d^2u(r)}{2!dr^2}\right|_{d_0} (r-d_0)^2 + \left.\frac{d^3u(r)}{3!dr^3}\right|_{d} (r-d_0)^3 \ldots \\ &= E_b(d_0) + \frac{k}{2}(r-d_0)^2 + \frac{k'}{6}(r-d_0)^3 + \ldots \\ &= E_b(d_0) + E_V(T) = \begin{cases} 0, & (Evaporation) \\ E_C, & (Critical - T_C) \end{cases} \end{aligned} \quad (37)$$

The term with index $n = 0$ corresponds to the minimal binding energy at T = 0 K, $E_b(d_0) < 0$. The term $n = 1$ is the force [ $\partial u(r)/\partial r|_{d_0} = 0$] at equilibrium state and the terms $n \geq 2$ correspond to the thermal vibration energy, $E_V(T)$. The $T_C$ can be any critical temperature for event such as melting point, $T_m$, or phase transition, $T_C$. By definition, the thermal vibration energy of a single bond is,

$$\begin{aligned} E_V(T) = \eta_1 T &= \mu\omega^2 (r-d_0)^2/2 = q_v (r-d_0)^2/2 \\ &= \sum_{n\geq 2} \left( \frac{d^n u(r)}{n!dr^n} (r-d_0)^{n-2} \right)_{d_0} (r-d_0)^2 \end{aligned} \quad (38)$$

where $r\text{-}d_0$ is the magnitude of lattice vibration. $\mu$ is the reduced mass of a dimer of concern. The $q_v = \mu\omega^2$ is the force constant for lattice vibration with an angular frequency of $\omega$.

The physical ground for the BOLS iteration is that if one wishes to peel off or loosen an atom in the solid thermally, one must supply sufficient thermal energy to overcome the cohesion that bind the specific atom to its surrounding neighbors. The thermal energy required to loosen or break one bond is the separation between the minimal bond energy $E_b$ and the $E_C$ (or $E_m$ for melting), as illustrated in section 2. If the $E_V(T)$ is sufficiently large, all the bonds of the specific atom will break and this atom will get rid of the solid. At the evaporating point of any kinds of solids, $E_{total} = 0$; at the critical point, $E_{total} = E_C$. We may consider step-by-step the energies required for melting (or dissociating) a single bond, a single atom, and then shell-by-shell of a nanosolid of radius lined with $K_j$ atoms.

The thermal energy required for loosening a single bond of an atom in the ith atomic layer by raising the temperature from T to $T_C$ is given as,

$$E_T = E_c(T_C) - E_b(T) = \eta_1(T_C - T) \propto E_b(0)$$

(39)

The energy required melting the entire atom in a bulk is proportional to the $E_B(0)$, which is a sum of the single bond energy over all the atomic *CN*.

Melting a nanosolid comprising $N_j$ atoms requires thermal energy that is proportional to the cohesive energy of the entire solid:

$$T_m(K_j) \propto E_{coh}(K_j) = \left[N_j z_b E_b + \sum_{i\le 3} N_i (z_i E_i - z_b E_b)\right]$$

(40)

It is reasonable to assume homogenous bond nature of the solid. The $E_{coh}$ may vary from material to material but for a specific sample, the portion of $E_{coh}$ needing for the phase transition should be fixed for a specific process occurring to the specimen.[284] The relative change of $T_m(K_j)$ is then:

$$\frac{\Delta T_m(K_j)}{T_m(\infty)} = \frac{\Delta E_B(K_j)}{E_B(\infty)} = \sum_{i\le 3}\gamma_{ij}\left(z_{ib}c_i^{-m} - 1\right) = \sum_{i\le 3}\gamma_{ij}(\alpha - 1) = \Delta_B$$

(41)

It is not surprising that the temperature is always the same throughout the small specimen in operation whereas the intrinsic $T_m$ may vary from site to site if the sample contains atoms with different *CN*, such as atoms at the surface, grain boundary, or sites surrounding voids or stacking faults. This mechanism may explain why the latent energy of fusion of a solid was measured to be a broad hump rather than a sharp peak.[279,290] For a solid with numerous randomly distributed defects, the mechanism of random fluctuation melting[82] could dominate because the energy required for breaking one bond and hence the energy needed to melt an individual atom with different *CN* is different. This mechanism may also explain the glass transition of an amorphous state as the random atomic CN-imperfection distributed in the solid, which happens in a range of temperatures and process condition dependent.[291]

On the other hand, from classical thermodynamic point of view, the thermal energy $E_T$ required for the liquid-solid phase transition can be estimated by integrating the specific heat over the entire solid with and without *CN*-imperfection from zero to the $T_m$:

$$\frac{\Delta E_T(K_j)}{E_T(\infty)} = \frac{\int_0^{T_m(K_j)} C_p(K_j, T)dT}{\int_0^{T_m(\infty)} C_p(\infty, T)dT} - 1 \cong \frac{\Delta T_m(K_j)}{T_m(\infty)} = \Delta_B ,$$

(42)

with an assumption of $C_P(K_j, T) \cong C_P(\infty, T) \cong C_V(\infty, T)$ = constant in the entire measured temperature range.[124] It is true in fact that $C_P(K_j, T) \neq C_P(\infty, T) \neq C_V(\infty, T) \neq$ constant. The Debye temperature and hence the specific heat $C_P$ is size and temperature dependent.[78,292] This effect may lead to 3 ~ 5% deviation of the $C_P$. Besides, $(C_P - C_V)/C_V \sim 3\%$.[124] However, compared with the precision in determining the size and shape of a nanosolid such errors are negligible. Actually, measurements[293,294,295] show that the $C_P$ varies insignificantly with the particle size in the measured temperature range. Therefore, it is acceptable to simplify the $C_P$ as a constant in the integration. Such simplification may lead to slight deviation in the integration in Eq (42) from the true value. Nevertheless, one should particularly note that the deviation of the integration from true value only affects the precision of the *m* value or the effective $z_{ib}$, it does no matter with the nature of the phenomenon. Actually, eq (41) extends the classical thermodynamics (42) in terms of atomistic approach.

5.2.3 Verification: liquidation and evaporation

Eq (41) indicates that the size-dependent $\Delta T_m(K_j)/T_m(\infty)$ originates from the relative change of the $E_{B,i}$ of a surface atom to the bulk value. The $\Delta T_m(K_j)/T_m(\infty)$ follows the scaling law given in eq (13) in section 2. Figure 17 compares predictions using parameters given in Table 5 with the measured size-dependent melting behavior of metals, semiconductors, inert gases and methyl-chloride polymer (m-Cl), as well as embedded systems showing superheating effects. The size dependent evaporating temperatures ($T_{eva}$) of Ag and CdS nanosolids are also compared.

It is interesting to note that *Al* nanosolids grown on *SiN* substrate are more plate-like ($\tau = 1$) throughout the measured size but *Sn* on *SiN* and *Au* on *C* are more spherical-like ($\tau = 3$) at the particle size smaller than 10 nm. The melting profiles show that at the smaller size range, the Au/W interface promotes more significantly the melting of Au (supercooling) than the Au/C interface. The silica matrix causes slightly superheating of the embedded Au solid compared with the curves for Au on the rest two substrates. This understanding also provides information about the fashion of epitaxial crystal growth and the bonding status between the nanosolid and the substrate. The deviation from theory and experiment may provide information about the difference in interfacial energy between the particles and the substrates, which is expected to be subject to the temperature of formation.

The current exercise indicates that the superheating of of In/Al ($T_{m,In}/T_{m,Al}$ = 530/932), Pb/Al (600/932), Pb/Zn (600/692), and Ag/Ni (1235/1726)[296] systems originates from the interfacial bond strengthening. It is understandable that an atom performs differently at a free surface from an atom at the interface. Although the coordination ratio at the interfaces suffers little change ($z_{ib} \sim 1$), formation of the interfacial compound or alloy alters the nature of the interatomic bond that should be stronger. In this case, we may replace the $z_{ib}c_i^{-m}$ with a parameter $\alpha$ to describe the interfacial bond enhancement due to interfacial effect, as indicated in panel (g). Although the coordination ratio at the interfaces suffers little change ($z_i/z \sim 1$), formation of the interfacial compound or alloy alters the nature of the interatomic bond that should be stronger. Numerical fit leads to a $\alpha$ value of 1.8, indicating that an interfacial bond is 80% stronger than a bond in the core/bulk interior. If we take the bond contraction, 0.90 ~ 0.92,[125] as determined from the As and Bi impurities in CdTe compound into consideration, it is ready to find that the m value is around 5.5 ~7.0. The high m value indicates that bond nature indeed evolves from a compound with m around 4 to covalent nature. Therefore, the deformed and shorten interfacial bond is much stronger. This finding means that electrons at an interface are deeply trapped and densified. Densification of energy and mass also happens as a result. Therefore, it is understandable that twins of nanograins[297] and the multilayered structures[298] are stronger and thermally more stable. It is anticipated that a thin insulating layer could form in a hetero-junction interface because of the interfacial bond nature alteration and the charge trapping effect. Interestingly, recent theoretical calculations confirmed by electron microscopy,[299] revealed that junction dislocations in aluminum can have compact or dissociated core interlayers. The calculated minimum stress ($\sigma_P$) required to move an edge dislocation is approximately 20 times higher for compact dislocations than for equivalent dissociated dislocations. As anticipated, this finding, provides new insights into the deformation of ultra-fine-grained metals. Density-functional simulations at temperatures near the $T_m(\infty)$ suggested that the solid-liquid phase-transition temperature at the semiconductor surfaces can be altered via a monolayer coating with a different lattice-matched semiconducting material. Results show that a single-monolayer GaAs coating on a Ge(110) surface could raise the Ge melting temperature (1211 K) with an association of a dramatic drop of the diffusion coefficient of the Ge atoms to prevent melting of the bulk Ge layers. In contrast, a single-monolayer coating of Ge on a GaAs(110) surface introduces defects into the bulk and induces melting of the top layer of GaAs atoms 300 K below the GaAs melting point (1540 K). Therefore, superheating is subject to the configuration of the hetero-junction interface and their respective $T_m(\infty)$ as well.

The measured $T_m$ of Si and CdS nanosolids appeared to be lower than the expected values compared with the predicted curve with m = 4.88 for Si. The definition of melting is different from source to source, which might be the reason of the deviation. For instance, molecular-dynamics calculations revealed that[300] coalescence occurs at temperatures lower than the cluster melting point, and that the difference between coalescence and melting temperatures increases with decreasing cluster size. In the normalization of the scaling relation, the coalescence temperature is lower than the $T_m$ and the coalescence T drops faster than $T_m$ with solid size. The size dependent $T_m$ of Kr, Ne, and O solids follow the curve of m = 4.88 as well, despite the accuracy of measurement. The Indium particle encapsulated in the controlled-pore silica exhibits superheating while the Indium embedded in Vycor glass shows no superheating effect. From the *RMSD* instability point of view, the interfacial binding constrains the *RMSD* of the interfacial atom to be smaller than that of a bulk atom.[85]

Equation (41) indicates that the quantity $\alpha = z_i / z_b c_i^{-m}$ dictates the process of superheating ($\alpha > 1$, $T_m$ elevation for chemically capped nanosolids) or supercooling ($\alpha < 1$, $T_m$ suppression of a freestanding nanosolid). For a capped nanosolid, $z_i/z_b \sim 1$, the $\alpha$ represents the interfacial bond strengthening as no apparent bond order loss can be recognized. For a freestanding nanosolid, there are two possibilities for $\alpha > 1$. One is that the m increases as $z_i$ is reduced and the other is that the $c_i$ is much lower than the prediction. No other sources contribute to the $\alpha$ value according to eq (41).

Figure 17 (Link) Agreement between predictions (solid lines) and experimental observations of the size-and-shape dependence of the $T_m$ suppression of (a) *Sn* and *Al* on $Si_3N_4$ substrate[124,295] (b) In and Pd, (c) *Au* on *C,*[69] W[68] and embedded in Silica,[301] (d) Ge and Si, (e) Bi and CdS, (f) Kr, Ne and O, and m-Cl, (g) superheating of embedded In and Pb, (g) $T_{eva}$ of Ag and PbS nanosolids.[214] Parameters and references are given in Table 5.

Table 5 Parameters used in calculations presented in Figure 17. For metals, m = 1. For embedded system, the $z_{ib}c_i^{-m}$ is replaced with a constant α that describes the bond strength enhancement due to the alloying at the interfaces. $T_m$ is the intercept of least-root-mean-square linearization of the experimental data, which calibrate the measurements. Atomic sizes are referred to Appendix A.

| Medium | $T_m(\infty)$ | $T_m$ intercept | Data sources (Ref.) |
|---|---|---|---|
| Al-01 (on SiN) | 933.25 | | 294 |
| Al-02 | | | 302 |
| Sn-01 (on SiN) | 505.06 | | 303 |
| Sn-02 | | | 304 |
| Au/C | 1337.33 | | 69 |
| $Au/SiO_2$ | | | 301 |
| Au/W | | | 68 |
| Ag | 1234 | *947* | |
| In-01 | 429.76 | 438.9 | 274 |
| In-02 | | 433 | 304 |
| In-03 | | 443 | 305 |
| Pb-01 | 600.6 | 632.6 | 306 |
| Pb-02 | 600.6 | 607 | 274 |
| Si-01 | 1685 | 1510 | 307 |
| Si-02 | | | 308 |
| Ge-01 (beginning) | 930 | 910 | 309 |
| Ge-02 (ending) | - | 1023.3 | |
| Ge-03 (Recrystallization) | - | 1260.8 | |
| CdS | 1678 | 1346 | 6 |
| Bi-01 | 544.52 | | 274 |
| Bi-02 | | 618.9 | 304 |
| Bi-03 | | 559.9 | 169 |
| Bi-04 | | 587.6 | 310 |
| Bi-05 | | 557.8 | 311 |
| Kr<br>O<br>Ne<br>Methyl chloride (m-Cl) | 116<br>54.4<br>24.6<br>175.6 | 109.2 | 312<br>313<br>307<br>312 |
| In/Al-01<br>In/Al-02<br>Pb/Al-01<br>Pb/Al-02<br>Pb/Al-03<br>Pb/Al-04<br>Pb/Zn | 429.76<br>429.76<br>933.25<br><br><br><br>692.73 | 433<br>423.8<br>613.2 | 314<br>304<br>269<br>315<br>316<br>317<br>313 |

As the particle size is reduced, the surface-curvature increases and the surface atomic *CN* further decreases and the bond should contract further. Increasing the particle size, the area of interface between the particle and the substrate increases. Atoms at the interface perform quite differently from atoms at a free surface. These artefacts may bring errors in the measurement that deviate from ideal modelling expectations. As we noted, the possible errors affect the accuracy of the *m* and the effective $z_{ib}$ ratio in eq (41) but not the nature of the phenomenon or the general trend of change. Compared with the accuracy of size determination, these artefacts may be negligible. From the perspective of equilibration between the thermal energy of melting and the cohesive energy of an atom at different sites, the proposed BOLS correlation mechanism could incorporate with the existing models including mechanism of random fluctuation melting[82] and could link all the competent factors involved to the effect of atomic *CN*-imperfection.

5.2.4 $T_m$ oscilation

- Observations

XRD in ultrahigh vacuum[282] confirmed that the $T_m$ of Pb nanosolids drops inversely with the crystallite size and favors the liquid-skin melting mechanism. Such melting is demonstrated via the reversible growth of a 0.5 nm thick liquid skin on 50 nm-sized crystallites. Tremendous works show that the surface shells are indeed harder than the bulk interior but the surface melts easier than the core of the solid. It is surprising, however, that sophisticated experimental[318,272] and theoretical[319,320,321] efforts have uncovered recently that a freestanding nanosolid at the lower end of the size limit, or clusters containing 10 ~ 50 atoms of $Ga^+$ or IV-an elements, melt at temperatures that are 10% ~100% or even higher than the bulk $T_m(\infty)$. For example, $Ga^+_{39-40}$ clusters were measured to melt at about 550 K, while a $Ga^+_{17}$ cluster does not melt even up to 700 K compared with the $T_m(\infty)$ of 303 K.[318] Small Sn clusters with 10 ~ 30 atoms melt at least 50 K above the $T_m(\infty)$ of 505 K.[272] Advanced *DFT* molecular dynamics simulations suggest that $Ga^+_{13}$ and $Ga^+_{17}$ clusters melt at 1400 and 650 K,[319] and $Sn_n$ (n = 6, 7, 10 and 13) clusters melt at 1300, 2100, 2000, and 1900 K, respectively.[321] For a $Sn_{10}$ cluster, the structural transition is calculated to happen at 500 and 1500 K and structural transition of a $Sn_{20}$ cluster occurs at 500 and 1200 K.[322] Recent calorimetric measurements[323] on unsupported $Sn^+_n$ particles clarified that $Sn^+_{10}$ and $Sn^+_{11}$ cluster can sustain till 1073 K while Sn clusters containing n >19 and n < 8 atoms are less thermally stable as the melt around 773 K or below. $Sn_{19}$ can sustain till 673 K while $Sn_{20}$ melts below 673 K. Calculations[320] suggested that the IV-a elements, $C_n$, $Si_n$, $Ge_n$, and $Sn_n$ (n ~ 13) clusters melt at temperature higher than their $T_m(\infty)$. The $C_{13}$ cluster prefers a monocyclic ring or a tadpole structure with the most probability to appear in the simulated annealing when the temperature is between 3000 and 3500 K. Although the $T_m$ may be overestimated to some extent for the smallest clusters,[321] the calculated $T_m$ elevation follows the trend of measurement.

The $T_m$ elevation of the smallest Ga and Sn nanosolid is attributed either to the bond nature alteration from covalent-metallic to pure covalent with slight bond contraction,[319,324] or to the heavily geometrical reconstruction as Ge, Si, and Sn clusters are found to be stacks of stable tricapped triagnal prism units.[325] However, consistent insight into the $T_m$ oscillation over the whole range of sizes (from single atom to the bulk crystal) is yet lacking though numerous outstanding models have been developed specifically for the $T_m$ elevation or suppression.

- Simulation

We may fit the measured $T_m$ oscillation over the whole range of sizes for Sn and $Ga^+$ clusters by varying the bond character parameter m with atomic CN. Optimization leads to the relation that transits the m value from seven at z = 2 to one at z > 4:

$$m(z) = 1 + 12/\{1 + \exp[(z-2)/1.5]\}$$

It is seen from Figure 18 that the $T_m$ curves drop generally with size and then bend up at $K_j > 3$ ($Log(K_j) > 0.5$, or $z_i > 3$) for higher m values. If the $T_m$ rise originates from the $c_i$ deviation with bond nature unchange, the bond will contract to $c_i = 0.7^7 = 0.082d$. A 92% bond contraction is strictly forbidden. Therefore, the m value, or bond nature, must change with CN reduction. As the smallest clusters are not in spherical shape, the equivalent size specified herein might be subject to some errors. It is surprising that evolution of m(z) matches closely to the measurement of $Ga^+_{17}$, $Ga^+_{39-40}$, $Sn_{19-31}$ and $Sn_{500}$ clusters and Sn nanosolids deposited on $Si_3N_4$ substrate as well.[290] Calculations[321] show that the $T_m$ transition for $Sn_{6-13}$ happens at $Sn_7$ though the estimated $T_m$ is subject to experimental confirmation.

Results indicate that the nature of the Sn-Sn and Ga-Ga bond indeed evolves from metallic-covalent to pure covalent as atomic CN reduces to much lower values, agreeing with that proposed by Chacko et al[319] This feature also complies with theoretical findings that the Al-Al bond for lower-coordinated or distorted Al atoms at grain boundaries[326] and at free surfaces[327] becomes shorter (~ 5%) and stronger with some covalent characteristics.[328] However, bond nature evolution in $Al^+_{49-60}$ clusters appears not as that significant as occurred in Sn and Ga, as the $T_m$ for $Al^+_{49-63}$ is 300 K lower than the $T_m(\infty)$ of bulk Al. The abrupt $T_m$ rise (~180 K) for $Al^+_{51-54}$, $Sn^+_{10-11}$ and $Sn^+_{19-20}$ clusters[**331**] may partly due to the closed shell structures that are highly stable.[329] Bond nature evolution should also cause conductor-insulator transition such as Pd solid containing $10^{1-2}$ atoms,[12] because of the depressed potential well of traps as all the involved atoms are lower-coordinated. As anticipated by the BOLS correlation, strong localization of charges in the surface-trapping region should be responsible for bond nature evolution/alteration and conductivity reduction of small specimen. Results show that bonding to two neighbors is stronger for an IV-a atom than to bond with three or more atoms due to the bond nature evolution. This mechanism may explain why a $C_{13}$ cluster prefers a ring or a tadpole structure with each atom two bonds, rather than the densely packed tetrahedron structure, as theoretically predicted by Ho and coworkers.[320] It is expected that for the covalent Si (m = 4.88) and C (m = 2.67) clusters should also show the $T_m$ elevation/bending at $K_j < 3$. For a pure covalent system, the bond strength increases as bond contracts without bond nature evolution. For Au, however, the value of m keeps unity throughout the course of CN reduction during monatomic chain formation.[116] Therefore, the bond nature evolution may be the unique property of the III-A and IV-A elements with larger number of electrons as compared with Al (m ~ 2), Ga(m = 6 ~ 7), C(m = 2.56), Si(m = 4.88), and Sn (m = 6 ~ 7).

Figure 18 (link) Comparison of the predicted $T_m$ ossification with those measured from $Ga^+_{13-17}$,[318,319] $Sn_{10-19}$,[323] $Sn_{19-31}$,[272] $Ga^+_{39-40}$,[318] $Sn_{500}$,[330] and Sn nanosolid on $Si_3N_4$ substrate.[290] The $T_m$ deviation of $Al^+_{50-60}$ clusters[331] from the predictions indicates that the bond nature alteration of Al is less significant compared to Sn and Ga bonds. Ideal fit is reached with a function of m(z) = 1+12/{1+exp(z-2)/1.5} to let m transit from 7 at z = 2 to 1 when z > 4.[332]

### 5.2.5 Remarks

Briefly, the BOLS correlation premise has enabled the observed suppression (supercooling), elevation (superheating), and oscillation of $T_m$ over the whole range of sizes of various specimens to be reconciled to the effect of atomic CN imperfection. The modified cohesive energy of the

lower–coordinated system also determines the geometrical reconstruction, surface lattice/phonon instability, and surface energy. Actually, the surface and interfacial energy, surface stress, the local mass density of liquid and solid are all functions of atomic separation and bond energy that are subject to the BOLS correlation. We relate the $T_m$ suppression directly to the atomic CN-imperfection and its effect on the strength of atomic bonding, and the $T_m$ enhancement of embedded system to the strengthening of the interfacial bond. The $T_m$ oscillation over the whole range of size results from the bond nature evolution with atomic CN.

The critical size at which the $T_m$ = 0 K for the current BOLS premise is z = 0, i.e., a single atom. In contrast, the $K_C$ varies from several atomic diameters to several nanometers in other models. The BOLS extends the zero melting to a single atom, which is the limitation of some other approaches. Compared with the existing models, the current BOLS premise is simpler and straightforward involving almost no assumptions or freely adjustable variables. The only parameter needed is the *m* that denotes the bond nature and can be determined from measurement of other properties.

If the nanosolid is involved randomly with distributed defects, the atomic CN surrounding the defects will be lower, and hence the $T_m$ of such nanosolid will follow the random fluctuation mechanism. The lattice vibration (both frequency and amplitude) or surface phonon instability depends on the bond strength (for the first order approximation, the force constant equals to the second order differentiation of the interatomic potential at equilibrium atomic separation) and the effective number of bonds. It is seen that in the present premise, the $\Delta T_m(K_j)$ originates from $z_{ib}c_i^{-m}$-1, compared with the *RMSD* wise in which $-(\beta-1)$ dominates. If $z_{ib}c_i^{-m} < 1$, then $\beta>1$. If the $E_B$ of a surface atom is weaker, its *RMSD* will be larger, and vice versa. For superheating, $z_{ib}c_i^{-m} > 1$, and $\beta < 1$. Therefore, the models of *RMSD* instability, *LSN, LNG,* surface phonon instability*,* and the current *BOLS* are in good accordance. All the modelling variables relating to the melting can be related functionally to the mean atomic $E_B$, which should provide natural link among the various models. The BOLS mechanism quantizes statistically the so-called 'liquid shell' structure as the contribution from individual atomic layers with different atomic *CN*s is different. The current model also supports the fluctuation for highly disordered system and the spontaneous melting at the lower end of the size limit. Therefore, all the proposed models are correct from a certain perspective.

### 5.3 Phase transition

#### 5.3.1 Observations

With reduction of a solid size, the phase stability of the solid becomes lower as well. The $T_C$ of ferromagnetic,[333,334,335] ferroelectric,[250,336,337] and superconductive[338,339,340] nanosolids can be modified by adjusting the shape and size of the nanosolid. The tunable $T_C$ will be an advantage for sensors or switches that can be functioning in a designed temperature range. However, understanding of the underlying mechanism for the $T_C$-tunability is yet primitive though numerous models have been developed.

- Ferromagnetic $T_C$

For ferromagnetic nanosolids, such as Fe, Co and Ni and their alloys or compounds,[341,342,343] the $T_C$ reduces with the particle size or with the thickness of the films.[335,344,333, 345, 346,347,348,349,350,351] According to the scaling theory,[352] a spin-spin correlation length (SSCL, or ξ) limitation model[352,353] defines the SSCL as the distance from a point beyond which there is no further correlation of a physical property associated with that point. Values for a given property at distances beyond the SSCL can be considered purely random. The SSCL, depends functionally on temperature as $\xi = \xi_0 (1-T/T_C)^{-v}$, where *v* is a universal critical exponent. The SSCL limitation premise indicates that the ξ is limited by the film thickness. If the ξ exceeds the film thickness $K_j$, the $T_C$ will be lower compared with the bulk value. The SSCL mechanism gives rise to the power-law form of $T_C(K_j)$ that involves two freely adjustable parameters, λ and C (or $C_0$). The λ value varies from unity to 1.59 for the mean field approximation and the three-dimensional Ising model, respectively:[345,352,354]

$$\frac{\Delta T_C(K_j)}{T_C(\infty)} = (C_0 K_j)^{-\lambda}$$

(43)

In order to match numerically to the measurement of ultra thin films, Eq (43) was modified by replacing the reference $T_C(\infty)$ with the $K_j$-dependent $T_C(K_j)$ for normalization:[333,355]

$$\frac{\Delta T_C(K_j)}{T_C(K_j)} = (C' K_j)^{-\lambda'}$$

However, convergence problem remains at the lower end of the size limit. An alternative non-continuous form was developed based on the mean field approximation to cover the thinner scales:[356]

$$\frac{\Delta T_C(K_j)}{T_C(\infty)} = \begin{cases} -(\frac{\xi+1}{2K_j})^{\lambda}, (K_j > \xi) \\ \frac{K_j - 1}{2\xi} - 1, (K_j < \xi) \end{cases}$$

(44)

Eq (44) shows that $T_C$ varies linearly with $K_j$ and approaches to zero at $K_j = 1$ (single atom). If $\lambda \neq 1$, there is a discontinuity at $K_j = \xi$.

Recently, Nikolaev and Shipilin[357] derived a simple model relating the $T_C$ change of a spherical nanosolid to the reduction of exchange bonds of surface atoms. It is assumed that, the number of exchange bonds per unit volume inside the bulk is equal to *z*. For the magnetically active surface atoms of a nanosolid, this number amounts to *z*/2 or less. The $T_C$ is assumed proportional to the mean number of exchange bonds per unit volume, and then the relative change of the $T_C$ due to size reduction is:

$$\frac{\Delta T_C(K_j)}{T_C(\infty)} = -\frac{\tau \Delta K_j}{2K_j}$$

(45)

where $\Delta K_j$ is the thickness of the layer with half-depleted exchange bonds. The quantity $\Delta K_j$ is an average that characterizes the features of the surface CN-deficient structure of a nanosolid. If $\Delta K_j$ is independent of the particle radius $K_j$, the $T_C$ drops inversely with $K_j$ and the critical $K_C$ at which $T_C$ is zero is $\tau \Delta K_j/2$. This relation characterizes qualitatively the interrelation between the degree of magnetic structure imperfection and the particle size for $Fe_3O_4$ spherical dots[358] by setting the critical thickness $\Delta K_j$ of half (larger size) and two (smallest size) atomic sizes with unclear mechanism for the $K_j$ dependence of $\Delta K_j$.

- Superconductive $T_C$

Highly dispersed superconducting nanosolids can be coupled due to the proximity effect when the interparticle spacing is of the order of twice the penetration length of the superconducting order parameter in the normal phase.[359,360] The electronic energy levels of the sample are discrete, with a mean level spacing of Kobo gap $\delta_K$ for fine metallic particles: [107,361] $\delta_K = 4E_F/3n \propto 1/V \propto K_j^{-3}$. As pointed out by Anderson,[362] superconductivity would not be possible when $\delta_K$ becomes larger than the bulk $E_G$. Based on this suggestion, the relation between the superconducting phase transition and the energy-level spacing for spherical granules is suggested to follow the relation:[363,364]

$$Ln\left(T_C(K_j)/T_C(\infty)\right) = \sum \left[2/(2m_j+1)\right] \times \left[\tanh\left[(\pi/2)\left((2m_j+1)2\pi k_B T_C/\delta_K\right)\right] - 1\right]$$

Index $m_j$ is the magnetic quantum number. Estimation using this relation yields a 2.5 nm critical size for the disappearance of superconductivity of Pb nanosolid. Experiments of Giaver and Zeller[340] on Sn confirmed the existence of a metastable energy gap only for particles of sizes larger than 2.5 nm. However, the $T_C$ for Pb is detectable when the grown Pb atomic layers on Si substrate are four more.[365] The $T_C$ suppression of Pb embedded in the Al-Cu-V matrix measured by Tsai et al [339] could not fit the above prediction. Instead, the data can be fitted to an empirical equation:

$$T_C(K_j) = T_C(\infty)\exp(-K_C / K_j)$$

with $T_C(\infty)$ = 7.2 K for Pb.[338] In the observed size region, due to the finite number of electrons in each particle (between 1000 and 64000 depending upon the grain size) the conventional BCS approach losses its validity because the bulk BCS theory of superconductivity assumes an infinite number of electrons. Small size implies fewer electrons at the Fermi surface and the discreteness of the Kubo levels. Additionally, energy-level spacing may be larger compared to thermal energy $k_BT$. Therefore, the assumption of metallic behavior of these particles may be subject to examination.[130]

Pogrebnyakov et al [338,366] found that the $T_C$ of superconductive $MgB_2$ thin films decreases and the residual resistance increases when the thickness of the epitaxial $MgB_2$ thin films is decreased. At sizes larger than 300 nm, the $T_C$ saturates at 41.8K. The resistivity also saturates to the bulk value of 0.28 Ωcm at 300 nm. The origin of the thickness dependence of the $MgB_2$ film properties is not clear at present. A possible explanation of higher $T_C$ is the strain in the film, while the grain size is not likely to be the direct cause of the thickness dependence of $T_C$. XRD measurement on a 230 nm-thick film shows a slight lattice expansion of $a$ = 0.3095 ± 0.0015 nm, compared with the value of 0.3086 nm for bulk $MgB_2$.[367] The measured $c$ lattice constant contracts from 0.3524 to 0.3515 nm. This contraction suggests that the films are under tensile strain in c-plane epitaxial growth. Hur *et al* [368] reported a higher-than-bulk $T_C$ in $MgB_2$ films on boron crystals and suggested that it is possibly due to tensile strain. Yildirim and Gulseren[369] have predicted an increase in $T_C$ by the *c*-axis compression in the first-principle calculations. Therefore, understanding of the size-induced $T_C$ suppression is still under debate.

- Ferroelectric $T_C$

Unlike ferromagnetic and superconductive nanosolids that show smaller critical sizes for $T_C$ = 0 K, a ferroelectric nanosolid often shows larger critical size at which the ferroelectric feature disappears.[370] Zhao et al [371] observed a progressive reduction of tetragonal distortion, heat of transition, $T_C$, and relative dielectric constant on dense $BaTiO_3$ ceramics with grain size decreasing from 1200 to 50 nm. The correlations between grain size, extent of tetragonal distortion, and ferroelectric properties strongly support the existence of an intrinsic size effect. The critical size for disappearance of ferroelectricity has been estimated to be 10–30 nm. The strong depression of the relative permittivity observed for the nanocrystalline ceramics can be ascribed to the combination of the intrinsic size effect and of the size dependent “dilution” effect of a grain boundary “dead” layer.

Different theoretical approaches have been developed including: (i) a microscopic pseudospin theory based on the Ising model in a transverse field, (ii) a classical and macroscopic Landau theory in which surface effects can be introduced phenomenologically, and, (iii) a polariton model appropriate for the very-long-wavelength region. Taking the surface and non-equilibrium energy into consideration, Zhong *et al* [38] extended the Landau-type phenomenological and classical theory by introducing a surface extrapolation length δ to the size dependent $T_C$ suppression of ferroelectric nanosolids expressed using the Ising model, $J_{ij} = J/r_{ij}^{\sigma}$. $\sigma = 0$ corresponds to an infinite-range interaction and $\sigma = \infty$ corresponds to a nearest-neighbour interaction.[372] However, the model shows limitations in explaining the thermal properties of $PbTiO_3$ and $PbZrO_3$ nanosolids. After that, Bursill *et al* [373] assumed the phenomenological Landau-Ginzburg-Devonshire (LGD) coefficients in the Gibbs energy to change with particle size to solve this problem. Huang et al [102] used the LGD phenomenological theory to study the size effect of ferroelectrics. The model assumes that the surface bond contraction occurs only within three outermost layers of unit cells of a spherical nanoferroelectric particle and the interior core of the particle is ferroelectric.[103] Recent experimental results seem to confirm this assumption. For example, barium titanate particles are found to consist of a shell of cubic material surrounding a core of tetragonal material.[374] Although such a core-shell structure has not been reported in the PZT system, an antiferrodistortive reconstruction of the (001) surface layer of $PbTiO_3$ has been found by *in situ* x-ray scattering measurements.[375] It is easy to understand that the rotation of the relatively rigid oxygen octahedra decreases the lattice parameter or the distance between neighbouring Pb-Pb ions.[376] A surface stress is therefore expected due to the antiferrodistortive reconstruction of the surface. In Huang's model, such a surface stress can be treated as a hydrostatic pressure for a nanoparticle or a two-dimensional stress for a thin film.[377]

An empirical equation that widely used to fit the $T_C$ suppression of a ferroelectric nanosolid is given as,[336]

$$\Delta T_C(K_j)/T_C(\infty) = C/(K_j - K_C) \tag{46}$$

where C and the critical $K_C$ are adjustable parameters. One may note that if $K_j = K_C$, eq (46) becomes singular. The proper form of the dividend seems to be $K_j + K_C$, instead, to shift the scaling relation towards larger critical size. More recently, Jiang *et al* [378] adopted their model for $T_m$ suppression to the size dependent $T_C$ of the ferroelectric nanosolids, which is expressed as:

$$\begin{cases} T_C(K_j)/T_C(\infty) = \exp\left\{-2S_0\left[3R_S(K_j/K_C - 1)\right]^{-1}\right\} \\ K_C = \alpha_{90}\left(2k_B\alpha\beta^2\right)^{-1} \end{cases}$$

where $S_0$ is the transition entropy and $R_S$ is the ideal gas constant. $\alpha$ is the thermal expansion coefficient and $\beta$ the compressibility. The constant $\alpha_{90}$ denotes the density of 90° domain walls. Numerical match for the $T_C$ suppression of $BaTiO_3$ and $PbTiO_3$ nanosolids has been realized with the documented $S_0$ values. Jiang et al found a relation that larger $K_C$ value corresponds to a smaller value of $S_0$ in the simulation.

- Antiferromagnetic transition

Zysler et al [379] investigated the size and temperature dependence of the spin-flop transition in antiferromagnetic $\alpha$-$Fe_2O_3$ nanosolids. When a sufficient large magnetic field is applied along the preferred axis, the so-called spin-flop reorientation occurs, i.e., a 90° rotation of the sublattice vectors. They found that both the spin-flop field, $H_{S-F}$ (T = 0), and the Morin transition temperature ($T_M$) decreases with particle size in a $K_j^{-1}$ way and tends to vanish below a diameter of 8 nm, for spherical particles.[380] Table 6 summarizes the size dependent $H_{S-F}$ and $T_M$. Zysler et al [379] related the change particularly to the distribution of CN for surface spins that determines a variety of reversal paths and in turn affects both the exchange and anisotropy fields. Therefore, the surface spins can undergo spin-flop instability at a field much lower than the bulk value. Weschke et al [381] measured the thickness dependence of the helical antiferromagnetic ordering temperature Néel temperature ($T_N$) for Ho films by resonant magnetic soft x-ray and neutron diffraction and found the $T_N$ to decrease with film thickness. The offset thickness is 11 ML for metallic Ho films in comparison with the value of 16 ML for Cr in sputtered, epitaxial Fe /Cr(001) superlattice.[382]

Table 6 Annealing temperature dependence of crystal size, Morin temperature ($T_M$), and spin-flop transition field at T = 0 ($H_{S-F,0}$) for the heminatite nanosolids.[379]

| Annealing T/K | D/nm | $T_M$/K | $H_{S-F,0}$/Tesla |
|---|---|---|---|
| 923 | 36.4 | 186 | 1.7 |
| 1023 | 40.0 | 200 | 2.5 |
| 1123 | 82.7 | 243 | 5.4 |
| 1273 | 159 | 261 | 6.6 |

Briefly, the SSCL theory considers the correlation length whereas the CN imperfection model considers the loss of exchange bonds of atoms in the ferromagnetic surface region for magnetic $T_C$ suppression. A model for the $T_C$ suppression of superconductive $MgB_2$ nanosolids[383] is yet lacking. Mechanisms for the ferroelectric $T_C$ suppression are under debate. Nevertheless, all the models developed insofar could have contributed significantly to the understanding of $T_C$ suppression from various perspectives. Consistent insight and a unification of size induced $T_C$ suppression of ferromagnetic, ferroelectric, and superconductive nanosolids as well as the $T_M$ and $H_{S-F}$ for antiferromagnetic heminatite is highly desirable. Here we extend the BOLS correlation into the Ising model that involves atomic cohesive/exchange energy, which has led to consistent insight, with a general expression, into the $T_C$ suppression of these nanosolids.

5.3.2 BOLS formulation

- Ising model

The Hamiltonian of an Ising spin system in an external field is expressed as:[356]

$$H_{ex} = \sum_{<i,j>} J_{ij} S_i S_j - g\mu_B B \sum_{i=1}^{N} S_i \propto z_i d_i^{-1}$$

The $H_{ex}$ is identical to the atomic $E_B$ if zero external field is applied, B = 0. $S_i$ and $S_j$ is the spin operator in site $i$ and site $j$, respectively. $J_{ij}$ is the exchange strength between spins at site $i$ and site $j$, which is inversely proportional to atomic distance. The sum is over all the possible coordinates, $z_i$. For phase transition, the thermal energy required is in equilibration with a certain portion of the exchange energy. This mechanism leads to the case being the same to $T_m$ suppression as described in Eq (41).

- High-order CN imperfection

For a ferroelectric system, the exchange energy also follows the Ising model, but the $S_j$ here represents the quanta of a dipole or an ion (may be called as quasi-dipole) that is responsible for the ferroelectric performance. The difference in the correlation length is that the dipole system is longer than that of a ferromagnetic spin-spin system. Usually, dipole-dipole Van der Vaals interaction follows the $r^{-6}$ type whereas the superparamagnetic interaction follows $r^{-3}$ relation. Hence, it is insufficient to count only the exchange bonds within the nearest neighbors for atoms with distant interaction in a ferroelectric system. A critical exchange correlation radius $K_C$ can be defined to count contribution from all atoms within the sphere of $K_C$ radius. Therefore, the sum in eq (41) changes from the $z_i$ neighbors to atoms within the $K_C$ sized volume.

For a ferroelectric spherical dot with $K_j$ radius, we need to consider the interaction between the specific central atom and its surrounding neighbors within the critical volume $V_C = 4\pi K_C^3/3$, in addition to the BOLS correlation in the surface region. The ferroelectric property drops down from the bulk value to a value smaller than 5/16 (estimated from Figure 19) when one goes from the central atom to the edge along the radius due to the volume loss. If the surrounding volume of the central atom is smaller than the critical $V_C$, the ferroelectric feature of this central atom attenuates; otherwise, the bulk value remains. For an atom in the *ith* surface layer, the number of the exchange bonds loss is proportional to the volume $V_{vac}$ that is the volume difference between the two caps of the $V_C$-sized sphere as illustrated in Figure 19a. Therefore, the relative change of the ferroelectric exchange energy of an atom in the ith atomic layer to that of a bulk atom due to volume loss becomes,

$$\frac{\Delta E_{exc,i}}{E_{exc}(\infty)} = \frac{V_C - V_{vac,i}}{V_C} - 1 = -\frac{V_{vac,i}}{V_C} = \delta_{V,i} \tag{47}$$

Figure 19 (Link) (a) Schematic illustration of the high-order exchange bonds lost of an atom in a spherical nanosolid with radius $K_j$. $K_C$ is the critical correlation radius. The $V_{vac}$ loss (the shaded portion) is calculated by differencing the volumes of the two spherical caps:

$$V_{vac,i}=\pi(K_C+K_i-K\cos\theta)^2(K_C-\frac{K_C+K_i-K\cos\theta}{3})$$
$$-\pi(K-K\cos\theta)^2(K-\frac{K-K\cos\theta}{3}),$$

where the angle θ is determined by the triangle $O_1O_2A$.

(b) Critical correlation radius $K_C$ dependence of the $T_C$ shift of ferroelectric and superconductive alloying nanosolids. For $K_C$ = 5 example, BOLS lowers the $T_C$ by -41.1% (follows the curve in Figure 1a) and the high-order bond loss contributes to the $T_C$ suppression by -53% and the over all $T_C$ shift is -94%. $K_C \le 4$, $T_C = 0$.[111]

- Generalization of $T_C$ suppression

Considering the BOLS correlation for the nearest neighbors and the volume loss of long-order CN imperfection, we have a generalized form for the $T_C$ suppression for ferromagnetic, ferroelectric, and superconductive nanosolids (*m* = 1 in the Ising model):

$$\frac{\Delta T_C(K_j)}{T_C(\infty)}=\frac{\Delta E_{exc}(K_j)}{E_{exc}(\infty)}=\begin{cases}\sum_{i\le 3}\gamma_{ij}\left(z_{ib}c_i^{-1}-1\right)=\Delta_B & (short-order-CN)\\ \sum_{i\le K_C}\gamma_{ij}\delta_{V,i}+\Delta_B=\Delta_{COH} & (long-order-CN)\end{cases} \tag{48}$$

For a short-order spin-spin interaction, it is sufficient to sum over the outermost three atomic layers whereas for a long-order dipole-dipole interaction the sum should be within the sphere of the critical volume $V_C$, in addition to the BOLS effect. It is understood now why $\Delta K_j$ in eq (45) is not a constant. In the current BOLS premise, the $\gamma_{ij}$ is not always proportional to the inverse of the radius, which drops instead from unity to infinitely small when the particle grows from atomic scale to macroscopic size. Meanwhile, the $z_i$ and the $c_i$ vary with the curvature of the sphere. Figure 19b shows the general $K_C$ dependence of the ferroelectric $T_C$ shift involving both volume loss and BOLS effect. For $K_C$ = 5 example, bond contraction lowers the $T_C$ by -41.1% and the volume loss contribution lowers the $T_C$ by -53% and the overall $T_C$ shift is -94%.

### 5.3.3 Verification: critical size

Least-root-mean-square linearization of the measured size dependent $T_C$ represented by eq (48) gives the slope B′ and an intercept that corresponds to the bulk $T_C(\infty)$. Compared with eq (48), one would find that $B' = K_j\Delta_{COH}$ for a ferroelectric system. For a ferromagnetic system, $B' = K_j\Delta_B$ = constant without needing numerical optimization. Calculations based on eq (48) were conducted using the average bond length (appendix A) and the $T_C(\infty)$ values as listed in Table 7.

Figure 20 (Link) Comparison of the predicted $T_C$ suppression with observations of (a) Ni thin films: data 1,[348] data 2, 3, and 4,[356] data 5,[333] (b) data 1[384] and data 2;[347] (c) Co films;[333] and, (d) $Fe_3O_4$ nanosolids.[358]

Figure 20 shows the match between the predicted curves and the measured $T_C$ suppression of ferromagnetic nanosolid. For ultra thin films, the measured data are closer to the predicted curve for a spherical dot. This coincidence indicates that at the beginning of film growth, the films prefer an island patterns that transform gradually into a continuous slab. For a ferroelectric system, we need to optimize the $K_C$ value in computation to match theoretical curves to the measured data. Figure 21 shows the $T_C$ suppression of ferroelectric $PbTiO_3$,[250] $SrBi_2Ta_2O_9$,[337] $BaTiO_3$[385] and anti-ferroelectric $PbZrO_3$[386] nanosolids. For ferroelectric and superconductive nanosolids, $T_C = 0$ happens at $V_{vac} = V_C$, which means that $K_C$ corresponds not to $T_C = 0$, but to a value that is too low to be detected. The difference in the optimized $K_C$ by different approaches, as compared in Table 7, lies in that the $\gamma_{ij}$ in the current approach is not a constant but changes with particle size.

Figure 21 (Link) Size-induced $T_C$ suppression of ferroelectric $PbTiO_3$,[250] $SrBi_2Ta_2O_9$,[337] $BaTiO_3$,[385] and anti-ferroelectric $PbZrO_3$,[386] nanosolids.

Comparing the BOLS prediction to the measured $T_C$ suppression of superconductive $MgB_2$ nanosolids in Figure 22 leads to an estimation of the critical radius $K_C = 3.5$, which agrees with that we determined recently ($R_C \sim 1.25$ nm).[387] For the smallest $MgB_2$ crystals, the relative Bragg intensities of the allowed reflections can only be successfully matched during Rietveld refinement by introducing statistically distributed B-vacancies, with the refined value falling from 1 to 2/3. This finding means that the average coordination of Mg to B falls from 12 to 8, which provides direct evidence for the loss of superconductivity of the lower-coordinated system.[388] Consistency of the BOLS prediction with the experiment data indicates that the long-range interaction dominates the superconductive $T_C$. For an Al-Cu-V embedded Pb nanosolid,[339] the $K_C$ is around 1, being the same as ferromagnetic solid. This finding may provide possible mechanisms for the origin of the superconductive $T_C$ suppression of compound $MgB_2$ and metallic Pb nanosolids. For the antiferromagnetic $\alpha$-$Fe_2O_3$ spin-flop transition with critical size of 8 nm can also be related to the high-order CN imperfection, which is the same to the ferroelectric and antiferroelectric compound nanosolids.

Figure 22 (Link) Comparison of the prediction with the measured size dependent $T_C$ suppression of (a) Pb particles embedded in Al-Cu-V matrix and (b) $MgB_2$ films.[338] (c) Size and temperature dependence of the magnetism of $MgB_2$ superconducting nanosolids and (d) size-induced $T_C$ and bond length suppression of $MgB_2$ nanosolids.[387]

Table 7 The bulk $T_C(\infty)$ and the critical correlation radius ($K_C$) at which the central atom lost its ferroelectricity attenuates. $K_C \le 4$ corresponds to $T_C = 0$. The critical radius $R'_C$ at which $T_C = 0$ is derived by other models.

| Materials | $T_C(\infty)$/K | $K_C/R_C$(nm) | $R'_C$/nm (Ref) |
|---|---|---|---|
| Fe | 1043 | 1 | 0[335] |
| Co | 1395 | 1 | 0[333] |
| Ni | 631 | 1 | 0[333] |
| $Fe_3O_4$ | 860 | 1 | 0[358] |
| $PbTiO_3$ | 773 | 4/1.04 | 6.3,[336] 4.5[250] |
| $SrBi_2Ta_2O_9$ | 605 | 4/1.0 | 1.3[337] |
| $PbZrO_3$ | 513 | 8/2.3 | 15[386] |
| $BaTiO_3$ | 403 | 100/24.3 | 24.5,[389] 55[385] |
| $MgB_2$ | 41.7 | 3.5/1.25 | 1.25[387] |
| Pb | 7.2 | 3.5/1.25 | 1.25[339] |

## 5.4 Other applications

### 5.4.1 Diffusivity and reactivity

- Diffusivity

The kinetics of diffusion processes occurring in nanostructured materials is a subject of intensive studies.[390] Although available database is ambiguous due to structural varieties during the process of diffusive experiments, the sharp acceleration of diffusion in these materials is not in any doubt.[391] The activation enthalpies for the interfacial diffusion are comparable to those for surface diffusion, which are much lower than those for diffusion along grain boundaries.[392,393] Measuring grain-boundary diffusion fluxes of Cu on creep behavior of coarse-grained and nanostructured Ni samples at 423 K and 573 K[394] revealed that the creep acceleration behavior is grain size dependent, which was attributed to higher diffusivity in the finer grain material. Experimental studies[395] of the Fe-tracer diffusion in submicrocrystalline Pd powders reveled that interfacial diffusion occurs at relatively low temperatures accompanied by a substantial recovery of grain growth. The recovery processes and the crystal growth occurring in a main recovery stage at 500 K are triggered by atomic defects. The lower-coordinated atoms surrounding the defects become mobile in this temperature regime, which is suggested to be responsible for the onset diffusion in the interfaces.

By means of surface mechanical attrition treatment (SMAT) to a pure iron plate, Wang et al [396] fabricated 5 μm-thick Fe surface layer composed of 10-25 nm sized grains without porosity and contamination on the Fe plate. They measured Cr diffusion kinetics within a temperature range of 573–653 K in the nano-Fe coated plate. Experimental results show that diffusivity of Cr in the nanocrystalline Fe is 7–9 orders higher in magnitude than that in Fe lattice and 4–5 orders higher than that in the grain boundaries (GBs) of α-Fe. The activation energy ($E_A$) for Cr diffusion in the Fe nanophase is comparable to that of the GB diffusion, but the pre-exponential factor is much higher. The enhanced diffusivity of Cr was suggested to originate from a large volume fraction of non-equilibrium GBs and a considerable amount of triple junctions in presence of the nanocrystalline Fe samples.

Under the given conditions, copper atoms were not detectable in the coarse-grained Ni even at a depth of 2 μm. However, the diffusive copper fluxes in nanostructured Ni penetrate into a depth greater than 25 and 35 μm at 423 and 573 K, respectively.[394] This information leads to the GB diffusion coefficients of copper in nanostructured nickel as derived as follows.

At 423 K, no migration of the GBs in nanostructured Ni was observed, and hence, the diffusion coefficient, $D_b$, could be determined using the equation describing the change of the GB impurity concentration versus time t of the diffusion annealing:[397]

$$c(x,t) = c_0 erfc\left[x/\left(2\sqrt{D_b t}\right)\right]$$

where $c_0$ is the concentration of copper at the surface. The depth $x$ is the distance from the surface at which log $c$ = -1 ($c$ = 0.1 %, which corresponds to the resolution limit of the SIMS unit). The value of $c_0$ was obtained by extrapolation of the experimental concentration curve at $x \to 0$. In this case, $D_b = 1\times10^{-14}$ m$^2$/s ($t$ = 3 hrs).

Grain growth occurs in nanostructured nickel when the sample is annealed at 573 K and the grain boundaries migration occurs at the velocity of $V \sim 7\times10^{-11}$ m$^2$/s. In this case, the $D_b$ follows:[397]

$$c(x,V,\beta) = c_0 \exp\left(-x\sqrt{V/D_b\beta_b}\right)$$

Considering the diffusion width of the boundary $\beta_b = 10^{-8}$ m, one can obtain that $D_b = 1.4\times10^{-12}$ m$^2$/s, which is two orders higher than that for the same sample annealed at 423 K. These experimental data demonstrate the increase in the GB diffusion coefficient of copper in nanostructured Ni in comparison with the coarse-grained nickel.

Systematic studies[39] of size-dependent alloy formation of silver-coated gold nanosolids in aqueous solution at the ambient temperature using XAFS reveal remarkable size dependent interdiffusion of the two metals at the ambient temperature. The diffusion between Ag and Au is enhanced when the Au particle size is reduced. For the very small particles (< 4.6 nm initial Au-core size), the two metals are almost randomly distributed within the particle. For larger particles, the diffusion boundary is only one monolayer. These results can be explained neither by enhanced self-diffusion that results from depression of the $T_m(K_j)$ nor by surface melting of the Au nanosolid. It was proposed that defects, such as vacancies, at the bimetallic interface enhance the radial migration (as well as atomic displacement around the interface) of one metal into the other.[39]

Kim et al [398] investigated the thickness dependence of Ag resistivity during temperature ramping as a means to access thermal stability of Ag thin films. *In situ* four-point probe analysis is used to determine the onset temperature at which the electrical resistivity deviates from linearity during the temperature ramping. At the deviation point, the Ag thin films become unstable due to void formation and growth during thermal annealing. In a vacuum, Ag thin films thicker than 85 nm on $SiO_2$ substrate are thermally stable. Using the Arrhenius relation in terms of onset temperature and film thickness, an $E_A$ of 0.326 ± 0.02 eV is obtained for the onset of Ag agglomeration ramped at a rate of 0.1 °C per second. This value is consistent with the $E_A$ for surface diffusion of Ag in a vacuum. Therefore, Ag agglomeration and surface diffusion share the same $E_A$, both of which should be related to the atomic cohesion.

The high diffusivity of nanosolid also enhances liquid to diffuse into the nanosolid, as observed firstly by Li and Cha.[399] Nanosolids surface adsorption and liquid diffusion are conventionally studied using planar electrodes in the field of catalytic chemistry. Cyclic voltammograms (CV) are often used to study surface adsorption of inorganic and organic molecules on metal nanosolids. However, conventional size of electrodes cannot study the fast electron transfer process due to the potential scan rate that is lower than 1 V/s. Solid ultramicroelectrodes can conduct fast CV but are not able to study powder nanosolids surface. Li and Cha[400] found much improved diffusion efficiency ($10\text{-}10^4$) in molding powder nanosolids as electrodes. Detailed studies on the surface chemistry including changes reduction or oxidation of the adsorbate, inhibition of oxygen adsorption, inhibition of hydrogen adsorption, and changes of electron-transfer rate show that the powder ultramicroelectrodes are capable of conducting CV with scan rate of 10 – 1000 V/s for study of high electron transfer process. Further, the powder ultramicroelectrode can significantly enhance the mass transportation rate from solution to the nanosolids surface. A mass transportation rate generated by a one-μm scale ultramicroelectrode is equivalent to that obtained at a conventional rotating disc electrode at a speed of 300,000 rounds per minute. The high efficiency of electronic behavior is general irrespective particular catalytic material.[401] Recently, the powder ultramicroelectrode has been used to biosensor that could enhance the enzymatic catalysis process.[402]

- Reactivity

It is known that nanocrystalline materials possess ultra fine grains with a large number of grain boundaries that may act as channels for fast atomic diffusion.[4] Greatly enhanced atomic diffusivities in nanocrystalline materials relative to their conventional coarse-grained counterparts have been experimentally confirmed.[5,403] A large number of grain boundaries with various kinds of nonequilibrium defects constitutes a high excess stored energy that may further facilitate their chemical reactivity. It has been demonstrated experimentally that chemical reaction (or phase transition) kinetics is greatly enhanced during mechanical attrition of solids, in which the grain size is significantly reduced to nanometer scale and large amount of structural defects are created by the severe plastic deformation,[404] which is associated with the actual temperature rise.

Nitriding of iron happens when Fe powders were ball-milled in a nitrogen-containing atmosphere at the ambient temperature.[405,406] Considerable transient temperature rise (as high as a few hundred degrees) always accompanies the impacts of the milling balls, which may contribute to enhancing the chemical reactivity. However, enhanced chemical reactivity at lower temperatures[38] can happen by converting the surface layer of the metal such as Fe into nanostructures. Observations show that surface nanocrystallization greatly facilitates the nitriding process of Fe, which happens in ammonia at much lower temperature (300 °C for 9 hrs) compared with nitriding of smooth Fe surface that occurs at 500 °C or higher for 20-80 hrs.[407] The much-depressed nitriding temperature is attributed to the enhanced diffusion of nitrogen in the nanocrystalline surface layer compared to the coarse grains. In conventional nitriding of coarse-grained Fe, diffusion in the Fe lattice dominates. In the nanocrystalline Fe specimen, however, nitrogen mostly diffuses along Fe grain boundaries because of the much smaller $E_A$, being proportional to $E_B$, compared with that for the lattice diffusion. Nitrogen diffusivity at nanostructured surface layers ($5.4 \times 10^{10}$ $cm^2/s$) at 300 °C is about two orders of magnitude higher than that in a α-Fe lattice ($3.8 \times 10^8$ $cm^2/s$) at 500°C. Therefore, the ultra fine-grained Fe phase in the surface layer provides a large number of defective grain boundaries (and other defects) that enhance the diffusion. Similarly, other surface chemical treatments by diffusing foreign atoms (such as chromium or aluminum) are useful in industry to improve the performance of engineering materials. Greatly enhanced diffusivity of chromium in the SMAT Fe has also been observed at 350°C that is about 300° to 400°C lower than the conventional exercise.

Pollution from automobiles is emitted in the first 5 min after startup. This is because Pt- or Pd-based catalysts currently used in automobile exhaust cleanup are inactive below about 200ºC. Interestingly, the low-temperature gold catalysts are very inactive unless the gold is in the form of particles smaller than ~8 nm in diameter. [408] Chen and Goodman [409] uncovered recently that a 1.33 ML (3×1 reconstructed surface with every third row of Au adding to the fully covered surface) coverage of Au on $TiO_2$ surface could improve the efficiency of CO oxidation at room temperature by ~50 folds compared with the fully Au covered surface. The catalytic activity was attributed entirely to the presence of neutral gold atoms on the gold nanoparticles.[408] These neutral atoms differ from atoms on bulk gold in three ways that might enhance their catalytic activity: (i) They have fewer nearest-neighbor atoms (that is, a high degree of coordinative unsaturation) and possibly a special bonding geometry to other gold atoms that creates a more reactive orbital. (ii) They exhibit quantum size effects that alter the electronic band structure of gold nanoparticles. (iii) They undergo electronic modification by interactions with the underlying oxide that cause partial electron donation to the gold cluster. Another proposal is that positively charged gold ions on the oxide support are the key to the catalytic activity of these gold catalysts. The finding of Chen and Goodman[409] provides direct, atomistic evidence for the significance of bond-order unsaturation on the catalytic effect of gold atoms.

Engineering alloys rely on the formation of protective oxide surfaces such as $Al_2O_3$ to resist corrosion at high-temperature. Unfortunately, relatively large (6 wt.% or higher) amount of Al or Cr are needed in bulk alloying to form a complete $Al_2O_3$ scale, which often renders the mechanical strength of the alloys.[410] If too-low Al content is added, complex oxides consisting of $Cr_2O_3$, $NiCr_2O_4$ and internal $Al_2O_3$ could form, which results in high reactivity and poor oxidation resistance. Gao et al [411] uncovered that diffusion and selective oxidation can be greatly enhanced by nano-structured coatings. With nano-crystal alloy coatings, the Al content can be substantially reduced to form a complete protective oxide scale. Practice reveals that when the grain size of Ni-20Cr-Al coatings are 60 nm or smaller, alloys containing ~2 wt% Al could form a complete α-$Al_2O_3$ scale at 1000°C in air. Numerical efforts suggest a mechanism in which simultaneous lattice and grain boundary diffusion dominates the selective oxidation process in the nanograined structures.[412]

- BOLS formulation

In order to gain consistent insight into the size enhanced diffusivity and reactivity, we extend the BOLS correlation to the $E_A$ that can be related to the atomic cohesion, leading to a conclusion that atomic CN imperfection suppressed atomic $E_B$ should be responsible for the $E_A$ for atomic diffusion, chemical reaction, and atomic agglomeration and glide dislocation as well.

The temperature dependence of the diffusion coefficient $D$ is expressed in the Arrhenius relation:

$$D(\infty,T) = D_0 \exp\left(-E_A(\infty)/k_B T\right) \tag{49}$$

where the activation enthalpy of diffusion is $E_A(\infty)$ = 1.76 eV and the pre-exponential factor is $D_0$ = 0.04 $cm^2s^{-1}$ for gold. It is possible to incorporate the BOLS premise to the interdiffusion and nano-alloying by letting $E_A \propto E_B$ and hence the $E_A$ is atomic CN dependent. It is understandable to diffuse an atom into the solid needing energy to loose partial of the bonds associated with atom dislocations. Applying Eq (40) to (49) by considering the size effect, one has,

$$\frac{D(K_j,T)}{D(\infty,T)} = \exp\left(-\frac{E_A(K_j)-E_A(\infty)}{k_BT}\right) = \exp\left(-\frac{E_A(\infty)}{k_BT}\left[\frac{E_B(K_j)}{E_B(\infty)}-1\right]\right)$$

$$= \exp\left(\frac{-E_A(\infty)}{k_BT}\Delta_B\right)$$

$$D(K_j,T) = D_0\exp\left(\frac{-E_A(\infty)}{k_BT}\frac{T_m(K_j)}{T_m(\infty)}\right) = D_0\exp\left(\frac{-E_A(\infty)}{k_BT}[1+\Delta_B]\right) \quad (50)$$

Therefore, the nano-diffusivity increases with the CN-imperfection reduced atomic $E_B$ and hence the $T_m(K_j)$ /$T_m(\infty)$ ratio drop in an exponential way. This understanding should provide a feasible mechanism for the nano-alloying, nano-diffusion, and nano-reaction in the grain boundaries where lower-coordinated atoms dominate. However, oxidation resistance of Si nanorod exhibits oscillation features.[36] At the lower end of size limit, the Si nanorod can hardly be oxidized, as oxide tetrahedron formation is strongly subject to the atomic geometrical environment. For instance, oxidation happens preferentially at the densely packed diamond {111} plane rather than the (110) surface.[413] The high surface curvature of the Si nanorod may provide an environment that resists the oxygen to form tetrahedron with Si atoms at the highly curved surface of a Si nanorod.

Figure 23 compares the measured size dependent $T_m$ suppression and diffusion-coefficient enhancement of silica-encapsulated gold particles,[301] and the CO oxidation catalytic reactivity of Au/$TiO_2$ monatomic chains and Au/oxide[414] nanosolid in comparison with BOLS prediction of diffusivity. The similarity in the trends of diffusivity and catalytic activity evidences the correlation between these two identities in terms of activation energy though the former is related to atomic dislocation while the latter to charge transportation. The actual link between the activation energy for atomic dislocation and charge transportation is a challenging topic for further study.

Figure 23 (Link) (a) Size dependence of the $T_m$ and the diffusion coefficient of silica-encapsulated gold particles. The solid curve (right-hand side axis) is the calculated Au self-diffusion coefficient. [301] (b) Atomic CN imperfection enhanced catalytic reactivity (~ 50 fold) of Au/$TiO_2$ monatomic chains for CO room temperature oxidation.[409] (c) Au/oxide particle size dependence of CO room temperature oxidation activity.[414] (d) BOLS prediction of the size-dependent coefficient of diffusion.

### 5.4.2 Crystal growth

- Liquid-solid epitaxy

Significant experimental and theoretical effort aiming at grasping factors controlling nucleation, growth and subsequent microstructural evolution of nanocrystalline materials such as silicon has been motivated by obtaining high-quality materials for electronic and optical applications. However, little is yet known about the initial stages of growth of nanometre-sized crystals from the molten or amorphous matrix. This important issue largely determines the resulting microstructure of a polycrystalline, which is extremely difficult to study experimentally due to the small size of the clusters and the small time scale involved. Intensive experiment and molecular dynamics (MD) simulations have been conducted towards understanding the kinetics and thermodynamics of the homoepitaxial melting and solidification of a material. Results on the homoepitaxial growth and melting of Si, for example, are well understood in terms of the transition-state theory of crystal growth.

According to transition-state theory, the driving force, $F_C$, for the movement of the liquid-crystal interface is the free energy difference between the liquid and bulk crystal. This difference is approximately proportional to the magnitude of the undercooling, $T_m - T$. The velocity of the moving interface, V, is proportional to the driving force $V = kF_C$, where k is the mobility of the liquid-crystal interface. This interfacial mobility determined by the movement of the atoms in the liquid phase as atoms residing in the crystalline phase are far less mobile. Therefore, it is usually assumed that this mobility is proportional to the thermally activated atomic diffusion in the liquid phase. As is well established, $T_m$ suppression happens to a cluster of finite-size due to atomic CN imperfection, which contributes to the free energy of the liquid-crystal interface. However, the kinetics of the highly curved liquid crystal interface is yet unclear.

Using the Stillinger-Weber (SW) empirical potential, Keblinski[308] studied temperature and size dependence of the growth and dissolution of Si nanosolids and found that there are actually no significant differences between the growth of nanosolids and planar interfaces. However, the $T_m$ of a cluster drops with solid size due to the reduced atomic $E_B$ and the mobility activation energy $E_A$ ($\propto$ atomic $E_B$) of the liquid-crystal interface is essentially the same as that for liquid diffusion. In the study of growth and melting of Si, the crystal front velocity was monitored using the fact that the SW potential consists of additive two-body and three-body energy terms. The three-body term is zero for the perfect-crystal structure at T = 0 K, but even at high temperature the three-body tern is assumed relatively low in the crystalline phase (e.g. the three-body energy is about 0.1 eV/atom at T = 1200 K). By contrast, the liquid phase is characterized by much larger three-body energy (-1 eV/atom). Using this large difference Keblinski calculated the amount of crystal and liquid phase present in the simulated cell simply by monitoring the total three-body energy and using as reference the corresponding values for the bulk liquid and bulk solid at the same temperature.

The size dependent $T_m(K_j)$ can be investigated by simulating the growth/melt behavior of clusters with various initial sizes as a function of temperature. The free energy of the cluster can be approximated by surface and bulk contribution. The surface contribution, $U_S$, is proportional to the surface area times the liquid-solid interracial free energy, $\gamma_{ls}$, such that $U_S = A\gamma_{ls}K_j^2$, where A is a geometrical constant of order one (for a spherical dot, $A = 4\pi d_0^2$). The bulk contribution, $U_B$, is proportional to the volume of the cluster and the difference between solid and liquid free energy densities, $\Delta u$, such that $U_B = B\Delta u K_j^3$, where B is another geometrical constant (for a spherical cluster $B = 4\pi d_0^3/3$). The difference between crystal and liquid free energy densities in the vicinity of the $T_m$ is proportional to the magnitude of undercooling (or overheating), $\Delta u = u_0(T - T_m(K_j))$, where $u_0$ is a constant (note that $\Delta u$ correctly vanishes at the $T_m$). For a given temperature, the critical cluster size corresponds to the maximum of the free energy $U = U_S + U_B$. By differentiation of the free energy with respect to the cluster size $K_j$, one finds the maximum at $T = T_m(K_j) - c\gamma_{sl}/K_j$, where c is a constant depending on A, B and $u_0$. The linear dependence of the $T_m$ on the inverse of the crystalline size implies that the interracial energy, $\gamma_{sl}$, does not change significantly with temperature, from the first order approximation. In reality, the interfacial energy varies with both size and temperature, as discussed in earlier sections.

In order to understand the temperature dependence of growth rate in terms of undercooling and thermally activated interfacial mobility, one may assume that,[308] in the classical nucleation theory, growth takes place on an atom-by-atom basis. Hence, the average rates of crystallization and dissolution are:

$$\nu_{\pm} = \nu_0 \exp\{\pm [(\Delta u - (A_{n+1} - A_n)\gamma_{sl})/(2k_B T)] - E_A/k_B T\}$$

where $A_{n+1} - A_n$ is an increase in the interfacial area due to the attachment of an atom to the crystal. The $\nu$ is the thermal vibration frequency of the interfacial atom. The cluster growth velocity resulting from the difference between $\nu_+$ and $\nu_-$ can be then written as:

$$\begin{aligned} V_{grow} &\sim \exp(-E_A/k_B T)\sinh\{[\Delta u - (A_{n+1} - A_n)\gamma_{sl}]/(2k_B T)\} \\ &\cong \{[\Delta u - (A_{n+1} - A_n)\gamma_{sl}]/(2k_B T)\}\exp(-E_A/k_B T) \end{aligned} \tag{51}$$

The argument of the hyperbolic sine is small near the $T_m$ (it is exactly zero at the $T_m(K_j)$). Eq (51) indicates that the rate of the growth/melting is driven by the lowering of the free energy, $\Delta u - (A_{n+1}-A_n)\gamma_{sl}$, while the interfacial mobility is determined by the $E_A$ for diffusion jumps of the interfacial atoms. Noting that $A_{n+1} - A_n$ is proportional to $K_j^{-1}$ and $\Delta u = u_0(T - T_m(K_j))$, and then the scaling law for melting complies: $\Delta T_m(K_j) \sim \gamma_{sl}/K_j$. ($T_m(K_j)$ is the temperature at which $V_{grow} = 0$). For planar growth the interfacial contribution to the free energy disappears; thus $V_{grow}$ is zero exactly at the $T_m(\infty)$ ($\Delta u = 0$).

For a given cluster size, the free energy term can be expanded around its $T_m(K_j)$ such that

$$V_{grow}(K_j) \sim [(T_m(K_j) - T)/T]\exp(-E_A/k_BT) \qquad (52)$$

This process describes the kinetics of liquid-nanosolid dissolution and growth. The $E_A$ obtained from the best fits are 0.75 ± 0.05 eV for 2.0 and 2.6 nm solids and 0.85 ± 0.05 eV for 3.5 nm solids, respectively. This result complies with the BOLS expectation that the mean atomic $E_B$ increases with solid size. Incorporating the BOLS correlation to the $T_m(K_j)$ and $E_A(K_j)$, eq (52) becomes,

$$\frac{\Delta E_A(K_j)}{E_A(\infty)} = \frac{\Delta T_m(K_j)}{T_m(\infty)} = \Delta_B$$

$$V_{grow}(D) \sim [(T_m(\infty)(1+\Delta_B) - T)/T]\exp\{-[E_A(\infty)(1+\Delta_B)]/k_BT\} \qquad (53)$$

The exponential part is the same to the reduced diffusivity (see eq (50)). Results in Figure 24a show the mobility of the liquid-solid interface that is determined by diffusion in the adjacent bulk liquid, which is exactly the case of homoepitaxial growth.

Figure 24 (Link) (a) Molecular dynamics simulation of size and temperature dependence of Si nanosolid melting (negative) and growth (positive).[308] (b) BOLS prediction of $T_S$ dependence of critical size of Sn ($T_m(\infty)$ = 505.1 K), Bi (544.5 K), Pd (600.6 K) nanosolids.

- Vapor phase deposition

The understanding of size dependent melting may provide guidelines for growing nanosolids on heated substrate in vapor deposition. For a given substrate temperature ($T_S$), there will be a critical size of the grown particle, thus, any particle larger than this size will be deposited as such. On the other hand, if the incident cluster size is smaller than the critical size, the particle will melt upon deposition and they will coagulate to produce clusters equal to the critical size or larger. If the $T_S$ is higher than the $T_m$, the arriving clusters may merge and then evaporate associated with size reduction of the coagulated solid.[214] This intuition implies that the deposition temperature should be as low as possible if one wants to obtain smaller particles. This mechanism also applies to the lower sinterability of nanosolids. As found by Hu et al ,[45,46] the solid size of oxide increases with annealing temperature and alogameration happens at a certain size range at room temperature in the process of ball-milling.

The $T_S$ dependence of the critical size $K_C$ can be estimated from the relation:

$$T_s(K_C) = T_m(\infty)(1+\Delta_B) \Rightarrow \Delta_B \cong \Delta'_B/K_C,$$

which gives the thermally stable crystal of critical size:

$$K_C = \frac{-\Delta'_B}{1 - T_S/T_m(\infty)} = \frac{\sum_3 \gamma_{ij}\left(1 - z_{ib}c_i^{-m}\right)}{1 - T_S/T_m(\infty)} \qquad (54)$$

where the constant $\Delta'_B$ = -2.96 for a spherical metallic dot (m = 1 and τ = 3). It is clear that the solid size and hence the number of atoms in the deposited nanosolid depends on the $T_S/T_m(\infty)$ ratio, which is in good agreement with the experimental observations.[415] This relation predicts that a monatomic layer of metals ($\tau = 1$) could only growth at $T_S = 0$ or nearby. The $R_C$ *($K_C d_0$)* at $T_S$ can be estimated with the known atomic size $d_0$ and $T_m(\infty)$ as illustrated in Figure 24b.

The growth of nanostructured multilayer thin films depends largely on how the adatoms aggregate in forming islands of various shapes in the submonolayers. However, little has been known about the detailed processes of nucleation and growth in the presence of surfactant at atomic scale until recently when Wang and co-workers[416,417] initiated and verified a reaction-limited aggregation mechanism, which forms one of the key competitive factors dominating the process of crystal growth. Using first-principle total-energy calculations, they show that an adatom can easily climb up at monatomic-layer-high steps on several representative fcc metal (110) surfaces via a place exchange mechanism. Inclusion of such novel processes of adatom ascending in kinetic Monte Carlo simulations of Al(110) homoepitaxy as a prototypical model system can lead to the existence of an intriguing faceting instability. A fractal-to-compact island shape transition can be induced by either *decreasing* the growth temperature or *increasing* the deposition flux, agreeing with experimental observations. Recent advance[418,419] in investigating the formation and decay of surface-based nanostructures and in identifying the key rate processes in kinetics-driven atomic processes have further confirmed the novel concept of adatom ascending at step edges and faceting on fcc metals.

The current BOLS correlation premise might provide a possible complementary mechanism from the perspective of bond making and breaking for the adatom climbing and position exchanging. The thermodynamic process of crystal growth is subject to the competition between atomic cohesion and thermal activation. Bombardment by the energetic deposition flux also supplies energy on the grown atom. The adatom tends to find a location with optimal total energy that is the sum of binding and heating and hence to ascend or descend crossing the step edge in the process of multilayer superlattice growth. At a given temperature, the magnitude of the total energy of an atom with z coordinate is described by (section 2):

$$E_{total}(T) = \sum_z \eta_{1l}(T_m - T) + \eta_{2l}$$

Relocation of the adatom from one site to another with a net gain in the $E_{total}(T)$ could be the force driving the process as such. The gain of $E_{total}(T)$ is subject to the difference of atomic CN, the specific heat per bond, $\eta_{1l}$, the melting point of the bond, and the 1/z fold entropy for atomization from molten phase, $\eta_{2l}$. Therefore, exchanging position happens because of nonequility of binding energy for the exchange specimens. Adatom ascending or descending across the step edge where the atomic CN is lower can be clear indicative that thermally activated bond broken is nonsimultaneous and that atoms with fewer bonds at the tip of edge are generally less stable. However, for Ga and Sn atoms, the $T_m$ becomes higher when the atomic CN is reduced,[318-321] as discussed in section 5.2.4. Therefore, it could be possible to observe that Ga and Sn atoms grow preferably at the tip of edge at a certain conditions, as the binding of such atoms with 2 ~ 3 bonds is stronger than even that in the bulk.

## 5.5 Summary

The BOLS correlation has enabled the thermodynamic behavior of a nanosolid to be consistently formulated and understood in terms of atomic CN imperfection and its effect on atomic cohesion. It is understood that the difference between the cohesive energy of an atom at the surface and that of an atom inside the solid determines the fall or rise of the $T_m$ of a surface and a nanosolid. The approach is in good accordance with existing models based on classical and molecular thermodynamics. Combination of these models should provide deeper insight into the physical origin and the general trends of the melting behavior of a nanosolid.

The Curie temperature suppression for ferromagnetic, ferroelectric, and superconducting nanosolids follows the same trend of $T_m$ suppression that is dictated by the BOLS correlation and the effect of high-order CN imperfection, as well as the criterion of thermal-vibration – exchange-interaction energy equilibration. At $T_C$, the atomic thermal vibration energy overcomes a portion of the atomic $E_B$, which triggers the order-disorder transition of the spin-spin exchange interaction.

Numerical match between predictions and measurements for a number of specimens reveals that the short spin-spin correlation dominates the exchange interaction in the ferromagnetic Fe, Co, Ni, and $Fe_3O_2$ nanosolids, whereas the long-range interaction dominates the exchange energy for the ferroelectric $PbTiO_3$, $PbZrO_3$, $SrBi_2Ta_2O_9$, and $BaTiO_3$, and the superconductive $MgB_2$ nanosolids. Consistency between predictions and experimental observations on the $T_C$ suppression of the considered nanosolids evidences the validity of current premise that has also been extended to the cases of nano-diffusivity, nano-reactivity and the $E_A$ for atomic dislocation and crystal growth.

## 6 Acoustic and optic phonons

### 6.1 Background

Vibration of atoms at a surface is of high interest because the behavior of phonons influence directly on the electrical and optical properties in semiconductor materials, such as electron-phonon coupling, photoabsorption and photoemission, as well as transport dynamics in devices.[420] With miniaturization of a solid down to nanometer scale, the transverse and the longitudinal optical (TO/LO) Raman modes shift towards lower frequency(or called as softening),[421] accompanied with generation of low-frequency Raman (LFR) acoustic modes at wave numbers of a few or a few tens $cm^{-1}$. The LFR peak shifts up (or called as hardening) towards higher frequency when the solid size is reduced.[422,423] Most of the theoretical studies of phonon modes are based on continuum dielectric mechanism.[424,425] A microscopic lattice dynamical calculation has already been developed.[423,426] However, the underlying mechanism behind the red and blue Raman shift is under debate with the possible mechanisms as summarized as follows.

The size dependent Raman shifts can be generalized empirically as,[423,421]

$$\omega(K_j) = \omega(\infty) + A_f \left(d_0/K_j\right)^{\kappa} ,$$

where $A_f$ and $\kappa$ are adjustable parameters used to fit the measured data. For the optical red shift, $A_f$ < 0. For Si example, $\omega(\infty)$ = 520 $cm^{-1}$ corresponding to wavelength of $2\times10^4$ nm. The index $\kappa$ varies from 1.08 to = 1.44. The $d_0$ is the lattice size that should contract with the solid dimension.[89] For the LFR blue shift, $A_f > 0$, $\kappa = 1$ and $\omega(\infty) = 0$. Therefore, the LFR results from nanosolid formation and the LFR should disappear for large particles.

#### 6.1.1 Acoustic phonon hardening

- Quadrupolar vibration

The LFR peaks were ever attributed to acoustic modes associated with the vibration of the individual nanoparticle as a whole. The phonon energies are size dependent and vary with materials of the host matrix. The LFR scattering from silver nanoclusters embedded in porous alumina[427] and $SiO_2$[428] was suggested to arise from the quadrupolar vibration modes that are enhanced by the excitation of the surface plasmas of the encapsulated Ag particles. The selection of modes by LFR scattering is due to the stronger plasmon-phonon coupling for these modes. For an Ag particle smaller than four nm, the size dependence of the peak frequency can be well explained by Lamb's theory,[429] which gives vibrational frequencies of a homogeneous elastic body with a spherical form. The mechanism for LFR enhancement is analogous to the case of surface-plasma enhanced Raman scattering from molecules adsorbed on rough metal surfaces. The surface acoustic phonons can be described as the eigenfrequencies of a homogeneous elastic sphere under stress-free boundary conditions, which give rise to a low frequency $\omega$ that is in the range of a few to a few tens of $cm^{-1}$ in the vibrational spectra. These modes are suggested to correspond to spheroidal and torsional modes of vibrations of a spherical or an ellipsoidal particle. Spheroidal motions are associated with dilation and strongly depend on the cluster material through $v_t$ and $v_l$, where $v_t$ and $v_l$ are the transverse and longitudinal sound velocities, respectively. The sound velocity in a medium depends functionally on the Young's modulus and the mass density, i.e., $v \sim (Y/\rho)^{0.5} \sim \sqrt{E_b}$ where $E_b$ is the cohesive energy per coordinate.[285] No volume change is assumed to the torsional motion of the particle. These modes are characterized by two indices $l$ and $n$, where $l$ is the angular-momentum quantum number and $n$ is the branch number. $n = 0$ represents the surface modes. It has been shown that spheroidal modes with $l = 0$ and 2 are Raman active and the torsional modes are Raman inactive.[430] The surface quadruple mode ($l = 2$) appears in both polarized and depolarized geometry, whereas the surface symmetrical mode ($l$ = 0) appears only in the polarized geometry. The relation between the particle size and the frequency of the polarized acoustic phonon can be established from:[431]

$$\sin(\xi) = 4n_{eff}^2 j_1(\xi)$$

with complex argument,

$$\xi = R(\omega + i\Gamma)/v_l \tag{55}$$

where ν and Γ are the phonon frequency and bandwidth, respectively, for the polarized-confined acoustic phonon of the first order. The term $j_l(\xi)$ is the spherical Bessel function of the first kind with order one, $v_l$ is the longitudinal sound velocity in the nanoparticle, and $n_{\rm eff}$ is an effective internal acoustic index given by,

$$n_{eff}^2 = n_p^2 - f_c\left\{n_m^2 - (k\xi)^2/\{4[1 - i(k\xi)]\}\right\} \tag{56}$$

where $n_p$ and $n_m$ are the ratios of transverse-to-longitudinal sound velocities in the particle and in the matrix, respectively, $k$ is the ratio between the longitudinal sound velocities in the particle and in the matrix, and $f_c$ is a coupling constant between the particle and the matrix, given by,

$$f_c = \rho_m/(k^2\rho_p)$$

with $\rho_m$ and $\rho_p$ being the mass densities for the matrix and for the particle, respectively. By substituting $n_{\rm eff}$ in Eq (56) into Eq (55), the relation between particle radius $R$ and the phonon frequency $v$ can be obtained from the real part of Eq (55). The eigenfrequencies for the torsional modes and the spheroidal modes with n = 0 can be written as:[432]

$$\begin{cases} \omega_t^1 = 0.815 v_t/Rc, \quad \omega_t^2 = 0.4 v_t/Rc & (Tortional) \\ \omega_s^0 = 0.18 v_l/Rc, \quad \omega_s^1 = 0.585 v_l/Rc, \quad \omega_s^2 = 0.42 v_l/Rc & (spheroidal) \end{cases}$$

where c is the velocity of light in vacuum. For bulk Ag, $v_t$ = 1660 m/s and $v_l$ = 3650 m/s, the ω is around $10^2$ $cm^{-1}$ level. This approach fits well the measured LFR data for Ag embedded in $Al_2O_3$ and $SiO_2$ matrix.[431,432]

- Lattice strain

The LFR blue shift was also attributed to the effect of lattice contraction induced stain. $CdS_xSe_{1-x}$ nanocrystals embedded in a borosilicate ($B_2O_3$-$SiO_2$) glass matrix[433] have been found to experience size-dependent compressive strain. The lattice strain causes the surface tension to increase when the crystal size is reduced. The observed blue shift of acoustic phonon energies was suggested to be consequence of the compressive stress overcoming the red shift caused by phonon confinement with negative dispersion. Liang et al [434] also presented a model for the Raman blue shift by relating the frequency shift to the bond strength and bond length that are functions of entropy, latent heat of fusion, and the melting point.

### 6.1.2 Optical phonon softening

The high-frequency Raman shift has ever been suggested to be activated by surface disorder,[435] and explained in terms of surface stress[436,437] or phonon quantum confinement,[438,439] as well as surface chemical passivation.[1] However, the effect of stress is usually ignorable for hydrogenated silicon,[440,441] in which hydrogen atoms terminate the surface dangling bonds, which reduce the bond strains and hence the residual stress. Phonon confinement model[438] attributes the red shift of the asymmetric Raman line to relaxation of the q-vector selection rule for the excitation of the Raman active phonons due to their localization. The relaxation of the momentum conservation rule arises from the finite crystalline size and the diameter distribution of nanosolid in the films. When the size is decreased, the rule of momentum conservation will be relaxed and the Raman active modes will not be limited at the center of the Brillouin zone.[436] A Gaussian-type phonon confinement model[439] that has been used to fit the experimental data indicates that strong phonon damping presents, whereas calculations[442] using the correlation functions of the local dielectric constant ignores the role of phonon damping in the nanosolid. The large surface-to-volume ratio of a nanodot strongly affects the optical properties mainly due to introducing surface polarization and surface states.[443] Using a phenomenological Gaussian envelope function of phonon amplitudes, Tanaka et al [444] show that the size dependence was originated from the relaxation of the q = 0 selection rule based on the phonon confinement argument with negative phonon dispersion. The phonon energies for all the glasses are reduced, the values of the phonon energies of CdSe nanodots are found to be quite different for different host glasses. The currently available models for the optical red shift are based on assumptions that the materials are homogeneous and isotropic, which is valid only in the long-wavelength limit. When the size of the nanosolid is in the range of a few nanometers, the continuum dielectric models exhibit limitations.

Hwang et al [260] indicated that the effect of lattice contraction must be considered to explain the observed differences in the red shift of phonon energies for CdSe nanodots embedded in different glass matrices. To obtain the phonon frequency as a function of the dot radius $K_j$ with contribution of lattice contraction, it was assumed that,

$$\omega(K_j) = \omega_L + \Delta\omega_D(K_j) + \Delta\omega_C(K_j) \tag{57}$$

where $\omega_L$ is the LO phonon frequency of the bulk. $\Delta\omega_D(K_j)$ is the peak shift due to phonon dispersion and $\Delta\omega_C(K_j)$ the peak shift due to lattice contraction. The $\Delta\omega_D(K_j)$ is given by,

$$\Delta\omega_D(K_j) = \left[\omega_L^2 - \beta_L^2\left(\frac{\mu_{np}}{K_j d_0}\right)^2\right]^{1/2} - \omega_L \cong -\left(\frac{\beta_L^2}{2\omega_L}\right)\left(\frac{\mu_{np}}{K_j d_0}\right)^2 \tag{58}$$

where the parameter $\beta_L$ describes the phonon dispersion assumed to be parabolic and $\mu_{np}$ is the nonzero $n_p$th root of the equation of $\tan(\mu_{np}) = \mu_{np}$. The phonon frequency shift due to the lattice contraction $\Delta\omega_C(K_j)$ is given as:[433]

$$\Delta\omega_C(K_j) = \omega_L\left[\left(1 + \frac{3\Delta d}{d}\right)^{-\gamma} - 1\right] \cong -3\gamma\omega_L\frac{\Delta d(K_j)}{d}$$

where,

$$\begin{aligned}\frac{\Delta d(K_j)}{d} &= (\alpha' - \alpha)(T - T_g) - \frac{2\beta_c}{3}\left(\frac{\sigma_\infty}{K_j d_0} + \frac{b}{2(K_j d_0)^2}\right)\\ &\cong (\alpha' - \alpha)(T - T_g) - \frac{\beta_c b}{3(K_j d_0)^2}\end{aligned} \tag{59}$$

$\gamma$ is the Grüneisen parameter, $\alpha'$ and $\alpha$ are the linear thermal-expansion coefficients of the host glass and the nanodot, respectively. $T$ and $T_g$ are the testing and the heat-treatment temperature, respectively. $\beta_c$ and $\sigma_\infty$ are the compressibility and the surface tension of the bulk, respectively, and $b$ is the parameter describing the size-dependent surface tension of the crystal. The surface tension for bulk crystals is assumed small. The first term describes the lattice contraction by thermal-expansion mismatch between the glass matrix and the crystal, and the second term arises from the increasing of surface tension with decreasing crystal size. Substituting eq (59) into (58), the phonon frequency change, is obtained,

$$\begin{aligned}\frac{\Delta\omega(K_j)}{\omega_L} &= -3\gamma(\alpha' - \alpha)(T - T_g) - \left[\frac{1}{2}\left(\frac{\beta_L \mu_{n_p}}{\omega_L}\right)^2 - \gamma\beta_C b\right](K_j d_0)^{-2}\\ &= A - BK_j^{-2}\end{aligned} \tag{60}$$

For a free surface, $\alpha' = \alpha$, and b = 0. There are some difficulties however to use this equation, as noted by Hwang et al [445] Since the thermal-expansion coefficient within the temperature range $T - T_g$ is hardly detectable, it is difficult to compare measurement with calculated values. The value of $B$ in eq (60) is given by the difference of the phonon negative dispersion and the size-dependent surface tension. Thus, a positive value of $B$ indicates that the phonon negative dispersion exceeds the size-dependent surface tension and consequently causes the red shift of phonon frequency. On the contrary, if the size-dependent surface tension is stronger than the phonon negative dispersion, blue shift of phonon frequency occurs. In case of balance of the two effects, i.e., $B = 0$, the size dependence disappears. Furthermore, the parameter $b$ introduced by the size-dependent surface tension is also unknown. At the lower end of the size limit, the $\omega(K_j)\to-\infty$ diverges in a $K_j^{-2}$ way. Therefore, the existing models could hardly reproduce with satisfactory the Raman frequency shifts near the lower end of the size limit.

6.2 BOLS formulation

Raman scattering is known to arise from the radiating dipole moment induced in a system by the electric field of incident electromagnetic radiation. The laws of momentum and energy conservation govern the interaction between a phonon and a photon. When we consider a solid containing numerous Bravais unit cells and each cell contains n atoms, there will be 3n modes of vibrations. Among the 3n modes, there will be three acoustic modes, LA, $TA_1$ and $TA_2$ and 3(n-1) optical modes. The acoustic mode represents the in-phase motion of the mass center of the unit cell or the entire solid. Therefore, the LFR should arise from the vibration of the entire nanosolid interacting with the host matrix. For a freestanding nanosolid, the LFR should correspond to intercluster interaction. The optical mode is the relative motion of the individual atoms in a complex unit cell that contains more than one atom. For elemental solids with a simple such as the fcc structure of Ag, there presents only acoustic modes. The structure for silicon or diamond is an interlock of two fcc structures that contains each cell two atoms in nonequivalent positions, there will be three acoustic modes and three optical modes.

The total energy $E_{total}$ due to the lattice thermal vibration and interatomic binding can be expressed in a Taylor's series, as given in eq (37) (section 5). When the atom is in equilibrium position, the bond energy is $E_b$. The second-order term corresponds to the Harmonic vibration energy, in which, the force constant $k = d^2u(r)/dr^2\big|_d \propto E_b/d_0^2$ and $k' = d^3u(r)/dr^3\big|_{d0} \propto E_b/d_0^3$ is in the dimension of stress. The vibration amplitude is $x = r - d_0$. The high-order terms correspond to the nonlinear contribution that can be negligible in the first order approximation. For a single bond, the $k$ and $k'$ are strengthened; for a lower-coordinated atom, the resultant $k$ could be lower because of the reduced CN. Since the short-range interaction on each atom results from its neighboring coordinating atoms, the atomic vibrating dislocation is the contribution from all the surrounding coordinates, $z$. Considering the vibration amplitude $x << d_0$, it is reasonable to take the mean contribution from each coordinate as a first order approximation, i.e., $k_1 = k_2 = \cdots = k_z = \mu_i(c\omega)^2$, and $x_1 = x_2 = \cdots = x_z = (r-d_0)/z$. Therefore, the total energy of a certain atom with $z$ coordinates follows:[97]

$$\begin{aligned} E_c &= -zE_b + \frac{z\mu c^2\omega^2}{2}\left(\frac{r-d_0}{z}\right)^2 + \ldots. \\ &= -zE_b + \frac{zd^2u(r)}{2!dr^2}\bigg|_{d_0}(r-d_0)^2 + \ldots \end{aligned} \tag{61}$$

where $\mu$ is the reduced mass of the dimer atoms, $c$ is the speed of light. The phonon frequency hence follows:

$$\omega = \frac{z}{c}\left[\frac{d^2u(r)}{\mu dr^2}\bigg|_{d_0}\right]^{\frac{1}{2}} \propto \frac{z(E_b)^{1/2}}{d_0}, \text{ and } \frac{\omega_i}{\omega} = z_{ib}c_i^{-(m/2+1)} \tag{62}$$

Considering the scaling law for the frequency [Q(∞) = ω(∞) - ω(1)], we have the size dependent Raman shift:

$$\omega(K_j) - \omega(1) = [\omega(\infty) - \omega(1)](1 + \Delta_R)$$

$$\Delta_R = \sum_{i \le 3} \gamma_{ij} \left( \frac{\omega_i}{\omega_b} - 1 \right) = \sum_{i \le 3} \gamma_{ij} \left( z_{ib} c_i^{-(m/2+1)} - 1 \right) \tag{63}$$

where $\omega_0$ and $\omega_i$ correspond to the vibration frequency of an atom inside the bulk and in the ith surface atomic shell. Combining Eqs (62) and (63) gives the size-dependent red shift of optical mode of a nanosolid:

$$\frac{\omega(K_j) - \omega(\infty)}{\omega(\infty) - \omega(1)} = \Delta_R < 0 \tag{64}$$

where $\omega(1)$ being the vibrational frequency of an isolated dimer is the reference point for the optical red shift upon nanosolid and bulk being formed. The frequency decreases from the dimer value with the number of atomic CN and then reaches the bulk value (z = 12) that is experimentally detectable.

## 6.3 Verification

### 6.3.1 Optical modes and dimer vibration

Incorporating the BOLS prediction with the least-root-mean-square linearization of measurements we have the scaling relation, with $\kappa = 1$:

$$\Delta\omega(K_j) = \begin{cases} \dfrac{-A'}{K_j} & (Measurement), \\ = \Delta_R[\omega(\infty) - \omega(1)] & (Theory). \end{cases} \tag{65}$$

Hence, the frequency shift from the dimer bond vibration to the bulk value, $\omega(\infty) - \omega(1) \equiv -A'/(\Delta_R K_j)$, is a constant as $\Delta_R \propto K_j^{-1}$. This relation provides an effective means allowing us to determine the vibrational frequency of an isolated dimer bond $\omega(1)$ and its bulk shift, which are beyond the scope of direct measurement. Figure 25 shows the match between the BOLS predictions with the theoretically calculated and the experimentally measured optical red shift of a number of samples. Derived information about the corresponding dimer vibration is given in Table 8.

Figure 25 (link) Comparison of the predictions with observations on the size-dependent TO shift of nano-silicon. Theoretical results: Si-1 was calculated using correlation length model;[446] Si-3 (dot) and Si-4 (rod) were calculated using the bulk dispersion relation of phonons; [447] Si-5 was calculated from the lattice-dynamic matrix;[423] Si-7 was calculated using phonon confinement model [448] and Si-8 (rod), and Si-9 (dot) were calculated using bond polarizability model.[421] Si-2,[449] Si-6;[450] Si-10 and Si-11 [436] are measured data. (b) $CdS_{0.65}Se_{0.35}$-1, $CdS_{0.65}Se_{0.35}$ (in glass)-$LO_2$, $CdS_{0.65}Se_{0.35}$-2, $CdS_{0.65}Se_{0.35}$ (in glass)-$LO_1$,[451] CdSe-1 CdSe(in $B_2O_3SiO_2$)-LO, CdSe-2 CdSe(in $SiO_2$)-LO and CdSe-3 CdSe(in $GeO_2$)-LO,[445] CdSe-4 CdSe(in $GeO_2$)-LO,[444] (c) $CeO_2$-1,[452] $SnO_2$-1,[453] $SnO_2$-2,[435] InP-1[454] are all measurement.

Table 8 Vibration frequencies of isolated dimers of various nanosolids and their red shift upon bulk formation derived from simulating the size dependent red shift of Raman optical modes, as shown in Figure 25.

| Material | Mode | $d_0$ (nm) | A′ | ω(∞) ($cm^{-1}$) | ω(1) ($cm^{-1}$) | ω(∞)-ω(1) ($cm^{-1}$) |
|---|---|---|---|---|---|---|
| $CdS_{0.65}Se_{0.35}$ | $LO_1$ CdSe-like | 0.286 | -23.9 | 203.4 | 158.8 | 44.6 |
| | $LO_2$ CdS-like | 0.286 | -24.3 | 303 | 257.7 | 45.3 |
| CdSe | LO | 0.294 | -7.76 | 210 | 195.2 | 14.8 |
| $CeO_2$ | | 0.22 | -20.89 | 464.5 | 415.1 | 49.4 |
| $SnO_2$ | $A_{1g}$ | 0.202 | -14.11 | 638 | 602.4 | 35.6 |
| InP | LO | 0.294 | -7.06 | 347 | 333.5 | 13.5 |
| Si | | 0.2632 | -5.32 | 520.0 | 502.3 | 17.7 |

### 6.3.2 Acoustic modes and intercluster interaction

Figure 26 shows the least-square-mean-root fitting of the size dependent LFR frequency for different nanosolids. The LFR frequency depends linearly on the inverse $K_j$. The zero intercept at the vertical axis indicates that when the $K_j$ approaches infinity, the LFR peaks disappear, which implies that the LFR peaks and their blue shifts originate from vibration of the entire nanosolid. However, it is hard to discriminate whether it arises from the quadruple motion or from the bond strain at the interface. The slope values for Si nanosolid are 97.77, 45.57 and 33.78 for the $A_1$, $T_2$ and $E$ modes, corresponding to the stretching (*LA*) and bending (*TA*) mode, respectively.

Figure 26 (link) Generation and blue shift of the LFR spectra where the solid, dotted and dashed lines are the corresponding results of the least squares fitting. (a) the Si-a ($A_1$ mode), Si-b ($T_2$ mode) and Si-c ($E$ mode) were calculated from the lattice-dynamic matrix by using a microscopic valence force field model;[423] the Si-d and Si-e are the experimental results.[422] (b) Ag-a (Ag in $SiO_2$),[455] Ag-b (Ag in $SiO_2$),[428] Ag-c (Ag in Alumina).[427] (c) $TiO_2$-a,[456] $TiO_2$-b,[456] $SnO_2$-a.[435] (d) CdSe-a (l = 0, n = 2), CdSe-b (l = 2, n = 1) and CdSe-c (l = 0, n = 1).[457] (e) $CdS_{0.65}Se_{0.35}$-a [$CdS_{0.65}Se_{0.35}$ (in glass)-LF2], and $CdS_{0.65}Se_{0.35}$-b [$CdS_{0.65}Se_{0.35}$ (in glass)-LF1][451] are all measured data.

Table 9 Linearization of the LFR acoustic modes of various nanosolids gives information about the sound velocity in the specific solid.

| Sample | A′ |
|---|---|
| Ag-a & Ag-b | 23.6 ± 0.72 |
| Ag-c | 18.2 ± 0.56 |
| TiO-a, TiO-b | 105.5 ± 0.13 |
| SnO-a | 93.5 ± 5.43 |
| CdSe-1-a | 146.1 ± 6.27 |
| CdSe-1-b | 83.8 ± 2.84 |
| CdSe-1-c | 46.7 ± 1.39 |
| CdSSe-a | 129.4 ± 1.18 |
| CdSSe-b | 58.4 ± 0.76 |
| Si | 97.77 |
| Si | 45.57 |
| Si | 33.78 |

### 6.3.3 Surface atom vibration

According to Einstein's relation, it can be derived that $\mu(c\omega x)^2/2z = k_B T$ . The vibrational amplitude of an atom is $x \propto z^{1/2}\omega^{-1}$ . The frequency and magnitude of vibration for an atom in the surface at room temperature are derived as:

$$\frac{\omega_1}{\omega_b} = z_{ib} c_1^{-(m/2+1)} = \begin{cases} 0.88^{-3.44}/3 = 0.517 & (Si) \\ 0.88^{-3/2}/3 = 0.404 & (Metal) \end{cases}$$

$$\frac{x_1}{x_b} = (z_1/z_b)^{1/2} \omega_b/\omega_1 = (z_b/z_1)^{1/2} c_1^{(m/2+1)}$$

$$= \begin{cases} \sqrt{3} \times 0.88^{3.44} = 1.09 & (Si) \\ \sqrt{3} \times 0.88^{3/2} = 1.43 & (Metal) \end{cases}$$

(66)

The vibrational amplitude of an atom at the surface is indeed greater than that of a bulk atom while the frequency is lower. The magnitude and frequency are sensitive to the m value and varies slightly with the curvature of a spherical dot when $K_j > 3$. This practice confirms for the first time the assumption made by Shi[29] and Jiang et al [85] that the vibration amplitude of a surface atom is always greater than the bulk value and it keeps constant at all particle sizes.

6.4 Summary

In summary, the BOLS correlation has enabled us to correlate the size-created and the size-induced blue shift in the LFR phonon frequency to the intergrain interaction and the red shift in the TO and LO phonon frequencies to the CN-imperfection reduced cohesive energy. Decoding the measured size-dependence of Raman optical shift has derived vibrational information of Si, InP, CdS, CdSe, $TiO_2$, $CeO_2$ and $SnO_2$ dimers and their bulk shifts and confirmation of the magnitude of surface atomic vibration that is indeed higher than the bulk value, which is beyond the scope of direct measurement.

## 7 Photon emission

7.1 Background

For the particular concern of the photoelectronic properties, nanostructured semiconductors exhibit the similar trends of change for the entire band-structure change with reducing the dimension of the solid.[32] The observable changes may be summarized as follows:

(i) The band gap expands with reducing particle size, which gives rise to the blue shift in the PL and photo-absorbance (PA) spectra of nanometric semiconductors such as Si oxides,[61,458,459,460] III-V[461] (GaN,[462,463] InAs,[464] GaP and InP,[465,466]) and II-VI (CdS,[467,468,469] ZnS,[470] CdSe,[471,472] ZnTe,[473] CdTe/CdZnTe[474]) compounds.

(ii) The energies of PL and PA involve the contribution from electron-phonon coupling that shifts the optical band gap $E_G$ from the true $E_G$ with the well-known amount of Stokes shift arising from electron-phonon interaction that also changes with solid size.[35]

(iii) The energy levels of the core bands and the adsorbate-induced chemical shifts move simultaneously towards higher binding energy (in absolute value). XPS measurements revealed that the main core-level peaks and the oxide satellites of Cu-$2p_{3/2}$,[475] (-932.1, -940.1 eV), Sn-3d (-484.4, -486.7 eV),[476] Sn-4d (-26, -31 eV), Ta-$4f_{5/2}$ (-23.4, -26.8eV) and Ta-$4f_{7/2}$ (-31.6, -36.5 eV),[476] move simultaneously up and the amounts of change depend on the original core-level position and particle size. These dedicated observations confirm that the particle-size and oxidation have important effects on the core-level shift of nanometric compounds, which is of great value in understanding the nature of nanometric system.

(iv) Because of $E_G$ expansion, the complex dielectric constant of a nanometric semiconductor is significantly suppressed,[477] which forms enormous impact in electronic and optical devices. The reduction of dielectric constant can enhance the Coulomb interaction among electrons, holes, and ionized shallow impurities in nanometric devices, and enhances the exciton binding energy.[104]

Therefore, it would be more appropriate to consider the simultaneous change of all the properties relating to the Hamiltonian rather than simply discriminate one phenomenon from others at a time in modelling practice of the Hamiltonian related properties.

## 7.2 Outstanding models

### 7.2.1 Quantum confinement

Among the numerous models for the PL blue shift, "quantum confinement (QC)" theory[57] has been elegantly accepted. Efros and Efros[58] firstly proposed, in 1982, this concept based on the experimental findings of the size effect on the blue shift in the main exciton absorption of CuCl (~ 3 nm across) nanocrystallite.[478] The confinement effect on the band gap, $E_G$, of a nanosolid of radius R was expressed as:[58]

$$E_G(R) = E_G(\infty) + \pi^2\hbar^2/(2\mu R^2) \tag{67}$$

where μ ($1/\mu = 1/m_h^* + 1/m_e^*$), being the reduced mass of an electron-hole (e-h) pair, is an adjustable parameter. Equation (67) indicates that the $E_G$ expansion arises from the kinetic energy of the e-h pairs that are separated by a distance of the particle dimension, R, or the quantum well size. In order to improve the simulations, Brus[59] and Kayanuma[60] further extended the QC theory by including the Coulomb interaction of an e-h pair of R separation and the correlation energy $E_R$ being the Rydberg (spatial correlation) energy for the bulk semiconductor. The modified form is given as:

$$\begin{aligned} E_G(R) &= E_G(\infty) + \pi^2\hbar^2/(2\mu R^2) - 1.786e^2/(\varepsilon_r R) + 0.284E_R \\ E_R &= \mu e^4/(2\varepsilon_r^2\varepsilon_0^2\hbar^2) = 13.56\mu/\varepsilon_r^2 m_e \text{ (eV)} \end{aligned} \tag{68}$$

The effective dielectric constant $\varepsilon_r$ and the effective mass, μ, describe the effect of the homogeneous medium in the quantum box, which is a mono-trapping central potential extended from that of a single atom by expanding atomic size to the dimension of the solid. The dictating factor for the QC convention is the production of e-h pairs as their kinetic energy and potential energy dominate the $E_G$ expansion. For CdS example,[60,479] $\varepsilon_r = 5.5$, $m_e = 0.19$, and $m_h = 0.8$.

According to the QC theory, electrons in the conduction band and holes in the valence band are confined spatially by the potential barrier of the surface, or trapped by the potential well of the quantum box. Because of the confinement of both the electrons and the holes, the lowest energy optical transition from the valence to the conduction band increases in energy, effectively increasing the $E_G$. The sum of kinetic and potential energy of the freely moving carriers is responsible for the $E_G$ expansion and therefore the width of the confined $E_G$ grows as the characteristic dimensions of the crystallite decrease.

Later development of the QC theory shows that the relation of $\Delta E_G \propto R^{-n}$ (n = 1.16,[480] 1.3,[471] 1.37[465]) fits better the size-dependent PL blue shift and the n values vary from source to source. Nevertheless, it is important to recognize that the QC premise is indeed a very helpful first order approximation, and can be used to estimate changes in energy levels, exciton and related energies, as a function of dot size. However, at the lower end of the size limit, the QC theoretical curve diverges from the true situation that the $E_G$ can never be larger than the separation of the involved energy levels of an isolated atom.

### 7.2.2 Other schemes

A free-exciton collision model[61] proposed for the PL blue shift suggests that the $E_G$ expansion arises from the contribution of thermally activated phonons in the grain boundaries rather than the QC effect. During PL measurement, the excitation laser heats the free excitons that then collide with the boundaries of the nanometer-sized fragments. The laser heating the free-excitons up to the temperature in excess of the activation energy required for the self-trapping gives rise to the extremely hot self-trapping excitons (STE's). Because the resulting temperature of the STE's is much higher than the lattice temperature, the cooling of the STE's is dominated by the emission of phonons. However, if the STE temperature comes into equilibrium with the lattice temperature, the absorption of lattice phonons becomes possible. As a result, the blue shift of the STE-PL band is suggested to originate from the activation of hot-phonon-assisted electronic transitions. The blue shift of the STE-PL band depends on the temperature of laser-heated free-excitons that in turn is determined by the size of nanometer-sized fragments. This event happens because the temperature (kinetic energy) of the laser-heated free-exciton increases with the number of collisions with the boundary of confined regions, which tends to be higher with decreasing size of the (silica was considered only) fragments in nanoscale materials. The energy gained from laser heating of the exciton increases with decreasing nanosolid diameter in an exp(1/R) way. Based on the analysis, Glinka et al [61] indicated that the size-dependent PL blue shift of a nanosolid in general does not need to be related always to the QC effect.

Other phenomenological models for the blue shift in PL of nanosolids include the impurity centers,[62] surface states,[481] surface alloying,[63] cluster interaction and oxidation effect.[65] However, all the models mentioned above are good for the blue shift in the PL and cover various possible sources. These models have their limitations, however, that could explain neither change of Hamiltonian as the origin nor other quantities relating to the Hamiltonain, such as the core level shift and dielectric suppression, which should be intrinsic to nanostructures.

## 7.3 BOLS formulation

### 7.3.1 Band formation

Figure 27 illustrates the evolution of the energy levels of a single atom to the energy bands of a bulk solid containing $N_j$ atoms. Electrons of a single atom confined by the intra-atomic trapping potential, $V_{atom}(r)$ = constant or $-\infty$, move around the central ion core in a standing-wave form inside the potential well. The corresponding eigen wave functions and the eigen energies are given as follows:

$$\phi_v(r) \propto \sin^2(2\pi nr/d_0), \text{ and } E(n) = 2(n\pi\hbar)^2/\left(m_e d_0^{\,2}\right), n = 1,2,3,...$$

the atomic diameter $d_0$ corresponds to the dimension of the potential well of the isolated atom. The branch numbers ($n$) correspond to different energy levels. The energy separation between the closest two levels depends on $(n+1)^2-n^2 = 2n+1$.

When a system contains two atoms, the single energy level splits into two separate sublevels and the separation between the sublevels is determined by the inter-atomic binding energy. Meanwhile, the presence of inter-atomic interaction shifts the center of the two levels down. Increasing the number of atoms up to $N_j$, the single energy level will expand into a band within which there are $N_j$ sublevels. The number of atoms $N_j$ in the solid determines the number of the sublevels in a particular energy band. What distinguishes a nanosolid from a bulk solid is that for the former the $N_j$ is accountable, while for the latter the $N_j$ is too large to be accounted. Therefore, the classical band theories are valid for a single nanometric solid that contains any number of atoms. As detected with XPS, the DOS of a core band for a nanosolid exhibits band-like features rather than the discrete spectral lines of a single atom. If the $N_j$ is sufficiently small, the separation between the sublevels is resolvable. The energy level spacing of the successive sublevels in the valence band, know as the Kubo gap ($\delta_K = 4E_F/3N_j$), decreases with increasing the number of valence electrons of the system.[12] For system contains 1000 silver atoms, the Kubo gap would be 5 ~ 10 meV. At room temperature, $k_BT \cong 25$ meV, a 3-nm particle containing 500 atoms or more would be metallic ($k_BT > \delta_K$). At low temperatures, however, the level spacings especially in a small particle may become comparable to $k_BT$ or higher, rendering them nonmetallic.[12] Because of the presence of the $\delta_K$ in an individual nanosolid, properties such as electron conductivity and magnetic susceptibility exhibit quantized features.[482] The resultant discreteness of energy sublevels also brings about fundamental changes in the characteristic spectral features of the nanosolids, especially those related to the valence band.

Figure 27 (link) Evolution of a single energy level into the band structure when particle grows from a single atom to a bulk solid that contains N atoms. Indicated is the work function$\phi$, band gap $E_G$, core level shift $\Delta E_v$, bandwidth $E_B$. The number of allowed sublevels in a certain band equals the number of atoms of the solid. The sublevel spacing is described by the Kubo gap, $4E_F/3N_j$, with $E_F$ being the Fermi level of the bulk.[100]

??

According to the band theory,[285] the Hamiltonian for an electron inside a solid is in the form:

$$\hat{H} = \hat{H}_0 + \hat{H}' = -\frac{\hbar^2\nabla^2}{2m} + V_{atom}(r) + V_{cry}(r + R_C) \tag{69}$$

where the $V_{atom}(r)$ is the intra-atomic trapping potential of an isolated atom and the $\hat{H}' = V_{cry}(r) = V_{cry}(r + R_C)$ is the periodic potential of the crystal, i.e., the inter-atomic binding potential or crystal field. $R_C$ is the lattice constant. According to the nearly-free-electron approximation, the $E_G$ originates from the crystal field and the width of the gap depends on the integral of the crystal field in combination with the Bloch wave of the nearly free electron, $\phi(k_l, r)$:

$$E_G = 2|V_1(k_l)|, \text{ and } V_1(k_l) = \langle \phi(k_l, r)|V(r + R_C)|\phi(k_l, r)\rangle \tag{70}$$

where $k_l$ is the wave-vector and $k_l = 2l\pi/R_C$. Therefore, the $E_G$ is simply twice of the first Fourier coefficient of the crystal field.

The energy dispersion of an electron in the ν th core band follows the relation:

$$\begin{aligned} E_v(k) &= E_v(1) + \Delta E_v(\infty) + \Delta E_B(k_l, R_C, z) \\ &= E_v(1) - (\beta + 2\alpha) + 4\alpha\Omega(k_l, R_C, z) \end{aligned} \tag{71}$$

where,

- $E_v(1) = \langle \phi_v(r)|\hat{H}_0|\phi_v(r)\rangle$ is the energy of the core electron of an isolated atom.
- $\beta = -\langle \phi_v(r)|V_{cry}(r)|\phi_v(r)\rangle$ is the crystal field effect on the specific core electron at site r.
- $\alpha = -\langle \phi_v(r - R_C)|V_{cry}(r - R_C)|\phi_v(r - R_C)\rangle$ is the crystal field effect on the coordinate neighbouring electrons.
- For an fcc structure example, the structure factor. $\Omega(k_l, R_C) = \sum_z \sin^2(k_l R_C/2)$.
- The sum is over all the contributing coordinates (z) surrounding the specific atom in the solid.

Eqs (70) and (71) indicate clearly that the $E_G$, the energy shift $\Delta E_v(\infty) = -(\beta + 2\alpha)$ of the $E_v(1)$ and the bandwidth $\Delta E_B$ (last term in Eq (71)) are all functions of the crystal field. Any perturbation to the crystal field will vary these quantities. Without the crystal field, neither the $E_G$ expansion nor the core-level shift would be possible; without the inter-atomic binding, neither a solid nor even a liquid would form.

### 7.3.2 Hamiltonian perturbation

Considering an assembly composed of *n* particles of mean size $K_j$ and with each particle, there are $N_j$ atoms, the total binding energy, $V(r, n, N_j)$:[15]

$$\begin{aligned} V(r, n, N_j) &= \sum_n \sum_{l \neq i} \sum_i v(r_{li}) \\ &= \frac{n}{2}\left[N_j \sum_{i=1} v(r_{li}) + \sum_{k \neq j} V(K_{kj})\right] \\ &\cong \frac{n}{2}\left[N_j^2 v(d_0) + nV(K_j)\right] \end{aligned} \tag{72}$$

$V(r, n, N_j)$ sums over all the $nN_j$ atoms and the n particles. The high order $r_{li}$ is a certain fold of the nearest atomic spacing, $d_0$. Besides, interaction between the nearest clusters, *k* and *j*, $V(K_{kj})$, should be taken into account. If $K_{kj}$ is considerably large (such as the case of porous Si, or highly dispersed particles), the last term is negligible, which is the case of an isolated particle. Normally, the intercluster interaction, $V(K_j)$, is much weaker than the interatomic interaction.[1] For example, if the cluster is treated as an electrical dipole or a magnetic dipole, the Van der Waals or the super-paramagnetic potential is much weaker. If the intercluster interaction cannot be neglected, Eq (72) becomes ($N_j = N_{shell} + N_{core}$):

$$
\begin{aligned}
V_{cry}(r,n,N_j) &= \frac{nN_j}{2}\left[N_{shell}v(d_s)+(N_j-N_{shell})v(d_0)+\frac{n}{N_j}V(K_j)\right] \\
&= \frac{nN_j^2 v(d_0)}{2}\left[\frac{N_{shell}v(d_s)}{N_j v(d_0)}+\left(1-\frac{N_{shell}}{N_j}\right)+\frac{nV(K_j)}{N_j^2 v(d_0)}\right] \\
&= V_{cry}(d_0,n,N_j)\{1+\gamma_{sj}[v(d_s)/v(d_0)-1]+\delta_{kj}\}
\end{aligned}
\tag{73}
$$

$N_{shell} = \sum_{i\le 3} N_i$ is the number of atoms in the outermost three atomic shells of the nanosolid. $V_{cry}(d_0, n, N_j)$ is the crystal potential of the system without the contribution from surface relaxation or from the intercluster interaction. The pair interatomic binding energy at equilibrium atomic separation, $v(d_i) \propto E_i = c_i^{-m}E_b$. Therefore, the perturbation to the crystal binding energy (the energy density in the relaxed region rather than the atomic cohesive energy) upon assembly of the nanosolids is,

$$
\begin{aligned}
\Delta_H(K_j) &= \frac{V_{cry}(r,n,N_j)}{V_{cry}(d_0,n,N_j)}-1 \\
&= \sum_{i\le 3}\gamma_{ij}\frac{\Delta v(d_i)}{v(d_0)}+\delta_{kj} \\
&= \sum_{i\le 3}\gamma_{ij}(c_i^{-m}-1)+\delta_{kj}
\end{aligned}
$$

$$
\text{and, } \delta_{kj} = \frac{nV(K_j)}{N_j^{\,2}v(d_0)}
\tag{74}
$$

The perturbation covers the weighted sum of contribution from the individual surface layers ($c_i^{-m}$-1) over the outermost three atomic layers of a nansolid, and the inter-cluster interaction, $\delta_{kj}$, that is negligible if the particle size is sufficient large.

The total potential in eq (69) becomes, $V(\Delta_H) = V_{atom}(r) + V_{cry}(r)[1+\Delta_H]$ . In conjunction with the corresponding Bloch wave functions, the atomic trapping potential, $V_{atom}(r)$, defines the discrete core-level energies of an isolated atom, $E_v(1)$. The crystal binding $V_{cry}(r)$ defines not only the $E_G$, but also the shift of the core-level energy away from the original position, $\Delta E_v(\infty) = E_v(\infty) - E_v(1)$, as well as other quantities such as the bandwidth and band tails. The dimensionless $\Delta_H$, being independent of the particular form of the interatomic potential, is the contribution from binding energy density in the relaxed surface region.

The perturbation to the Hamiltonian will cause the changes of $E_G$ and $E_v(K_j)$, which follows the scaling relation:

$$
\frac{\Delta E_G(K_j)}{E_G(\infty)} = \frac{\Delta E_v(K_j)-\Delta E_v(\infty)}{\Delta E_v(\infty)} = \Delta_H
\tag{75}
$$

We now turn to look at the possible mechanisms that could modify the crystal field, $V_{cry}$(r), that depends functionally on atomic distance and the nature of the chemical bond. Bond formation transports charge among the bonding constituents. Different types of interatomic potential describe different kinds of chemical bonds. If the atomic distance, or bond length, relaxes spontaneously, the crystal field will be enhanced. Chemical reaction in which charge transport dominates not only reduces the “screening” effect of the core electrons, but also alters the nature of the bond.

Therefore, shortened bond length and altered bond nature will enhance the binding energy and, consequently, the crystal field of the solid. On the other hand, chemical reaction will repopulates with electrons in the valence band, which will expand the $E_G$ extrinsically.[133,483]

### 7.3.3 Remarks

In the quantum theory for condensed matters, the key elements are the Hamiltonian and the Bloch wave functions. Nanosolid densification may modify the wave functions slightly as there no chemical reaction occurs. In the first order approximation, we may ignore the size effect on the wave function shrinkage. Therefore, the Hamiltonian becomes key important. The potential energy of the Hamiltonian in various modeling consideration is different, which is compared as follows:

$$V = \begin{cases} V_{atom}(\infty), & (r \le r_0), & isolated-atom \\ V_{dot}(= V_{atom}), & (r \le R), & QC-I \\ V_{dot} + V_{e-h} + V_R, & (r \le R), & QC-II \\ V_{atom} + V_{cry}, & (-\infty < r < \infty), & Extended-solid \\ V_{atom} + V_{cry}(1+\Delta_H), & (r \le R), & BOLS-perturbation \end{cases} \quad (76)$$

The major difference between the QC theory and the BOLS correlation for a nanosolid lies in that:

(i) The QC effect is dictated by the kinetic and the potential energy of the e-h pairs that are produced in the measurement. Production of e-h pairs is crucial to the trigger the function of the QC theory. The Coulomb potential energy for the e-h pair, $V_{e-h}$, is about $10^{-1}$ eV order,[500] which is negligibly small compared to the interatomic binding energy (1 ~ 7 eV). Furthermore, the radiation recombination of the e-h pair occurs depending on the overlap extent of wave functions of the e-h pair. The localization length of a carrier, $R_0$, is about several Bohr's radius.[134] The probability of e-h recombination is proportional to $\exp(-2r_{e-h}/R_0)$. If the e-h separation, $r_{e-h}$, is considerably larger than the localization length, $R_0$, the probability of the radiation recombination is extremely small. On the other hand, the involved dielectric constant $\varepsilon_r$ is no longer constant but it is size dependent.[104,477]

(ii) The BOLS correlation is dictated, however, by the fact of atomic CN-imperfection and its consequences on the Hamiltonian. The BOLS premise adds its perturbation to the crystal potential of an extended solid without e-h pair or the correlation energy being involved. The intra-atomic trapping, $V_{atom}$, is responsible for the discrete energy levels of an isolated atom. As trapping centers, the $V_{atom}$ localizes electrons to spend most time moving inside the $V_{atom}$ in the form of standing waves. The inter-atomic binding potential or crystal field, $V_{cry}$, is crucial to binding atoms to form a solid. Therefore, the $V_{cry}$ can never be removed, or replaced with otherwise other alternatives, in dealing with a system containing atoms more than one. Despite the numerical convergence of the $E_G$ at the lower end of the size limit, The BOLS premise is able to formulate not only the entire band structure change (band gap expansion, the core-level shift, the core bandwidth and band tails) but also other properties such as the strength of electron-phonon coupling.

Therefore, the traditional QC theory dictated by e-h production appears to be too ideal and the modification with dominance of crystal binding and electron-phonon coupling would be necessary.

## 7.4 Verification: photon emission and absorption

### 7.4.1 Electron-phonon coupling

Figure 28 illustrates the effect of electron-phonon (e-p) coupling and crystal binding on the $E_{PL}$ and $E_{PA}$. The energies of the ground state ($E_1$) and the excited state ($E_2$) are expressed as:[134]

$$\begin{cases} E_1(q) = Aq^2 \\ E_2(q) = A(q-q_0)^2 + E_G \end{cases} \quad (77)$$

Constant A is the slope of the parabolas. The $q$ is in the dimension of wave-vector. The vertical distance between the two minima is the real $E_G$ that depends functionally on the crystal potential. The lateral displacement ($q_0$) originates from the e-p coupling that can be strengthened by enhancing lattice vibration. Therefore, the blue shift in the $E_{PL}$ and in the $E_{PA}$ is the joint contribution from crystal binding and e-p coupling. At a surface, the CN-imperfection-enhanced bond strength affects both the frequency and magnitude[85,29] of lattice vibration. Hence, at a surface, the e-p coupling and hence the Stokes shift will be enhanced.

Figure 28 (link) Mechanisms for $E_{PA}$ and $E_{PL}$ of a nano-semiconductor, involving crystal binding ($E_G$) and electron-phonon coupling (W). Insertion illustrates the Stokes shift from $E_{PA}$ to $E_{PL}$.[134] Electron is excited by absorbing a photon with energy $E_G$+W from the ground minimum to the excited state and then undergoes a thermalization to the excited minimum, and then transmits to the ground emitting a photon with energy $E_G$-W.[134]

In the process of carrier formation and recombination, an electron is excited by a photon with $E_G$+W energy from the ground minimum to the excited state with creation of an electron-hole pair. The excited electron then undergoes a thermalization and moves to the minimum of the excited state, and eventually transmits to the ground combining with the hole. The carrier recombination is associated with emission of a photon with energy $E_{PL} = E_G - W$. The transition processes (e-h pair production and recombination) follow the rule of momentum and energy conservation though the conservation law may be subject to relax for the short ordered nanosolid. Such conservation law relaxation is responsible for the broad peaks in the PA and PL.

The insertion illustrates the Stokes shift, $2W = 2Aq_0^2$, from $E_{PL}$ to $E_{PA}$. The $q_0$ is inversely proportional to atomic distance $d_i$, and hence, $W_i = A/(c_i d_i)^2$, in the surface region. Based on this premise, the blue shift of the $E_{PL}$, the $E_{PA}$, and the Stokes shift can be correlated to the CN-imperfection-induced bond contraction:[110]

$$\left.\begin{array}{c} \dfrac{\Delta E_{PL}(K_j)}{E_{PL}(\infty)} \\ \dfrac{\Delta E_{PA}(K_j)}{E_{PA}(\infty)} \end{array}\right\} = \frac{\Delta E_G(K_j) \mp \Delta W(K_j)}{E_G(\infty) \mp W(\infty)} \cong \sum_{i\le 3} \gamma_i \left[(c_i^{-m}-1) \mp B(c_i^{-2}-1)\right]$$

$$\left(B = \frac{A}{E_G(\infty) d^2};\quad \frac{W(\infty)}{E_G(\infty)} \approx \frac{0.007}{1.12} \approx 0\right) \quad (78)$$

Compared with the bulk $E_G(\infty)$ = 1.12 eV, the $W(\infty)$ ~ 0.007 eV obtained using empirical tight-binding calculations[484] is negligible. One can easily calculate the size dependent $E_{PL}$, $E_{PA}$, and $E_G = (E_{PL} + E_{PA})/2$ as well using Eq (78). Fitting the measured data gives the values of $m$ and $A$ for a specific semiconductor.

### 7.4.2 $E_G$ expansion

The size dependence of both the $E_{PL}(K_j)$ and $E_{PA}(K_j)$ of porous Silicon (p-Si) fabricated using electrochemical method has been obtained.[110] The room temperature reflectivity (Figure 29a) varies with size in the photon energy range of 200-900 nm wavelength, which is related to the change of dimension and geometry of columns and voids on the p-Si surface. The absorption coefficient was obtained by fitting the reflection spectra using the Scout software package.[485] The $E_{PA}$ values were extracted from the absorption spectra (Figure 29b) using the Tauc plot method.[486, 487] The PL and XPS E-2p energy shift profiles are given in Figure 29c and d, compared with the predicted size-dependence.

Figure 29 (link) (a) Reflection and (b) absorption spectra of PS samples with different particle sizes measured at ambient temperature. $E_{PA}$ is obtained with the Tauc plot fitting of the reflection and absorption data. Size dependence of (c) PL spectra, and (d) E-2p core-level shift in particle size range of R = 1.4 nm to 2.1 nm.[110]

Figure 30 (link) Comparison between predictions (solid lines) and the measured size dependence:

(a) The $E_{PA}$ blue shift of PS with Data-1,[486] Data-2,[488] Data-3,[489] Data-4,[490] and Data-5.[33]
(b) The $E_{PL}$ blue shift of nano-Si. Data-1,[440] Data-2,[491] Data-3,[484] Data-4,[492] Data-5,[493] Data-6,[494] Data-7,[495] Data-8,[490] are calculation results. Data-9,[490] Data-10,[488] Data-11,[458] Data-12,[496] and Data-13[33] are measurements.
(c) The $E_G$-expansion measured using STS[36] and optical method, Data –1 ($E_G = E_{PA} – W$),[439] Data –2 ($E_G = (E_{PL} + E_{PA})/2$).[33]
(d) The core level shift of Si.[35]

Matching the predictions in Eq (78) with the measured $E_{PA}$ and $E_{PL}$ data (Figure 30a) gives coefficient $B$ = 0.91 and $m$ = 4.88 that refines the original value $m$ = 4 of which the e-p interaction was not considered.[33] The refined form is able to discriminate the effect of e-p coupling (B = 0.91) from the effect of crystal binding (m = 4.88) on the $E_{PL}$ and $E_{PA}$.

Most strikingly, without triggering electron-phonon interaction or electron-hole production, STM/S varies from 1.1 to 3.5 eV with decreasing the rod diameter from 7.0 to 1.3 nm and that the surface Si-Si bond contracts by ~12% from the bulk value (0.263 nm) to ~0.23 nm. The STS findings concur excitingly with the BOLS premise: CN-imperfection shortens the remaining bonds of the lower-coordinated atoms spontaneously associated with $E_G$ expansion, consequently. It is important to note that STS collects *localized* $E_G$ information without needing any energetic stimulus. The bias ($|V_b| < 2$ eV) between the tip and the sample surface is not sufficiently large to break the Si-Si bond. What happens upon being biased is that the tip introduces holes or electrons into the sample rather than excites electron-hole pairs inside the specimen. As such, neither electron excitation from the ground to the excited states nor electron-hole pair production or carrier recombination occurs during STS/M measurement. What contribute to the STS-$E_G$ are states occupied by the covalent bonding electrons and the empty states that are strongly localized at the probed site rather than the Coulomb interaction between the exited electron-hole or kinetic energies of the mobile carriers moving inside, or being confined by, the nanosolid. Without triggering the dictating QC factors,[100] STS-$E_G$ continues expanding upon the size being reduced. Surface hydride may form upon the sample being passivated. However, hydride formation reduces the mid-gap impurity DOS and hence to improve the quantum efficiency in the irradiation recombination, and hence, the surface hydride formation could never expand the $E_G$ at all.[134] As shown in Figure 30b, the size-enlarged $E_G$ of Si nanorods (STS derived)[36] and Si nanodots (mean value of $E_{PA}$ and $E_{PL}$) follows the BOLS prediction which involves no events of electron-hole interaction, e-p coupling or quantum confinement.

### 7.4.3 Nanocompound photoluminescence

Copper-doped zinc oxide nanowires ranges from 30 to 100 nm in diameter and tens to hundreds of microns in length show broad and continuous PL spectra extending from the ultraviolet to the red region at room temperature, depending on the excitation wavelength, which is different from that of the bulk. The mechanism of the excitation wavelength dependence of the PL emission of Cu-ZnO2 is complicated, which should be the joint effect of size and oxidation98 and Cu doping may add new levels for transition.

**Table 10** lists the parameters used in simulating the size-dependence of the PL blue shift of nanometric compound semiconductors. The bond length in the bulk takes the values of covalent bond of the corresponding materials (appendix A). Figure 31 compares the predictions with the relative PL shift observed for InP, InAs, CdS, and CdSe nanosolids. It can be seen that the curve of m = 4 gives generally better fit of the PL spectra of these compounds without involving the e-p coupling. This trend also agrees with the $E_G$ expansion determined with an XPS from the Si:H nanosolids.[13] The deviation of theory from experiment may arise from the accuracy in determining the shape and size of the particles or from the uncertainty of chemical reaction. The extent of reaction determines the $E_G$ which was used as a scale to normalize the entire set of the PL data. The scattered and broad distribution of the measured data for InP, InAs, CdS, and CdSe may be due to the same reason. However, all the data follow the similar trend of m = 4 ~ 6. The cluster interaction appears to play an insignificant role in the PL blue shift, which coincide with BOLS anticipation, as discussed in Section 7.3.2. The general trends of the simulated PL peak shift show that the size-induced frequency shift varies little with the materials or with the particular crystal structures (Wurtzite and zinc-blend structures in Figure 31b), as noted by Yoffe.[14]

It should be noted that the $E_G$ of the bulk compound varies with the extent of chemical reaction.[497,498] For example, the $E_G$ for the $SiO_x$ varies from 1.12 (Si) to 9.0 ($SiO_2$) eV. Therefore, it is not realistic to fit the measured data perfectly without considering the possible errors in experiment and the effect of surface passivation. Our attention, however, should focus on the trends of change and their origins. We may compare predictions with experimental observations on the *PL* blue shift of nanometric semiconductors near the lower end of the size as shown in Figure 30 and Figure 31. Agreement with *PL* shift of Si, *CdS*-I,[467] *CdSe*-I,[499] *CdSe*-II,[500] *CdSe*-III,[501] *CdSe*-IV,[502] and *CdSe*-V[503] nanosolids ($D$ < 5 nm) has been realized. The QC theoretical curves of $D^{-\lambda}$ ($1155\times D^{-1}$ and $80850\times D^{-2}$) are also compared, which diverge at the lower end of critical size though they match the *PL* data at larger particle size. Figure 31 shows the agreement between prediction and the measured PL shift of InP nanosolids.[504,465,505] The data also match curves of $R^{-1.04}$ and $100\times (5.8D^2+27.2D+10.4)^{-1}$ forms as well.[506]

The extent of bond contraction varies slightly from *CdS* to *CdSe*. The difference should be due to the difference in electronic configuration or the covalent bond length between S ($3p^4$, 0.104 nm) and Se ($4p^4$, 0.114 nm), which is beyond the scope of BOLS correlation.

Figure 31 (link) Comparison of the modeling predictions with the measured PL peak shifts of (a) InAs,[464] InP-01.[466,504] (b) Zn-blende-I and Wurtzite structure-II of CdS and CdSe,[461] and *CdS*-I,[467] *CdSe*-I,[499] *CdSe*-II,[500] *CdSe*-III,[501] *CdSe*-IV,[502] and *CdSe*-V[503] nanosolids. (c) InP[465,505] and CdSe rod measured using STS and PL.[507] The $R^{-1}$ and $R^{-2}$ curves diverge at the lower end of critical size.

Copper-doped zinc oxide nanowires ranges from 30 to 100 nm in diameter and tens to hundreds of microns in length show broad and continuous PL spectra extending from the ultraviolet to the red region at room temperature, depending on the excitation wavelength, which is different from that of the bulk.[508] The mechanism of the excitation wavelength dependence of the PL emission of Cu-$ZnO_2$ is complicated, which should be the joint effect of size and oxidation[98] and Cu doping may add new levels for transition.

Table 10 Summary of the simulating parameters for the $E_G$ expansion of nanometric semiconductors.

| | $E_G$(bulk) (eV) | $E_G$ (∞) (eV) |
|---|---|---|
| Si | 1.12 | 1.12 |
| InP | 1.45/1.34 | 1.45 |
| InAs | 0.35 | 0.9 |
| CdS | - | 2.2 |
| CdSe | 1.75 | 1.75 |

## 7.5 Bandwidth and band tails

### 7.5.1 Bandwidth

The predicted size-dependent bandwidth derived in Section 7.3.2 indicates that the bandwidth is determined by both the crystal field and the effective atomic CN(z) and the bandwidth shrinks with reducing particle size:

$$\frac{\Delta E_B(K_j)}{E_B(\infty)} = \sum_{i \le 3} \gamma_{ij} \left( \frac{\Delta \alpha_i}{\alpha} + \frac{\Delta z}{z} \right), \tag{79}$$

which agrees with the trends measured using XPS from CuO surface.[509] The observed peak intensity increases and the peak-base width (rather than the full width at half maximum that describes the distribution of the occupied DOS in the core band) decreases with reducing the particle size. It is understandable that the number of electrons is conservative in the deeper band, as the core electrons do not involve the charge transportation in a process of chemical reaction.[1] If the z reduces to one or two, the bandwidth will approach to the single energy level of an isolated atom.

### 7.5.2 Band tails and surface states

For an isolated nanosolid or a surface, there are two kind surface states. One is the dangling bonds or surface impurities, which add impurity states within the $E_G$ of semiconductors. Termination of the dangling bonds by H adsorption could minimize the impurity states. The other is the contracted bonds in the relaxed surface region, which offsets the entire band structure associated with $E_G$ enlargement and the presence of band tails.

The difference between an assembly of nanosolids and a bulk solid in amorphous state is the distribution of defects. In the amorphous phase, the randomly distributed CN deficiency causes the bond length and angle of the specific atom to distort in a disordered way, which adds traps randomly in depths inside the bulk. In an amorphous solid, the number of the lower-coordinated atoms is hardly controllable as the amorphous state depends heavily on the processing conditions. For a nanosolid or nanocrystallite, CN-deficiency only happens orderly at the surface and the number of sites of CN-deficiency is controllable by adjusting the shape and size of the nanosolids.

The effect of CN-deficiency in both amorphous and nanosolid states bends the energy near the conduction and the valence band edges with production of band tails occupied by the localized states. The resultant of the two band tails gives the Urbach edge appearing in the photoabsorption spectra.[134] According to the BOLS premise, the Urbach edge of a nanosolid resulting from bond contraction due to the CN-imperfection in the surface region is comparable to the random traps inside the amorphous bulk solid. The deepened potential traps near the surface edges are responsible for the localization of carriers in the band tails of nanosolids. Therefore, the CN-imperfection enhanced interatomic interaction near the surface edge of a nanosolid should also produce such band tails that are identical to the band tails of amorphous solid though the tail states are originated from different sites in real space. As expected, such Urbach edges have been identified from the photo-absorption spectra of InAs,[464] InP,[465] and the XPS measurement of Si:H nanosolids.[13]

## 7.6 Summary

We have thus developed a consistent understanding of the factors dominating the entire band-structure change of nanostructured solids by incorporating the BOLS correlation to the Hamiltonian of an extended solid of which the Hamiltonian contains the intra-atomic trapping interaction and the inter-atomic binding interaction. Introducing the effect of CN-imperfection to the convention of an extended solid has led to a new Hamiltonian that enlarges the $E_G$ of nanometric semiconductors. This approach allows us to discriminates the contribution from crystal binding from the effect of e-p coupling in determining the $E_G$ expansion and PL blue shift. In addition, we have shown that the conventional band theories are still valid for a nanosolid that contains numerous atoms in the form of multiple trapping centers in the energy box. It is anticipated that the spontaneous contraction of chemical bond at surface originates the size dependency of a nanosolid as all the detectable quantities are functions of interatomic binding energy. Therefore, the CN-imperfection induced bond contraction and the rise in the surface-to-volume ratio with reducing particle size originate the change of the band features of nanometric semiconductors and the performance of electrons, phonons and photons in the small particles. Agreement between modeling predictions and the observed size-dependency in the PL of Si and some nanometric III-V and II-VI semiconductors evidence further the significance of atomic CN imperfection and the BOLS correlation.

## 8 Electronic energy

### 8.1 Core bands: intra-atomic trapping and crystal binding

#### 8.1.1 Observations

Unlike the valence DOS that provides direct information about charge transportation kinetics during reaction,[1] the energy shift of a core level of an isolated atom gives profound information about the intensity of crystal binding that is dominated by inter-atomic interaction. Alteration of bond nature and variation of bond length will affect the crystal field and hence shift the core level by a certain extent towards normally higher binding energy if the processes are spontaneous. Being able to discriminate the crystal binding (core level shift) from the atomic trapping (core level of an isolated atom) of a core electron under various physical and chemical environment is a great challenge, which is beyond the scope of direct measurement using currently available probing technologies. Combining the most advanced laser cooling technology and XPS, one can measure the energy separation between different energy levels of the slowly-moving gaseous atoms trapped by the laser beams but yet the individual core-level energy of an statically isolated atom.[109] What one can measure using XPS are the convoluted broad peaks of the core-bands contributing from atomic trapping, crystal binding, crystal orientation, surface relaxation or nanosolid formation and the effect of surface passivation.

Figure 32 (link) Illustration of the positive shift ($S_1$, $S_2$, …, B) of the core-band components with respect to the energy level of an isolated atom, $E_v(1)$. $\Delta E_v(S_i) = \Delta E_v(\infty)[1 + \Delta_i]$. Measurements show that the intensities of the low-energy bulk component often decrease with incident beam energy and with the increase of the angle between the incident beam and surface normal.[107]

In addition to the well-known chemical shift caused by the core-hole ‘screening’ due to charge transportation in reaction, relaxed atomic layer spacings at a surface can split the core-level of a specimen into a few components, as illustrated in Figure 32. However, the assignment for the components induced by surface relaxation is quite confusing, as summarized in Table 11, due to the lack of guidelines for determining which peak arises from the surface and which one comes from the bulk. With the widely used sign convention, a positive shift relates the high-energy component to the surface contribution ($S_i$, i = 1, 2, …, B) while the low-energy component to the bulk origin (B) (Figure 32). The resultant peak is often located in between the components and the exact position of the resultant peak varies with experimental conditions, which is perhaps why the recorded values for the core-level energy of a specimen vary from source to source. XPS measurements[533,475,510,511,512] reveal that the intensity of the low-energy component often increases with the incident beam energy or with decreasing the angle between the incident beam and the surface normal in the XPS measurement (Figure 32). The intensity of the low-energy component also increases with decreasing the surface atomic density under the same beam conditions (energy and incident angle). For example, at 390 eV beam energy, two $3d_{5/2}$ components at 334.35 and 334.92 eV have been identified from Pd(110, 100, 111) surfaces. The lower 334.35 eV peak intensity decreases with the variation of the surface geometry from (110) to (111)[513] (with atomic density $n_{110} : n_{100} : n_{111} = 1/\sqrt{2} : 1 : 2/\sqrt{3}$). The 306.42 eV component of the Rh(111) $3d_{5/2}$ level measured under 380 eV beam energy is relatively higher than the same peak of Rh(110) measured using 370 eV beam energy compared with the high-energy component at 307.18 eV.[513] The energy of individual component should be intrinsic disregarding the surface atomic density but the resultant peak changes with crystal orientation due to the contribution from the individual component. The dependence of the low-energy-component intensity on the beam conditions and atomic density implies that the surface-relaxation induces most likely positive shift in the XPS measurement due to the varied penetration depth of the incident beams.

Table 11 Specifications and the possible origins of the surface-induced core-level splitting.

| Specification ($\lvert E_v \rvert$: high $\rightarrow$ low) | Samples |
|---|---|
| Positive shift:<br>$S_1$, $S_2$, …, and B | Nb(001),[475,510] graphite,[533]<br>Tb(0001)-4f,[511] Ta(001)-4f,[512]<br>Ta(110),[514] Mg($10\bar{1}0$),[515]<br>Ga(0001)[516] |
| Negative shift:<br>B, $S_4$, $S_3$, $S_2$ and $S_1$<br>B, $S_2$, $S_3$, $S_4$ and $S_1$ | Be(0001),[517] Be($10\bar{1}0$),[515,518,519]<br>Ru($10\bar{1}0$),[520] Mo(110),[521]<br>Al(001),[522] W(110),[523]<br>W(320),[524] Pd(110,100, 111)[513] |
| Mixed shift:<br>$S_1$, B, $S_{dimer\text{-}up}$,$S_{dimer\text{-}down}$<br>$S_2$, B, $S_1$<br>$S_1$, B, $S_2$<br>$S_2$, $S_3+S_4$, $S_1$, B | Si(111),[525] Si(113)[526]<br>Ge(001)[527]<br>Ru(0001)[528]<br>Be($10\bar{1}0$)[529] |

Upon reacting with electronegative elements such as oxygen, the core-level also splits with a production of high-energy satellite. This well-known 'chemical shift' arises from core-hole production due to bond formation that weakens the 'screening' of the crystal field acting on the specific core electrons. Interestingly, the effects of surface relaxation and chemical reaction on the core-level shift can be distinguished easily. For instances, two distinct Ru-$3d_{5/2}$ core-level components were resolved from a clean Ru(0001) surface due to the relaxation. Both components then shift up simultaneously further by up to 1.0 eV upon oxygen adsorption.[525] The Rh-$3d_{5/2}$ core-level of Rh(100) surface has a split of 0.65 eV relative to the main peak of the bulk, while with oxygen addition both of the components shift 0.40 eV further towards high binding energy.[530] XPS spectra in Figure 33 (c) shows the Ta-4*f* spectra taken after removing about 30% (upper) and 50% (lower) of the surface nanoparticles of a gate device by sputtering method,[476] respectively. The first pair of doublets (Ta-$4f_{5/2}$ and Ta-$4f_{7/2}$) at (23.4 and 26.8 eV) arises from $TaSi_x$ and $Ta_2O_5$, respectively. The second pair of doublets at (31.6 and 34.5 eV) is the corresponding satellites due to the size effect. Size effect that can be weakened by removing the nanoparticles causes a simultaneous shift of both the oxide coated nanoparticle and its metallic environment by about eight eV (bold arrow). These observations confirm that both surface relaxation and catalytic reaction could shift the core-level positively by an amount that may vary depending not only on the original core-level position but also on the extent of reaction.

When a solid reduces its size down to nanometer scale, the entire core-level features (both the main peak and the chemical satellites) move simultaneously towards higher binding energy and the amounts of shift depend on not only the original core-level position but also the shape-and-size of the particle. This trend has been confirmed with XPS on the size-dependence of the main core-level peaks and the oxide satellites, as introduced in section 8.1. The trend of Au-4f core-level shift coincides with the change of the inverse capacitance of the Au particles measured using STS.[531] Compared with the mono-peak of S-2p and S-2s core bands of a bulk solid, ZnS and CdS nanosolids exhibit three components of each the S-2p and S-2s core band.[178,532] These components have been ascribed as the contribution, from high to low binding energy, from the outmost capping layer (0.2 ~ 0.3 nm thick), surface layer (0.2 ~ 0.3 nm thick), and the core of the nanosolid, as shown in Figure 33 (a) and (b). This specification is in accordance with the surface positive shift. The energy value of each component changes insignificantly with particle size but the resultant peak varies considerably with the atomic portions of the capping, surface and the core of the nanosolid. For example, when the particle size is reduced, the intensity of the core component decreases while the capping component increases, which follows the size-dependence of the surface-to-volume ratio of a nanosolid. This convention has enabled an effective method of determining the particle size to be developed,[178,532] which is competent with transition electron microscopy (TEM) and XRD.

Figure 33 (link) XPS spectra (a) S-2p and (b) S-2s of CdS nanosolids show the core (1), surface (2) and capping (3) features. The intensity of feature (1) decreases whereas the intensity of feature (2) and (3) increases with the decreases of particle size. XPS profiles in (c) show both the size and oxidation effect on the core level shift of Ta oxide. [178,532]

Generally, the core level shift of a nanosolid follows the scaling relation with the slope *B* that changes depending on surface treatment, particle dimensionality and particle-substrate interaction.[99] Photoemission from highly oriented pyrolytic graphite (HOPG)[533] shows two $C_{1s}$ components separated in binding energy by 0.12 eV. The higher binding-energy component of the $C_{1s}$ is ascribed to electrons of atoms in the outermost atomic layer and the other to the bulk. The Cu-$2p_{3/2}$ peak of Cu nanosolids deposited on HOPG and CYLC (polymer) substrates,[534] the Au-4f peak of Au nanosolids deposited on Octanedithiol,[531] $TiO_2$[535] and Pt(001)[536] substrates as well as the Pd/HOPG[482] follow exactly the scaling relation. Therefore, as physical origin (without charge transport being involved), surface relaxation and nanosolid formation play the equivalent yet unclear role in splitting and shifting the core-levels of a specimen.

### 8.1.2 Outstanding models

The underlying mechanism for the surface- and size-induced core-level shift is under debate with the following major arguments:

(i) The high-energy component of the core level shift was attributed to the surface interlayer contraction.[533,512] For Nb(001)-$3d_{3/2}$ example, the first layer spacing was found to contract by 12% with an association of 0.50 eV core level shift.[475] A (10 ± 3)% contraction of the first layer spacing has caused the Ta(001)-$4f_{5/2(7/2)}$ level to shift by 0.75 eV.[512] The corresponding positive shift was explained as the enhanced interlayer charge density and the enhanced resonant diffraction of the incident irradiation light due to the surface bond contraction. [512,475,533]

(ii) The size-induced Cu-2p core level shift of CuO nanosolid was ascribed as the size-enhanced *ionicity* of copper and oxygen.[509] This suggestion means that an oxygen atom bonds more strongly to the Cu atoms in a nanosolid than does the oxygen atom to the Cu atoms inside the bulk.

(iii) The size-enhanced Sn-3d, Sn-4d and Ta-4f core level shift of the O-Sn and O-Ta covered metallic clusters was considered as the contribution from the *interfacial dipole* formation between the substrate and the particles.[476] The number of dipoles or the momentum of the dipoles was expected to increase with reducing particle size.

(iv) The thermo-chemical or the 'initial (neutral, un-ionized specimen with n electrons) - final (radiation beam ionized specimen with n-1 electrons) states' model[510,511] defines the core-level shift as the difference in cohesive energy that is needed to remove a core electron either from a surface atom or from a bulk atom. The surface atom is assumed as a 'Z+1 impurity' sitting on the substrate metal of Z atomic number. The final states of atoms at a flat surface or at the curved surface of a nanosolid were expected to increase/decrease while the initial states to decreases/increase when the particle size is reduced. Although often derives the negative or the mixed surface shift in theoretical calculations this model has been elegantly accepted.

(v) Experimental investigations[537,538] have shown that the 'initial-final states' effects cannot explain all the observations and that a metal-to-nonmetal transition mechanism was suggested to occur with a progressive decrease in cluster size.[12] The increase in the core-level binding energy in small particles was also attributed to the poor screening of the core-hole and hence a manifestation of the size-induced metal-nonmetal transition that happens at particle size in the range of 1-2 nm diameter consisting of 300 ± 100 atoms.[482] However, the metal-insulator transition for Au nanoparticles deposited on diamond is excluded based on an XPS observation and the occurrence of the band offset was assigned to the range of cluster sizes.[539]

(vi) Yang and Wu[540] investigated the core-level shifts in sparse Au clusters on oxides, Au/MgO(001) and Au/$TiO_2$(110), with a varying coverage and in the presence of surface oxygen vacancies, by using the DFT full-potential-linearized augmented plane-wave method. The final-state effects are treated self-consistently by moving one core electron to the valence band. They concluded that it is not the final-state contribution but the presence of surface O vacancies that causes the positive core-level shifts in Au nanosolids.

Briefly, signs show that surface relaxation and nanosolid formation share indeed common yet unclear origin in splitting and shifting the core-level to higher binding energy. However, definition of the components is quite confusing and the origin for the surface- and size-induced core-level shift is under debate. Therefore, consistent understanding of the effect of surface relaxation and nanosolid formation on the core-level shift is therefore highly desirable. The BOLS correlation mechanism allows us to unify the core-level shift to the origin of atomic CN-imperfection and the associated rise of binding energy density in the relaxed surface region on the electronic properties of a surface and a nanosolid.

### 8.1.3 BOLS formulation

According to the band theory and the BOLS correlation, the surface relaxation induced and the size-induced shift of the energy level of an isolated atom $E_v(1)$ follows the same relation (l = i, j) to the $E_G(\infty)$ expansion:

$$\begin{cases} V(\Delta_l) = V_{atom}(r) + V_{cry}(r)[1+\Delta_l] & (a) \\ E_v(\Delta_l) = E_v(1) + [E_v(\infty) - E_v(1)](1+\Delta_l), & or \\ E_v(\Delta_l) = E_v(\infty) + [E_v(\infty) - E_v(1)]\Delta_l & (b) \end{cases} \tag{80}$$

where $E_v(\infty) - E_v(1) = \Delta E_v(\infty)$ being equivalence of $E_G(\infty)$ is independent of crystal size, surface relaxation, or chemical reaction. $\Delta_l$ can be expressed as

$$\Delta_l = \begin{cases} \Delta_i(S_i) = \dfrac{\varepsilon_i - \varepsilon_0}{\varepsilon_0} = c_i^{-m} - 1 & (surface) \\ \Delta_j(K_j) = \sum_{i\le 3} \gamma_{ij}\Delta_i & (Nanosolid) \end{cases} \tag{81}$$

$\Delta_l$ is the contribution from interlayer bond contraction ($\Delta_i$) or its sum over the outmost two or three atomic layers ($\Delta_j$). At the lower end of the size limit, the perturbation to the Hamiltonian of a nanosolid relates directly to the behavior of a single bond, being the cases of the outermost surface layer and a monatomic chain. Thus, we have the relation for the relaxed surface,

$$\frac{E_v(\Delta_l)-E_v(1)}{E_v(\Delta_{l'})-E_v(1)}=\frac{1+\Delta_l}{1+\Delta_{l'}},(l'\neq l),or$$

$$\frac{E_v(\Delta_l)-E_v(\infty)}{E_v(\Delta_{l'})-E_v(\infty)}=\frac{\Delta_l}{\Delta_{l'}},(l'\neq l)$$

(82)

Not surprisingly, given an XPS profile with clearly identified $E_v(\Delta_i)$ and $E_v(\infty)$ components of a surface ($l$ = i = 1, 2, …B), or a set XPS data collected from a certain type of nanosolid of different sizes ($l$ = j = 1, 2, …), one can calculate easily the energy level of an isolated atom, $E_v(1)$, and the bulk shift, $\Delta E_v(\infty)$ as well, with the following relations derived from Eq (82):

$$\begin{cases} E_v(1) & =\dfrac{(1+\Delta_{l'})E_v(\Delta_l)-(1+\Delta_l)E_v(\Delta_{l'})}{\Delta_{l'}-\Delta_l}, \quad (l\neq l') \\ \Delta E_v(\infty) & =E_v(\infty)-E_v(1) \end{cases}$$

or,

$$E_v(1)=E_v(\infty)-\frac{E_v(\Delta_l)-E_v(\infty)}{\Delta_l}$$

(83)

If $l$ (> 2) components are given, the $E_v(1)$ and the $\Delta E_v(\infty)$ should take the mean value of the $C_l^2 = l!/[(l-2)!2!]$ possible combinations with a standard deviation σ as both of the $E_v(1)$ and the $\Delta E_v(\infty)$ are independent of particle dimension or surface relaxation. Chemical reaction changes neither these two quantities. Accuracy of determination is subject strictly to the XPS data calibration and the bond length that may not always follow exactly the BOLS specification (section 2). Nevertheless, furnished with this approach, we would be able to elucidate, in principle, the core level positions of an isolated atom and the strength of bulk crystal binding using the conventional XPS measurement.

8.1.4 Verification: single energy level

- Surfaces

The $E_v(1)$ and $\Delta E_v(\infty)$ values of several surfaces have been derived based on the XPS database and Eq (83).[107] As listed in Table 12, the small deviation σ values evidence that the BOLS correlation describes adequately the real situations and that the parameters of $m$ and $z_i$ represent the true situations. Interestingly, a slight refinement of the mid-component $E_v(S_2)$ within the XPS resolution reduces the σ values to less than 0.1%, which indicates that the XPS precision is critical and the developed method is sensitive and reliable.

Results show that the crystal binding is stronger to the electrons in the outer shells than to the electrons in the inner ones. For example, the binding to the C-1s electrons is weaker (~0.8 eV) than the binding to the Be-1s (~5.6 eV) electrons. The former is screened by the four $2s^2 2p^2$ electrons and the latter by the two $2s^2$ electrons only.

Table 12 Calculated atomic $E_v(1)$, bulk shift $\Delta E_v(\infty)$, and the standard deviation σ for different surfaces based on available XPS database.[107] For elemental surface, m = 1. $z_1$ = 4, $z_2$ = 6 and $z_3$ = 8 are used in calculation. Refinement of the $E_v(S_2)$ within XPS resolution reduces the σ to < 0.1%, indicating the importance of accuracy in XPS calibration.

| Surface | XPS components | | | Calculation results | | |
|---|---|---|---|---|---|---|
| | $E_v(S_1)$ | $E_v(S_2)$-refined | $E_v(\infty)$ | $E_v(1)$ | $\Delta E_v(\infty)$ | σ (%) |
| Poly C 1s [533] | 284.42 | - | 284.30 | | | - |
| Pd-$3d_{5/2}$[513] | 334.92 | - | 334.35 | 330.34 | 4.01 | - |
| Rh-$3d_{5/2}$[513] | 307.18 | - | 306.42 | 301.17 | 5.35 | - |
| Ru(0001) | 280.21 | 279.955 | 279.73 | 276.344 | 3.3856 | 0.003 |
| $3d_{1/2}$[528] | | | | | | |
| W(110) $4f_{7/2}$[523] | 31.50 | 31.335 | 31.19 | 29.006 | 2.1835 | 0.003 |
| Nb(100) $3d_{5/2}$[475] | 202.80 | 202.54 | 202.31 | 198.856 | 3.4544 | 0.002 |
| Be(10$\bar{1}$0) 1s [518] | 111.85 | 111.475 | 111.1 | 105.817 | 5.2835 | 0.002 |
| Be(0001) 1s [517] | 111.9 | 111.48 | 111.1 | 105.465 | 5.6350 | 0.007 |

- Nanosolids

BLOS prediction yields a simpler form for elucidating the $E_v(1)$ from a set of data of size dependent core-level shift, which follows the scaling law (Eq (13) in section 2). The $Q(\infty) = \Delta E_v(\infty) - \Delta E_v(1) = B/(\Delta_j \times K_j)$ varies simply with the parameter *m* and the given dimensionality ($\tau$) and size ($K_j$) of the solid because $\Delta_j \propto K_j^{-1}$. There are only two independent variables, *m* and $\Delta E_v(\infty)$, in calculations. If a certain known quantity $Q(\infty)$ in the scaling law, such as the $T_m(\infty)$ or the $E_G(\infty)$, and the measured size dependent $Q(K_j)$ of the considered system are given, the *m* can be readily obtained by equilibrating both the theoretical and experimental scaling law. With the determined *m*, any other unknown quantities $Q(\infty)$ such as the crystal binding intensity, $\Delta E_v(\infty)$, of the same system, and hence the energy level of an isolated atom, $E_v(1)$, can be determined uniquely with the above relations.

The $\Delta E_v(\infty)$ and $E_v(1)$ for Cu-2p, Au-4f and Pd-3d were calculated by using Eq (13). Figure 34 compares the predicted (solid) curves with the measured size dependence of the core levels shifts of these samples (scattered data). In order to find the intercepts and slopes in the scaling relation, all the experimental results were linearized with the Least-root-mean-square optimization method. The intercepts provide calibration of the measurement as the intercepts reflect the space charging effect or the system error. The slopes are the major concern in the current decoding exercises. The $E_v(1)$ and $\Delta E_v(\infty)$ of Cu-2p can be obtained by calculating the Cu/HOPG system with m = 1 using Eq (13). The reason to take m = 1 is that Cu atoms react hardly with the carbon surface at room temperature,[541] and that m = 1 always holds for elemental metallic solid. Decoding gives rise to the atomic trapping energy $E_{2p}(1) = -931.0$ eV for the Cu-2p electrons of an isolated Cu atom and the bulk crystal binding energy $\Delta E_{2p}(\infty) = -1.70$ eV for an extended Cu solid. Taking the obtained Cu-$\Delta E_{2p}(\infty)$ value as reference in simulating the measured size dependent $\Delta E_{2p}(K_j)$ for Cu on CYCL gives m = 1.82, which adds the contribution from the reactivity of Cu to CYCL polymer substrate to the m = 1. Therefore, the change of m value provides means for information about particle-substrate interaction.

For the Au nanosolid, m = 1 has been confirmed in decoding the size dependent melting temperature of Au on C and on W substrates.[68] Fitting the measured $\Delta E_{4f}(K_j)$ of Au on Octan with m = 1 gives the $E_{4f}(1) = -81.50$ eV for an isolated Au atom and $\Delta E_{4f}(\infty) = -2.86$ eV for the Au bulk bonding. Simulations with the derived $\Delta E_{4f}(\infty) = -2.86$ eV as reference reveal that Au growth on $TiO_2$ and on Pt(001) substrates proceeds in a layer-by-layer mode, agreeing with the growth modes as reported by the initial practitioners.[535,536] Simulating the XPS data of both Pd surfaces[513] and Pd/HOPG nanosolids[482] led to the value of $\Delta E_{Pd\text{-}3d}(\infty) = -4.00 \pm 0.02$ eV and $E_{Pd\text{-}3d}(1) = -330.34$ eV.[106] Incorporating m = 4.88 value into the simulation of the measured size dependence of the Si-2p level shift gives the $E_{2p}(1) = -96.74$ eV for a Si atom and $\Delta E_{2p}(\infty) = -2.46$ eV for Si bulk.

Calculation results from counting the capping, surface and the core of the ZnS and CdS nanosolids show that the crystal binding to S-2p of ZnS is stronger than that of CdS, as compared in Table 13, because the Zn-S bond[30] is shorter than the Cd-S[48] bond. The $E_v(1)$ should not change under any circumstances. However, the crystal binding to the same levels of an atom may offset when the atom forms compounds with different elemental atoms. Surface charging also affect the measurement. Therefore, the measured S-2s and S-2p peaks of CdS should shift up or down consistently against the same peaks of ZnS. Compared with the measured S-2s and S-2p peaks from CdS, it can be found that the S-2p peak from ZnS goes slightly down while the S-2s peak floats up with respect to those of CdS. This may cause the $E_{2p}(1)$ values of S in the two samples to vary slightly.

Assuming m = 4 for O-Cu, we can use the measured data[509] to estimate the $E_{2p}(1)$ and the $\Delta E_{2p}(\infty)$ of bulk Cu and bulk CuO, as given in Table 13. The estimated values seem to be too large to be reasonable compared with those obtained from Cu/HOPG or Cu/CYCL. As mentioned earlier, the accuracy is strictly subject to the precision of the XPS data. The modeling predictions agree also with the trends of the core-level shift for O-Sn and O-Ta compound nanosolids, of which both the satellites and the main peaks in the XPS profiles shift towards higher binding energy with reducing particle size.

Figure 34 (link) Comparison of the BOLS prediction with the measured size dependence of the core level shift. (a) Thiol-caped Au[199] and Au on Octan[531] shows three-dimensional features while Au on $TiO_2$[535] and on Pt[536] show plate pattern of formation. (b) Pd on HOPG substrate.[482] The different m values in (c) of Cu on HOPG and CYCL[534] indicate the contribution from the reaction between Cu nanosolid and polymer CYCL substrate. (d) Core-level shift and bandwidth of CuO nanoparticles.

Table 13 Calculated atomic $E_v$, bulk shift $\Delta E_C(\infty)$ and the standard deviation σ for different nanosolids based on available XPS database. For compounds, m = 4, $z_1 = 4$, $z_2 = 6$, and $z_3 = 8$ are used in calculation.

| Nanosolid | XPS measurement | | | Calculated | | |
|---|---|---|---|---|---|---|
| | $E_v$(Cap) | $E_v$ (Surf) | $E_v$ (core) | $E_v(1)$ | $\Delta E_v(\infty)$ | σ |
| CdS-S $2p_{3/2}$[532] | 163.9 | 162.7 | 161.7 | 158.56 | 3.14 | 0.002 |
| ZnS-S 2p [178] | 164.0 | 162.4 | 161.4 | 157.69 | 3.71 | 0.002 |
| CdS-S 2s [532] | 226.0 | 224.7 | 223.8 | 220.66 | 3.14 | 0.001 |
| ZnS-S 2s [178] | 229.0 | 227.3 | 226.3 | 222.32 | 3.92 | 0.001 |
| CuO-Cu $2p_{3/2}$ [509] | 936.0/934.9 (4/6 nm) | 932.9 (25nm) | 932.1 (Bulk) | 919.47 | 12.63 | 0.36(2.8%) |
| CuO-Cu $2p_{3/2}$ Refined data | 935.95/934.85 (4/6 nm) | 932.95 (25 nm) | 932.1 (Bulk) | 919.58 | 12.52 | 0.30 (2.4%) |

Table 14 The $E_v(1)$ of an isolated Au, Cu and Si atom and the crystal binding energy of $\Delta E_v(\infty)$ obtained from decoding the size-dependent $E_v(K_j)$ of the corresponding nanosolids.[107]

| | Au/Octan | Au/$TiO_2$ | Au/Pt | Cu/HOPG | Cu/CYCL | Si | Pd |
|---|---|---|---|---|---|---|---|
| m | 1 | | | | 1.82 | 4.88 | 1 |
| τ | 3 | 1 | 1 | 3 | 3 | 3 | 3 |
| $d_0$/nm | 0.288 | | | 0.256 | | 0.263 | 0.273 |
| $E_v(\infty)$/eV | -84.37(4f) | | | -932.7(2p) | | -99.20(2p) | -334.35(3d) |
| $E_v(1)$/eV | -81.504 | -81.506 | -81.504 | -931.0 | | -96.74 | -330.34 |
| $\Delta E_v(\infty)$/eV | -2.866 | -2.864 | -2.866 | -1.70 | | -2.46 | -3.98 |

- Conductor-insulator transition

In the current modeling approach, we have found that the interfacial bond nature (character m) changes with atomic CN. For instance, the m value for Sn and Ga nanosolids increases from on to seven when the solid size decreases from the bulk to the lower end of the size limit, as shown in section 5.3. For an isolated metallic nanosolid, the metallic bond suffers from relaxation due to CN imperfection but no nature alteration if chemical process is without being involved. Metal-nonmetal transition may happen at a certain critical size, 1 ~ 2 nm. Such transition was suggested to originate from the Kubo-gap expansion in which no bond character is involved. However, from the bond relaxation perspective, the bonds near the surface region become shorter and stronger and the trapping potential wells become deeper. The BOLS correlation indicates that it is the deepened trapping potential well that confines the moving electrons to be more localized, and hence, the conductivity of the metallic nanosolid becomes lower.[542] At the lower end of the size limit (1 ~ 2 nm) there is no core exists, all the bonds will contract by 20 ~ 30% associated with 30 ~ 50% deepening of the trapping potential wells, see Figure 5 in section 2. As a complementary mechanism to the Kubo gap expansion, the BOLS correlation may provide a scenario in real space to explain why the conductivity is reduced and how the conductor-insulator transforms for a metallic nanosolid, though the local change density (work function) is reduced.

## 8.2 Work function

### 8.2.1 Chemical modulation

The work function (Φ) of a specimen is the energy required to get an electron from or to add an electron into the surface. The Φ or the threshold in cold-cathode field emission of materials such as diamond, diamond like carbon (a-C) or carbon nanotubes (CNTs), can be chemically modulated/enhanced by doping the materials with proper amount of properly selected impurities besides the geometric enhancement of the emitters.[543] It has been realized that co-doping O or N with low-Φ metals to form metal dipoles at the surface could be promising route[139,544,545] in lowering the Φ at the surface. For example, ZnO nanopins and Ga-doped ZnO nanorods show a low field emission threshold ~2.0 V/mum at a current density of 0.1 μAcm$^{-1}$.[546,547] The lone-pair induced antibonding dipole states that are located at energy levels higher than the Fermi level are responsible for the Φ reduction. However, the production of a H-like bond at the surface due to O or N over-dosing may have detrimental effects on the Φ reduction,[1] such as the case of carbon nanotubes with over-doped oxygen.[548] Lower doses of oxygen to the tubes improve significantly the field emission characteristics while overdosing with oxygen makes electron emission difficult. Therefore, appropriate amounts of impurity density are necessary[545] to avoid H-like bond formation that narrows the antibonding band. [139,549] The chemical effect on the Φ has been intensively discussed in a previous report on the electronic process of oxidation and nitridation.[1]

### 8.2.2 Geometric modulation

The current BOLS correlation argument indicates that the bond contraction not only deepens the atomic potential well but also enhances the charge density in the relaxed surface region. The confined electrons near the surface edge are denser and more localized. For an isolated nanosolid of size $K_j$, the Φ satisfies ($V \propto d^{\tau}$):[139]

$$\Phi = E_0 - E_F;\ \text{and}\ E_F \propto n^{2/3} = \left(N_e/V\right)^{2/3} \propto \left(\overline{d}\right)^{-2\tau/3} \tag{84}$$

The total number of electrons $N_e$ of a nanosolid is conserved. At the lower end of the size limit of a spherical or semispherical dot (R ~ 1 nm, τ = 3), the average bond length is around 20% shorter than the bulk value and hence the Φ will reduce from the original value by 30% ($E_F$ shifts up by $0.8^{-2}$-1), according to Eq (84). Using He-II ultraviolet beam source of 21.2 eV, Abbot et al [550] measured the Φ of diamond {111} surface to be about 4.8 eV at grain size of 108 μm, as shown in Figure 35a. The Φ of the diamond decreases with particle size to a minimum of 3.2 eV at an average grain size of about 4 μm, and then the Φ recovers to a maximum of 5.1 eV at diamond particle size of 0.32 μm. Rouse et al [272] measured at room temperature that the field-emission threshold decreases from 3.8 to 3.4 V/μm of polycrystalline diamond films on molybdenum tips as the diamond average grain size increases from 0.25 to 6 μm. They related the Φ change to the increases of negative electron affinity within the grain due to increased surface hydrogen bonding and with perhaps a contribution from surface defect states. The Φ of Na particles around 0.4 ~ 2.0 nm size was measured to vary inversely with the size R and lowered the bulk value from 2.75 to 2.25 eV (by 18%).[551] Majority of the nanotubes have a Φ of 4.6–4.8 eV at the tips, which is 0.2–0.4 eV lower than that of carbon (graphite) bulk. A small fraction of the nanotubes have a Φ of ~5.6 eV, about 0.6 eV higher than that of carbon (graphite). This discrepancy is thought to arise from the metallic and semiconductive characteristics of the nanotubes. The average Φ of porous Si with different crystalline columnar dimensions was measured using a retarding field diode method to increase as the crystalline size decreases.[552] The variation of the Φ was attributed to the etching effect and the formation of impurity Si-H, Si-O and Si-H-O bonds at the surface.[552]

It appears that the measured size-dependent Φ change for diamond is in conflicting with the BOLS prediction. However, one needs to note that if the emitters are packed too closely, the system is identical to a smooth surface. It has been found[553] that hydrogen-rich or oxygen-containing CVD precursors cold promote electron emission from discrete diamond particles and *non-continuous* diamond films but not for high quality and continuous diamond films, nanocrystalline diamond, and glassy carbon coatings even if they contain conductive graphitic carbon. The Φ at the tips of individual multi-walled carbon nanotubes was measured using a TEM to show no significant dependence on the diameter of the nanotubes in the range of 14 – 55 nm.[554] Although the calibrated diamond particles are much larger the curvature of the tips should be much higher. The particle size corresponds only to the separation of the sharp emitters. This phenomenon indicates the significance of CN-imperfection on the Φ reduction that is subject to the separation between the nanoparticles, and surface chemical states.

Figure 35 (link) Size modulated work function of (a) diamond[550] and (b) Na[551] nanocrystals. Recovery of the work function is due to the geometrically flatness.[553]

### 8.2.3 Hydrophobic-hydrophilic transition

The wettability that governs the surface chemical states and geometric structures is an important factor influencing the properties of functional materials. Special wettabilities, such as superhydrophilicity and superhydrophobicity, have aroused great interest in recent years because of their advantages in applications, such as anti-contamination, anti-oxidation, and prevention of current conduction. Superhydrophobicity and superhydrophobicity have been observed by Jiang et al  on as grown and alkylfluorosilane (NaF) modified carbon nanotubes[555] and other aligned nanostructures.[556] Interestingly, reversible switching between superhydrophilicity and superhydrophobicity through constructing special surface structures on the respective surfaces becomes possible by surface conditioning.[557,558] On a poly(N-isopropylacrylamide)-modified rough silicon substrate,[557] switching from superhydrophobicity to superhydrophilicity can be achieved at temperatures 302 - 313 K because the inverse competition between intermolecular and intramolecular hydrogen bonding in the polymer chains. While on an aligned ZnO nanorod surface,[558] switching can be achieved by UV irradiation (365 ± 10 nm) and dark aging, which is considered the result of the reversible generation and annihilation of the photo-generated surface oxygen vacancies.

The combination of the BOLS and the BBB correlation[1] premises may provide a possible complementary mechanism in terms of bond formation and relaxation for the superhydrophilicity and superhydrophobicity as identified by Jiang and coworkers.[555-558] As demonstrated, charge densification in the relaxed flat or curved surface region could lower the local work function by as high as 30% at the expense of raising the chemical potential ($E_F$) of the nanosolid. Furthermore, sp-orbital hybridization of electronegative elements such as N, O, and F will produce nonbonding lone pairs that polarize electrons of the neighboring atoms to form antibonding dipoles.[1] Dipole formation lowers the local work function by about 1.2 eV for N and O involvement. The joint effect of nanostructures and antibonding states could be responsible for the superhydrophobicity arising from the raised chemical potential. Warming up the passivated samples to a certain temperature,[559] irradiated by light of a certain wavelength, or bombarded by energetic beams, dehybridization occurs. External stimulus such as heating breaks the lone pairs and hence the antibonding dipoles, as observed for O-Cu(001) surface of which the lone pair DOS feature disappears upon annealing at “dull red color”. The lone pair DOS feature of the annealed O-Cu(001) surface is restored after cooling down and aging for some while. If overdosed with electronegative additives to the surface, H-like bond may form rendering the antibonding dipoles and restore the work function, as the dipoles become positive ions due to charge transport. This happening is the case of carbon nanotubes dosed with oxygen.[548] Small amount of oxygen lowers the work function while overdosing with oxygen raises the work function to a value that is even higher than the undoped case. Changing the dosage of electronegative elements could be a new manner in which to switch the superhydrophilicity and superhydrophobicity nature, and further experimental verification would be required.

### 8.2.4 Mechanical modulation

Amorphous carbon (a-C) films have an uniquely intrinsic stress (~12 GPa) that is almost one order of magnitude higher than those found in other amorphous materials such as *a*-Si, *a*-Ge, or metals (<1 GPa).[560] Although it is known from theoretical studies[561] that by applying pressure to a material one can modify its electronic properties, e.g., band structure, resistivity, work function, etc, the influence of the intrinsic stress on the electron emission properties of a material has not been clear so far. Poa et al [202,560] investigated electron emission from highly compressive carbon films obtained by bombardment of noble gas plasma and found correlation between the stress and the threshold field for electron emission, as shown in Figure 36. By carefully controlling deposition conditions, they vary the internal stresses from 1 to 12 GPa, which is associated with suppression of the electron emission threshold field. The lowering of the threshold field is related to the enhanced stress that pushes the π and the π* bands together with a reduced gap between them to even an overlap by gathering the $sp^2$ clusters closer to each other. Such a band overlap increases the electron conductivity and hence the drop of the threshold. On the other hand, the "*c*-axis" spacing of the $sp^2$ clusters is likely to be smaller than that of crystalline graphite under the intrinsic stress. The reduced lattice spacing will densify the charge in the shrieked region, which suppresses internally the threshold field. Applying an external stress by bending a-C films or carbon nanotubes has the same effect on reducing the threshold for electron emission.[562] However, the threshold will restore when it reduces to a certain value if further stress is applied.

Figure 36 (link) Correlation between the threshold field and intrinsic stress of amorphous carbon as a function of the assisting energy with different noble gases.[202]

Using extended near-edge XAFS, Lacerda et al [563] investigated the effect of trapping noble gases (Ar, Kr, and Xe) in an a-C matrix on the internal stress of the a-C films. When one to 11 GPa internal stress is generated by controlling the size of the pores within which noble gases are trapped, they found ~ 1 eV lower of the core level binding energy of the entrapped gases associated with 0.05 nm expansion of the atomic distance of the noble gases. For Ar (Xe), the first interatomic separation varies from 0.24 (0.29) nm to 0.29(0.32) nm in the 1–11-GPa pressure range. This enhancement indicates clearly that the gas entrapped pores expand and the interfacial C-C bonds contract. An external pressure around 11 GPa could suppress the interplanar distance of microcrystalline graphite by ~15%,[564] gathering the core/valence electrons and carbon atoms closer together. The resistivity decreases of *a*-C films when the external hydrostatic pressure is increased.[565] These results are in agreement with the recent work of Umemoto *et al* [566] who proposed a dense, metallic, and rigid form of graphitic carbon with characteristics being very similar to the findings of Poa et al However, the spontaneous lattice contraction could raise the resistivity, instead, as the densified charges are strongly trapped within the lowered potential well though both the intrinsic and extrinsic pressure could densify the mass, charge, and the stress (energy) of a highly $sp^2$ rich *a*-C film.

We may suggest a possible mechanism for the intrinsic and extrinsic stress enhanced threshold field of carbon films. Nanopore formation creates lower-coordinated atoms at the interfaces between the gas-trapped pores and the a-C matrix. The CN-imperfection induced lattice contraction of the host matrix will apply to the pores and hence act to expand atomic distance between the inter-trapped atoms. The bond expansion is associated with weakening of interatomic binding of the noble gas atoms confined in the pores, as observed. The interfacial C-C bond contraction leads to simultaneous enhancement of both the charge density and the internal stress, associated with a drop of the local work function by as high as 30%, as derived in the previous section. Therefore, the internal stress affects the work function by enhancing the local charge density trapped in the deep potential well. From this perspective, a $sp^2$ cluster with a shortened bond (≤0.142 nm) would be beneficial to the field emission properties when compared with a $sp^3$ cluster (0.154 nm bond length) despite the less localized van der walls bond electrons that should add a DOS feature in the mid gap. Therefore, atomic CN imperfection enhances the charge density and hence the magnitude of the N(E).

In contrast, external stress could raise the atomic binding and the total energy between a pair of atoms, being the same in effect as to heating and thus weakening the bond. Therefore, heating or pressing should raise the N(E) higher, and as a consequence, minimize the gap between the π and π* bands, as proposed by Poa et al  Overstressing the specimen bonds tend to break and dangling bonds formed, which adds DOS features in the midgap despite the enlarged gap between the emitter and the grid in measurement. This understanding may provide a possible mechanism for the threshold recovery upon being overstressed. Therefore, mechanism for the external pressure lowered Φ should differ from that of intrinsic pressure though the effects are the same. The intrinsic stress amplifies the N(E) magnitude, raises the resistivity; whereas the external one "pumps" the N(E) up, and raises the conductivity. Lu and coworkers[5,51] have demonstrated that for as grown nanosolid, resistivity increases with the inverse of solid size but under stretching, electrical conductivity comparable retains that of bulk copper, which could be evidence for the recommended effect of intrinsic and extrinsic stress on the conductivity behavior of a metallic nanosolid.

### 8.3 Summary

The BOLS correlation premise has enabled us to unify the core-level physical shift induced by surface-relaxation and nanosolid-formation into the same origin of atomic CN-imperfection. The mechanism of surface interlayer relaxation[533,512] induced positive shift and the specification of the capping and surface layers in CdS and ZnS nanosolid are highly favored. Atomic CN-imperfection also enhances the iconicity of the constituent atoms such as oxygen and metals.[509] The artifacts added to the XPS spectrum due to photovoltaic effect in experiment and the excited final states could be removed by proper calibration of the data. The CN-imperfection enhanced binding intensity acts on the core electrons of an atom disregarding the atomic states whether it is in the neutral initial or the ionized final.

Besides, we have developed an effective yet straightforward method to determine the core-level energies of an isolated atom and hence to discriminate the contribution of crystal binding from the effect of atomic trapping to the core electrons at energy levels shifted by bulk formation, surface relaxation or nanosolid formation. The developed method not only allows the predicted size-dependence of core-level shift to match with observations but also enables the conventional XPS to provide comprehensive information about the behavior of electrons in the deeper shells of an isolated atom and the influence of crystal formation.

Understanding of the effect of intrinsic and extrinsic stress and factors controlling work function, resistivity, and the intrinsic stress should provide guidelines for materials design and fabrication for applications of electron emission and superhydrophilicity – superhydrophobicity transition.

## 9 Dielectric suppression

### 9.1 Background

The complex dielectric constant, $\varepsilon_r(\omega) = \mathrm{Re}[\varepsilon_r(\infty)] + i\mathrm{Im}[\varepsilon'_r(\omega)]$, is a direct measure of electron polarization response to external electric field, which has enormous impact on the electrical and optical performance of a solid and related devices. For example, low $\varepsilon_r(\infty)$ media are required for the replacement of Al with Cu in microelectronic circuitry to prevent the 'cross-talk' between connections while media of higher $\varepsilon_r(\infty)$ are required for the miniaturized conductor-metal-oxide-semiconductor gate devices. Miniaturizing a semiconductor solid to nanometer scale often causes the $\varepsilon_r(K_j)$ to decrease. [104,567,568] The $\varepsilon_r(K_j)$ reduction enhances the Coulomb interaction between charged particles such as electrons, holes, and ionized shallow impurities in nanometric devices, leading to abnormal responses. The increase of exciton activation energy in nano-semiconductors due to $\varepsilon_r(K_j)$ reduction would significantly influence optical absorption and transport properties of the devices. Both the ac conductivity and dielectric susceptibility of amorphous Se films drop with thickness in the range of 15-850.[569] The complex dielectric constant decreases when the frequency is increased and the temperature is decreased in the range of 300 and 350 K. Carrier motion is suggested to be the dominant mechanism in both ac polarization and dc conduction.

The relative change of the dielectric susceptibility, $\chi = \varepsilon_r - 1$, can be modeled as:

$$\Delta\chi(K_j)/\chi(\infty) = \begin{cases} -\left[1 + (K_j/\alpha)^{\lambda}\right]^{-1} & (Penn) \\ \dfrac{-2\Delta E_G(K_j)}{E_G(\infty)} & (Tsu) \\ \dfrac{-2}{1-(E/E_G(\infty))^2}\left(\dfrac{\Delta E_G(K_j)}{E_G(\infty)}\right) & (Chen) \end{cases}$$

where $\alpha$ and $\lambda$ in the Penn's empirical model[570] are freely adjustable parameters that vary from situation to situation as listed in Table 15. Tsu et al [571] related the susceptibility change directly to the offset of $E_G(K_j)$. Considering the contribution from incident photon energy, E = $\hbar\omega$, Chen et al [572,573] modified Tsu's model and studied the dielectric response of nanosolid Si embedded in $SiO_2$ matrix using elipsometry. They suggested that the dielectric suppression varies with the photon beam energy that should be lower than the intrinsic $E_G(\infty)$ of Si. Delerue et al [574] deposited PbSe nanocrystals of a few nanometers in height on an Au(111) substrate and measured the thickness-dependent dielectric function. Compared with electronic structure calculations of the imaginary part of the dielectric function of PbSe nanocrystals they suggested that the size-dependent variation of the dielectric function is affected by quantum confinement at well-identifiable points in the Brillouin zone, instead of the band-gap transition. The size-induced decrease of the average dielectric response is also suggested to be mainly due to the breaking of the polarizable bonds at the surface[575] rather than the $E_G$ expansion or quantum confinement effect. A recent theoretical study[576] of the third-order susceptibility for Ag dielectric composite suggests the saturation of optical transitions between discrete states of conduction electrons in metal dots. Saturation effects lead to a decrease of the local field enhancement factor that is of particular importance for surface-enhanced phenomena, such as Raman scattering and nonlinear optical responses.

One may note that the modified models[571,572] suite only cases of which the $\Delta E_G(K_j)/E_G(\infty) < 0.5$, otherwise, $\chi < 0$, which is physically forbidden, as commented by Chen. Generally, the $E_G$ often expands beyond this critical value such as the case of Si nanorods with $E_G$ = 3.5 eV.[36] Therefore, understanding of dielectric suppression of nanosolid semiconductors is still under debate. Furthermore, the size dependence of the imaginary part of the dielectric constant and of the photoabsorption coefficient needs yet to be established. Therefore, deeper and consistent insight into the origin and a clearer and complete expression for the size dependence of the complex dielectric constant of a nanosolid semiconductor is necessary.

Table 15 Simulation results in Penn's model.

| | $\varepsilon_r$ (bulk) | $\alpha$ /nm | $\lambda$ |
|---|---|---|---|
| CdSe[173] | 6.2 | 0.75 | 1.2 |
| Si-a[477,571] | 11.4 | 2.2 | 2 |
| Si-b[577] | 11.4 | 1.84 | 1.18 |
| Si-c[567] | 10.38 | 0.85 | 1.25 |
| Si-d[567] | 9.5 | 0.69 | 1.37 |

This section presents analytical expressions for the size dependent complex dielectrics by incorporating the BOLS correlation to the Kramers–Kronig relation. The dielectric performance of a nanosolid Si is examined experimentally by measuring the effective dielectrics of p-Si. Consistency between predictions and observations reveals that the complex dielectrics of a nanosolid depends on the crystal binding and electron–phonon coupling that are subject to the BOLS correlation and chemical passivation.[1]

## 9.2 BOLS formulation

### 9.2.1 Electron polarization

Electronic polarization through a process of transition from the lower ground states (valence band, or the mid-gap impurity states) to the upper excited states in the conduction band takes the responsibility for complex dielectrics. This process is subject to the selection rule of energy and momentum conservation, which determines the optical response of semiconductors and reflects how strongly the electrons in ground states are coupling with the excited states that shift with lattice phonon frequencies.[490] Therefore, the $\varepsilon_r$ of a semiconductor is directly related to its band gap $E_G$ at zero temperature, as no lattice vibration occurs at zero Kelvin.

Since the involvement of electron–phonon coupling, electron excitation from the ground states to the excited upper states is complicated, as illustrated in Figure 28. The energy for photon absorption, or energy difference between the upper excited state $E_2(q)$ and the lower ground state $E_1(q)$, at $q$ is given as:

$$\begin{aligned} \hbar\omega &= E_2(q) - E_1(q) \\ &= E_G - Aq_0^2 + 2Aqq_0 \\ &= E_{PL} + 2Aqq_0. \end{aligned} \tag{85}$$

The imaginary part, $\varepsilon'_r(\omega)$, describes the electromagnetic wave absorption and is responsible for the energy loss of incident irradiation through the mechanism of electron polarization. The $\varepsilon'_r(\omega)$ can be obtained by inserting the gradient of eq (85) into the relation,[285,578]

$$\begin{aligned} \varepsilon_r'(\omega) &= \frac{F}{\omega^2}\int ds \frac{f_{CV}}{|\nabla[E_C(q) - E_V(q)]|} \\ &= \frac{\pi F f_{CV}}{A\omega^2} q \\ &= \frac{\pi F f_{CV}}{2A^2} \frac{\hbar\omega - E_{PL}}{q_0\omega^2} \propto \frac{\hbar\omega - E_{PL}}{q_0\omega^2}, \end{aligned}$$

where the gradient and the elemental area for integral are derived as follows:[94]

$$\begin{aligned} \nabla[E_C(q) - E_V(q)] &= 2Aq_0 \\ ds &= 2\pi q_0 dq \end{aligned} \tag{86}$$

The $s$ is the area difference of the two curved surfaces in q space of the upper excited band and the lower ground band. $F$ is a constant. $f_{CV}$, the probability of inter–subband (Kubo gap) transition is size dependent. However, the size–induced change of transition probability between the sublevels is negligibly small, and for the first order approximation, $f_{CV}$ is taken as constant.

### 9.2.2 Complex dielectrics

- Dielectric susceptibility

The Kramers–Kronig relation correlates the real part to the imaginary part by,[579]

$$\begin{aligned}\varepsilon_r(\infty)-1=\chi &= \frac{2}{\pi}\int_{\omega_0}^{\infty}\frac{\varepsilon_r'(\omega)}{\omega}d\omega \qquad (\omega_0 = E_{PL}/\hbar)\\ &= \frac{Ff_{CV}}{A^2 q_0}\int_{\omega_0}^{\infty}\frac{\hbar\omega - E_{PL}}{\omega^3}d\omega\\ &= \frac{G}{q_0 E_{PL}}, \qquad (G=\hbar^2 Ff_{CV}/2A^2)\end{aligned} \tag{87}$$

where $\hbar\omega - E_{PL} = 2Aqq_0$ as given in eq (85). Hence, the size–depressed dielectric susceptibility depends functionally on the characteristics of e-p interaction and the photoluminescence energy. Using the relation of $\Delta E_{PL}(K_j)/E_{PL}(\infty) = \Delta_H - B\Delta_{e-p}$ (section 7), the size-induced relative change of both the $\chi$ and the $\varepsilon'_r(\omega)$ can be obtained as:[94]

$$\begin{aligned}\frac{\Delta\chi(K_j)}{\chi(\infty)} &= -\frac{\Delta E_{PL}(K_j)}{E_{PL}(\infty)} - \frac{\Delta q_0(K_j)}{q(\infty)}\\ &= -\frac{\Delta E_{PL}(K_j)}{E_{PL}(\infty)} + \frac{\Delta d_i(K_j)}{d_0}\\ &= -(\Delta_H - B\Delta_{e-p}) + \Delta_d,\\ \frac{\Delta\varepsilon_r'(K_j,\omega)}{\varepsilon_r'(\infty)} &= \frac{-E_{PL}(\infty)}{\hbar\omega - E_{PL}(\infty)}\frac{\Delta E_{PL}(K_j)}{E_{PL}(\infty)} + \frac{\Delta d_i(K_j)}{d_0}\\ &= \frac{-E_{PL}(\infty)}{\hbar\omega - E_{PL}(\infty)}(\Delta_H - B\Delta_{e-p}) + \Delta_d\end{aligned} \tag{88}$$

where $B$ is the e-p coupling coefficient. $\Delta_H$ and $\Delta_{e-p}$ represent the contribution from the CN–imperfection perturbed Hamiltonian and the e-p coupling in the relaxed region. The last term is the bond length change ($q \propto d^{-1}$). They are given as:[33]

$$\begin{cases}\Delta_H &= \sum_{i\le 3}\gamma_{ij}(c_i^{-m}-1), \quad (Hamiltonain - perturbation)\\ \Delta_{e-p} &= \sum_{i\le 3}\gamma_{ij}(c_i^{-2}-1), \quad (e-p-coupling)\\ \Delta_d &= \sum_{i\le 3}\gamma_{ij}(c_i-1), \quad (bond-contraction)\end{cases} \tag{89}$$

For a spherical silicon dot, $B = 0.91$, $m = 4.88$, $z_2 = 6$, and $z_3 = 12$. Compared with the relations given in (88), the complex dielectric performance of a nanosolid semiconductor depends functionally on crystal binding and e-p coupling. The imaginary dielectric constant depends also functionally on the photon energy. Both components drop with solid size, which follow the BOLS correlation.

- Direct and indirect band transition

For direct and indirect band-gap optical transition, the $\varepsilon'_r(\omega)$ can be traditionally simplified as: [104,580]

$$\begin{aligned}\varepsilon_r'(\omega) &= \begin{cases}\frac{B'}{\omega^2}(\hbar\omega - E_G)^{1/2} & (direct - E_G)\\ A'(T)(\hbar\omega - E_G)^2 & (indirect - E_G)\end{cases}\\ B' &= \pi(2\mu/\hbar^2)^{\frac{3}{2}} f_{cv} A \qquad ,\\ &(\hbar\omega > E_G)\end{aligned}$$

(90)

where A'(T) containing parameters for band structure and temperature describes the momentum contribution of phonons to the indirect $E_G$ transition. The probability of interband transition, $f_{cv}$ and *A'(T)*, should also vary with the particle size. It would be reasonable to assume that the size-induced transition-probability change is negligibly small despite the availability of the exact correlation of the transition probability to the Kubo gaps.

Compared with eq (88), the traditional form of size dependent $\varepsilon_r'$ varies with the $E_G$ and the incident beam energy:

$$\frac{\Delta[\varepsilon_r'(K_j,\omega)]}{\varepsilon_r'(\infty,\omega)} = \frac{\alpha' E_G(\infty)}{E_G(\infty)-\hbar\omega}\left(\frac{\Delta E_G(K_j)}{E_G(\infty)}\right) = \frac{\alpha' E_G(\infty)}{E_G(\infty)-\hbar\omega}\Delta_H \qquad (91)$$

where α' = ½ and 2 correspond to direct and indirect $E_G$ transition, respectively. The traditional at a certain optical energy, $\hbar\omega > E_G(\infty)$, decreases with $E_G$ expansion, $\Delta_H$, without involvement of bond contraction and e-p interaction.

- Photon absorption

The absorption coefficient, α, the refractive index, n (= $\sqrt{\varepsilon_r}$), and the complex dielectric function are correlated as: $\alpha(\omega) = 2\pi\varepsilon'_r(\omega)/n\lambda$ , and the transmittance of light is given as $T \propto \exp(-\alpha x)$, where x is the thickness of the medium for light transmission. This relation leads to the size-induced change of α as:

$$\begin{aligned}
\frac{\Delta\alpha(K_j,\omega)}{\alpha(\infty,\omega)} &= \frac{\Delta\varepsilon_r'(K_j,\omega)}{\varepsilon_r'(\infty,\omega)} - \frac{\Delta\varepsilon_r(K_j)}{2\varepsilon_r(\infty)} \\
&= -\left[\frac{\chi(\infty)}{\chi(\infty)+1} + \frac{\alpha' E_G(\infty)}{\hbar\omega - E_G(\infty)}\right]\Delta_H \qquad (convention) \qquad (92) \\
or \quad &= \left[\frac{\chi(\infty)}{2[\chi(\infty)+1]} - \frac{E_{PL}(\infty)}{\hbar\omega - E_{PL}(\infty)}\right]\times(\Delta_H - B\Delta_{e-p}) + \frac{\chi(\infty)+2}{2[\chi(\infty)+1]}\times\Delta_d \qquad (BOLS)
\end{aligned}$$

The traditional form [ $\Delta\chi(K_j)/\chi(\infty) = -2\Delta_H$ ] discriminates the direct and indirect $E_G$ transition by the α' while the BOLS form [ $\Delta\chi(K_j)/\chi(\infty) = \Delta_d - (\Delta_H - B\Delta_{e-p})$ ] counts the contribution from e-p coupling, lattice relaxation, and crystal binding.

9.3 Verification

9.3.1 Dielectric suppression

It is possible to discriminate the dielectric contribution of the nanosoild Si backbone from the measured effective $\varepsilon_{eff}$ of p-Si by matching the prediction with the measured impedance spectra. P-Si samples were prepared and their impedance was measured at ambient temperature in the frequency range of 50 Hz -1.0 MHz under 100 mV potential.[94] Silver paste was used for an ohmic contact. The samples were then dried at 353 K for 1 h to make the experimental data reproducible.

The impedance behaviour can be described by Debye's formula for a serial-parallel *RC* circuit[581] with elements that correspond to the dielectric behaviour of different components. The high temperature impedance behaviour can be described by a series of triple parallel RC circuit elements[581] that correspond to the dielectric behaviour of grain interior, grain boundary and electrode/film interface, respectively, as shown in Figure 37. The complex impedance response commonly exhibits semicircular forms in the measured Cole–Cole plot[582] as shown in Figure 38. At higher temperatures, two or more semicircles present corresponding different transition mechanisms.[95] The grain boundary resistance is normally higher than the grain interior and the electrode/film interface resistance is higher that that of the boundary. The larger radius of the Cole-Cole plot in frequency space corresponds to contribution from constituent of lower resistance.

Therefore, the first semicircle in the high frequency region can be attributed to the behaviour of grain interior while the intermediate and tertiary semicircles in the lower frequency region correspond to the grain boundary and the electrode/film interface, respectively.

The fitting procedure used here is the same as the one described by Kleitz and Kennedy.[583] The complex impedance $Z^*$ measured by RCL meter can be expressed as:

$$\begin{cases} Z^* = Z' - jZ'' \\ Z' = \sum_l \frac{R_l}{1+\omega^2 R_l^2 C_l^2}; \quad Z'' = \sum_l \frac{\omega R_l^2 C_l}{1+\omega^2 R_l^2 C_l^2} \end{cases} \tag{93}$$

where $\omega$ is the angular frequency. The resistance $R_l$ represents ionic or electronic conduction mechanisms, while the capacitance $C_l$ represents the polarizability of the sample from different components labeled $l$, which are related to grain interior, grain boundary, and electrode/interface.[94] Curves A–E in Figure 38 denote the responses of different samples (Table 16) measured at the ambient temperature. The complex impedance plots show only one depressed single semicircular arc, indicating that only one primary mechanism, corresponding to the bulk grain behavior, dominates the polarization and easy path for conductance within the specimen. The second intercept on the lateral real axis made by the semicircle corresponds to the resistance in the bulk grain. As it is seen, the intercept of the semicircles shifts away from the origin as the solid size decreases, indicating an increase of the Nan grain resistance, due to the lowering of the atomic potential well that trap the electrons in the surface region.

Figure 37 (link) Effective circuits for the impedance measurement of sample containing several components.

Figure 38 (link) Simulated and measured size dependence of Cole–Cole plots of p-Si and the *RC* parallel circuit model (inset) for typical dielectric materials.[94]

The capacitance and dielectric constant is extracted by using the relation: $Z''=1/(\omega C)$ from the data measured in high frequency ranging of $10^5$–$10^6$ Hz.[584] The bulk grain capacitance $C$ of the sample is given by the slope of the straight line determined by the variation of $Z''$ as a function of $1/\omega$. Then, the effective dielectric constant $\varepsilon_{eff}$ of the porous structure is calculated based on the equation: $\varepsilon_{eff} = Cx/(\varepsilon_0 S)$. With the measured $\varepsilon_{eff}$, we can calculate the $\varepsilon_{nano\text{-}Si}$ based on the Looygenga approximation:[585]

$$\varepsilon_{eff}^{1/3} = (1-p)\varepsilon_{nano-Si}^{1/3} + p\varepsilon_{air}^{1/3},$$

where $\varepsilon_{air}$ ($\approx 1$) is the dielectric constant of air and $p$ is the porosity of the p-Si. Results in Table 16 show that the $\varepsilon_{nano\text{-}Si}$ decreases with solid size.

Figure 39 compares the $\varepsilon_{nano\text{-}Si}$ derived herein and other sophisticated calculations of nanosolid Si and the third order dielectric susceptibility of Ag nanodots. Although the dielectric susceptibility does not follow the BOLS prediction but it shows the suppressed trend. Consistency in trends between BOLS predictions and the measured results evidences that the BOLS correlation describes adequately the true situation in which the $\varepsilon_{nano\text{-}Si}$ suppression is dictated by atomic CN imperfection. Other factors may contribute to dielectric suppression, which makes the prediction deviate from measurement compared with other simulations reported in previous sections. The apparent factors are the accuracy and uniformity of the shape and size of porous Si and the porosity. Atomic CN at a negatively curved surface of a pore is higher than that at the positively curved surface of a dot. As the numerical solution sums the contribution from crystal binding ($E_G$ expansion), electron-phonon coupling, and bond contraction, which accumulate the errors from the three aspects, contributing to the observed deviation. However, from physical and chemical insight point of view, the first (main) order approximation would be acceptable as other artifacts from measurement or form impurities are hardly controllable.

Table 16 Summary of the D-dependent $\varepsilon_{nano\text{-}Si}$ derived from the measured $E_{PL}$, porosity, and $\varepsilon_{eff}$, p-Si.

| Sample | *D* (nm) | $E_{PL}$ (eV) | Porosity (%) | $\varepsilon_{eff}$ | $\varepsilon_{nano\text{-}Si}$ |
|---|---|---|---|---|---|
| A | 1.7 | 2.08 | 85 | 1.43 | 6.27 |
| B | 2.0 | 1.82 | 76 | 1.84 | 7.29 |
| C | 2.1 | 1.81 | 71 | 2.11 | 7.7 |
| D | 2.2 | 1.79 | 68 | 2.28 | 7.86 |
| E | 2.4 | 1.76 | 66 | 2.45 | 8.29 |

Figure 39 (link) Comparison of the BOLS predictions with the sophisticated calculation and measurement results on the size–dependent dielectric constants of (a) silicon nanosolids with calculated Data–1, 2, 3;[577] Data -4 and 5;[567] and Data–6;[94] and (b) the third-order dielectric susceptibility of Ag nanosolid.[576] Note that a logarithmic *y*-axis has been used for clarity.

9.3.2 Blue shift of photoabsorption

The coefficient of photon absorption is calculated based on the relation:

$$\begin{aligned}\frac{\Delta\alpha(K_j,\omega)}{\alpha(\infty,\omega)} &= \frac{\Delta\varepsilon_r'(K_j,\omega)}{\varepsilon_r'(\infty,\omega)} - \frac{\Delta\varepsilon_r(K_j)}{2\varepsilon_r(\infty)} \\ &= -\left[\frac{\chi(\infty)}{\chi(\infty)+1} + \frac{\alpha' E_G(\infty)}{\hbar\omega - E_G(\infty)}\right]\Delta_H \qquad (convention) \quad (92) \\ or \quad &= \left[\frac{\chi(\infty)}{2[\chi(\infty)+1]} - \frac{E_{PL}(\infty)}{\hbar\omega - E_{PL}(\infty)}\right]\times(\Delta_H - B\Delta_{e-p}) + \frac{\chi(\infty)+2}{2[\chi(\infty)+1]}\times\Delta_d \qquad (BOLS)\end{aligned}$$

by taking $\chi(\infty)$ =10.4 and $E_{PL}(\infty) \sim E_G(\infty)$ = 1.12 eV. It is surprising that, as shown in Figure 40, a blue shift of the absorption edge takes place for the nano-Si. The threshold of absorption for the indirect band gap is slightly higher than that of the direct band gap materials. Such a blue shift of absorption edges should be advantageous in designing devices for optical communication of nanometer-scaled wires, tubes or superlattice structures. The lowered absorption coefficient and refractive index makes a nanometer-sized adsorbate more transparent, which may form the basis of quantum lasers, as observed at room-temperature from nanostructured ZnO tubes which emits ultraviolet laser at 393 ± 3 nm under 355 nm optical excitation.[586]

Figure 40 (link) Energy dependence of (a) imaginary dielectrics and (b) photoabsorption coefficient in conventional and BOLS approaches. Spherical size $K_j$= 5 is used corresponding to $\Delta_H$ = 0.506, $\Delta_{e\text{-}p}$ = 0.182 and $\Delta_d$ = -0.083.

9.4 Summary

The BOLS correlation has enabled us to derive numerical solutions for the first time to unifying the complex dielectric constants and the coefficient of photoabsorption of nanosemiconductors to the often-overlooked event of atomic CN imperfection and its effect on crystal binding and electron-phonon coupling. The solution applies to the whole range of measuring energies. The dielectric constant drops dramatically at the surface edge of the solid due to atomic CN imperfection. This mechanism can be used to trap and amplify light within the nanosolid by internal reflection, which may form a possible mechanism for random lasers. Understanding could be of use in designing photonic crystals with thermally and electrically tunability for optical switches and in fabricating wave guides for light trapping and amplifying device applications.[587]

Besides, the BOLS correlation has also allowed us to formulate and understand the dielectric suppression, dispersion, and conductivity and dielectric transition of nanosemiconductors. Effect of temperature and frequency on the dielectric transition and relaxation of nanosolid Si and nanodiamond were also examined, which derives activation energy for conductivity and dielectric transition of both nanodiamond and nanosilicon, giving information about the impurity mid-gap states of the corresponding systems. Interested readers may be referred to Refs.[94,95]

# 10 Magnetic modulation

## 10.1 Background

### 10.1.1 Observations

When a ferromagnetic solid is reduced in the nanometer regime, the magnetic properties of the solid will change. The Curie temperature $T_C$ drops with size.[93,111,347] For a nanograined solid the coercivity ($H_C$) increases whereas for an isolated nanosolid, the $H_C$ drops.[588,589,590,591] The saturation magnetization ($M_S$) increases at low temperature with quantized features, whereas the $M_S$ drops at ambient temperatures when the solid size is decreased.[592,593,594,595] Generally, the exchange bias field and the blocking temperature decrease, whereas the coercivity increases, as the size of the ferromagnetic (Pt/Co)-antiferromagnet (FeMn) coupled nanostructure is reduced.[596,597] Figure 41 (a) and (b) show the magnetic oscillation of small Ni and Rh particles at temperatures closing to zero Kelvin.

- Surface magnetron

In the case of surfaces and thin films, it has been clear that the magnetic moment of an atom ($\mu_i$) in the surface region is larger than the corresponding bulk value ($\mu_b$).[598,599] Theoretical calculations suggest that the $\mu_i$ for a surface Fe atom increases by 15% to 2.54 $\mu_B$ for 1 ML(monolayer) Fe on 5 ML W (110). The $\mu_i$ for a two-ML Fe atom is 29% higher than the bcc bulk moment of 2.2 $\mu_B$ of Fe, where $\mu_B$ is the Bohr magnetron. The significant surface relaxation (-12%) of Fe (310)[188] and Ni (210)[189] surfaces leads to enhancement of the atomic $\mu_i$ by up to 27%.

- Nanosolid at low temperature

Surface effects become stronger in the case of a nanosolid since larger fraction of atoms of the system is located at the curved surface. Furthermore, additional unexpected features present arising from the finite size effect. However, controversy remains in the measured trend of the $M_S(K_j)$.[592,593,594,595,600,601,602,603]

One trend in measurement shows that at temperatures below 200 K, the $M_S(K_j)$ increases with the inverse of size.[603,604,605,606,607] For example, the $M_S$ per atom of Fe, Co and Ni (at 78 –120 K) was measured[607] to increase until the value of a free atom when the solid size is reduced to a cluster that contains 30 atoms or less; as the size is increased up to 700 atoms, the magnetic moment approaches to the bulk limit. The $M_S$ of Ni clusters also increase inversely with size at temperature between 73 and 198 K.[339] $Co_n$ particles of 1.8 - 4.4 nm sizes carry magnetic moments that are ~20% higher than the bulk value.[339] The moment of a Co surface atom is enhanced by 32% compared to the bulk value of 1.73 $\mu_B$.[608] In the temperature range of 77 K and 570 K, the $M_S$ of Fe-Ni alloy films increase gradually[609] when the film thickness is decreased from 75 nm to 35 nm. Using a laser vaporization of an iron rod inside the throat of a high-pressure pulsed nozzle, Cox *et al*[601] firstly measured the magnetic properties of isolated iron-atom clusters containing 2 to 17 atoms and Fe monoxides and dioxides clusters at 22 K. The metal cluster beam passes through a Stern-Gerlach magnet. The deflected beam was detected by spatially resolved time-of-flight photo ionization mass spectrometry. It was found that the spin per atom of iron clusters was larger than that of bulk iron. It was therefore widely accepted that size reduction could enhance the magnetization of the small ferromagnetic particles.

Figure 41 (Link) Size dependence of magnetic moments of (a) $Ni_n$[610] and (b) $Rh_n$[611] particles measured at low temperature shows the size-enhanced and quantized $M_S(N_j)$ with oscillating features; (c) Cobalt particles[595] and (d) Ni thin films[590] measured at room temperature show size-tailed $M_S(K_j)$, instead.

- Nanosolid at ambient temperature

Since the observation of Cox in 1985,[601] overwhelming experiments have been conducted on various ferromagnetic nanosolids. Repeating the same Stern-Gerlach deflections of Fe clusters in a molecular beam, Heer *et al* [592] found instead that the average magnetic moments for small iron clusters (50~230 atoms) decrease with size. This trend also holds for $Pd_{96}Fe_4$,[594,593] $Pd_{97.1}Fe_{2.9}$,[594] $NiFe_2O_4$,[20] and $Ni_3Fe$[600] alloying nanosolids. Similarly, a remarkable reduction of magnetization at room temperature for Fe-Ni invar alloy (<40 nm)[612] and Ni thin films has been reported.[590,613] The $M_S$ for $Fe_3O_4$ thin films[614] drops rapidly when the film is 70 nm or thinner. Figure 41c and d shows the magnetic suppression of Co clusters[595,600] and Ni films. Small $Pd_{100-x}Fe_x$, grains with $x = 4$, 6, 8, and 12, with a radii of approximately 5 nm at 4.2, 100, and 295 K show a typical superparamagnetic features with $M_S$ values that are substantially smaller than the $M_S$ of the bulk.[593] However, for MnBi films,[615] the magnetism changes with neither thickness nor chemical composition. Therefore, it was surprisingly conflicting that some measurements show magnetic elevation whereas some show suppression without taking the operating temperature into condition.

- Coercive performance

An isolated magnetic domain or highly dispersed ones often show no hysteresis at any temperature. When the size of an isolated ferromagnetic solid is reduced to a certain critical size, the coercivity of the isolated nanosolid will approach to zero.[616] $Fe_{69}Ni_9CO_2$ powders of 10-15 nm grain sizes show almost no hysteresis, being indicative of supperparamagnetic characteristics.[617] However, when the particles get closer together, the supperparamagnetic behavior vanishes and the coercivity presents.[616,618] The coercivity increases with the inverse of grain size, which follows a $H_C \sim 1/K_j$ relation.[589,616,619] Investigation on the $Fe_{74.5-x}Cu_xNb_3Si_{13.5}B_9$ ($x = 0 \sim 1$ at.%) ribbons with grain sizes between 10 and 300 nm suggests that the $H_C$ increases following a $K_j^6$ relation and then drops in a $1/K_j$ fashion at the critical size of 50 nm. Similar trend of transition holds for Fe, Ni, and Co metal films, with corresponding critical sizes of 20, 40, and 30 nm.[620] Figure 42a shows the size-enhanced $H_C$ of Ni thin films consisting of 3-10 nm grains. Panel b shows the CN imperfection enhanced magnetization of Ni, Fe, and Co particles and panel c the Monte Carlo simulated $M_S(T, N_j)$ profiles.[621]

Figure 42 (Link) (a) Size-enhanced $H_C$ of Ni films[590] measured at room temperature and (b) CN dependence of the magnetic moment in (a) Fe, (b) Co, and (c) Ni as a function of nearest neighbor coordination (in various structures).[622] (c) Monte Carlo simulated $M_S(T, N_j)$ profiles.[621]

10.1.2 Possible mechanisms

- Magnetization

Puzzles of some of these observations have not been well understood in particular the oscillatory behavior at low-T as a function of size (see Figure 41a, b)[603,607] and the inconsistent trends of $M_S$ measured at different temperature ranges. Several shell structural models have been proposed for this purpose.[607,623] Jensen and Bennemann[623] firstly suggested that the $\mu_b$ of an atom is determined by their local atomic CN. By assuming bulk-like structures (such as fcc and bcc) and different global cluster shapes (cube, octahedron, cube octahedron), they found that the average $\mu_b$ oscillates with the cluster size, and that this magnetic "shell structure" reflects the progressive formation of concentric atomic layers.

Without considering the effect of temperature, the magnetic properties of transition metals are described by using a simple rectangular d-band approximation[605] together with the second moment approximation, as a first order approximation.[624] It was assumed that the d-band splitting between the major and the minor spin caused by exchange interaction is invariant for the cluster to the bulk solid, leading to the following expression:[599]

$$\frac{\mu_i}{\mu_b} = \begin{cases} \mu_{\dim}/\mu_b, & if \quad z_i \le z_b(\mu_b/\mu_{\dim})^2 \\ (z_b/z_i)^{1/2}, & otherwise \end{cases} \tag{94}$$

where $\mu_{dim}$ is the magnetic moment of one atom with one neighbour.[624] In the case of Fe, $\mu_b = 2.22$ $\mu_B$[130] and $\mu_{dim} = 3.2\ \mu_B$.[622] If $z_b = 12$, the step function transits at $z_i = 5.775 \sim 6$. The magnetic moment of an atom will take the dimer value if its CN is smaller than six. Considering the geometrical arrangement of atoms in different lattice structures of various shapes, the oscillation features could be reproduced using the shell structure.[599,625] Calculations using the tight binding theory[622] also show that the magnetic moment of Fe, Co, and Ni atoms increases towards the atomic value when the CN is reduced, as shown in Figure 42b.

The $M_S$ suppression at mid-T was explained with the following mechanisms:

(i) Surface spins are weakly coupled and more disordered at ambient temperatures compared to the bulk spins. The magnetization is then dominated by the interior bulk spins that drop in number when the solid size is decreased.[626]

(ii) $T_C$ suppression of the nanosolid lowers the $M_S$. In the shell structure, the surface layer is magnetically melted, which contributes little to the total $M_S$ of the system.[621]

(iii) In contrast, Monte Carlo simulations, as shown in Figure 42c, suggest that the $M_S$ of a small cluster is never higher than the bulk value due to the reduction of exchange bonds of the surface atoms. Based on an assumption that the clusters undergo a super-paramagnetic relaxation, Khanna and Linderoth[627] derived that the effective $M_S$ of small Fe and Co clusters decrease with size, which was explained as a consequence of fluctuations due to thermal vibration and rotation effects on the giant spinner.

Here we show that all the proposed mechanisms are correct and consistent provided with the BOLS as complementary origin.

- Coercivity

It is known that both the inter-spin interaction within a domain and the inter-grain interaction within a solid composed of nanograins can be described using the same Ising model, or other approaches such as the mean field approximation as well. We prefer using the Ising model, as it is sufficient for the first order approximation. Considering a domain as a giant spinner with a moment $J$, the exchange energy of the spinner ($E_{exc,\,i}$) interacting with its $z$ nearest giant spin neighbors follows the Ising relation. The $d_i$ is then replaced with grain diameter, $D_j$ (also structural correlation length), if uniform grain size is assumed. The $H_C(D_j)$ transition from $D_j^6$ to $D_j^{-1}$ can be expressed as:[628]

$$H_C \sim \frac{1}{20}\frac{K_2^4 D_j^6}{A^3 M_0} \quad \Rightarrow H_C \propto \frac{1}{2} H_S \propto E'_{exc} \propto z D_j^{-1}$$

where $K_2$ is the strength of local uniaxial anisotropy and $M_0$ is the magnitude of the local magnetization vector. $A$ is the exchange stiffness parameter. The former corresponds to the random anisotropy mechanism of domain-wall pinning at grain boundaries; the latter relates herewith to the inter-grain and grain-substrate interaction, which dominates the anisotropy energy.[628]

## 10.2 BOLS formulation

### 10.2.1 Charge localization

It is known that the CN-imperfection-enhanced bond-energy deepens the atomic trapping potential well of the lower–coordinated atom from one unit to $c_i^{-m}$. Electrons inside the trap are then more localized. If the localization probability is proportional to the trapping well depth, then the densely localized electrons contribute to the $\mu_i$ of the lower-coordinated atom. The corresponding change of the mean $\mu(K_j)$ varies monotonically with the coefficient of bond contraction:

$$\begin{cases} \mu_i(z_i) = c_i^{-m}\mu_b \\ \dfrac{\Delta\mu(K_j)}{\mu_b} = \sum_{i\le 3} r_{ij}\left(c_i^{-m}-1\right) \end{cases}$$

(95)

For a dimer Fe atom ($z_i$ = 2, $c_i$ ~ 0.7), $\mu_i = 0.7^{-1}\mu_b = 3.25\ \mu_B$, which is 1.43 times the bulk value, agreeing with measured value of $\mu_{dim} = 3.2\ \mu_B$. Compared with the model given in eq (94), here we use a smooth function rather than a step transiting at $z_i$ ~ 6, albeit the difference in physical origin. As the effective CN of an atom at a flat or a curved surface is 4 or lower, the BOLS premise predicts a $0.88^{-1}$ = 12% or higher magnetic enhancement of a surface atom at zero Kelvin, agreeing with theoretical predictions.[189,187] Therefore, it is reasonable to suggest that it is the very atomic CN-imperfection deepened atomic potential that traps the electrons with high probability of localization to contribute to the $\mu_i$ of the lower-coordinated atom.

10.2.2 Brillouin function

The inter-spin interaction dominates the order of the spin system and hence the $M_S$ and $T_C$. At low temperatures, the total angular moment of an atom changes its direction in a quantum tunneling process.[629] At higher temperatures, the spin direction will fluctuate due to thermal agitation. The easiness of fluctuation is determined by the strength of inter-spin coupling that varies with atomic CN as well. Because of fluctuation, the magnetic momentum will reduce and eventually vanish at the $T_C$. In the first order approximation to the size and temperature dependence of the $\mu_S(T, K_j)$, we use the concept of "molecular field",[630] to describe the spontaneous magnetization at $T$ in terms of Brillouin function, $B_J(y)$:

$$\begin{cases} \mu(T) & = g_J J \mu_B B_J(y) \\ B_J(y) & = \dfrac{2J+1}{2J}\coth\dfrac{2J+1}{2J}y - \dfrac{1}{2J}\coth\dfrac{y}{2J} \\ y & = \dfrac{J g_J \mu_B}{k_B T}H_m \end{cases}$$

(96)

$g_J$ is the Lande's g-factor, $J$ is the total angular momentum and $E_{exc} \sim E_{coh}$ is the molecular field. When $T$ approaches to 0.8 $T_C$, $\mu(T) \approx \mu_S(T)$.[630] Therefore, the $\mu_S(K_j, T)$ can be obtained by replacing the bulk $J$ and $H_m$ with the size dependent $J(K_j)$ and $E_{exc}(K_j)$ that are given as:

$$\begin{aligned} E_{exc}(K_j) &= E_{exc}(\infty)\left[1+\sum_{i\le 3}\gamma_{ij}(z_{ib}c_i^{-m}-1)\right] \\ J(K_j) &= J\left[1+\sum_{i\le 3} r_{ij}\left(c_i^{-m}-1\right)\right] \end{aligned}$$

(97)

Differentiating eq (96) against $E_{exc}(K_j)$ leads to the size and temperature dependent $\mu_S(K_j, T)$:

$$\begin{aligned} \frac{\Delta\mu_S(K_j,T)}{\mu_S(\infty,T)} &= \left\{\frac{1}{2J}\operatorname{csc}h^2\left[\frac{2g_J\mu_B}{2k_BT}AE_{exc}(\infty)\right] - \frac{2J+1}{2J}\operatorname{csc}h^2\left[\frac{(2J+1)g_J\mu_B}{2k_BT}AE_{exc}(\infty)\right]\right\} \\ &\quad \times\frac{\Delta E_{exc}(K_j)}{E_{exc}(\infty)} \\ &= \alpha(T)\sum\nolimits_{i\le 3}\gamma_i(z_{ib}c_i^{-m}-1) \end{aligned}$$

(98)

parameter $\alpha(T)$ is T and material dependent. Eq (98) indicates that for a specific ferromagnetic solid and at a given temperature, the $\mu_S(K_j, T)$ changes with the atomic cohesive energy. One needs to note that eq (98) does not apply to an isolated atom without exchange interaction being involved though the isolated atom possesses intrinsically higher magnetic momentum.

## 10.3 Verification

### 10.3.1 Monte Carlo simulation

- Structures

In order to examine the model consideration, Monte Carlo simulation was carried out based on the BOLS incorporated Ising convention in comparison with the modified Brillouin function. The atomic CN imperfection enhanced magnetic moment was taken into consideration by varying the spin value $S_i'$ for each atom. We employed six kinds of nanosolids to investigate the size, shape, and crystal structure effects on the $\mu_S$ at various temperatures.

The fcc spherical dots are formed in such a way that layers of successive atoms are added to the initial central atom. Figure 43 (a) shows, for example, the fcc spherical dot containing $N_{141}$ atoms with $S = 9$ shells and $K_j = 3.3$ atomic size. Here we only consider those clusters with completely closed outermost shells as a convention. The rod and the plate systems are also formed based on the fcc lattice along the <100> direction. The length of the rod is maintained at $K_j = 28.3$ of which variation has insignificant effect on the result. The radius of the rod ranges from $S = 1$ to 11 ($K_j = 0.5 \sim 3.66$). The width and length of the plate are maintained at $K_j = 28.3$. The thickness ranges from $S = 1$ to 14 ($K_j = 0.5 \sim 5.1$). Figure 43 (b) and (c) illustrate an fcc rod and plate with $S = 3$, $K_j = 1.9$ and $S = 2$, $K_j = 1.7$, respectively.

Figure 43 (link) Illustration of atomic configurations of (a) an fcc dot of 9 shells with $K_j = 3.3$, (b) an fcc rod of 3 shells with $K_j = 1.9$, and (c) an fcc plate of $K_j = 1.7$ thick, (d) an Icosahedron with $N_{147}$ atoms, (e) a Marks decahedron with $N_{101}$ and (f) an fcc truncated octahedron with $N_{201}$ atoms.[101]

Calculations were also conducted using the ordered structures of icosahedra, decahedra and the close-packed fcc truncated octahedra that are favored from the energetic point of view. Figure 43 (d) ~ (f) show the close-packed structures with a total number of $N_{101}$, $N_{147,}$ and $N_{201}$ atoms, respectively. Icosahedra and decahedra are noncrystalline structures that cannot be found in bulk crystals because of the fivefold symmetry. Icosahedra are quasispherical, where atoms are arranged in the concentric shells. Marks-truncated decahedra have reentrant (111) facets that are introduced via a modified-Wulff construction. Fcc truncated-octahedra own the crystalline structure and have the open (100) facets.

- MC algorithm

To compute effectively, we use cool state initialization at low temperature ($k_BT/J_{exc} < 6$) and hot state initialization at relatively high temperature ($k_BT/J_{exc} \geq 6$).[101] In the hot state, spins orientate randomly; and in the cool state, spins align parallel to the applied magnetic field. For a certain spin system, the value of the Hamiltonian $H_{ex,k-1}$ was calculated. A spin $S_i'$ was chosen randomly and the orientation was flipped from $S_i'$ to $S_{i,trial}$. The $H_{ex,k}$ was optimized to satisfy the Metropolis criterion:[631]

$$\begin{cases} \exp\left(-\Delta H_{ex}/kT\right) > \delta_{mc} \\ \Delta H_{ex} = H_{ex,k} - H_{ex,k-1} \end{cases}$$

$\Delta H_{ex}$ is the energy change for a spin re-orientation and $\delta_{mc}$ is a uniform random deviate. After several MC steps of sweeping over all the lattice sites of the spin system, the spin system of a specific size at a specific temperature reaches to thermal equilibrium. In simulation, each atom is taken as an independent spin with $\mu_i$ in unit of the bulk $\mu_b$. For the bulk value, S takes the values of +1 or –1 for the up and down flip. The energy change is calculated for the spin flip from *k-1* to *k* step due to thermal vibration: $\Delta H_{ex} = H_{ex,k} - H_{ex,k-1}$. At a given temperature, the system will reach to a stable state after sufficient steps of operation. The magnetization is then calculated:

$< M >= \left[ \sum_N M(s_1^{(i)}, s_2^{(i)}, ....s_N^{(i)}) \right] / N$ with 5000 thermalization steps for each spin to reach thermal equilibrium state.

### 10.3.2 Measurement

Ni films with grain sizes in the range of 3 ~10 nm were grown on Si(100) substrates using physical vapor deposition. The grain size was calibrated using XRD profiles and Scherrer's equation. The in-plane magnetic properties were measured using vibrational sample magnetometer at room temperature.[590]

## 10.4 Findings

### 10.4.1 Ni films at ambient temperature

Figure 41d and Figure 42a show the agreement between the predicted with m = 1 and the measured size dependence of the $M_S$ and the $H_C$ for the Ni films. The match of $M_S(K_j)$ at 300 K is realized with $\alpha(J, T) = 4.0$. When the particle size is reduced to $K_j = 5$ ($D_j = 2.5$ nm), $M_S = 0$. This result is consistent with the findings of the size induced $T_C$ suppression of ferromagnetic nanosolids, as discussed in 5.3.3.[93,111] For a Ni particle of $K_j = 5$, the $T_C$ drops by ~ 51% from 631 K to 309 K and the $M_S$ is not detectable.[111]

Figure 44 (link) The magnetic hysteretic loops of Ni films of different grain sizes measured at room temperature.[590]

### 10.4.2 Numerical findings

Figure 45 shows the MC simulated $M_S(K_j, T)$ curves at zero applied magnetic field for (a) an fcc dot, (b) an fcc rod, (c) an fcc plate and (d) an Icosahedra spin system. Generally, at very low temperature region ($k_BT/J_{exc} < 3$), the $M_S(K_j, T)$ increases with oscillatory features as the solid size is reduced. At mid-$T$ region ($k_BT/J_{exc} \sim 6$), the $M_S$ drops with size. In the paramagnetic region, the residual $M_S$ increases as the size is reduced. These features are intrinsically common depending less on the shape and the crystal structure of the specimen. Figure 46 shows a BOLS predicted $M_S(K_j, T)$ counterplot for a spherical dot. No oscillatory features are given as we used a smooth function for the surface-to-volume ratio. Therefore, the strong discrepancy of $M_S$ between low and mid temperatures is unified with the BOLS correlation.

Figure 45 (Link) MC simulated temperature and size dependence of the $M_S$ for (a) an fcc dot, (b) an fcc rod, (c) an fcc plate and (d) an Icosahedral spin system.

Figure 46 (Link) Counterplot of the BOLS predicted $M_S(T, K_j)$, which shows that $M_S$ increases with inverse size at low temperature and decreases with size at mid temperature. The $M_S$ is normalized by $M_S(T = 0, K_j = \infty)$ and $T$ is normalized by $AE_{exc}(\infty)$.

- $M_S(K_j, T \sim 0$ K) enhancement

It is seen from Figure 45 (a) that for a specific size $K_j$, the $M_S$ of the fcc dot is higher than that of the fcc plate because a spherical dot has higher $\gamma_{ij}$ value. It is understood that when T → 0, y → ∞, and then $B_J(y) \to 1$, eq (96) can then be approximated as: $\mu_S(T \to 0) = Jg_j\mu_B$. Using a shell structure in the BOLS correlation that calculates the magnetic moment of every atom layer-by-layer leads to the size-enhanced $M_S$ for a nanosolid at very low temperature, which follows equation (95).

Figure 47 compares the BOLS predictions with the measured low-temperature $M_S(K_j)$ of Fe, Ni and Co particles.[607] As the measured data are much scattered, it is hard to conclude though the trends generally match; however, the close match of $T_C(K_j)$ suppression and lattice contraction, as shown earlier in this report evidences sufficiently the validity of the BOLS correlation as complementary origin of the unusual magnetic behavior of the ferromagnetic nanosolids at different temperatures.

Figure 47(Link) Comparison of BOLS predictions with measured size dependence of $M_S$ at low T.

Figure 48 (Link) Size dependence of the $M_S$ at temperature (a) $k_BT/J_{exc}$ = 1, (b) $k_BT/J_{exc}$ = 6 and (c) $k_BT/J_{exc}$ = 12 for fcc nanosolids shows three outstanding regions, where $|M|$ shift = $[|M|(K)-|M|(\infty)] / |M|(\infty) \times$ 100%.

- $M_S(K_j$, T ~$T_C)$ suppression

Figure 48 (b) shows the matching between predictions with $\alpha(J, T)$ = 1.4 and the Monte Carlo simulative results at mid T. The calculated trend is consistent with the measurement with $\alpha(J, T)$ value that is different from Ni sample. The calculation takes the surface CN to the half number of the bulk (12) but in the BOLS premise, the effective surface CN is 4 or less. In the paramagnetic phase as shown in Figure 48(c), the remnant magnetism is higher for smaller particles, which has been attributed to the slower temperature decay in the Monte Carlo study and to the increasing fluctuations with decrease in cluster size.[621]

- $M_S(K_j)$ oscillation and structural stability

The oscillation behavior of $M_S$ at smaller sizes, as shown in Figure 49, depends less on the crystal structures. This relation suggests that the oscillatory originates from the surface-to-volume ratio because some particles may have fewer atoms at the surface with smaller $\gamma_{ij}$ value than those of the adjacent larger or smaller sizes, as illustrated in section 2 for the fcc and bcc structures.[610] Therefore, it is not surprising to resolve the $M_S$ oscillation in the low temperature measurement of smaller nanostructures with quantized surface-to-volume ratios.

The physical properties of nanosolids in molecular regime typically exhibit a very irregular dependence on their aggregate size, namely, magic numbers, while they behave in a regular way in the mesoscopic regime. The icosahedron, Marks-decahedron and the fcc truncated-octahedron have lower $M_S$ in the low-$T$ region, especially, the small icosahedral particles of $N_{55}$ and $N_{147}$ atoms, compared to other structures. An icosahedron has fewer low-CN atoms at the surface with most compact structures. The mass spectra of nanosolids usually exhibit especially abundant sizes that often reflect particularly stable structures, especially reactive nanosolids, or closed electronic shells.[632] These "magic number" sizes are of great theoretical interest since many of them correspond to compact structures that are especially stable. The simulative results presented herein show that the magnetic number of $N_{13}$, $N_{55}$, $N_{147}$ for icosahedron magnetic nanoparticles.[606] However, when the $N$ is larger than 300, the fcc truncated-octahedron is magnetically most stable compared with the decahedra, icosahedra and the fcc spherical dot. The MC simulation results here are consistent with experimental findings that the icosahedral structure transits at ~3.8 nm to the fcc truncated structure when the particle size is increased.[633] The competition between the surface energy reduction and the strain energy enhancement determines the structural stability. Therefore, icosahedra are the most stable at small sizes due to its low surface energy and good quasispherical structures, while decahedra are favorable at intermediate sizes, and regular crystalline structures are restore of large objects.

Figure 49 (Link) Magnetic oscillation features of different crystal structures at different temperatures.

### 10.5 Summary

Incorporating the BOLS correlation to the Ising premise and the Brillouin function, we have conducted the MC simulations, and BOLS predictions to examine the size, shape, structural and temperature dependence of the magnetization of a ferromagnetic nanosolid with experimental verification. MC simulations and BOLS predictions have produced all the observable features at various temperatures, including the oscillatory with clear physical insight into the origin of the changes. Conclusion is drawn as follows:

(i) For a ferromagnetic nanosolid, the magnetic moment at very low temperature increases with the inverse of size compared with the bulk value due to the deepening of the intra-atomic potential well that trap the surface charges contributing to the angular momentum of the lower-coordinated atoms of a nanosolid.

(ii) The $M_S$ at temperature around $T_C$ reduces, which is dominated by the decrease of exchange energy that dominates the thermal stability of a nanosolid.

(iii) The $M_S$ oscillates with the total number of atoms arise from nothing more than the surface-to-volume ratio of the solid.

(iv) The $H_C$ is dominated by inter-grain interaction.

(v) Structure transition from icosahedron to fcc truncated-octahedron happens at size containing 300 atoms, which is common to observations using other means.

Consistency in the Monte Carlo calculations, BOLS predictions and experimental observations not only clarifies for the first time the long-standing confusion on magnetic behavior of a ferromagnetic nanosolid at various temperatures. The joint contribution from the CN-imperfection and the associated bond energy rise lowers the exchange energy that tailors the temperature of phase transition. Therefore, it is not surprising that some measurements show the enhanced $M_S$ at low temperature while some observed the tailoring of the $M_S$ at temperature closing to the $T_C$.[634,635,636]

## 11 Concluding remarks

### 11.1 Attainment

As demonstrated, the impact of the often-overlooked event of atomic CN imperfection is indeed tremendous, which has enabled us to view the performance of a surface, a nanosolid and a solid in amorphous state consistently in a way from the perspective of bond relaxation and its consequences on bond energy. Progress made insofar can be summarized as follows:

(i) The unusual behavior of a surface and a nanosolid in various aspects has been consistently understood and systematically formulated as functions of atomic CN imperfection and its derivatives on the atomic trapping potential, crystal binding intensity, and electron-phonon coupling. These properties include the lattice contraction (nanosolid densification and surface relaxation), mechanical strength (resistance to both elastic and plastic deformation), thermal stability (phase transition, liquid-solid transition, and evaporation), and lattice vibration (acoustic and optical phonons). They also cover photon emission and absorption (blue shift), electronic structures (core level disposition and work function modulation), magnetic modulation, dielectric suppression, and activation energies for atomic dislocation, diffusion, and chemical reaction.

(ii) The new freedom of dimension has enabled us to elucidate information such as single energy levels of an *isolated* Si, Pd, Au, Ag and Cu atoms and their shift upon bulk and nanosolid formation by matching predictions to the observed size and shape dependence of the XPS data, or simply the components of the surface and the bulk XPS signatures. The new freedom also allows us to gain quantitative information about dimer vibration and electron-phonon interaction by matching predictions to the measured shape and size dependence of Raman and photoemission/absorption spectra of Si and other III-V and II-VI compounds.

(iii) New approaches enhance in turn the capability of Raman and XPS and provide a new way towards discriminating the contribution of intra-atomic trapping from the contribution of crystal binding and the electron-phonon coupling to the behavior of specific electrons, phonons and photons inside a nanosolid.

(iv) The bonding identities such as the length, strength, extensibility, and thermal and chemical stability, in metallic monatomic chains and in the carbon nanotubes have been determined. Understanding could be extended to the mechanical strength and ductility of metallic nanowires. Further investigation is in progress.

(v) In combination with the bond-band-barrier correlation for chemical reaction, the BOLS premise has also enabled us to discriminate the extent of oxidation and the effect of fluorine passivation on the performance of nanostructured silicon. The latter two topics would form the subject of nanometrology, and further pursue is in progress.

Consistency between the BOLS prediction and the measurements evidences not only the essentiality and validity of the BOLS correlation premise but also the significance of atomic CN imperfection to the low-dimensional and disordered systems that are dominated by atomic CN deficiencies. Understanding gained insofar should be able to help us in predicting nanosolid performance and hence provide guideline in designing process and fabricating materials with desired functions.

## 10.2 Advantages and limitations

The significance of the approaches is that it covers the whole range of sizes from a dimer bond to the bulk solid and covers the states of surface, amorphous, and nanosolid of various shapes to bulk solid with defects inside, with few adjustable parameters and almost no assumptions. Almost all of the imaginable and detectable quantities are consistently related to the BOLS correlation and the population of the lower-coordinated atoms as well. No multiscale model is necessary. For instance, the surface energy, interfacial energy, surface stress, the local mass density of liquid and solid are all functions of atomic separation and bond energy that are subject to the effect of atomic CN-imperfection. The original BOLS premise overcomes the convergence difficulties in numerical efforts faced by other documented approaches. The difficulties in describing the photoluminescence blue shift at the lower end of the size limit and the melting point oscillation over the whole range of sizes have been completely resolved. The parameters involved are just the bond nature represented by the parameter m and the corresponding bulk values of quantities of concern, which is independent of the particularity of element, crystal structures, or the form of interatomic potentials.

One may wonder that there is often competition between various origins for a specific phenomenon. As demonstrated in the context, the atomic CN imperfection affects almost all the aspects of concern, and therefore, the atomic CN imperfection should dominate the performance of a nanosolid through the competiting factors. For instances, the atomic cohesive energy dictates the phase transition or melting while the binding energy density dominates angular momentum and mechanical strength. These two competition factors determine the unusual behavior of a nanosolid in magnetism and mechanical strength under various conditions.

One may also wonder the effect of impurities such as surface oxidation on the measurement. Although XRD and XPS revealed no impurities in the Ni samples, for instance, we cannot exclude the existence of trace impurities. However, if all the samples were prepared and measured under the same conditions and we use the relative change of the quantities, artifacts caused by impurities should be minimized, and the results are purely size dependent.

The BOLS premise does not apply to the so-called dangling bond, as a dangling bond is not a real bond that forms between two neighboring atoms. It is true that the concept of localized bond is not applicable to metallic systems due to the demoralized valence electrons whose wave function often extends to the entire solid. However, the demoralized valence electrons are often treated as a Fermi sea and the metal ions are arranged regularly in the Fermi-sea background. As a standard practice, the metallic bond length corresponds to the equilibrium atomic separation and the bond energy is defined as the division of the atomic cohesive energy $E_B$ by the atomic CN in a real system. For the tetrahedral bond of diamond and Si, the full CN is not four as the tetrahedron is an interlock of fcc structures. Therefore, the BOLS premise holds for any solid disregarding the nature of the chemical bond. The pair interatomic potential for metallic interatomic interaction also holds, as the pair potential represents the resultant effect of various orders of coordination and the charge-density distribution. Density functional theory calculations on the dimer bond contraction and bond strength gain of Ni, Cu, Ag, Au, Pt and Pd contracts evidence sufficiently the validity of the current BOLS correlation for metallic systems.

Stimuli in measurement may affect the data acquisited. For example, in mechnical strength detection, the stress-strain profiles of a nanosolid may not be symmetric under tension and compression, and the flow stress is strain rate, loading mode, and materials compactness as well as size distribution dependent. However, one could not expect to cover the fluctuations of mechanical (strain rate, stress direction, loading mode, etc), thermal (self-heating during process and electron bombardment in TEM), crystal structure orientation, or grain-size distributions in a theoretical model, as these fluctuations add random artifacts that are hardly controllable. These effects can be minimized in the present approach by using relative changes that are intrinsic in physics.

As this practice is a first (but main) order approximation, there are still rooms for improvement by involving other high-order effects that contribute to the physical properties. If counting atom-by-atom in a specific crystal structure, the theoretical curves at the lower end of the size limit should show oscillation features with “magic number” of atoms due to the surface-to-volume ratio. For illustration purpose, it would be adequate to employ the smooth function for the surface-to-volume ratio in the present approach, as one should focus, in the first place, on the nature, trend, and origins for the size-induced changes and to grasp with factors controlling the property change.

I would like to indicate that all the models mentioned in the context are successful from different physical perspectives, and with the BOLS correlation as complementary origin, they would be complete and in good accordance.

## 10.3 Future directions

Although the imaginable and detectable quantities of a nanosolid have been preliminarily formulated and verified with experimental measurements in terms of the BOLS correlation, there are still more challenges ahead of us:

(i) Further attention is needed to pay on the joint effect of physical size and chemical reaction. At an interface, no significant CN imperfection is expected but chemical bond may evolve when alloy or compound is formed. Chemical effect revises the nature of the bond while the physical size causes the bond contraction. Both will modify the atomic trapping, crystal binding, electron-phonon coupling, which should be origin for the detectable physical properties of a solid including transport properties. Switching between superhydrophilicity and superhydrophobicity of chemically treated nanostructure could be successful samples for the joint effect of the BOLS and BBB proposals.

(ii) Traditional practice in theoretical calculations may be subject to modification at the lower end of the size limit to involve the CN imperfection effect. Consideration of the real boundary conditions with atomic CN imperfection instead of the ideal periodic boundary conditions would be necessary. As demonstrated, the atomic CN imperfection and the large portion of surface/interface atoms play the key roles determining the performance of small structures. Recent tight-binding potential molecular dynamics calculations of tetrahedron carbon (t-C) graphitization by Zheng et al [637] reveal that the graphitization of t-C cluster with hundreds atoms happens at temperature that is 10% higher than measurement (1100 ~ 1200 K). A first-principle calculation[638] predicted that the hardness of the optimal $BC_2N$ structure is lower than the measured extreme hardness of $BC_2N$ nanocomposites. It is suggested that the effects of the nanocrystalline size and the bonding with the amorphous carbon matrix in $BC_2N$ anocomposites likely play a crucial role in producing the extreme hardness measured in experiments. If the effect of bond-order loss on the atomic cohesive energy and biding energy density in the contracted surface region may be considered for the large number of surface atoms instead of the ideally periodic boundary condition is considered, the transition temperature would be lowered to a value closing to the measurement and the calculated hardness with approach the measured values.

Excitingly, *In situ* $T_C$ and valence DOS measurement of atomic–layered growth of superconductive Pb on stepped Si substrate by Guo et al[365] revealed oscillation of both $T_C$ and the valence DOS peak near the $E_F$ when the film thickness was increased by one atomic layer at a time. The $T_C$ increases gradually to the bulk value at about 30 layers with saw-tooth like fashion of oscillatory with 0.5 K magnitude in a period of every other layer. The two DOS peaks at 0 and 0.3 eV below $E_F$ dominate alternatively with the layer-by-layer growth. These discoveries provide direct evidence for the BOLS premise which indicates that the atomic cohesive energy (and $T_C$) drops associated with deepened potential well of trapping as the atomic CN decreases. Therefore, the $T_C$ valley and the dominance of the DOS peak away from the $E_F$ are suggested to arise from the large number of lower-coordinated atoms, which correspond to the deepened potential well (DOS peak away from $E_F$) and the lowered atomic cohesive energy ($T_C$). The $T_C$ peak and the dominance of the DOS peak near the $E_F$ are consequence of the small number of lower-coordinated atoms.

(iii) Transport in thermal conductivity, electric conductivity plays important role in the performance of nanostructured devices, which would be more challenging for studies. Employing the BOLS crystal potential for a nanosolid and for an assembly of nanosolids could improve the understanding on the kinetic and dynamic performance of a nanosolid under external stimuli.

(iv) The new freedom of size allows us to tune the physical properties of a nanosolid by simply changing the shape and size. Far beyond that, the new freedom provides us with opportunities to gain information such as dimer vibration, single energy level of an isolated atom, which could form important impact in basic science.

(v) Application of the BOLS and its derivative to process and materials design is important in practical applications. If we know what is intrinsic and what the limit is, we may save our spirit and resources in fabricating devices and materials. For example, one who is working in microelectronics often expects to expand the limit of dielectrics to the lower end for interconnection and to the higher end for Gate devices by changing the grain size. The BOLS derivative is able to tell that it is unlikely possible to raise the dielectrics by reducing the particle size and one has to seek other chemical route for the objectives. One cannot expect proper functioning of a ferromagnetic, ferroelectric and a superconductive nanosolid when the solid size is smaller than 2.5 nm, as derived in the present work.

These topics would form new branches of study towards profound knowledge and practical applications and this report just scratches the skin of this vast filed and further investigation is in progress.

**Acknowledgment**

I would like to express my sincere thanks to Professors Stan Viprek, Chunli Bai, John S. Colligon, S.R.P. Silva, David S. Y. Tong, F.M. Ashby, Philip J. Jennings, Andris Stelbovics, Xi Yao, A. Bhalla, E. Sacher, and En-Yong Jiang, for illuminating communications and encouragement. I also thank Drs Haitao Huang, Yongqing Fu, Likun Pan, and Weihua Zhong for their assistance. Permission of reprinting diagrams from Elsevier, IOP, APS, ACS, and AIP is also acknowledged.

**Appendix A**

Link

**Bibliography**

[1] Sun CQ, Oxidation electronics:bond-band-barrier correlation and its applications, Prog Mater Sci 2003;48:521-685.

[2] Ashby MF, Lu TJ, Metal foams: A survey. Sci China B 2003;46:521-32.

[3] Edelstein AS, Cammarata RC, Nanomaterials: Synthesis, Properties and Applications. Inst Phys Bristol, 1996.
[4] Gleiter H, Nanocrystalline materials. Prog Mater Sci 1989;33:223-315.
[5] Lu L, Sui ML, Lu K, Superplastic extensibility of nanocrystalline copper at room temperature. Science 2000;287:1463-6.
[6] Goldstein N, Echer CM, Alivistos AP, Melting in semiconductor nanocrystals. Science 1992;256:1425-7.
[7] Christenson HK, Confinement effects on freezing and melting. J Phys:Condens Matt R 2001;13: 95-133.
[8] Zhang Z, Li ZC, and Jiang Q, Modelling for size-dependent and dimension-dependent melting of nanocrystals. J Phys D 2000;33:2653-6.
[9] Mintova S, Olson NH, Valtchev V, and Bein T, Mechanism of zeolite a nanocrystal growth from colloids at room temperature. Science 1999;283:958-60.
[10] Horch S, Lorensen HT, Helveg S, Laegsgaard E, Stensgaard I, Jacobsen KW, Norskov JK, and Besenbacher F, Enhancement of surface self-diffusion of platinum atoms by adsorbed hydrogen. Nature (London) 1999;398:134-6.
[11] Tan OK, Zhu W, Yan Q, Kong LB, Size effect and gas sensing characteristics of nanocrystalline $SnO_{2-(1-x)}\alpha$-$Fe_2O_3$ ethanol sensors. Sensor Actuat B 2000;65:361-5.
[12] Rao CNR, Kulkarni GU, Thomas PJ, Edwards PP, Size-dependent chemistry:Properties of nanocrystals. Chem Euro J 2002;8:28-35.
[13] van Buuren T, Dinh LN, Chase LL, Siekhaus WJ, and Terminello LJ, Changes in the Electronic Properties of Si Nanocrystals as a Function of Particle Size. Phys Rev Lett 1998;80:3803-6
[14] Yoffe AD, Semiconductor quantum dots and related systems:electronic, optical, luminescence and related properties of low dimensional systems. Adv Phys 2001;50:1-208.
[15] Sun CQ, Gong HQ, Hing P and Ye H, Behind the quantum confinement and surface passivation of nanoclusters. Surf Rev Lett 1999;6:L171-6.
[16] Modrow H, Tuning nanoparticle properties - the X-ray absorption spectroscopic point of view. Appl Spec Rew 2004;39:183-290.
[17] Brauman JI, Small clusters hit the big time. Science 1996;271:889-9.
[18] Service RF, Semiconductor clusters, nanoparticles, and quantum dots. Science 1996;271:890.
[19] Collins RT, Faucher PM, and Tischler MA, Porous Sillicon:from luminescence to LEDs. Phys Today:1997;25-31
[20] Kodama RH, Berkowitz AE, McNiff EJ Jr, and Foner S, Surface Spin Disorder in $NiFe_2O_4$ Nanoparticles. Phys Rev Lett 1996;77:394-97.
[21] Berkowitz AE, and Takano K, Exchange anisotropy. J Magn Magn Mater 1999;200:552-70.
[22] Gerberich WW, Mook WM, Perrey CR, Carter CB, Baskes MI, Mukherjee R, Gidwani A, Heberlein J, McMurry PH, and Girshick SL, Superhard silicon nanospheres. J Mech Phys Solids 2003;51:979-92.
[23] Veprek S, Electronic and mechanical properties of nanocrystalline composites when approaching molecular size. Thin Solid Films 1997;297:145-53.
[24] Veprek S, Reiprich S, A concept for the design of novel superhard coatings. Thin Solid Films 1995;268:64-71.
[25] Zhao M, Li JC, and Jiang Q, Hall-Petch relationship in nanometer size range. J Alloys Comp 2003;361:160-4.
[26] Wong EW, Sheehan PE, Lieber CM, Nanobeam mechanics:elasticity, strength and toughness of nanorods and nanotubes. Science 1997;277:1971-5.
[27] Ren Y, Fu YQ, Liao K, Li F and Cheng HM, Fatigue failure mechanisms of single-walled carbon nanotube ropes embedded in epoxy. Appl Phys Lett 2004;84:2811-3.
[28] An B, Fukuyama S, Yokogawa K, Yoshimura M, Surface superstructures of carbon nanotubes on highly oriented pyrolytic graphite annealed at elevated temperatures. Jpn J Appl Phys 1998;137:3809-11.
[29] Shi FG, Size dependent thermal vibrations and melting in nanocrystals. J Mater Res 1994;9:1307-13.
[30] Allan G, Delerue C, Lannoo M, and Martin E, Hydrogenic impurity levels, dielectric constant, and Coulumb charging effects in silicon crystallites. Phys Rev B 1995;52:11982-8.
[31] Chen XS, Zhao JJ, Wang GH, and Shen XC, The effect of size distributions of nanoclusters on photoluminescence from ensembles of Si nanoclusters. Phys Lett A 1996;212:285-9.
[32] Alivisatos AP, Semiconductor clusters, nanocrystals, and quantum dots. Science 1996;271:933-7.
[33] Pan LK, Sun CQ, Tay BK, Chen TP, and Li S, Photoluminescence of Si nanosolids near the lower end of the size limit. J Phys Chem B 2002;106:11725-7.
[34] Sun CQ, Li S, Tay BK, Chen TP, Upper limit of blue shift in the photoluminescence of CdSe and CdS nanosolids. Acta Mater 2002;50:4687-93.

[35] Sun CQ, Pan LK, Fu YQ, Tay BK and Li S, Size dependence of the 2p-level shift of nanosolid Silicon. J Phys Chem B 2003;107:5113-5.
[36] Ma DDD, Lee CS, Au FCK, Tong SY, and Lee ST, Small-diameter silicon nanowire surfaces. Science 2003;299:1874-7.
[37] Shi J, Gider S, Babcock K, and Awschalom DD, Magnetic Clusters in molecular beams, metals, and semiconductors. Science 1996;271:937-41.
[38] Tong WP, Tao NR, Wang ZB, Lu J, and Lu K, Nitriding Iron at Lower Temperatures. Science 2003;299:686-8.
[39] Shibata T, Bunker BA, Zhang ZY, Meisel D, Vardeman CF, and Gezelter JD, Size-dependent spontaneous alloying of Au-Ag nanoparticles. J Am Chem Soc 2002;124:11989-96.
[40] Kennedy MK, Kruis FE, Fissan H, Mehta BR, Stappert S, and Dumpich G, Tailored nanoparticle films from monosized tin oxide nanocrystals: Particle synthesis, film formation, and size-dependent gas-sensing properties. J Appl Phys 2003;93:551-60.
[41] Wang YM, Ma E, Chen MW, Enhanced tensile ductility and toughness in nanostructured Cu. Appl Phys Lett 2002;80:2395-7.
[42] Champion Y, Langlois C, Guerin-Mailly S, Langlois P, Bonnentien JL, and Hytch MJ, Near-perfect elastoplasticity in pure nanocrystalline copper. Science 2003;300:310-1.
[43] Lu L, Shen YF, Chen XH, Qian LH, Lu K, Ultrahigh strength and high electrical conductivity in copper. Science 2004;304:422-6.
[44] Weissmuller J, Viswanath RN, Kramer D, Zimmer P, Wurschum R, and Gleiter H, Charge-Induced Reversible Strain in a Metal. Science, 2003;300:312-5.
[45] Hu Y, Tan OK, Cao W, Zhu W, Fabrication and characterization of nano-sized $SrTiO_3$-based oxygen sensor for near-room temperature operation. IEEE Sensors J 2004;in press
[46] Hu Y, Tan OK, Pan JS, Yao X, A new form of nano-sized $SrTiO_3$ material for near human body temperature oxygen sensing applications. J Phys Chem B 2004;108:11214-8.
[47] Kennedy M. K., Kruis F. E., and Fissan H., Mehta B.R., Stappert S. and Dumpich G., Tailored nanoparticle films from monosized tin oxide nanocrystals: Particle synthesis, film formation, and size-dependent gas-sensing properties, J Appl Phys 2003;93:552-
[48] Balamurugan B., Mehta B.R. and Shivaprasad S.M., Surface-modified CuO layer in the size-stabilized $Cu_2O$ nanoparticles, Appl Phys Lett 2001;79:3176-8.
[49] Zhou J, Sun CQ, Pita K, Lam YL, Zhou Y, Ng SL, Kam CH, Li LT, Gui ZL, Thermally tuning of the photonic band-gap of SiO2 colloid-crystal infilled with ferroelectric $BaTiO_3$. Appl Phys Lett 2001;78:661-3.
[50] Li B, Zhou J, Hao LF, Hu W, Zong RL, Cai MM, Fu M, Gui ZL, Li LT, Li Q, Photonic band gap in $(Pb,La)(Zr,Ti)O_3$ inverse opals. Appl Phys Lett 2003;82:3617-9.
[51] Lu K, Nanocrystalline metals crystallized from amorphous solids:nanocrystallization, structure, and properties. Mater Sci Eng R 1996;16:161-221.
[52] Agraït N, Yeyati AL and van Ruitenbeek JM, Quantum properties of atomic-sized conductors. Phys Rep 2003;377:81-279.
[53] Halporin WP, Quantum size effects in metal particles. Rev Mod Phys 58:1966;533-97
[54] Binns C, Nanoclusters deposited on surfaces. Surf Sci Rep 2001;44:1-49.
[55] Baletto F and Ferrando R, Structural properties of nanoclusters: energetic, thermodynamic, and kinetic effects, Rev Mod Phys 2005; 77:371-423.
[56] Baretzky B, Baro MD, Grabovetskaya GP, Gubicza J, Ivanov MB, Kolobov YR, Langdon TG, Lendvai J, Lipnitskii AG, Mazilkin AA, Nazarov AA, Nogues J, Ovidko IA, Protasova SG, Raab GI, Revesz A, Skiba NV, Sort J, Starink MJ, Straumal BB, Surinach S, Ungar T, Zhilyaev AP, **Fundamentals of interface phenomena in advanced bulk nanoscale materials,** REV ADV MATER SCI 2005; 9 (1): 45-108.
[57] Trwoga PF, Kenyon AJ, and Pitt CW, Modeling the contribution of quantum confinement to luminescence from silicon nanoclusters. J Appl Phys 1998;83:3789-94.
[58] Efros AL and Efros AL, Interband absorption of light in a semiconductor sphere. Sov Phys Semicond 1982;16:772-5
[59] Brus JE, On the development of bulk optical properties in small semiconductor crystallites. J Lumin 1984;31:381-4
[60] Kayanuma Y, Quantum-size effects of interacting electrons and holes in semiconductor microcrystals with spherical shape. Phys Rev 1988;B38:9797-805.
[61] Glinka, YD Lin SH, Hwang LP, Chen YT, Tolk NH, Size effect in self-trapped exciton photoluminescence from $SiO_2$-based nanoscale materials. Phys Rev B 2001;64:085421.
[62] Qin GG, Song HZ, Zhang BR, Lin J, Duan JQ, and Yao GQ, Experimental evidence for luminescence from silicon oxide layers in oxidized porous silicon. Phys Rev B 1996;54:2548-55.

[63] Koch F, Petrova-Koch V, Muschik T, Nikolov A and Gavrilenko V, in Micrystalline Semiconductors:Materials Science and Devices:Materials research Society, Pittsburgh, PA, 1993:283:197.
[64] Prokes SM, Surface and optical properties of porous silicon. J Mate Res 1996;11:305-19.
[65] Iwayama TS, Hole DE, and Boyd IW, Mechanism of photoluminescence of Si nanocrystals in $SiO_2$ fabricated by ion implantation:the role of interactions of nanocrystals and oxygen. J Phys Condens Matter 1999;11:6595-604.
[66] Lindemann FA, Z Phys1910;11:609.
[67] Born M, J Chem Phys 1939;7:591.
[68] Buffat P and Borel JP, Size effect on the melting temperature of gold particles. Phys Rev A 1976;13:2287-98.
[69] Pawlow P, Z Phys Chem Munich 1909;65:1.
[70] Reiss H, Mirabel P, and Whetten RL, Capillarity theory for the "coexistence" of liquid and solid clusters. J Phys Chem 1988;92:7241-6.
[71] Sakai H, Surface-induced melting of small particles. Surf Sci 1996;351:285-91.
[72] Ubbelohde AR, The molten State of Materials. Wiley, New York, 1978.
[73] Wronski CRM, J Appl Phys 1967;18:1731.
[74] Hanszen KJ, Z Phys 1960;157:523.
[75] Couchman PR and Jesser WA, Nature (London) 1977;269:481.
[76] Vanfleet RR and Mochel JM, Thermodynamics of melting and freezing in small particles. Surf Sci 1995;341:40-50.
[77] Jiang Q, Liang LH, and Li JC, Thermodynamic superheating and relevant interface stability of low-dimensional metallic crystals. J Phys Condens Matt 2001;13:565-71.
[78] Jiang Q, Tong HY, Hsu DT, Okuyama K, and Shi FG, Thermal stability of crystalline thin films. Thin Solid Films 1998;312:357-61.
[79] Jiang Q, Shi HX, and Li JC, Finite size effect on glass transition temperatures. Thin Solid Films 1999;354:283-6.
[80] Wen Z, Zhao M, Jiang Q, The melting temperature of molecular nanocrystals at the lower bound of the mesoscopic size range. J Phys Condens Matt 2000;12:8819-24.
[81] Jiang Q, Liang LH, and Zhao M, Modelling of the melting temperature of nano-ice in MCM-41 pores. J Phys Condens Matt 2001;13:L397-401.
[82] Vekhter B and Berry RS, Phase coexistence in clusters:an "experimental" isobar and an elementary model. J Chem Phys 1997;106:6456-9.
[83] Nanda KK, Sahu SN, and Behera SN, Liquid-drop model for the size-dependent melting of low-dimensional systems. Phys Rev A 2002;66:013208.
[84] Wautelet M, Estimation of the variation of the melting temperature with the size of small particles, on the basis of a surface-phonon instability model. J Phys D 1991;24:343-6.
[85] Jiang Q, Zhang Z, and Li JC, Superheating of nanocrystals embedded in matrix. Chem Phys Lett 2000;322:549-52.
[86] Vallee R, Wautelet M, Dauchot JP, and Hecq M, Size and segregation effects on the phase diagrams of nanoparticles of binary systems. Nanotechnology 2001;12:68-74.
[87] Goldschmidt VM, Ber Deut Chem Ges 1927;60:1270.
[88] Pauling L, Atomic radii and interatomic distances in metals. J Am Chem Soc 1947;69:542-53.
[89] Sun CQ, Li S, and Tay BK, Laser-like mechanoluminescence in ZnMnTe-diluted magnetic semiconductor. Appl Phys Lett 2003;82:3568-9.
[90] Sun CQ, The lattice contraction of nanometersized Sn and Bi particles. J Phys Condens Matt 1999;11:4801-3.
[91] Sun CQ, Li S, Li CM, Impact of bond-order loss on surface and nanosolid mechanics. J Phys Chem B 2005;109: 415-23.
[92] Zeng XT, Zhang S, Sun CQ, and Liu YC, Nanometric-layered CrN/TiN thin films:mechanical strength and thermal stability. Thin Solid Films 2003;424:99-102.
[93] Zhong WH, Sun CQ, Tay BK, Li S, Bai HL, and Jiang EY, Curie temperature suppression of ferromagnetic nanosolids. J Phys Condens Matt 2002;14:L399-405.
[94] Pan LK, Huang HT, Sun CQ, Dielectric transition and relaxation of nanosolid silicon. J Appl Phys 2003;94:2695-700.
[95] Ye HT, Sun CQ, Huang HT, Hing P, Dielectric transition of nanostructured diamond films. Appl Phys Lett 2001;78:1826-8.
[96] Sun CQ, Wang Y, Tay BK, Li S, Huang H, Zhang YB, Correlation between the melting point of a nanosolid and the cohesive energy of a surface atom. J Phys Chem B 2002;106:10701-5.

[97] Pan LK, Sun CQ and Li CM, Elucidating Si-Si dimer vibration from the size-dependent Raman shift of nanosolid Si. J Phys Chem B 2004;108:L3404-6.
[98] Sun CQ, Sun XW, Gong HQ, Huang H, Ye H, Jin D, and Hing P, Frequency shift in the photoluminescence of nanometric SiOx: surface bond contraction and oxidation. J Phys Condens Matt 1999;11:L547-50.
[99] Sun CQ, Pan LK, Bai HL, Li ZQ, Wu P, and Jiang EY, Effects of surface passivation and interfacial reaction on the size-dependent 2p-level shift of supported copper nanosolids. Acta Mater 2003;51:4631-6.
[100] Sun CQ, Chen TP, Tay BK, Li S, Huang H, Zhang YB, Pan LK, Lau SP, Sun XW, An extended quantum confinement theory:surface-coordination imperfection modifies the entire band structure of a nanosolid. J Phys D 2001;34:3470-9.
[101] Zhong WH, Sun CQ, Li S, Bai HL, and Jiang EY, Impact of bond order loss on surface and nanosolid magnetism, Acta Materialia, in press.
[102] Huang HT, Sun CQ, Zhang TS, Hing P, Grain-size effect on ferroelectric $Pb(Zr_{1-x}Ti_x)O_3$ solid solutions induced by surface bond contraction. Phys Rev B 2001;63:184112.
[103] Huang HT, Sun CQ, Hing P, Surface bond contraction and its effect on the nanometric sized lead zirconate titanate. J Phys Condens Matt 2000;12:L127-32.
[104] Sun CQ, Sun XW, Tay BK, Lau SP, Huang H, and Li S, Dielectric suppression and its effect on photoabsorption of nanometric semiconductors. J Phys D 2001;34:2359-62.
[105] Pan LK, Ee YK, Sun CQ, Yu GQ, Zhang QY, and Tay BK, Band-gap expansion, core-level shift and dielectric suppression of porous Si passivated by plasma fluorination. J Vac Sci Technol 2004;22:583-7.
[106] Sun CQ, Atomic-coordination-imperfection-enhanced Pd-$3d_{5/2}$ crystal binding energy. Surf Rev Lett 2003;10:1009-13.
[107] Sun CQ, Surface and Nanosolid Core-level Shift: Impact of Atomic Coordination Number Imperfection. Phys Rev B 2004;69:045105.
[108] Sun CQ, Tay BK, Fu YQ, Li S, Chen TP, Bai HL and Jiang EY, Discriminating crystal bonding from the atomic trapping of a core electron at energy levels shifted by surface relaxation or nanosolid formation. J Phys Chem B 2003;107:L411-4.
[109] Phillips WD, Laser cooling and trapping of neutral atoms. Rev Mod Phys 1998;70:721-41
[110] Pan LK and Sun CQ, Coordination imperfection enhanced electron-phonon interaction. J Appl Phys 2004;95:3819-21.
[111] Sun CQ, Zhong WH, Li S, Tay BK, Coordination imperfection suppressed phase stability of ferromagnetic, ferroelectric, and superconductive nanosolids. J Phys Chem B 2004;108:1080-4.
[112] Pan LK, Sun CQ, Li CM, Estimating the extent of surface oxidation by measuring the porosity dependent dielectrics of oxygenated porous silicon. Appl Surf Sci 2005;240:19-23.
[113] Pan LK, Sun CQ, Yu GQ, Zhang QY, Fu YQ, and Tay BK, Distinguishing the effect of surface passivation from the effect of size on the photonic and electronic behavior of porous silicon. J App Phys 2004;96:1074-5.
[114] Sun CQ, Bai HL, Li S, Tay BK, Jiang EY, Size effect on the electronic structure and the thermal stability of a gold nanosolid. Acta Mater 2004;52:501-5.
[115] Sun CQ, Bai HL, Li S, Tay BK, Li CM, Chen TP, Jiang EY, Length, strength, extensibility and thermal stability of an Au-Au bond in the gold monatomic chain. J Phys Chem B 2004;108:2162-7.
[116] Sun CQ, Li CM, Li S and Tay BK, Breaking limit of atomic distance in an impurity-free monatomic chain. Phys Rev 2004;B67:245402.
[117] Sun CQ, Bai HL, Tay BK, Li S, Jiang EY, Dimension, strength, and chemical and thermal stability of a single C-C bond in carbon nanotubes. J Phys Chem B 2003;107:7544-6.
[118] Wales DJ, Structure, dynamics, and thermodynamics of clusters:tales from topographic potential surfaces. Science 1996;271:925-9.
[119] Sun CQ, Oxygen-reduced inner potential and work function in VLEED. Vacuum 1997;48:865-9
[120] Sun CQ, Spectral sensitivity of the VLEED to the bonding geometry and the potential barrier of the O-Cu(001) surface, Vacuum 1997;48:491-8.
[121] Sun CQ and Bai CL, Modelling of non-uniform electrical potential barriers for metal surfaces with chemisorbed oxygen, J Phys Condens Matt 1997;9:5823-36.
[122] Jones RO, Jennings PJ, and Jepsen O, Surface barrier in metals:A new model with application to W(001). Phys Rev B 1984;29:6474-80.
[123] Jennings PJ, Sun CQ, Low energy electron diffraction, In the Surface Analysis Methods in Materials Science, Eds O'Connor DJ, Sexton BA and Smart RC, Berlin Springer-Verlag, New York 2th ed 2003.
[124] Sinnott MJ, The solid state for engineers, Wiley and Sons, New York, 1963.

[125] Shpyrko OG, Grigoriev AY, Steimer C, Pershan PS, Lin B, Meron M, Graber T, Gerbhardt J, Ocko B, Deutsch M, Phys Rev B 70, 224206 (2004)
[126] Mahnke H.-E., Haas H., Holub-Krappe E., Koteski V., Novakovic N., Fochuk P. and Panchuk O, Lattice distortion around impurity atoms as dopants in CdTe, *Thin Solid Films 480-481, (2005) 279-282.*
[127] Feibelman PJ, Relaxation of hcp(0001) surfaces: A chemical view. Phys Rev 1996;B53:13740-6.
[128] Bahn SR and Jacobsen KW, Chain Formation of Metal Atoms. Phys Rev Lett 2001;87:266101.
[129] Kocks UF, Argon AS, and Ashby AS, Prog Mater Sci 1975;19:1.
[130] Kittel C, Introduction to Solid State Physics. 6th Ed, Wiley, New York, 1986.
[131] Harrison WA, Electronic Structure and the Properties of Solids. Freeman, San Francisco, 1980.
[132] Sun CQ and Li S, Oxygen derived DOS features in the valence band of metals. Surf Rev Lett 2000;7:L213-7.
[133] Sun CQ, A model of bonding and band-forming for oxides and nitrides. Appl Phys Lett 1998;72:1706-8.
[134] Street RA, Hydrogenated amorphous silicon. Cambridge University Press, Cambridge, 1991.
[135] Sun CQ, O-Cu(001): I Binding the signatures of LEED, STM and PES in a bond-forming way. Surf Rev Lett:2001;8:367-402.
[136] Sun CQ, O-Cu(001): II VLEED quantification of the four-stage $Cu_3O_2$ bonding kinetics. Surf Rev Lett 2001;8:703-34.
[137] Sun CQ, The sp hybrid bonding of C, N and O to the fcc(001) surface of nickel and rhodium. Surf Rev Lett 2000;7:347-63.
[138] Sun CQ, Jin D, Zhou J, Li S, Tay BK, Lau SP, Sun XW, Huang HT, Hing P, Intense and stable blue-light emission of $Pb(Zr_xTi_{1-x})O_3$. Appl Phys Lett 2001;79:1082-4.
[139] Zheng WT, Sun CQ, and Tay BK, Modulating the work function of carbon by O and N addition and nanotip formation. Solid State Commun 2003;128:381-4.
[140] Li J, Zheng WT, Gu C, Jin Z, Zhao Y, Mei X, Mu Z, Dong C, Sun CQ, Field emission enhancement of amorphous carbon films by nitrogen-implantation. Carbon 2004;42:2309–14.
[141] Sun CQ, Fu YQ, Yan BB, Hsieh JH, Lau SP, Sun XW, Tay BK, Improving diamond-metal adhesion with graded TiCN interlayers. J Appl Phys 2002;91:2051-4.
[142] Sun CQ, Tay BK, Lau SP, Sun XW, Zeng XT, Bai H, Liu H, Liu ZH, Jiang EY, Bond contraction and lone pair interaction at nitride surfaces. J Appl Phys 2001;90:2615-7.
[143] Sun CQ A model of bond-and-band for the behavior of nitrides. Mod Phys Lett B 1997;11:1021-9.
[144] Jiang EY, Sun CQ, Li JE and Liu YG, The structures and magnetic properties of FeN films prepared by the facing targets sputtering method. J Appl Phys 1989;65:1659-63.
[145] Fu YQ, Sun CQ, Yan BB, and Du HJ, Carbon turns the tensile surface stress of Ti to be compressive. J Phys D 2001;34:L129-32.
[146] Fu YQ, Sun CQ, Du HJ, and Yan BB, Crystalline carbonitride formation is harder than the hexagonal Si-carbonitride crystallite. J Phys D 2001;34:1430-5.
[147] Tománek D, Mukherjee S, and Bennemann KH, Simple theory for the electronic and atomic structure of small clusters. Phys Rev B 1983;28:665-73.
[148] Duan Y, Li J, Structure study of nickel nanoparticles. Mater Chem Phys 2004;87:452–4.
[149] Montano PA, Schulze W, Tesche B, Shenoy GK, and Morrison TI, Extended x-ray-absorption fine-structure study of Ag particles isolated in solid argon. Phys Rev B 1984;30:672–7.
[150] Lamber R, Wetjen S, and Jaeger NI, Size dependence of the lattice parameter of small palladium particles. Phys Rev B 1995;51:10968–71.
[151] Mi WB and Bai HL, private communications. Tianjin, June 2004.
[152] Yu XF, Liu X, Zhang K, and Hu ZQ, Lattice contraction of nanometre-sized Sn and Bi particles produced by an electrohydrodynamic technique. J Phys Condens Matt 1999;11:937-44.
[153] Reddy DR, and Reddy BK, Laser-like mechanoluminescence in ZnMnTe-diluted magnetic semiconductor. Appl Phys Lett 2002;81:460-2.
[154] Montano PA, Shenoy GK, Alp EE, Schulze W, and Urban J, Structure of Copper Microclusters Isolated in Solid Argon. Phys Rev Lett 1986;56:2076–9.
[155] Hansen LB, Stoltze P, and Nørskov JK, Is there a contraction of the interatomic distance in small metal particles? Phys Rev Lett 1990;64:3155–8.
[156] Weissker HC, Furthmuller J, Bechstedt F, Structural relaxation in Si and Ge nanocrystallites: Influence on the electronic and optical properties. Phys Rev B 2003;67:245304.
[157] Buttard D, Dolino G, Faivre C, Halimaoui A, Comin F, Formoso V, and Ortega L, Porous silicon strain during in situ ultrahigh vacuum thermal annealing. J Appl Phys 1999;85:7105-11.
[158] Mays CW, Vermaak JS, and Kuhlmann-Wilsdorf D, On surface stress and surface tension II Determination of the surface stress of gold. Surf Sci 1968;12:134-40.

[159] Stoneham AM, The lattice contraction of nanometersized Sn and Bi particles. J Phys Condens Matt 1999;11:8351-2.
[160] Wasserman HJ, and Vermaak JS, On the determination of a lattice contraction in very small silver particles. Surf Sci 1970;22:164-72.
[161] Nanda KK, Behera SN, and Sahu SN, The lattice contraction of nanometre-sized Sn and Bi particles produced by an electrohydrodynamic technique. J Phys Condens Matt 2001;13:2861-4
[162] Jiang Q, Liang LH, and Zhao DS. Lattice Contraction and Surface Stress of fcc Nanocrystals. J Phys Chem B 2001;105:6275-7.
[163] Liang LH, Li JC and Jiang Q, Size-dependent melting depression and lattice contraction of Bi nanocrystals. Physica B: Condens Matt 2003;334:49-53.
[164] Kara A and Rahman TS, Vibrational Properties of Metallic Nanocrystals. Phys Rev Lett 1998;81:1453-6.
[165] Toyama T, Adachi D and Okamoto H, Electroluminescent devices with nanostructured ZnS:Mn emission layer operated at 20 V0-p. Mater Res Symp Proc 2000;621:q441-6.
[166] Woltersdorf J, Nepijko AS, Pippel E, Dependence of lattice parameters of small particles on the size of the nuclei. Surf Sci 1981;106:64-9.
[167] Zhao M, Zhou XH and Jiang Q, Comparison of different models for melting point change of metallic nanocrystals. J Mater Res 2001;16:3304-8.
[168] Müller H, Opitz C, Strickert K, Skala LZ, Phys Chemie Leipzig 1987;268:634.
[169] Kellermann G and Craievich AF. Structure and melting of Bi nanocrystals embedded in a $B_2O_3$-$Na_2O$ glass. Phys Rev B 2002;65:134204.
[170] Apai G, Hamilton JF, Extended x-ray absorption fine structure of small copper and nickel clusters:binding energy and bond length changes with cluster size. Phys Rev Lett 1979;43:165-8.
[171] Wasserman HJ, and Vermaak JS, On the determination of the surface stress of copper and platinum. Surf Sci 1972;32:168-74.
[172] Liu JP, Zaumseil P, Bugiel E, and Osten HJ, Epitaxial growth of $Pr_2O_3$ on Si(111) and the observation of a hexagonal tocubic phase transition during postgrowth $N_2$ annealing. Appl Phys Lett 79:2001;671-3.
[173] Wang LW, Zunger A, Pseudopotential calculations of nanoscale CdSe quantum dots. Phys Rev 1996;53:9579-82. B
[174] Leung K, Whaley KB, Surface relaxation in CdSe nanocrystals. J Chem Phys 1999;110:11012-22.
[175] Rabani E, Structure and electrostatic properties of passivated CdSe nanocrystals. J Chem Phys 2001;115:1493-7.
[176] Puzder A, Williamson AJ, Reboredo FA, Galli G, Structural Stability and Optical Properties of Nanomaterials with Reconstructed Surfaces. Phys Rev Lett 2003;91:1574051-4.
[177] Gilbert B, Huang F, Zhang H, Waychunas GA, Banfield JF, Nanoparticles:Strained and Stiff. Science 2004;305: 651-4.
[178] Nanda J, and Sarma DD, Photoemission spectroscopy of size selected zinc sulfide nanocrystallites. J Appl Phys 2001;90:2504-10.
[179] Baraldi G, Silvano L, Comelli G, Oxygen adsorption and ordering on Ru(1010). Phys Rev B 2001;63:115410.
[180] Over H, Kleinle G, Ertl G, Moritz W, Ernst KH, Wohlgemuth H, Christmann K, and Schwarz E, LEED structural analysis of the Co(1010) surface. Surf Sci 1991;254:L469-74.
[181] Davis HL, Hannon JB, Ray KB, and Plummer EW, Anomalous interplanar expansion at the (0001) surface of Be. Phys Rev Lett 1992;68:2632-5.
[182] Halicioglu T, Calculation of surface energies for low index planes of diamond. Surf Sci 1991;259:L714-8.
[183] Sun CQ, Exposure-resolved VLEED from the O-Cu(001): Bonding dynamics, Vacuum 1997;48:535-41.
[184] Sun CQ, Origin and processes of O-Cu(001;and the O-Cu(110) biphase ordering, Int J Mod Phys B 1998;12:951-64.
[185] Nautiyal T, Youn SJ, and Kim KS, Effect of dimensionality on the electronic structure of Cu, Ag, and Au. Phys Rev B 2003;68:033407.
[186] Zhao YH, Zhang K, and Lu K, Structure characteristics of nanocrystalline element selenium with different grain sizes. Phys Rev B 1997;56:14322-9.
[187] Qian X and Hübner W, First-principles calculation of structural and magnetic properties for Fe monolayers and bilayers on W(110). Phys Rev B 1999;60:16192-7.
[188] Geng WT, Freeman AJ, and Wu RQ, Magnetism at high-index transition-metal surfaces and the effect of metalloid impurities: Ni(210). Phys Rev B 2001;63:064427.

[189] Geng WT, Kim M, and Freeman AJ, Multilayer relaxation and magnetism of a high-index transition metal surface: Fe(310). Phys Rev B 2001;63:245401.
[190] Batra IP, Lattice relaxation in aluminum monolayers. J Vac Sci Technol A 1985;3:1603-6.
[191] Banerjee R, Sperling EA, Thompson GB, Fraser HL, Bose S, and Ayyub P, Lattice expansion in nanocrystalline niobium thin films. Appl Phys Lett 2003;82:4250-2.
[192] Walko DA and Robinson IK, Structure of Cu(115): Clean surface and its oxygen-induced facets. Phys Rev B 1999;59:15446-58.
[193] Durukanolu S and Rahman TS, Structure of Ag(410) and Cu(320). Phys Rev B 2003;67:205406.
[194] Nascimento VB, Soares EA, de Carvalho VE, Lopes EL, Paniago R, and de Castilho CM, Thermal expansion of the Ag(110) surface studied by low-energy electron diffraction and density-functional theory. Phys Rev B 2003;68:245408.
[195] Galanakis I, Papanikolaou N, Dederichs PH, Applicability of the broken-bond rule to the surface energy of the fcc metals. Surf Sci 2002;511:1–12.
[196] Jona F, Marcus PM, Zanazzi E, and Maglietta M, Structure of Ag(410). Surf Rev Lett 1999;6:355-9
[197] Sklyadneva YI, Rusina GG, and Chulkov EV, Vibrational states on vicinal surfaces of Al, Ag, Cu and Pd. Surf Sci 1998;416:17-36.
[198] Kara A, Durukanoglu S, and Rahman TS, Local thermodynamic properties of a stepped metal surface: Cu(711). Phys Rev B 1996;53:15489-92.
[199] Zhang P and Sham TK, X-Ray Studies of the Structure and Electronic Behavior of Alkanethiolate-Capped Gold Nanoparticles: The Interplay of Size and Surface Effects. Phys Rev Lett 2003;90:245502.
[200] Ibach H, Role of surface stress in reconstruction, epitaxial growth and stabilization of mesoscopic structures. Surf Sci Rep 1997;29:193-263.
[201] Haiss W, Surface stress of clean and adsorbate-covered solids. Rep Prog Phys 2001;64:59-648.
[202] Poa CHP, Lacerda RG, Cox DC, Silva SRP, and Marques FC, Stress-induced electron emission from nanocomposite amorphous carbon thin films. Appl Phys Lett 2002;81:853-5.
[203] de Boer FR, Boom R, Mattens WCM, Miedema AR and Niessen AK, Cohesion in Metals. Amsterdam, North-Holland Publishing Company, 1998, Vol 1.
[204] Vitos L, Ruban AV, Skriver HL, and Kollár J, Surface energy of metals. Surf Sci 1998;411:186-202.
[205] Mattsson AE and Jennison DR, Computing accurate surface energies and the importance of electron self-energy in metal/metal-oxide adhesion. Surf Sci 2002;520:L611-8.
[206] Methfessel M, Hennig D, and Scheffler M, Trends of the surface relaxations, surface energies, and work functions of the 4d transition metals. Phys Rev B 1992;46:4816-29.
[207] Skriver HL, and Rosengaard NM, Surface energy and work function of elemental metals. Phys Rev B 1992;46:7157-68.
[208]Mehl MJ and Papaconstantopoulos DA, Applications of a tight-binding total-energy method for transition and noble metals: Elastic constants, vacancies, and surfaces of monatomic metals. Phys Rev 1996;B54:4519-30.
[209] Rodríguez AM, Bozzolo G, and Ferrante J, Multilayer relaxation and surface energies of FCC and BCC metals using equivalent crystal theory. Surf Sci 1993;289:100-26.
[210] Alden M, Skriver HL, Mirbt S, and Johansson B, Calculated surface-energy anomaly in the 3d metals. Phys Rev Lett 1992;69:2296-8.
[211] Moriarty JA, and Phillips R, First-principles interatomic potentials for transition-metal surfaces. Phys Rev Lett 1991;66:3036-9.
[212] Thomson W (Kelvin), Philos Mag 1871;42:448.
[213] Rytkönen A, Valkealahti S, and Manninen M, Melting and evaporation of argon clusters. J Chem Phys 1997;106:1888-92.
[214] Nanda KK, Maisels A, Kruis FE, Fissan H, and Stappert S, Higher Surface Energy of Free Nanoparticles. Phys Rev Lett 2003;91:106102.
[215] Lu HM, and Jiang Q, Comment on: Higher Surface Energy of Free Nanoparticles. Phys Rev Lett 2004;92:179601.
[216] Desjonquères MC and Spanjaard D, Concepts in Surface Physics. Springer Series in Surface: Berlin, Heidelberg:Springer-Verlag, 1993, Vol 30.
[217] Jiang Q, Lu HM, and Zhao M, Modelling of surface energies of elemental crystals. J Phys Condens Matter 2004;16:521–30.
[218] Baskes MI, Many-Body Effects in fcc Metals: A Lennard-Jones Embedded-Atom Potential. Phys Rev Lett 1999;83:2592-5.
[219] Shi X, Tay BK, Flynn DL, and Sun Z, Tribological properties of tetrahedral carbon films deposited by filtered cathodic vacuum arc technique. Mat Res Symp Proc 1997;436:293-8.

[220] Caceres D, Vergara I, Gonzalez R, Monroy E, Calle F, Munoz E, Omnes F, Nanoindentation on AlGaN thin films. J Appl Phys 1999;86:6773-8.
[221] Mirshams RA, Parakala P, Nanoindentation of nanocrystalline Ni with geometrically different indenters. Mater Sci Eng A 2004;372:252–60.
[222] Wang YH, Moitreyee MR, Kumar R, Wu SY, Xie JL, Yew P, Subramanian B, Shen L and Zeng KY, The mechanical properties of ultra-low-dielectric-constant films, Thin Solid Films 2004;462-463:227-230.
[223] Liu E, Shi X, Tan HS, Cheah LK, Sun Z, Tay BK, Shi JR, The effect of nitrogen on the mechanical properties of metrahedral amorphous carbon films deposited with a filtered cathodic vacuum arc. Surf Coat Technol 1999;120–121:601–6.
[224] Zhao M, Slaughter WS, Li M, Mao SX, Material-length-scale-controlled nanoindentation size effects due to strain-gradient plasticity. Acta Mater 2003;51:4461–9.
[225] Dodson BW, Many-Body Surface Strain and Surface Reconstructions in fcc Transition Metals,. Phys Rev Lett 1988;60:2288-91.
[226] Streitz FH, Cammarata RC, and Sieradzki K, Surface-stress effects on elastic properties: I Thin metal films. Phys Rev 1994;B49:10699-706.
[227] Liu M, Shi B, Guo J, Cai X and Song H, Lattice constant dependence of elastic modulus for ultrafine grained mild steel. Scripta Mater 2003;49:167-71.
[228] Horstemeyer MF, Baskes MI, and Plimpton SJ, Length scale and time scale Dects on the plastic Cow of fcc metals. Acta Mater 2001;49:4363-74.
[229] Kim JJ, Choi Y, Suresh S, Argon AS, Nanocrystallization during nanoindentation of a bulk amorphous metal alloy at room temperature. Science 2002;295:654-7.
[230] Baker SP, Vinci RP, and Arias T, Elastic and anelastic behavior of materials in small dimensions. MRS Bull 2002;27:26-9.
[231] Murayama M, Howe JM, Hidaka H, Tokai S, Atomic Level Observation of Disclination Diposes in Mechanically Milled, Nanocrystalline Fe. Science 2002;295:2433-5.
[232] Veprek S, Argon AS, Mechanical properties of superhard nanocomposites. Surf Coat Technol 2001;146–147:175–82.
[233] Ferro D, Teghil R, Barinov SM, D'Alessio L, DeMaria G, Thickness-dependent hardness of pulsed laser ablation deposited films of refractory carbides. Mater Chem Phys 2004;87:233–6.
[234] Yamakov V, Wolf D, Salazar M, Phillpot SR, and Gleiter H, Length-scale Dects in the nucleation of extended dislocations in nanocrystalline Al by molecular-dynamics simulation. Acta Mater 2001;49:2713–22.
[235] Cantor B, Chang ITH, Knight P, Vincent AJB, Microstructural development in equiatomic multicomponent. Mater Sci Eng 2004;375-377:213-8.
[236] Gao F, He J, Wu E, Liu S, Yu D, Li D, Zhang S, and Tian Y, Hardness of Covalent Crystals. Phys Rev Lett 2003;91:015502.
[237] Cohen ML, and Bergstresser TK, Band Structures and Pseudopotential Form Factors for Fourteen Semiconductors of the Diamond and Zinc-blende Structures. Phys Rev 1966;141:789-96.
[238] Philips JC, Covalent Bonding in Crystals, Molecules and Polimers: Chicago University Press, Chicago, 1969.
[239] Liu AY and L CohenM, Prediction of new low compressibility solids. Science 1989;245:841-2
[240] Korsunskii BL and Pepekin VP, Usp Khim 1977;66:1003.
[241] Litovchenko V, Analysis of the band structure of tetrahedral diamondlike crystals with valence bonds: Prediction of materials with superhigh hardness and negative electron affinity. Phys Rev B 2002;65:153108.
[242] Wang EG, Research on Carbon Nitrides. Prog Mater Sci 1997;41:241-300.
[243] Wang EG, New development in covalently bonded carbon-nitride and related materials. Adv Mater 1999;11:1129-33.
[244] Zheng WT, Sjostrom H, Ivanov I, Xing KZ, Broitman E, Salaneck WR, Greene JE, Sundgreen JE, Reactive magnetron sputter deposited CNx: Effects of $N_2$ pressure and growth temperature on film composition, bonding, and microstructure. J Vac Sci Technol A 1996;14:2696-2701.
[245] Veprek S, Conventional and new approaches towards the design of novel superhard materials. Surf Coat technol 1997;97:15-22.
[246] Arzt E, Ashby MF, Threshoul stresses in materials containing dispersed particles. Scripta Metallurgica 1982;16:1285-90.
[247] Van Swygenhoven H, Derlet PM, and Hasnaoui A, Atomic mechanism for dislocation emission from nanosized grain boundaries. Phys Rev B 2002;66:024101.

[248] Cheng S, Spencer JA, and Milligan WW, Strength and tension/compression asymmetry in nanostructured and ultrafine-grain metals. Acta Mater 2003;51:4505-18.
[249] Schiøtz J, and Jacobsen KW, A maximum in the strength of nanocrystalline copper. Science 2003;301:1357-9.
[250] Zhong WL, Jiang B, Zhang PL, Ma JM, Cheng HM, Yang ZH, and Li LX, Phase transition in $PbTiO_3$ ultrafine particles of different sizes. J Phys Condens Matt 1993;5:2619-24.
[251] Bata V and Pereloma EV, An alternative physical explanation of the Hall-Petch relation. Acta Mater 2004;52:657-65.
[252] Wang GF, Feng XQ, Yu SW, Interfacial effects on effective elastic moduli of nanocrystalline Materials. Mat Sci Eng A 2003;363:1C8.
[253] Wolf D, Yamakov V, Phillpot SR, and Mukherjee AK, Deformation mechanism and inverse Hall-Petch behavior in nanocrystalline materials. ZEITSCHRIFT FUR METALLKUNDE 2003;94:1091-7.
[254] Yamakov V, Wolf D, Phillpot SR, Mukherjee AK, and Gleiter H, Deformation-mechanism map for nanocrystalline metals by molecular-dynamics simulation. Nature Mater 2004;3:43-7.
[255] Kumar KS, Van Swygenhoven H, and Suresh S, Mechanical behavior of nanocrystalline metals and alloys. Acta Mater 2003;51:5743-74.
[256] Qi WH and Wang MP, Size effect on the cohesive energy of nanoparticle. J Mater Sci Lett 2002;21:1743-5.
[257] Xie D, Wang MP and Qi WH, A simplified model to calculate the surface-to-volume atomic ratio dependent cohesive energy of nanocrystals. J Phys Condens Matter 2004;16:L401-5.
[258] Qi WH, Wang MP, Xu GY, The particle size dependence of cohesive energy of metallic nanosolids. Chem Phys Lett 2003;372:632-4.
[259] Jiang Q, Li JC, Chi BQ, Size-dependent cohesive energy of nanocrystals. Chem Phys Lett 2002;366:551-4
[260] Kim HK, Huh SH, Park JW, Jeong JW, Lee GH, The cluster size dependence of thermal stabilities of both molybdenum and tungsten nanoclusters. Chem Phys Lett 2002;354:165-72
[261] Gorecki T, Z Metallkde 1974;65:426.
[262] Finis MW, The Harris functional applied to surface and vacancy formation energies in aluminium. J Phys Condens Mat 1990;2:331-42.
[263] Brooks H, Impurities and Imperfection. American Socienty for Metals, Cleveland, 1955.
[264] Qi, WH and Wang MP, Size dependence of vacancy formation energy of metallic nanosolids. Physica B 2003;334:432-35.
[265] Brands EA, Smithells Metals Reference Book. 6th Ed, Butterworths, London, 1983, 15–22.
[266] Miedema AR, Surface energies of solid metals. Zeitschrift fur Metallkunde 1978;69:287-92.
[267] Dash JG, History of the search for continuous melting. Rev Mod Phys 1999;71:1737-43.
[268] Penfold J, The structure of the surface of pure liquids. Rep Prog Phys 2001;64:777–814.
[269] Sheng HW, Ren G, Peng LM, Hu ZQ, and Lu K, Superheating and melting-point depression of Pb nanosolids embedded in Al matrices. Philos Mag Lett 1996;73:179-86.
[270] Sheng HW, Ren G, Peng LM, Hu ZQ, and Lu K, Epitaxial dependence of the melting behavior of In nanosolids embedded in Al matrices. J Mater Res 1997;12:119-23.
[271] Lereah Y, Deutscher G, Cheyssac P, and Kofman R, A direct observation of low-dimensional effects on melting of small lead particles. Europhys Lett 1990;12:709-13.
[272] Rouse AA, Bernhard JB, Sosa ED, and Golden DE, Variation of field emission and photoelectric thresholds of diamond films with average grain size. Appl Phys Lett 1999;75:3417-9.
[273] Hamada N, Sawada A, and Oshiyama A, New one-dimensional conductors: Graphitic microtubules. Phys Rev Lett 1992;68:1579-81.
[274] Skripov VP, P KoverdaV, and Skokov VN, Size effect on melting of small particles. Phys Status Solidi A 1981;66:109-18.
[275] Pcza JF, Barna A, and Barna PB, Formation Processes of Vacuum-Deposited Indium Films and Thermodynamical Properties of Submicroscopic Particles Observed by In Situ Electron Microscopy. J Vac Sci Technol 1969;6:472-5.
[276] Santucci SC, Goldoni A, Larciprete R, Lizzit S, Bertolo M, Baraldi A, and Masciovecchio C, Calorimetry at Surfaces Using High-Resolution Core-Level Photoemission. Phys Rev Lett 2004;93:106105.
[277] Bergese P, Colombo I, Gervasoni D, and Depero LE, Melting of Nanostructured Drugs Embedded into a Polymeric Matrix. J Phys Chem B 2004;108:15488-15493.
[278] Hwang I-S, Chang S-H, Fang C-K, Chen L-J, and Tsong TT, Observation of Finite-Size Effects on a Structural Phase Transition of 2D Nanoislands. Phys Rev Lett 2004;93:106101.

[279] Zhang M, Yu Efremov M, Schiettekatte F, Olson EA, Kwan AT, Lai SL, Wisleder T, Greene JE, and Allen LH, Size-dependent melting point depression of nanostructures: Nanocalorimetric measurements. Phys Rev B 2000;62:10548-57.
[280] Jin ZH, Gumbsch P, Lu K, and Ma E, Melting Mechanisms at the Limit of Superheating. Phys Rev Lett 2001;87:055703.
[281] Defay R and Prigogine I, Surface Tension and adsorption. Wiley, New York, 1951.
[282] Peters KF, Cohen JB, and Chung YW, Melting of Pb nanocrystals. Phys Rev B 1998;57:13430-8.
[283] Rose JH, Smith JR, and Ferrante J, Universal features of bonding in metals. Phys Rev B 1983;28:1835-45.
[284] Tateno J, An empirical relation for the melting temperature of some ionic crystals. Solid State Commun 1972;10:61-2.
[285] Omar MA, Elementary Solid State Physics: Principles and Applications. Addison-Wesley, New York, 1975.
[286] Pluis B, Frenkel D, van der Veen JF, Surface-induced melting and freezing: II A semi-empirical Landau-type model. Surf Sci 1990;239:282-300.
[287] Wautelet M, Dauchot JP and Hecq M, Phase diagrams of small particles of binary systems:a theoretical approach. Nanotechnology 2000;11:6–9
[288] Wautelet M, Phase stability of electronically excited Si nanoparticles. J Phys Condens Matter 2004;16:L163–6.
[289] Wang J, Chen X, Wang G, Wang B, Lu W, and Zhao J, Melting behavior in ultrathin metallic nanowires. Phys Rev B 2002;66:085408
[290] Lai SL, Guo JY, Petrova V, Ramanath G, and Allen LH, Size-Dependent Melting Properties of Small Tin Particles: Nanocalorimetric Measurements. Phys Rev Lett 1996;77:99-102.
[291] Jiang Q, Lang XY, Glass transition of Low-dimensional Polystyrene. Macromol Rapid Comm 2004;25:825-8.
[292] Ding XZ and Liu XH, The Debye temperature of nanocrystalline titania measured by two different methods. Phys Status Solidi A 1996;158:433-9.
[293] Schmidt M, Kusche R, von Issendorff B, and Haberland H, Irregular variations in the melting point of size-selected atomic clusters. Nature (London) 1998;393:238-40.
[294] Lai SL, Carlsson JRA, and Allen LH, Melting point depression of Al clusters generated during the early stages of film growth: Nanocalorimetry measurements. Appl Phys Lett 1998;72:1098-100.
[295] Bottani CE, Bassi AL, Tanner BK, Stella A, Tognini P, Cheyssac P, and Kofman R, Melting in metallic Sn nanoparticles studied by surface Brillouin scattering and synchrotron-x-ray diffraction. Phys Rev B 1999;59:R15601-4.
[296] Zhong J, Zhang LH, Jin ZH, Sui ML, Lu K, Superheating of Ag nanoparticles embedded in Ni matrix. Acta Mater 2001;49:2897-2904.
[297] Lu L, Schwaiger R, Shan ZW, Dao M, Lu K, Suresh S, Nano-sized twins induce high rate sensitivity of flow stress in pure copper, ACTA MATERIALIA 2005;53:2169-79.
[298] Veprek S, Veprek-Heijman MGJ, Karvankova P, Prochazka J, Different approaches to superhard coatings and nanocomposites, THIN SOLID FILMS 2005;476:1-29.
[299] Srinivasan, S. G.; Liao, X. Z.; Baskes,M. I.; McCabe, R. J.; Zhao, Y. H. and Zhu, Y. T. Compact and Dissociated Dislocations in Aluminum: Implications for Deformation, Phys Rev Lett 2005, **94**, 125502.
[300] Ding F, Rosén A, and Bolton K, Size dependence of the coalescence and melting of iron clusters:A molecular-dynamics study. Phys Rev B 2004;70:075416.
[301] Dick K, Dhanasekaran T, Zhang Z, and Meisel D, Size-dependent melting of silica-encapsulated gold nanoparticles. J Am Chem Soc 2002;124:2312-7.
[302] Eckert J, Holzer JC, Ahn CC, Fu Z and Johnson WL, Melting behavior of nanocrystalline alminum powders. Nanostruc Mater 1993;2:407.
[303] Lai SL, Ramanath G, Allen LH, and Infante P, Heat capacity measurements of Sn nanostructures using a thin-film differential scanning calorimeter with 0.2 nJ sensitivity. Appl Phys Lett 1997;70:43-5.
[304] Allen GL, Gile WW, Jesser WA, Melting temperature of microcrystals embedded in a matrix. Acta Metall 1980;28:1695-701.
[305] Unruh KM, Huber TE, and Huber CA, Melting and freezing behavior of indium metal in porous glasses. Phys Rev B 1993;48:9021-7.
[306] David TB, Lereah Y, Deutscher G, Kofmans R, Cheyssac P, Solid-liquid transition in ultra-fine lead particles. Phil Mag A 1995;71:1135.
[307] Goldstein N, The melting of silicon nanocrystals submicro thin–film structures derived from nanocrystal precursors. Appl Phys A 1996;62:33-7.

[308] Kenlinski P, Thermodynamics and kinetics of melting and growth of crystalline silicon clusters. Mat Res Soc Symp Proc 1999;536:311-6.
[309] Wu Y, and Yang P, Melting andWelding Semiconductor Nanowires in Nanotubes. Adv Mater 2001;13:520-521
[310] Peppiat SJ, Proc R Soc London Ser A 1975;345:401.
[311] Itoigawa H, Kamiyama T, and Nakamura Y, Bi precipitates in $Na_2O–B_2O_3$ glasses. J Non-Cryst Solids 1997;210:95-100.
[312] Morishige K, Kawano K, Freezing and melting of methyl chloride in a single cylindrical pore:anomalous pore-size dependence of phase-transition temperature. J Phys Chem B 2000;104:2894-900.
[313] Molz E, Wong AP, Chan MHW, Beamish JR, Freezing and melting of fluids in porous glasses. Phys Rev B 1993;48:5741-50.
[314] Saka H, Nishikawa Y, Imura T, Temperature dependence of the stacking fault energy in silver-base alloys. Phil Mag A 1983;57:859-68.
[315] Gråback L, Bohr J, Superheating and supercooling of lead precipitates in aluminum. Phys Rev Lett 1990;64:934-7.
[316] Zhang L, Jin ZH, Zhang LH, Sui ML, Lu K, Superheating of confined Pb thin films. Phys Rev Lett 2000;85:1484-7.
[317] Chattopadhyay K, Goswami R, Melting and superheating of metals and alloys. Prog Mater Sci 1997;42:287-300.
[318] Breaux GA, Benirschke RC, Sugai T, Kinnear BS, and Jarrold MF, Hot and solid gallium clusters: too small to melt. Phys Rev Lett 2003;91:215508.
[319] Chacko S, Joshi K, and Kanhere DG, Why Do Gallium Clusters Have a Higher Melting Point than the Bulk. Phys Rev Lett 2004;92:135506.
[320] Lu ZY, Wang CZ, and Ho KM, Structures and dynamical properties of $C_n$, $Si_n$, $Ge_n$, and $Sn_n$ clusters with n up to 13. Phys Rev B 2000;61:2329-34.
[321] Chuang FC, Wang CZ, SerdarÖüt, Chelikowsky JR, and Ho KM, Melting of small Sn clusters by ab initio molecular dynamics simulations. Phys Rev B 2004;69:165408.
[322] Joshi K, Kanhere DG, and Blundell SA, Abnormally high melting temperature of the $Sn_{10}$ cluster. Phys Rev B 2002;66:155329.
[323] Breaux GA, Neal CM, Cao B, and Jarrold MF, Tin clusters that do not melt: Calorimetry measurements up to 650 K ,PHYS REV 2005, B 71, 073410.
[324] Jones RO, Simulated annealing study of neutral and charged clusters: AlN and GaN. J Chem Phys 1993;99:1194-206.
[325] Shvartsburg AA, Liu B, Lu ZY, Wang CZ, Jarrold MF, and Ho KM, Structures of germanium clusters:where the growth patterns of silicon and germanium clusters diverge. Phys Rev Lett 1999;83:2167-70.
[326] Lu GH, Deng S, Wang T, Kohyama M, and Yamamoto R, Theoretical tensile strength of an Al grain boundary. Phys Rev B 2004;69:134106.
[327] Carling K, Wahnstrüm G, Mattsson TR, Mattsson AE, Sandberg N, and Grimvall G, Vacancies in metals:from first-principles calculations to experimental data. Phys Rev Lett 2000;85:3862-5.
[328] Ogata S, Li J, and Yip S, Ideal pure shear strength of aluminum and copper. Science 2002;298: 807-11.
[329] Liu, H.H.; Jiang, E. Y.; Bai, H.L; Wu, P.; Li Z. Q.; and Sun, C. Q. Possible paths of magic clusters growth, J Theor Chem. in press.
[330] Bachels T, Guntherodt HJ, and Schafer R, Melting of Isolated Tin Nanoparticles. Phys Rev Lett 2000;85:1250-3.
[331] Breaux, G. A.; Neal, C. M.; Cao, B.; and Jarrold, M. F. Melting, Premelting, and Structural Transitions in Size-Selected Aluminum Clusters with around 55 Atoms, Phys. Rev. Lett. 2005, 94, 173401.
[332] Sun CQ, Li CM, Bai HL and Jiang EY, Melting point oscilation over the whole range of sizes, Nanotechnology, in press.
[333] Huang FF, Mankey GJ, Kief MT, and Willis RF, Finite-size scaling behavior of ferromagnetic thin films. J Appl Phys 1993;73:6760-2.
[334] Kenning GG, Slughter JM, and Cowen JA, Finite-Size Effects in a CuMn Spin-Glass. Phys Rev Lett 1997;59:2596-9.
[335] Qiu ZQ, Person J, and Bader SD, Asymmetry of the spin reorientation transition in ultrathin Fe films and wedges grown on Ag(100). Phys Rev Lett 1993;70:1006-9.
[336] Ishikawa K, Yoshikawa K, Okada N, Size effect on the ferroelectric phase transition in $PbTiO_3$ ultrafine particles. Phys Rev 1988;B37:5852-5.

[337] Yu T, Shen ZX, Toh WS, Xue JM, and Wang JJ, Size effect on the ferroelectric phase transition in $SrBi_2Ta_2O_9$ nanoparticles. J Appl Phys 2003;94:618-20.
[338] Pogrebnyakov AV, Redwing JM, Jones JE, Xi XX, Xu SY, Li Q, Vaithyanathan V, Schlom DG, Thickness dependence of the properties of epitaxial $MgB_2$ thin films grown by hybrid physical-chemical vapor deposition. Appl Phys Lett 2003;82:4319-21.
[339] Tsai AP, Chandrasekhar, Chattopadhyay N, Size effect on the superconducting transition of embedded lead particles in an Al-Cu-V amorphous matrix. Appl Phys Lett 1999;75:1527-8.
[340] Giaever I, Zeller HR, Superconductivity of Small Tin Particles Measured by Tunneling. Phys Rev Lett 1968;20:1504-7.
[341] Stampanoni M, Vaterlaus A, Aeschlimann M, and Meier F, Magnetism of Epitaxial bcc Iron on Ag(001) Observed by Spin-Polarized Photoemission. Phys Rev Lett 1987;59:2483-5.
[342] Baibich MN, Broto JM, Fert A, Van Dau FN, Petrou F, Eitenne P, Creuzet G, Friederich A, and Chazelas J, Giant Magnetoresistance of (001)Fe/(001)Cr Magnetic Superlattices. Phys Rev Lett 1988;61:2472-5.
[343] Liu C, Moog R, and Bader SD, Polar Kerr-Effect Observation of Perpendicular Surface Anisotropy for Ultrathin fcc Fe Grown on Cu(100). Phys Rev Lett 1998;60:2422-5.
[344] Schneider CM, Bressler P, Schuster P, Kirschner J, de Miguel JJ, and Miranda R, Curie temperature of ultrathin films of fcc-cobalt epitaxially grown on atomically flat Cu(100) surfaces. Phys Rev Lett 1990;64:1059-62.
[345] Hu X and Kawazoe Y, Mean-field theory for critical phenomena in bilayer systems. Phys Rev B 1994;50:12647-58.
[346] Ou JT, Wang F, Lin DL, Critical behavior of magnetic films in the Ising model. Phys Rev E 1997;56:2805-10.
[347] Li Y and Baberschke K, Dimensional crossover in ultrathin Ni(111) films on W(110). Phys Rev Lett 68:1992;1208-11.
[348] Tischer M, Arvanitis D, Yokoyanma T, Lederer T, Troger L, and Baberschke K, Temperature dependent MCXD measurements of thin Ni films on Cu(100). Surf Sci 1994;307-309:1096-101.
[349] Tjeng LH, Idzerda YU, Rudolf P, Sette F, and Chen CT, Soft-X-ray magnetic circular dichroism A new technique for probing magnetic properties of magnetic surfaces and ultrathin films. J Magn, Magn Mater 1992;109:288-92.
[350] Jiang JS, and Chien CL, Magnetization and finite-size effects in Gd/W multilayers. J Appl Phys 1996;79:5615-7.
[351] Jiang JS, Davidovic D, Reich DH, and Chien CL, Oscillatory Superconducting Transition Temperature in Nb/Gd Multilayers. Phys Rev Lett 1995;74:314-7.
[352] Fisher ME and Barber MN, Scaling Theory for Finite-Size Effects in the Critical Region. Phys Rev Lett 1972;28:1516-9.
[353] Richie DS and Fisher ME, Finite-Size and Surface Effects in Heisenberg Films. Phys Rev B 1973;7:480-94.
[354] Barber MN, in Phase Transitions and Critical Phenomena, Ed. Domb C and Lebowita J, Academic, New York, 1983, Vol 8.
[355] Binder K and Hohenberg PC, Surface effects on magnetic phase transitions. Phys Rev B 1974;9:2194-214.
[356] Zhang R and Willis RF, Thickness-Dependent Curie Temperatures of Ultrathin Magnetic Films:Effect of the Range of Spin-Spin Interactions. Phys Rev Lett 2001;86:2665-8.
[357] Nikolaev VI, and Shipilin AM, The Influence of Breaking of Exchange Bonds on the Curie Temperature. Phys Solid State 2003;45:1079–80.
[358] Sadeh B, Doi M, Shimizu T, and Matsui MJ, Dependence of the Curie temperature on the diameter of Fe3O4 ultra-fine particles. J Magn Soc Jpn 2000;24:511-4.
[359] Degennes PG, Superconductivity of Metals and Alloys. Benjamin, New York, 1966.
[360] Coswami R, Banerjee S, Chattopadhyay K, and Raychaudhuri AK, Superconductivity in rapidly quenched metallic systems with nanoscale structure. J Appl Phys 1993;73:2934-40.
[361] Kubo R, J Phys Soc Jpn 1962;17:975.
[362] Anderson PW, Theory of dirty superconductors. J Phys Chem Solids 1959;11:26-30.
[363] Strongin M, Thompson RS, Karnmerer OF, Crow JE, Destruction of Superconductivity in Disordered Near-Monolayer Films. Phys Rev B 1970;1:1078-91.
[364] Muhlschlegel B, Scalapino DJ, and Denton R, Thermodynamic properties of small superconducting particles. Phys Rev B 1972;6:1767-77.
[365]Guo Y, Zhang YF, Bao XY, Han TZ, Tang Z, Zhang LX, Zhu WG, Wang EG, Niu Q, Qiu ZQ, Jia JF, Zhao ZX, Xue QK. Superconductivity Modulated by Quantum Size Effects. Science 2004;306:1915-7.

[366] Pogrebnyakov AV, Redwing JM, Raghavan S, Vaithyanathan V, Schlom DG, Xu SY, Li Q, Tenne DA, Soukiassian A, Xi XX, Johannes MD, Kasinathan D, Pickett WE, Wu JS, and Spence JCH, Enhancement of the Superconducting Transition Temperature of $MgB_2$ by a Strain-Induced Bond-Stretching Mode Softening. Phys Rev Lett 2004;93:147006.
[367] Nagamatsu J, Nakagawa N, Muranaka T, Zenitani Y, and Akimitsu J, Superconductivity at 39 K in magnesium diboride. Nature (London) 2001;410:63-4.
[368] Hur N, Sharma PA, Guha S, Cieplak MZ, Werder DJ, Horibe Y, Chen CH, and Cheong SW, High-quality $MgB_2$ films on boron crystals with onset Tc of 417 K. Appl Phys Lett 2001;79:4180-2.
[369] Yildirim T and Gulseren O, A simple theory of 40 K superconductivity in $MgB_2$:first-principles calculations of Tc, its dependence on boron mass and pressure. J Phys Chem Solids 2002;63:2201-6.
[370] Honebecq V, Huber C, Maglione M, Antonietti M, and Elissalde C, Dielectric properties of pure(BaSr)$TiO_3$ and composites with different grain sizes ranging from the nanometer to the micrometer. Adv Func Mater 2004;19:899-904.
[371] Zhao Z, Buscaglia V, Viviani M, Buscaglia MT, Mitoseriu L, Testino A, Nygren M, Johnsson M, and Nanni P, Grain-size effects on the ferroelectric behavior of dense nanocrystalline $BaTiO_3$ ceramics. Phys Rev B 2004;70:024107.
[372] Wang CL, Xin Y, Wang XS, Zhong WL, Size effects of ferroelectric particles described by the transverse Ising model. Phys Rev B 2000;62:11423-7.
[373] Jiang B, Bursill LA, Phenomenological theory of size effects in ultrafine ferroelectric particles of lead titanate. Phys Rev B 1999;60:9978-82.
[374] Tanaka M and Makino Y, Ferroelec Lett 1998;24:13.
[375] Munkholm A, Streiffer SK, Murty MVR, Eastman JA, Thompson C, Auciello O, Thompson L, Moore JF, and Stephenson GB, Antiferrodistortive reconstruction of the $PbTiO_3$(001) surface. PhysRevLett 2002;88:016101.
[376] Huang H, Zhou LM, Guo J, Hng HH, Oh JT, and Hing P, F spots and domain patterns in rhombohedral $PbZr_{0.9}Ti_{0.1}O_3$. Appl Phys Lett 2003;83:3692-4.
[377] Huang H, Sun CQ, Z Tianshu, Hong Z, Oh JT, and Hing P, Stress effect on the pyroelectric properties of lead titanate thin films. Integrated Ferroelectrics 2003;51:81-90.
[378] Jiang Q, Cui XF, Zhao M, Size effects on Curie temperature of ferroelectric particles. Appl Phys A 2004;78:703-4.
[379] Zysler ED, Firani D, Testa AM, Suber L, Agostnelli E, and Godinho M, Size dependence of the spin-flop transition in hematite nanosolids. Phys Rev B 2003;68:212408
[380] Amin A, and Arajs S, Morin temperature of annealed submicronic α -$Fe_2O_3$ particles. Phys Rev B 1987;35:4810-1.
[381] Weschke E, Ott H, Schierle E, Schüßler-Langeheine C, Vyalikh DV, Kaindl G, Leiner V, Ay M, Schmitte T, Zabel H, and Jensen PJ, Finite-Size Effect on Magnetic Ordering Temperatures in Long-Period Antiferromagnets: Holmium Thin Films. Phys Rev Lett 2004;93:57204.
[382] Fullerton EE, Riggs KT, Sowers CH, and Bader SD and Berger A, Suppression of Biquadratic Coupling in Fe /Cr(001) Superlattices below the Néel Transition of Cr. Phys Rev Lett 1995;75:330-3.
[383] Von Braun F and Delft J, Superconductivity in ultrasmall metallic grains. Phys Rev B 1999;59:9527-44.
[384] Bergholz R and Gradmann U, Structure and magnetism of oligatomic Ni(111)-films on Re(0001). J Magn Magn Mater 1984;45:389-98.
[385] Uchina K, Sadanaga Y, Hirose T, J Am Ceram Soc 1999;72:1555.
[386] Chattopadhyay S, Ayyub P, Palkar VR, Gurjar AV, Wankar RM, Multani, M Finite-size effects in antiferroelectric $PbZrO_3$ nanoparticles. J Phys Condens Matt 1997;9:8135-45.
[387] Li S, White T, Plevert J and Sun CQ, Superconductivity of nano-crystalline $MgB_2$. Supercond Sci Technol 2004;17:S589-94
[388] Li S, White T, Sun CQ, Fu YQ, Plevert J and Lauren K, Discriminating Lattice Structural Effects from Electronic Contributions to the Superconductivity of Doped $MgB_2$ with Nanotechnology. J Phys Chem B 2004;108: 6415-9.
[389] Schlag S, Eicke HF, Stern WB, Size driven phase transition and thermodynamic properties of nanocrystalline $BaTiO_3$. Ferroelectrics 1995;173:351-69.
[390] Valiev RZ, Islamgaliev RK, and Alexandrov IV, Bulk nanostructured materials from severe plastic deformation. Prog Mater Sci 2000;45:103–89.
[391] Razumovskii IM, Kornelyuk LG, Valiev RZ, and Sergeev VI, Diffusion along nonequilibrium grain boundaries in a nickel-base superalloy. Mater Sci Eng A 1993;167:123-7.
[392] Horvath J, Diffusion in nanocrystalline materials. Diffusion and Defect Data - Solid State Data, Part A: Defect and Diffusion Forum 1989;66-69:207-27.

[393] Mütschele T and Kirchheim R, Segregation and diffusion of hydrogen in grain boundaries of palladium. Scripta Metall 1987;21:135-40.
[394] Kolobov YR, Grabovetskaya GR, Ratochka IV, Kabanova ER, Naidenkin EV and Lowe T, Effect of grain-boundary diffusion fluxes of copper on the acceleration of creep in submicrocrystalline nickel. Ann Chim Sci des Materiaux 1996;21:483-91.
[395] Wurschum A, Kubler, Gruss S, Acharwaechter P, Frank W, Valiev RZ, Mulyukov RR and Schaeffer HE, Tracer diffusion and crystallite growth in ultra-fine-grained Pd prepared by severe plastic deformation. Ann Chim Sci Mater 1996;21:471-82.
[396] Wang ZB, Tao NR, Tong, WP Lu J, Lu K, Diffusion of chromium in nanocrystalline iron produced by means of surface mechanical attrition treatment. Acta Mater 2003;51:4319–29.
[397] Mishin YM and Razumovskii IM, Development of boundary diffusion models. Scripta Metall 1991;25:1375-80.
[398] Kim HC, Alford TL, and Allee DR, Thickness dependence on the thermal stability of silver thin films. Appl Phys Lett 2002;81:4287-9.
[399] Li CM and Cha CS, Electrodes with surface intercalated powder of catalysts: II Microelectrodes with surface-intercalated powder of catalysts. Acta Chimica Sinica 1988;1:14-20.
[400] Li CM and Cha CS, Powder microelectrodes: II Irreversible electrode system. Acta Physico-Chimica Sinica 1988;4:273-7.
[401] Cha CS, Li CM, Yang HX and Liu PF, Powder Microelectrodes. J Electroanalytical Chem 1994;368:47-51.
[402] Li CM and Cha CS, Porous carbon/Teflon composite enzyme glucose sensors. Front Biosci 2004;9:3324-30.
[403] Schumacher S, Birringer R, Strauss R, and Gleiter H, Diffusion of silver in nanocrystalline copper between 303 and 373 K. Acta Metall 1989;37:2485-8.
[404] Suryanarayana C, Mechanical alloying and milling. Prog Mater Sci 2001;46:1-184.
[405] Calka A, Nikolov JI, and Williams JS, Formation, structure and stability of iron nitrides made by reactive ball milling. Mater Sci Forum 1996;225-227:527-32.
[406] El-Eskandarany MS, Sumiyama K, Aoki K, Suzuki K. Mater Sci Forum 1992;88-90:801-6.
[407] Mongis J, Peyre JP, and Tournier C, Nitriding of nicroalloyed steels. Heat Treat Metals 1984;3:71-5.
[408] Campbell CT, The Active Site in Nanoparticle Gold Catalysis. Science 2004;306:234-5.
[409] Chen MS and Goodman DW, The Structure of Catalytically Active Gold on Titania. Science 2004;306:252-5.
[410] Liu Z, Gao W, Dahm K, and Wang F, Oxidation Behaviour of Sputter-Deposited Ni-Cr-Al Micro-Crystalline Coatings. Acta Metallurg Mater 1998;46:1691-700.
[411] Gao W, Liu Z and Li Z, Nano- and Micro-Crystal Coatings and Their High-Temperature Applications. Adv Mater 2001;13:1001-4.
[412] Liu Z, Gao W, and He Y, Surface Nano-Crystallisation of 310s Stainless Steel and Its Effect on the Oxidation Behaviour. J Mater Eng Perf 1998;7:88-92.
[413] Sun CQ, Xie H, Zhang W, Ye H and Hing P, Preferential oxidation of diamond {111}. J Phys D 2000;33:2196-9.
[414] Lopez N, Janssens TVW, Clausen BS, Xu Y, Mavrikakis M, Bligaard T and Nørskov JK, On the origin of the catalytic activity of gold nanoparticles for low-temperature CO oxidation. J Catalysis 2004;223:232-5.
[415] Röder H, Hahn, E Brune H, Bucher JP, and Kern K, Building one- and two- dimensional nanostructures by diffusion-controlled aggregation at surfaces. Nature (London) 1993;366:141-3.
[416] Liu BG, Wu J, Wang EG, and Zhang Z, Two-dimensional pattern formation in surfactant-mediated epitaxial growth. Phys Rev Lett 1999;83:1195-8.
[417] Li MZ, Wendelken JF, Liu BG, Wang EG, and Zhang Z, Decay Characteristics of Surface Mounds with Contrasting Interlayer Mass Transport Channels. Phys Rev Lett 2001;86:2345-8.
[418] Zhu WG, Mongeot FB, Valbusa U, Wang EG, and Zhang Z, Adatom Ascending at Step Edges and Faceting on fcc Metal (110) Surfaces. Phys Rev Lett 2004;92:106102.
[419] Wu J, Wang EG, Varga K, Liu BG, Panfelides ST, and Zhang Z, Island Shape Selection in Pt(111) Submonolayer Homoepitaxy without or with CO as Adsorbates. Phys Rev Lett 2002;89:146103.
[420] Takagahara T, Electron-phonon interactions and excitonic dephasing in semiconductor nanocrystals. Phys Rev Lett 1993;71:3577-80.
[421] Zi J, Büscher H, Falter C, Ludwig W, Zhang KM, and Xie XD, Raman shifts in Si nanocrystals. Appl Phys Lett 1996;69:200-2.
[422] Fujii M, Kanzawa Y, Hayashi S, and Yamamoto K, Raman scattering from acoustic phonons confined in Si nanocrystals. Phys Rev B 1996;54:R8373-6.

[423] Cheng W and Ren SF, Calculations on the size effects of Raman intensities of silicon quantum dots. Phys Rev B 2002;65:205305.
[424] Klein MC, Hache F, Ricard D and Flytzanis C, Size dependence of electron-phonon coupling in semiconductor nanospheres: The case of CdSe. Phys Rev B 1990;42:11123-32.
[425] Trallero-Giner C, Debernardi A, Cardona M, Menendez-Proupi E, and Ekimov AI, Optical vibrons in CdSe dots and dispersion relation of the bulk material. Phys Rev 1998;B57:4664-9.
[426] Hu XH and Zi J, Reconstruction of phonon dispersion in Si nanocrystals. J Phys Condens Matter 2002;14:L671-7.
[427] Palpant B, Portales H and Saviot L, Lerme J, Prevel B, Pellarin M, Duval E, Perez A, Broyer M, Quadrupolar vibrational mode of silver clusters from plasmon-assisted Raman scattering. Phys Rev B 1999;60:17107-11.
[428] Fujii M, Nagareda T, Hayashi S, and Yamamoto K, Low-frequency Raman scattering from small silver particles embedded in $SiO_2$ thin films. Phys Rev B 1991;44:6243-8.
[429] Lamb H, Proc London Math Soc 1882;13:189.
[430] Duval E, Far-infrared and Raman vibrational transitions of a solid sphere:Selection rules. Phys Rev B 1992;46:5795-7.
[431] Verma P, Cordts W, Irmer G, and Monecke J, Acoustic vibrations of semiconductor nanocrystals in doped glasses. Phys Rev B 1999;60:5778-85.
[432] Ferrari M, Gonella F, Montagna M, and Tosello C, Detection and size determination of Ag nanoclusters in ion-exchanged soda-lime glasses by waveguided Raman spectroscopy. J Appl Phys 1996;79:2055-9.
[433] Scamarcio G, Lugara M, and Manno D, Size-dependent lattice contraction in $CdS_{1-x}Se_x$ nanocrystals embedded in glass observed by Raman scattering. Phys Rev B 1992;45:13792-5.
[434] Liang LH, Shen CM, Chen XP, Liu WM and Gao HJ, The size-dependent phonon frequency of semiconductor nanocrystals. J Phys Condens Matt 2004;16:267–72.
[435] Dieguez A, Romano-Rodrýguez A, Vila A, and Morante JR, The complete Raman spectrum of nanometric $SnO_2$ particles. J Appl Phys 2001;90:1550-7.
[436] Iqbal Z and Vepřek S, Raman scattering from hydrogenated microcrystalline and amorphous silicon. J Phys C 1982;15:377-92.
[437] Anastassakis E and Liarokapis E, Polycrystalline Si under strain:.Elastic and lattice-dynamical considerations. J Appl Phys 1987;62:3346-52.
[438] Richter H, Wang ZP, and Ley L, One phonon Raman spectrum in microcrystalline silicon. Solid State Commun 1981;39:625-9.
[439] Campbell IH, and Fauchet PM, Eeffects of microcrystal size and shape on the phonon Raman spectra of of crystalline semiconductors. Solid State Commun 1986;58:739-41.
[440] Wang X, Huang DM, YeL, Yang M, Hao PH, Fu HX, Hou XY, and Xie XD, Pinning of photoluminescence peak positions for light-emitting porous silicon:An evidence of quantum size effect. Phys Rev Lett 1993;71:1265-7.
[441] Andújar JL, Bertran E, Canillas A, Roch C, and Morenza JL, Influence of pressure and radio frequency power on deposition rate and structural properties of hydrogenated amorphous silicon thin films prepared by plasma deposition. J Vac Sci Technol A 1991;9:2216-21.
[442] Ohtani N and Kawamura K, Theoretical investigation of Raman scattering from microcrystallites. Solid State Commun 1990;75:711-5.
[443] Banyai L, and Koch SW, Semiconductor Quantum Dots. World Scientific, Singapore, 1993.
[444] Tanaka A, Onari S, and Arai T, Raman Scattering from CdSe microcrystals embedded in germanate glass matrix. Phys Rev B 1992;45:6587-92.
[445] Hwang YN, Shin S, Park HL, Park SH, Kim U, Jeong HS, Shin EJ, and Kim D, Effect of lattice contraction on the Raman shifts of CdSe quantum dots in glass matrices. Phys Rev 1996;B54:15120-4.
[446] Viera G, Huet S, and Boufendi L, Crystal size and temperature measurements in nanostructured silicon using Raman spectroscopy. J Appl Phys 2001;90:4175-83.
[447] Fauchet PM and Campell IH, Crit Rev Solid State Mater Sci 1988;14:S79.
[448] Sood AK, Jayaram K, and Victor D, Muthu S, Raman and high-pressure photoluminescence studies on porous silicon. J Appl Phys 1992;72:4963-5.
[449] Ossadnik C, Vepřek S, and Gregora I, Applicability of Raman scattering for the characterization of nanocrystalline silicon. Thin Solid Films 1999;337:148-51.
[450] Cheng GX, Xia H, Chen KJ, Zhang W, and Zhang XK, Raman measurement of the grain size for silicon crystallites. Phys Status Solidi A 1990;118:K51-4.
[451] Verma P, Gupta L, Abbi SC, and Jain KP, Confinement effects on the electronic and vibronic properties of $CdS_{0.65}Se_{0.35}$ nanoparticles grown by thermal annealing. J Appl Phys 2000;88:4109-16.

[452] Spanier JE, Robinson RD, Zhang F, Chan SW, and Herman IP, Size-dependent properties of $CeO_2$ nanoparticles as studied by Raman scattering. Phys Rev B 2001;64:245407.
[453] Shek CH, Lin GM and Lai JKL, Effect of oxygen deficiency on the Raman spectra and hyperfine interactions of nanometer $SnO_2$. Nanosructured Mater 1999;11:831–5.
[454] Seong MJ, Micic OI, Nozik AJ, Mascarenhas A, and Cheong HM, Size-dependent Raman study of InP quantum dots. Appl Phys Lett 2003;82:185-7.
[455] Gangopadhyay P, Kesavamoorthy R, Nair KGM, and Dhandapani R, Raman scattering studies on silver nanoclusters in a silica matrix formed by ion-beam mixing. J Appl Phys 2000;88:4975-9.
[456] Gotić M, Ivanda M, Sekulić A, Music S, Popović S, Turković A, Furić K, Microstructure of nanosized $TiO_2$ obtained by sol-gel synthesis. Mater Lett 1996;28:225-9.
[457] Saviot L, champagnon B, Duval E, Kudriavtsev IA, Ekimov AI, Size dependence of acoustic and optical vibrational modes of CdSe nanocrystals in glasses. J Non-Crystal Solids 1996;197:238-46.
[458] Schuppler S, Friedman SL, Marcus MA, Adler DL, Xie YH, Ross FM, Chabal YJ, Harris TD, Brus LE, Brown WL, Chaban EE, Szajowski PF, Christman SB, and Citrin PH, Size, shape, and composition of luminescent species in oxidized Si nanocrystals and H-passivated porous Si. Phys Rev B 1995;52:4910-25.
[459] Heikkila L, Kuusela T, Hedman HP, Electroluminescence in Si/$SiO_2$ layer structures. J Appl Phys 2001;89:2179-84.
[460] Garrido B, Lopez M, Gonzalez O, Perez-Rodrýguez A, Morante JR, and Bonafos C, Correlation between structural and optical properties of Si nanocrystals embedded in $SiO_2$: The mechanism of visible light emission. Appl Phys Lett 2000;77:3143-5.
[461] Tomasulo A and Ramakrishna MV, Quantum confinement effects in semiconductor clusters II. J Chem Phys 1996;105:3612-26.
[462] Jain SC, Willander M, Narayan J, and van Overstraeten R, III–nitrides: Growth, characterization, and properties. J Appl Phys 2000;87:965-1006.
[463] Ramvall P, Tanaka S, Nomura S, Riblet P, and Aoyagi Y, Observation of confinement-dependent exciton binding energy of GaN quantum dots. Appl Phys Lett 1998;73:1104-6.
[464] Guzelian AA, Banin U, Kadavanich AV, Peng X, and Alivisatos AP, Colloidal chemical synthesis and characterization of InAs nanocrystal quantum dots. Appl Phys Lett 1996;69:1432-4.
[465] Micic OI, Sprague J, Lu Z, and Nozik AJ, Highly efficient band-edge emission from InP quantum dots. Appl Phys Lett 1996;68:3150-2.
[466] Ferreyra JM, and Proetto CR, Quantum size effects on excitonic Coulomb and exchange energies in finite-barrier semiconductor quantum dots. Phys Rev B 1999;60:10672-5.
[467] Wang Y and Herron N, Quantum size effects on the exciton energy of CdS clusters. Phys Rev B 1990;42:7253-5.
[468] Krishna MVR and Friesner RA, Exciton spectra of semiconductor clusters. Phys Rev Lett 1991;67:629-32.
[469] Lippens PE, and Lannoo M, Comparison between calculated and experimental values of the lowest excited electronic state of small CdSe crystallites. Phys Rev B 1989;41:6079-81.
[470] Chu DS and Dai CM, Quantum size effects in CdS thin films. Phys Rev B 1992;45:11805-10.
[471] Albe V, Jouanin C, and Bertho D, Confinement and shape effects on the optical spectra of small CdSe nanocrystals. Phys Rev B 1998;58:4713-20.
[472] Bawendi MG, Wilson WL, Rothberg L, Carroll PJ, Jedju TM, Steigerwald ML, and Brus LE, Electronic structure and photoexcited-carrier dynamics in nanometer-size CdSe clusters. Phys Rev Lett 1990;65:1623-6.
[473] Hayashi S, Sanda H, Agata M, and Yamamoto K, Resonant Raman scattering from ZnTe microcrystals: Evidence for quantum size effects. Phys Rev B 1989;40:5544-8.
[474] Gourgon C, Dang LS and Mariette H, Optical properties of CdTe/CdZnTe wires and dots fabricated by a final anodic oxidation etching. Appl Phys Lett 1995;66:1635-7.
[475] Fang BS, Lo WS, Chien TS, Leung TC and Lue CY, Chan CT, and Ho KM, Surface band structures on Nb(001). Phys Rev B 1994;50:11093-101.
[476] Schmeißer D, Böhme O, Yfantis A, Heller T, Batchelor DR, Lundstrom I, and Spetz AL, Dipole moment of nanoparticles at interfaces. Phys Rev Lett 1999;83:380-3
[477] Tsu R, Babic D, and Loriatti L Jr, Simple model for the dielectric constant of nanoscale silicon particle. J Appl Phys 1997;82:1327-9.
[478] Ekimov AI and Onushchenko AA, Quantum size effect in the optical spectra of semiconductor microcrystals. Sov phys Semicon 1982;16:775-8.
[479] Steigerwald ML, and Brus LE, Semiconductor crystallites:a class of large molecules. Acc Chem Res 1990;23:183-8

[480] Fu H, and Zunger A, Local-density-derived semiempirical nonlocal pseudopotentials for InP with applications to large quantum dots. Phys Rev B 1997;55:1642-53.
[481] Wang X, Qu L, Zhang J, Peng X, and Xiao M, Surface-Related Emission in Highly Luminescent CdSe Quantum Dots. Nanolett 2003;3:1103-6.
[482] Aiyer HN, Vijayakrishnan V, Subbanna GN, Rao CNR, Investigations of Pd clusters by the combined use of HREM, STM, high-energy spectroscopies and tunneling conductance measurements. Surf Sci 1994;313:392-8.
[483] Yang DQ, Kabashin AV, Pilon-Marien VG, Sacher E, and Meunier M, Optical breakdown processing:Influence of the ambient gas on the properties of the nanostructured Si-based layers formed. J Appl Phys2004;95:5722-8.
[484] Sanders GD, and Chang YC, Theory of optical properties of quantum wires in porous silicon. Phys Rev 1992;B45:9202-13.
[485] Theiss W, Scout Thin Film Analysis Software Handbook, Hard and Software: M Theiss, Aachen, Germany, 2001, www.mtheiss.com).
[486] Furukawa S and Miyasato T, Quantum size effects on the optical band gap of microcrystalline Si:H. Phys Rev B 1988;38:5726-9.
[487] Guha S, Steiner P, and Lang W, Resonant Raman scattering and photoluminescence studies of porous silicon membranes. J Appl Phys 1996;79:8664-8.
[488] Kanemitsu Y, Uto H, Masumoto Y, Matsumoto T, Futagi, T and Mimura H, Microstructure and optical properties of free-standing porous silicon films:Size dependence of absorption spectra in Si nanometer-sized crystallites. Phys Rev B 1993;48:2827-30.
[489] von Behren J, van Buuren T, Zacharias M, Chimowitz EH, and Fauchet PM, Quantum confinement in nanoscale silicon:the correlation of size with bandgap and luminescence. Solid State Commun 1998;105:317-22.
[490] Canham L, Properties of Porous Silicon:INSPEC, London, 1997, p213.
[491] Dorigoni L, Bisi O, Bernardini F, and Ossicini S, Electron states and luminescence transition in porous silicon. Phys Rev B 1996;53:4557-64.
[492] Hybertsen MS and Needels M, First-principles analysis of electronic states in silicon nanoscale quantum wires. Phys Rev B 1993;48:4608-11.
[493] Ohno T, Shiraishi K, and Ogawa T, Intrinsic origin of visible light emission from silicon quantum wires: Electronic structure and geometrically restricted exciton. Phys Rev Lett 1992;69:2400-3.
[494] Yeh CY, Zhang SB, and Zunger A, Confinement, surface, and chemisorption effects on the optical properties of Si quantum wires. Phys Rev B 1994;50:14405-15.
[495] Read AJ, Needs RJ, Nash KJ, Canham LT, Calcott PDJ, and Qteish A, First-principles calculations of the electronic properties of silicon quantum wires. Phys Rev Lett 1992;69:1232-5.
[496] Pavesi L, Giebel G, Ziglio F, Mariotto G, Priolo F, Campisano SU, and Spinella C, Nanocrystal size modifications in porous silicon by preanodization ion implantation. Appl Phys Lett 1994;65:2182-4.
[497] Hasegawa S, Matsuda M, and Kurata Y, Bonding configuration and defects in amorphous SiNx:H films. Appl Phys Lett 1991;58:741-3.
[498] Brongersma ML, Polman A, Min KS, Boer E, Tambo T, and Atwater HA, Tuning the emission wavelength of Si nanocrystals in $SiO_2$ by oxidation. Appl Phys Lett 1998;72:2577-9.
[499] Nogami N, Suzuki S, and Nagasaka K, Sol-gel processing of small-sized CdSe crystal-doped silica glasses. J Non-Cryst Solids 1991;135:182-8.
[500] Bawendi MG, Carroll PJ, Wilson WL, and Brus LE, Luminescence properties of CdSe quantum crystallites: Resonance between interior and surface localized states. J Chem Phys 1992;96:946-54.
[501] Katari JEB, Colvin VL, and Alivisatos AP, X-ray photoelectron spectroscopy of CdSe nanocrystals with applications to studies of the nanocrystal surface. J Phys Chem 1994;98:4109-17.
[502] Hoheisel W, Colvin VL, Johnson CS, and Alivisatos AP, Threshold for quasicontinuum absorption and reduced luminescence efficiency in CdSe nanocrystals. J Chem Phys 1994;101:8455-60.
[503] Alivisatos AP, Harris RD, Brus, LE and Jayaraman A, Resonance Raman scattering and optical absorption studies of CdSe microclusters at high pressure. J Chem Phys 1988;89:5979-82.
[504] Micic OI, Cheong HM, Fu H, and Zunger A, Sprague JR, Mascarenhas A, and Nozik AJ, Size-dependent spectroscopy of InP quantum dots. J Phys Chem B 1997;101:4904-12.
[505] Micic OI, Jones KM, Cahill A, and Nozik AJ, Optical, electronic, and structural properties of uncoupled and close-packed arrays of InP quantum dots. J Phys Chem B 1998;102:9791-6.
[506] Sapra S, Viswanatha R and Sarma DD, An accurate description of quantum size effects in InP nanocrystallites over a wide range of sizes J Phys D 2003;36:1595–8.
[507] Katz D, Wizansky T, Millo O, Rothenberg E, Mokari T, Banin U, Size-Dependent Tunneling and Optical Spectroscopy of CdSe Quantum Rods. Phys Rev Lett 2002;89:086801.

[508] Xu CX, Sun XW, Zhang XH, Ke L, and Chua SJ, Photoluminescent properties of opper-doped zinc oxide nanowires. Nanotechnology 2004;15:856–61.
[509] Borgohain K, Singh JB, Rao MVR, Shripathi T, Mahamuni S, Quantum size effects in CuO nanoparticles. Phys Rev B 2000;61:11093-6.
[510] Aldén M, Skriver HL, and Johansson B, Ab initio surface core-level shifts and surface segregation energies. Phys Rev Lett 1993;71:2449-52.
[511] Navas EE, Starke K, Laubschat C, Weschke E, and Kaindl D, Surface core-level shift of 4f states for Tb(0001). Phys Rev B 1993;48:14753-5.
[512] Bartynski RA, Heskett D, Garrison K, Watson G, Zehner DM, Mei WN, Tong SY and Pan X, The first interlayer spacing of Ta(100) determined by photoelectron diffraction. J Vac Sci Technol A 1989;7:1931-6.
[513] Andersen JN, Hennig D, Lundgren E, Methfessel M, Nyholm R, and Scheffler M, Surface core-level shifts of some 4d-metal single-crystal surfaces:Experiments and ab initio calculations. Phys Rev B 1994;50:17525-33.
[514] Riffe DM, and Wertheim GK, Ta(110) surface and subsurface core-level shifts and $4f_{7/2}$ line shapes, Phys Rev B 1993;47:6672-9.
[515] Cho JH, Kim KS, Lee SH, Kang MH, and Zhang Z, Origin of contrasting surface core-level shifts at the Be(10**Error! Objects cannot be created from editing field codes.**0) and Mg(10**Error! Objects cannot be created from editing field codes.**0) surfaces. Phys Rev B 2000;61:9975-8.
[516] Fedorov AV, Arenholz E, Starke K, Navas E, Baumgarten L, Laubschat C, and Kaindl G, Surface shifts of 4f electron-addition and electron-removal states in Gd(0001). Phys Rev Lett 1994;73:601-4.
[517] Johansson LI, Johansson HI, Andersen JN, Lundgren E, and Nyholm R, Three surface-shifted core levels on Be(0001). Phys Rev Lett 1993;71:2453-6.
[518] Lizzit S, Pohl K, Baraldi A, Comelli G, Fritzsche V, Plummer EW, Stumpf R, and Hofmann P, Physics of the Be(10**Error! Objects cannot be created from editing field codes.**0) Surface Core Level Spectrum. Phys Rev Lett 1998;81:3271-4.
[519] Johansson HIP and Johansson LI, Lundgren E, Andersen JN, and Nyholm R, Core-level shifts on Be(10**Error! Objects cannot be created from editing field codes.**0). Phys Rev B 1994;49:17460-3.
[520] Baraldi A, Lizzit S, Comelli G, Goldoni A, Hofmann P, Paolucci G, Core-level subsurface shifted component in a 4d transition metal: Ru(1010). Phys Rev B 2000;61:4534-7.
[521] Lunsgren E, Johansson U, Nyholm R, and Anderson JN, Surface core-level shift of the Mo(110) surface. Phys Rev B 1993;48:5525-29.
[522] Nyholm R, Andersen JN, van Acker JF, and Qvarford M, Surface core-level shifts of the Al(100) and Al(111) surfaces. Phys Rev B 1991;44:10987-90.
[523] Riffe DM, Kim B, Erskine JL, and Shinn ND, Surface core-level shifts and atomic coordination at a stepped W(110) surface. Phys Rev B 1994;50:14481-8.
[524] Cho JH, Oh DH, and Kleinman L, Core-level shifts of low coordination atoms at the W(320) stepped surface. Phys Rev B 2001;64:115404.
[525] Karlsson CK, Landemark E, Chao YC, and Uhrberg RIG, Atomic origins of the surface components in the Si 2p core-level spectra of the Si(111)7 x 7 surface. Phys Rev B 1994;50:5767-70.
[526] Scholz SM and Jacobi K, Core-level shifts on clean and adsorbate-covered Si(113) surfaces. Phys Rev B1995;52:5795-802.
[527] Pi TW, Wen JF, Ouyang CP, and Wu RT, Surface core-level shifts of Ge(100)-2×1. Phys Rev B 2001;63:153310.
[528] Lizzit S, Baraldi A, Groso A, Reuter K, Ganduglia-Pirovano MV, Stampfl C, Scheffler M, Stichler M, Keller C, Wurth W, and Menzel D, Surface core-level shifts of clean and oxygen-covered Ru(0001). Phys Rev 2001;B63:205419.
[529] Glans PA and Johansson LI, Balasubramanian T, Blake RJ, Assignment of the surface core-level shifts to the surface layers of Be(10$\bar{1}$0). Phys Rev B 2004;70:033408.
[530] Zacchigna M, Astaldi C, Prince K, Sastry M, Comicioli C, Rosei R, Quaresima C, Ottaviani C, Crotti C, Antonini A, Matteucci M, Perfetti P, Photoemission from atomic and molecular adsorbates on Rh(100). Surf Sci 1996;347:53-62.
[531] Ohgi T and Fujita D, Consistent size dependency of core-level binding energy shifts and single-electron tunneling effects in supported gold nanoclusters. Phys Rev B 2002;66:115410.
[532] Nanda J, Kuruvilla A, and Sarma DD, Photoelectron spectroscopic study of CdS nanocrystallites. Phys Rev 1999;B59:7473-9.
[533] Balasubramanian T, Andersen JN, and Wallden L, Surface-bulk core-level splitting in graphite. Phys Rev B 2001;64:205420.

[534] Yang DQ, and Sacher E, Initial- and final-state effects on metal cluster/substrate interactions, as determined by XPS: Copper clusters on Dow Cyclotene and highly oriented pyrolytic graphite. Appl Surf Sci 2002;195:187-195.
[535] Howard A, Clark DNS, Mitchell CEJ, Egdell RG, and Dhanak VR, Initial and final state effects in photoemission from Au nanoclusters on $TiO_2$(110). Surf Sci 2002;518:210-24.
[536] Salmon M, Ferrer S, Jazzar M and Somojai GA, Core- and valence-band energy-level shifts in small two-dimensional islands of gold deposited on Pt(100): The effect of step-edge, surface, and bulk atoms. Phys Rev B 1983;28:1158-60.
[537] Vijayakrishnan VA, Chainani A, Sarma DD, and Rao CNR, Metal-insulator-transitions in metal clusters A high-energy spectroscopy study of Pd and Ag clusters. J Phys Chem 1992;96:8679-82.
[538] Mason MG, in Cluster Models for Surface and Bulk Phenomena. Ed G Pacchioni, Plenum, New York, 1992.
[539] Boyen HG, Herzog T, Kastle G, Weigl F, Ziemann P, Spatz JP, Moller M, Wahrenberg R, Garnier MG, and Oelhafen P, X-ray photoelectron spectroscopy study on gold nanoparticles supported on diamond. Phy Rev B 2002;65:075412.
[540] Yang ZN and Wu R, Origin of positive core-level shifts in Au clusters on oxides. Phys Rev B 2003;67:081403.
[541] Egelhoff WF Jr and Tibbetts GG, Growth of copper, nickel, and palladium films on graphite and amorphous carbon. Phys Rev B 1979;19:5028-35.
[542] Kästle G, Boyen HG, Schröder A, Plettl A, and Ziemann P. Size effect of the resistivity of thin epitaxial gold films. Phys Rev B 2004;**70**:165414.
[543] Amaratunga GAJ and Silva SRP, Nitrogen containing hydrogenated amorphous carbon for thin-film field emission cathodes. Appl Phys Lett 1996;68:2529-31.
[544] Pickett WE, Negative electron affinity and low work function surface Cesium on oxygenated diamond (100),. Phys Rev Lett 1994;73:1664-7.
[545] Lin LW, The role of oxygen and fluorine in the electron emission of some kinds of cathodes. J Vac Sci Technol A 1998;6:1053-7.
[546] Xu CX, Sun XW, Field emission from zinc oxide nanopins. Appl Phys Lett 2003;83:3806-8
[547] Xu CX, Sun XW, Chen BJ, Field emission from gallium-doped zinc oxide nanofiber array. Appl Phys Lett 2004;84:1540-2.
[548] Wadhawan A, Stallcup RE, Stephens KF, Perez JM, and Akwani IA, Effects of $O_2$, Ar, and $H_2$ gases on the field-emission properties of single-walled and multiwalled carbon nanotubes. Appl Phys Lett 2001;79:1867-9.
[549] Zheng WT, Li JJ, Wang X, Li XT, Jin ZS, Tay BK and Sun CQ, Mechanism for the nitrogen-lowered threshold in carbon nitride cold-cathode. J Appl Phys 2003;94:2741-5.
[550] Abbott P, Sosa ED, and Golden DE, Effect of average grain size on the work function of diamond films. Appl Phys Lett 2001;79:2835-7.
[551] Kappes MM and Schuhmacher E, Bunsenges B, Physik Chem 1984;88:220.
[552] Bhave TM and Bhoraskar SV, Surface work function studies in porous silicon. J Vac Sci Technol B 1998;16:2073-8.
[553] Tzeng Y, Liu C, and Hirata A, Effects of oxygen and hydrogen on electron field emission from microwave plasma chemically vapor deposited microcrystalline diamond, nanocrystalline diamond, and glassy carbon coatings. Diamond Rel Mater 2003;12:456-63.
[554] Gao R, Pan Z, and Wang ZL, Work function at the tips of multiwalled carbon nanotubes. Appl Phys Lett 2001;78:1757-9.
[555] Li H, Wang X, Song Y, Liu Y, Li Q, Jiang L, Zhu D, Super-"Amphiphobic" Aligned Carbon Nanotube Films. Angew Chem Int Ed 2001;40:1743-6.
[556] Feng L, Li S, Li H, Zhai J, Song Y, Jiang L, Zhu D, Super-Hydrophobic Surface of Aligned Polyacrylonitrile Nanofibers. Angew Chem Int Ed 2002;41:1221-3.
[557] Sun T, Wang G, Feng L, Liu B, Ma Y, Jiang L, Zhu D, Reversible Switching between Superhydrophilicity and Superhydrophobicity. Angew Chem Int Ed 2004;43:357-60.
[558] Feng X, Feng L, Jin M, Zhai J, Jiang L, Zhu D, Reversible Super-hydrophobicity to Super-hydrophilicity Transition of Aligned ZnO Nanorod Films. J Am Chem Soc 2004;126:62-3.
[559] Sun CQ, Time-resolved VLEED from the O-Cu(001):Atomic processes of oxidation. Vacuum 1997;48:525-30.
[560] Poa CHP, Lacerda RG, Cox DC, Marques FC, Silva SRP, Effects of stress on electron emission from nanostructured carbon materials. J Vac Sci Technol B 2003;21:1710-4.
[561] Uher C, Hockey RL, and Ben-Jacob E, Pressure dependence of the c-axis resistivity of graphite. Phys Rev B 1987;35:4483-8.

[562] Silva SRP, Private communication Singapore, 2003.
[563] Lacerda RG, dos Santos MC, Tessler LR, Hammer P, Alvarez F, and Marques FC, Pressure-induced physical changes of noble gases implanted in highly stressed amorphous carbon films Phys Rev B 2003;68:054104.
[564] Lynch RW and Drickamer HG, Effect of High Pressure on the Lattice Parameters of Diamond, Graphite, and Hexagonal Boron Nitride J Chem Phys 1966;44:181-4.
[565] Bhattacharyya S and Subramanyam SV, Metallic conductivity of amorphous carbon films under high pressure Appl Phys Lett 1997;71:632-4.
[566] Umemoto K, Saito S, Berber S, and Tomanek D, Carbon foam: Spanning the phase space between graphite and diamond. Phys Rev B 2001;64:193409.
[567] Wang LW and Zunger A, Dielectric Constants of Silicon Quantum Dots. Phys Rev Lett 1994;73:1039-42.
[568] Walter JP and Cohen ML, Wave-Vector-Dependent Dielectric Function for Si, Ge, GaAs, and ZnSe. Phys Rev B 1970;2:1821-6.
[569] Deger D, Ulutas K, Conduction and dielectric polarization in Se thin films. Vacuum 2003;72:307-12.
[570] Penn DR, Wave-Number-Dependent Dielectric Function of Semiconductors**.** Phys Rev 1962;128: 2093-7.
[571] Tsu R and Babic D, Doping of a quantum dot. Appl Phys Lett 1994;64:1806-8.
[572] Chen TP, Liu Y, Tse MS, Tan OK, Ho PF, Liu KY, Gui D, and Tan ALK, Dielectric functions of Si nanocrystals embedded in a $SiO_2$ matrix. Phys Rev B 2003;68:153301.
[573] Chen TP, Liu Y, Tse MS, Ho PF, Dong G and Fung S, Depth Profiling of Si Nanocrystals in Si-Implanted SiO2 Films by Spectroscopic Ellipsometry. Appl Phys Lett 2002;81:4724-6.
[574] Hens Z, Vanmaekelbergh D, Kooij ES, Wormeester H, Allan G, and Delerue C, Effect of Quantum Confinement on the Dielectric Function of PbSe. Phys Rev Lett 2004;92:026808.
[575] Delerue C, Lannoo M, and Allan G, Concept of dielectric constant for nanosized systems. Phys Rev B 2003;68:115411-4.
[576] Drachev VP, Buin AK, Nakotte H, and Shalaev VM, Size Dependent $\chi^{(3)}$ for Conduction Electrons in Ag Nanoparticles. Nano Lett 2004;4 (8):1535-1539.
[577] Lannoo M, Delerue C, and Allan G, Screening in Semiconductor Nanocrystallites and Its Consequences for Porous Silicon. Phys Rev Lett 1995;74:3415-8.
[578] Greenway DL and Harbeke G, Optical Properties and Band Structure of Semiconductors. Pergamon Press, New York, 1968.
[579] Brown FG, The Physics of Solids. Benjamin Press, New York, 1968.
[580] Blatt FJ, Physics of Electronic Conduction in Solids. McGraw-Hill, New York, 1967
[581] Macdonald JR, Impedance Spectroscopy. Wiley, New York, 1987, Chap 4.
[582] Orton JW and Powell MJ, The Hall effect in polycrystalline and powdered semiconductors. Rep Prog Phys 1980;43:1263-307.
[583] Kleitz M and Kennedy JH, Fast Ion Transport in Solids, edited by Vashishta P, Mundy JN, and Shenoy GK. Elsevier, Noth Holland, 1979, 185.
[584] Lanfredi S, Carvalho JF, and Hernandes AC, Electric and dielectric properties of $Bi_{12}TiO_{20}$ single crystals. J Appl Phys 2000;88:283-7.
[585] Looyenga H, Physica 1965;31:401.
[586] Sun XW, Yu SF, Xu CX, Yuen C, Chen BJ, and Li S, Room-Temperature Ultraviolet Lasing from Zinc Oxide Microtubes. Jpn J Appl Phys 2003;42:L1229–31.
[587] Zhou J, Zhou Y, Ng SL, Zhang X, Que WX, Lam YL, Chan YC, and Kam CH, Three-dimensional photonic band gap structure of a polymer-metal composite. Appl Phys Lett 2000;76:3337-9.
[588] Tronc E, Noguès M, Chanéac C, Lucari F, D'Orazio F, Grenèche JM, Jolivet JP, Fiorani D, Testa AM, Magnetic properties of $\gamma$-$Fe_2O_3$ dispersed particles:Size and matrix effects, J Magn Magn Mater 2004;272-276:1474-5.
[589] Seehra MS and Punnoose A, Particle size dependence of exchange-bias and coercivity in CuO nanoparticles. Solid State Commun 2003;128:299-302.
[590] Zhong WH, Sun CQ, and Li S, Size effect on the magnetism of nanocrystalline Ni films at ambient temperature. Solid State Commun 2004;13:603-6.
[591] Li J, Qin Y, Kou X and Huang J, The microstructure and magnetic properties of Ni nanoplatelets. Nanotechnology 2004;15:982-6.
[592] De Heer WA, Milani P, and Chatelain A, Spin relaxation in small free iron clusters. Phys Rev Lett 1990;65:488-91.

[593] Guzman M, Delplancke JL, Long CJ, Delwiche J, Hubin-Franskin MJ, Grandjean F, Morphologic and magnetic properties of $Pd_{1-x}Fe_x$ nanoparticles prepared by ultrasound assisted electrochemistry. J Appl Phys 2002;92:2634-60.
[594] Taniyama T, Ohta E, Sato T, and Takeda M, Magnetic properties of Pd–29 at% Fe fine particles. Phys Rev B 1997;55:977-82.
[595] Qiang Y, Sabiryanov RF, Jaswal SS, Liu Y, Haberland H, and Sellmyer DJ, Manetism of Co nanocluster films. Phys Rev B 2002;66:064404.
[596] Sort J, Dieny B, Fraune M, Koenig C, Lunnebach F, Beschoten B, Guntherodt G, Perpendicular exchange bias in antiferromagnetic-ferromagnetic nanostructures. Appl Phys Lett 2004;84:3696-8.
[597] Baltz V, Sort J, Rodmacq B, Dieny B, Landis S, Size effects on exchange bias in sub-100 nm ferromagnetic-antiferromagnetic dots deposited on prepatterned substrates. Appl Phys Lett 2004;84:4923-5.
[598] Falicov LM, Pierce DT, Bader SD, Gronsky R, Hathaway KB, Hopster HJ, Lambeth DN, Parkin SP, Prinz G, Salamon M, Schuller IK, Victora RH, Surface, interface, and thin-film magnetism. J Mater Res 1990;5:1299-340.
[599] Aguilera-Granja F, Moran-Lepez JL, Ising model of phase transitions in ultrathin films. Solid State Comm 1990;74:155-8.
[600] Chinnasamy CN, Narayanasamy A, Ponpandian N, Chattopadhyay K and Saravanakumar M, Order-disorder studies and magnetic properties of mechanically alloyed nanocrystalline $Ni_3Fe$ alloy. Mater Sci Eng A 2001;304:408-12.
[601] Cox GM, Trevor DJ, Whetten RL, Rohifing EA, and Kaldor A, Magnetic behavior of free-iron and iron oxide clusters. Phys Rev B 1985;32:7290-8.
[602] Bucher JP, Douglas DC, Xia P, Haynes B, Bloomfield LA, Magnetic properties of free cobalt clusters. Phys Rev Lett 1991;66:3052-5.
[603] Billas IML, Becker JA, Chatelain A, de Herr WA, Magnetic moments of iron clusters with 25 to 700 atoms and their dependence on temperature. Phys Rev Lett 1993;71:4067-70.
[604] Ohnishi S, Freeman AJ, and Weinert M, Surface magnetism of Fe(001). Phys Rev B 1983;28:6741-8.
[605] Pastor GM, Dorantes-Dvaila J, and Bennemann KH, Size and structural dependence of the magnetic properties of small 3d-transition-metal clusters. Phys Rev B 1989;40:7642-54.
[606] Yang CY, Johnson KH, Salahub DR, Kaspar J, and Messmer RP, Iron clusters: electronic structure and magnetism. Phys Rev B 1981;24:5673-90.
[607] Billas IML, Chatelain A, De Heer WA, Magnetism from the Atom to the Bulk in Iron, Cobalt, and Nickel Clusters. Science 1994;265:1682-4.
[608] Ney A, Poulopoulos P, Baberschke K, Surface and interface magnetic moments of Co/Cu(001). Europhys Lett 2001;54:820-5.
[609] Konno M, Anomalous thickness dependence of the saturation magnetization in Fe-Ni Invar alloy films. J Phys Soc Japan 1980;49:1185-6.
[610] Apsel SE, Emmert JW, Deng, J and Bloomfield LA, Surface-Enhanced Magnetism in Nickel Clusters. Phys Rev Lett 1996;76:1441-4.
[611] Cox AJ, Louderback JG, Apsel SE, and Bloomfield LA, Magnetism in 4d-transition metal clusters. Phys Rev B 1994;49:12295-8.
[612] Sumiyama K, Sato T, Graham GM, Thickness dependence of magnetization in Fe-Ni Invar alloy film. Solid State Commun 1976;19:403-4.
[613] Wedler G, Schneck H, Galvanomagnetic and magnetic properties of evaporated thin nickel films: II Thickness dependence of the Hall coefficients, the magnetoresistivity and the saturation magnetization. Thin Solid Films 1977;47:147-53.
[614] Ohta S, Terada A, Ishii Y, and Hattori S, Thickness dependence of magnetic properties and read-write characteristics for iron oxide thin films. Trans Inst Electr Commun Eng Jap 1985;E68:173
[615] Kempter K, Maurer I, Harms H, Saturation magnetization of MnBi films with different thicknesses and composition. Appl Phys 1975;7:7-9.
[616] Manaf A, Buckley RA, Davies HA, and Leonowicz M, Enhanced magnetic properties in rapidly solidified Nd-Fe-B based alloys. J Magn Magn Mater 1991;101:360-2.
[617] Shafi KVPM, Gedanken A, Prozorov R, Revesz, A and Lendvai J, Preparation and magnetic properties of nanosized amorphous ternary Fe-Ni-Co alloy powders. J Mater Res 2000;15:332-7.
[618] Degauque J, Astié B, Porteseil JL, and Vergne R, Influence of the grain size on the magnetic and magnetomechanical properties of high-purity iron. J Magn Magn Mater 1982;26:261-3.
[619] Herzer G, Grain size dependence of coercivity and permeability in nanocrystalline ferromagnets. IEEE Trans Mag 1990;26:1397-402.

[620] Sato F, Tezuka N, Sakurai T, Miyazaki T, Grain diameter and coercivity of Fe, Ni, and Co metals. IEEE Trans Mag Jpn 1994;9:100-6.
[621] Merikoski J, Timonen J, and Manninen M, Ferromagnetism in small clusters. Phys Rev Lett 1991;66:938-41.
[622] Liu F, Press MR, Khanna SN, and Jena P, Magnetism and local order: ab initio tight-binding theory. Phys Rev 1989;B39:6914-24.
[623] Jensen PJ, Bennemann KH, Theory for the atomic chshell structure of the cluster magnetic-moment and magnetoresistance of a cluster ensemble. Z Phys D 1995;35:273-8.
[624] Zhao J, Chen X, Sun Q, Liu E, Wang G, A Simple d-band model for the magnetic property of ferromagnetic transition-metal clusters. Phys Lett A 1995;205:308-12.
[625] Montejano-Carrizales JM, Aguilera-Granja F, Moran-Lσpez JL, Direct enumeration of the geometrical characteristics of clusters. Nanostruct Mat 1997;8:269-87.
[626] Han DH, Wang JP, and Luo HL, Crystallite size effect on saturation magnetization of fine ferrimagnetic particles. J Magn Mang Mater 1994;136:176-82.
[627] Khanna SN and Linderoth S, Magnetic behavior of clusters of ferromagnetic transition metals, Phys Rev Lett 1991;67:742-45.
[628] Alben R, Becker JJ, and Chi MC, Random anisotropy in amorphous ferromagnets. J Appl Phys 1978;49:1653-8.
[629] Sangregorio C, Ohm T, Paulsen C, Sessoli T, and Gatteschi D, Quantum tunneling of the magnetization in an iron cluster nanomagnet. Phys Rev Lett 1997;78:4645-8.
[630] Dai DS and Qian KS, Ferromagnetism. Beijing, Sci Press, 2000.
[631] Metropolis N, Rosenbluth A, Rosenbluth M, Teller A, Teller E. J Chem Phys 1953;21:1087.
[632] Sakurai M, Watanabe K, Sumiyama K, and Suzuki K, Magic numbers in transition metal (Fe, Ti, Zr, Nb, and Ta) clusters observed by time-of-flight mass spectrometry. J Chem Phys 1999;111:235-8.
[633] Reinhard D, Size-Dependent Icosahedral-to-fcc Structure Change Confirmed in Unsupported Nanometer-Sized Copper Clusters. Phys Rev Lett 1997;79:1459-62.
[634] Du Y, Wu J, Lu H, Wang T, Qiu ZQ, Tang H, and Walker JC, Magnetic properties of fine iron particles. J Appl Phys 1987;61:3314-6.
[635] Gangopandhyay S, Hahjipanayis GC, Sorensen CM, and Klabunde KJ, Magnetic properties of ultrafine Co particles. IEEE Trans Magn 1992;28:3174-6.
[636] Sánchez RD, Rivas J, Vaqueiro P, López-Quintela MA, and Caeiro D, Particle size effects on magnetic garnets prepared by a properties of yttrium iron sol-gel method. J Magn Magn Mater 2002;247:92-8.
[637] Zheng WT, Private communication, Changchun, June 2004.
[638] Zhang Y, Sun H, and Chen C. Superhard Cubic BC2N Compared to Diamond. Phys Rev Lett 2004;93:195504.